%% file: MWD.tex
\documentclass[a4paper]{aa}
\usepackage{txfonts,graphicx,natbib}
\usepackage{aalongtable}
\bibpunct{(}{)}{;}{a}{}{,}
\newcommand{\fo}{\ensuremath{f_\parallel}}
\newcommand{\fe}{\ensuremath{f_\perp}}
\newcommand{\pv}{\ensuremath{P_V}}
\newcommand{\nnv}{\ensuremath{N_V}}

\newcommand{\teff}{\ensuremath{T_{\rm eff}}}
\newcommand{\bz}{\ensuremath{\langle B_z \rangle}}
\newcommand{\nz}{\ensuremath{\langle N_z \rangle}}
\newcommand{\cz}{\ensuremath{C_z}}
\newcommand{\bs}{\ensuremath{\langle \vert B \vert \rangle}}
\newcommand{\sz}{\ensuremath{\sigma_{\langle B_z \rangle}}}
\newcommand{\snr}{\ensuremath{S/N}}

\begin{document} 
   \title{Searching for the weakest detectable magnetic fields\\ in white dwarfs}
   \subtitle{Highly-sensitive measurements from first VLT and WHT surveys\thanks{Tables and reduced data are available in electronic form at the CDS via anonymous ftp to cdsarc.u-strasbg.fr (130.79.128.5)}}
   \author{S. Bagnulo          \inst{1}
          \and
          J.D. Landstreet      \inst{1,2}
          }
   \institute{Armagh Observatory and Planetarium, College Hill, Armagh BT61 9DG, U.K.
              \email{stefano.bagnulo@armagh.ac.uk}
         \and
             University of Western Ontario, London, Ontario, Canada N6A 3K7
             \email{jlandstr@uwo.ca}
             }
   \date{Received April 15, 2018; accepted 18 July 2018}

   \abstract{

Our knowledge of the magnetism in white dwarfs is based on an
observational dataset that is biased in favour of stars with very
strong magnetic fields. Most of the field measurements available in
the literature have a relatively low sensitivity, while current
instruments allow us to detect magnetic fields of white dwarfs with
sub-kG precision. With the aim of obtaining a more complete view of
the incidence of magnetic fields in degenerate stars, we have started
a long-term campaign of high-precision spectropolarimetric
observations of white dwarfs. Here we report the results obtained so
far with the low-resolution FORS2 instrument of the ESO VLT and the
medium-resolution ISIS instrument of the WHT. We have considered a
sample of 48 stars, of which five are known magnetic or
suspected magnetic stars, and obtained new longitudinal magnetic field
measurements with a mean uncertainty of about 0.6\,kG. Overall, in the
course of our survey (the results of which have been partially published in
papers devoted to individual stars) we have discovered one new weak-field
magnetic white dwarf, confirmed the magnetic nature of another, found
that a suspected magnetic star is not magnetic, and suggested two new
candidate magnetic white dwarfs. Even combined with data previously
obtained in the literature, our sample is not sufficient yet to reach
any final conclusions about the actual incidence of very weak magnetic fields
in white dwarfs, but we have set the basis to achieve a homogeneous
survey of an unbiased sample of white dwarfs.
As a by-product, our survey has also enabled us to carry out
a detailed characterisation of the ISIS and the FORS2
instruments for the detection of extremely weak magnetic fields in white dwarfs, and in
particular to relate the {\it S/N} to measurement uncertainty for white dwarfs
of different spectral types. This study will help the optimisation of
future observations.
}


\maketitle

%
%________________________________________________________________

\section{Introduction}

During the past century, observations have gradually established that
magnetic fields can be directly detected, usually through the Zeeman
effect, in some (but not all) stars in most of the major phases of
stellar evolution. Evolutionary stages in which stellar magnetic
fields have been detected include the T Tau and Herbig AeBe phases,
the main sequence (upper and lower), and the red giant, AGB, white
dwarf, and neutron star phases \citep{DonLan09,BagLan15}. Thus, the
potential importance of magnetic fields extends over most of the
observable HR diagram. 

The incidence and the typical strength and
morphology of the magnetic field are different for different kinds of
stars. In most of the cases, the origin of the magnetic field is not
understood, and we do not know how fields evolve as stars evolve.  At
an even more basic level, the situation is that we simply do not
understand why magnetic fields occur in some stars but not in others.
In this situation, exploratory observations can play an important role
by establishing clearly the circumstances in which magnetic fields are
found, the strength and surface geometry of the fields, and the
statistics of field occurence as a function of such parameters as
stellar mass, age, and rotation, during various evolutionary
phases.

In this and forthcoming papers we focus on the incidence of
magnetic fields in the most common final stage of stellar evolution,
that of white dwarfs (WDs). During this phase, we observe that a
rather small fraction of stars \cite[of the order of 10\,\%,
  e.g.][]{Lanetal12} exhibit surface magnetic fields. These fields
range in global field strength from a few kG to nearly
$10^3$\,MG. Most of the known fields have strengths between roughly
1.5\,MG and 75\,MG (which may be due to observational bias, see
Sect.~\ref{Sect_Previous}).

The fields observed in WDs do not appear to change their intrinsic
structure on an observable time scale, and seem to be of fossil nature
(that is, fields inherited from an earlier stage of evolution; for a review
  on the possible mechanisms that may have generated a magnetic field,
  including the merging of a binary system, see \citeauthor{Feretal15}
  \citeyear{Feretal15}). In principle,
the rotation period of a magnetic white dwarf (MWD) may be determined
from variation of the appearance of the field on the visible
hemisphere of the star as it rotates; and modelling of the shape of
spectral lines, particularly of the effect of the magnetic fields on
these lines, may make it possible to obtain an approximate map of the
surface structure of the stellar magnetic field. This has been
attempted for a number of MWDs with some, but not complete, success
\citep{Jordan92,Eucetal02,Eucetal06}.

At present, the observational situation for WD magnetism still leaves
a number of questions for which the answers should help to stimulate
and improve our theoretical ideas about how the observed fields arise
and evolve. (1) We do not yet have a clear picture of the frequency of
occurence of magnetic fields of various strengths in either magnitude
or volume limited WD samples. In particular, is the relative deficit
of large and small fields relative to the $1.5 - 75$\,MG range of
field strengths \citep{Feretal15} real, or is it an artefact of
selection effects in the discovery process?  (2) We do not know if
fields occur that are larger than $10^9$\,G or less than about
5\,kG. (3) We do not know how the frequency of occurence of fields
varies with WD mass or age, although there are hints that magnetic WDs
may be more massive than the average, and that magnetic fields may be
more common in cooler, older WDs than in hot, young stars
\citep{LiebertSion79,Lieetal03}. (4) We have no information on how the surface
structure of magnetic fields may vary with WD age or mass.

The observational data currently available do not provide sufficient
constraints to any of these issues \cite[see e.g.][]{Feretal15},
mainly because of two reasons: (1) the weak-field regime is probed
only by a small fraction of the relevant measurements, or, in simpler words, the
large majority of magnetic field strength measurements available in
the literature have rather large uncertainties; and (2) the database
of available measurements is quite inhomogeneous in both space
distribution and stellar parameters.

To improve this situation, ideally we need a large survey of WD
magnetism aimed at a complete coverage of a volume-limited region of
space, with uniform field detection thresholds. The survey volume
should be large enough to establish with statistical precision the
incidence of magnetism as a function of mass and age.

In practice, there are major problems with completing such a
survey. Even within a distance of 20\,pc from the Sun it is thought
that the known sample of about 140 WDs is still missing about 20
undiscovered stars \citep{Holetal08,Holetal16}, whose absence will
clearly affect statistical conclusions. A certain number of the WDs
that are within such a volume are nevertheless old enough (more than
about 5\,Gyr) and cool enough ($T_{\rm eff} < 5000$\,K) that some are
fainter than $V \sim 17$. Such stars require the use of the largest
telescopes to acquire the necessary normal or polarised spectra, but
the field detection threshold would still be substantially higher than
in case of $V=13$ or 14 stars. A third problem is that the sensitivity
of (spectro-)polarimetry to the detection and measurement of fields
varies considerably from WDs with a strong line spectrum (DA, DB, DZ
stars) for which uncertainties smaller than 1\,kG can be obtained, to
stars lacking any optical atomic spectral lines (DC, DQ stars), for
which the practical uncertainties are of the order of 1\,MG or larger.
To the best of our knowledge, the only attempt to conduct a volume-limited
sample study of the incidence of magnetism was made by
\citet{Kawetal07}, who considered a volume within 13\,pc thought to be
complete by \citet{Holetal02}.

An alternative approach would be to observe a magnitude-limited
sample, which, however, would very strongly favour hot, young WDs, and
certainly would be a poor representation of the local WD population
(and obviously would still have the problem of insensitivity to the
fields of DC stars). In practice, all the surveys undertaken so far
have effectively been magnitude-limited surveys, and the best possible
approach, especially now that the Gaia parallaxes are made available,
will be to collect new data that complement the existing ones towards
the completion of a volume-limited sample, while being aware fully
aware of the various observational biases.

To inform the strategy of future surveys, we need first to assess the degree
of completeness of the available observations of WDs. In
Sect.~\ref{Sect_Previous} we present a short review the main
characteristics and the results of the surveys carried out so far,
which will be a starting point to establish our criteria for target
selection in this and forthcoming survey papers.
Regarding the choice of the instrument for our surveys, over the past
50 years or so, several different techniques have been used to measure
magnetic fields in WDs. Their detection limits depend not only on
\snr\ but also on the spectral characteristics of the star, and this
will be discussed in Sect.~\ref{Sect_Techniques}.  For our survey we
have used the FORS2 instrument of the European Southern Observatory's
Very Large Telescope (ESO VLT), the ISIS instrument of the William
Herschel Telescope (WHT), a low-resolution and a mid-resolution
spectropolarimeter, respectively, and the high-resolution
spectropolarimeter ESPaDOnS of the Canada-France-Hawaii Telescope
(CFHT). In this paper we present the results obtained so far with the
FORS2 and ISIS spectropolarimeters. More details on the FORS2 and ISIS
instruments and the settings used in this survey are given in
Sect.~\ref{Sect_Instruments}. Our observing strategy was aimed in part
at assessing the reliability of the instruments (in particular of ISIS,
which is less commonly used in spectropolarimetric mode than FORS2),
and our experiments are described in
Sect.~\ref{Sect_Observations}. Data reduction is described in
Sect.~\ref{Sect_Data_Reduction}. The results of our tests, including
quality checks, are given in Sect.~\ref{Sect_QC}.  The results of our
observations of scientific targets are given in
Sect.~\ref{Sect_Results}. In Sect.~\ref{Sect_Efficiencies} we consider
the relative sensitivity of magnetic field detections as obtained with
the two different instruments, and as a function of the spectral types
of the WDs. In Sect.~\ref{Sect_Discussion_Science} we discuss our
scientific results, and in Sect.~\ref{Sect_Conclusions} we summarise our
conclusions.

\section{Previous surveys of magnetic fields in WDs}\label{Sect_Previous}
Techniques employed for detection and modelling of magnetic fields in
WDs are optical spectroscopy, broad-band polarimetry, and
spectropolarimetry, and the quantitative data interpretation depends
on the field strength. The simplest case is when field strength is
$\la 1$\,MG. In this regime, the continuum is not detectably polarised, and the
Stokes profiles of spectral lines may be interpreted in terms of
the linear Zeeman effect. In this regime, detection and modelling of
magnetic fields in WDs is very similar to that carried out for
non-degenerate stars \citep[e.g.][]{Mathys89,DonLan09,BagLan15}. For
field strengths in the range $\sim 1$ to 50\,MG, spectral lines are
formed in the quadratic Zeeman regime, and line polarisation and
splitting may be interpreted in terms of field strength with the aid
of numerical computations of the atomic structure of H and He \citep[e.g.][]{Kemic74}. For field strengths $\ga
50$\,MG, the magnetic field polarises the continuum \citep{Kemp70},
and the various components of the spectral lines may be shifted by
several hundred \AA, or are washed out so completely as to become indistinguishable
from the continuum. The estimate of the field strength rely again on
numerical atomic computations \citep{Wunetal85}. In all regimes, as a general
rule, unpolarised spectroscopy is sensitive to the field strength
averaged over the visible stellar disk, or mean field modulus \bs; circular
polarisation is sensitive to the longitudinal component of the
magnetic field, again averaged over the visible stellar disk, and called the mean
longitudinal field \bz. Linear polarisation is sensitive to the field
transverse components, but is far less commonly employed as a diagnostic tool
than the other techniques. Below we summarise the outcome of the main
surveys for magnetic fields in WDs carrried out in the last 60 years.

\subsection{Early non-detections of WD fields}\label{Sect_Earliest}
During the 1960's, extensive spectroscopy of WDs \citep[particularly
  that carried out by J.\,L.\,Greenstein, for example][]{EggGre65}, made it
clear that WDs with H or He line spectra (DA or DB white
dwarfs) show in general no sign of magnetic splitting. Based on a
sample of more than 100 DA and DB stars observed with a typical
spectral resolving power of several hundred \citep[typically with
  dispersion of 190\,\AA\,mm$^{-1}$,][]{EggGre67}, no indication was
found that large fields having \bs\ of the order of a MG or more occur
in WDs. The threshhold of this general absence of fields was quantitatively estimated 
by \citet{Pre70}, who showed that, based on a sample of about
20 DA WDs, ``few if any WDs \ldots\ have surface magnetic fields
as large as $5\,\times\,10^5$\,G''.

The first survey of WDs for still weaker magnetic fields (a
project suggested by L.\ Woltjer) was made by \citet{AngLan70a}, using
interference filters with a photoelectric polarimeter to isolate the wings of the H$\beta$ line in DA
stars, searching for Zeeman-effect induced circular polarisation in
these line wings. This survey reached uncertainties in \bz\ of the
order of 30\,kG for nine WDs, but detected no fields.

\subsection{First discoveries and early successful surveys}
The discovery of the first magnetic WD was made by \citet{Kemetal70}
by means of broad-band circular polarimetry. This approach was
stimulated by Kemp's theory \citep{Kemp70} that a field of the order
of $10^7$\,G at the surface of a WD would cause broad-band
continuum circular polarisation of the optical radiation. This idea
turned out to be qualitatively correct, and led to the discovery of a
field of many MG in the white dwarf Grw+70\,8247 (= WD\,1900+705). It
was quickly discovered that the radiation of this MWD is also
linearly polarised \citep{AngLan70b}, and eventually it was shown that
the field of this star is in the 100s of MG range \citep{Angetal85}.
Grw+70\,8247 still has one of the strongest MWD fields known.

Further surveys, mostly looking for broad-band circular polarisation,
gradually uncovered roughly 1--2 MWDs per year: G195--19
\citep{AngLan71-Second}, G99-37 \citep{LanAng71}, G99-47
\citep{AngLan72}, etc. The first clear indication of magnetic
variability was observed in the 1.33\,d periodic variation of circular
polarisation in the light of G195-19 \citep{AngLan71-Periodic}.  For
the first 20-25 years of magnetic investigations of WDs, most MWDs were
detected and studied using broad-band optical circular polarisation
\citep[see e.g.][]{Angetal81,Landstreet92}, but it was also realised
that Zeeman splitting of Balmer lines by magnetic fields is rare in
WDs, but not absent \citep[e.g.\ G99-47 and Feige
  7,][]{Lieetal75,Lieetal77}.  Combining all field detections and
non-detection of the first decade together, \citet{Angetal81}
concluded that the probability of finding a magnetic field of between
$3\,10^6$ and $3\,10^8$\,G in a WD is at least 3\,\%.

\subsection{More recent spectropolarimetric surveys}\label{Sect_SPOL_Surveys}
\citet{SchSmi95} carried out a spectropolarimetric survey of 170 DA
stars brighter than $B=15$ to search for fields below roughly
1\,MG. The method used was to search for the circular polarisation
produced in the wings of Balmer lines by the presence of a non-zero
mean longitudinal field \bz. This method is very similar in principle to
that of \citet{AngLan70a}; however, the use of a low-resolution ($R
\sim 700$) spectropolarimeter allowed both H$\alpha$ and H$\beta$ to
be observed simultaneously. The mean error bar \sz\ was 8.6\,kG, but 7
targets were re-observed with long exposure times to achieve 
$\sz \la 2$\,kG. This survey discovered four new MWDs,
bringing the total number of MWDs known at that time to 42 \citep[see 
Table~2 of][]{SchSmi95}.

\citet{Putney97} observed 46 isolated WDs classified as DCs in the WD
catalogue by \citet{McCSio87}  for spectrally resolved circular polarisation. 
Her survey used spectral coverage
from 3700 to 8000\,\AA\ with spectral resolving power $R \sim 400$. 
She found that many of her targets were misclassified:
of the 46 DC WDs, only 22 are genuine DC stars. Most of
the remaining ones are DA stars with very weak H$\alpha$ and almost no
other visible Balmer series lines. Fields
were detected in five faint stars ($V \sim 16 - 17$), two of which
still need to be confirmed by further observations.

A spectropolarimetric survey of 61 bright DA white dwarfs in the southern
hemisphere, with similar field measurement uncertainties as those of
\citet{SchSmi95}, was reported by \citet{Kawetal07}. This survey
reported marginal evidence of a field in WD\,0310-688, which was not
confirmed by later observations. At the time of their survey,
\citet{Kawetal07} were able to list approximately 170 known MWDs. The abrupt
increase in the total number of known MWDs was
due to the first impact of wholesale discovery of WD fields in the range
of $\bs \sim 2$ to 80\,kG by the Sloan Digital Sky Survey (SDSS; see
Sec.~\ref{Sect_SDSS} below.)

\citet{KawVen12} carried out a magnetic survey of some 58 high proper
motion DA white dwarfs using ESO FORS1 and FORS2 with uncertainties
\sz\ of typically a few kG. These stars tend to be relatively cool 
(the median value of $T_{\rm eff}$ is about 6500\,K). \citet{KawVen12}
discovered a previously unknown magnetic field in one DAZ star of
$T_{\rm eff} \approx 6\,000$\,K, and may have found a weak field in a
cool DA star (NLTT\,347); they also discovered magnetic variability in two
previously unknown cool, MG-field WDs.

\citet{Voretal10, Voretal13} carried out a small survey of 11 DQ
(C-rich) WDs for magnetic fields, and found a field of $\bz \approx
1.5$\,MG in WD\,2153$-$512 = GJ\,841B, a rare C-rich WD with CH
bands. It appears that cool WDs with atomic or molecular metal lines
may be a fruitful category of WD to search for weak fields, but such
stars are usually rather faint, and the molecular Swan bands of C$_2$
in the visible are extremely insensitive to magnetic fields.

\subsection{Searching for extremely-weak fields in WDs}\label{Sect_Weak}

Two papers based on small spectro-polarimetric surveys of WDs with the
VLT brought attention to the possibility that a large number of WDs
may have a magnetic field that is not strong enough to be detected with
in the previous low-resolution spectroscopic
surveys. \citet{Aznetal04} and \citet{Joretal07} used FORS1 to observe
samples of 12 and 10 WDs respectively, with typical \bz\ error bars of
1\,kG, and found respectively 3 and 1 new MWDs, each of which has a
longitudinal field of only a few kG. The conclusion from these papers
was that the rate of weak-field MWD was actually relatively high,
perhaps up to 25\,\%. However, one of the MWDs detected by these two
studies was previously identified as a candidate magnetic star in a
high-resolution spectroscopic survey (see Sect.~\ref{Sect_High_Res}),
and one was not confirmed. When these two stars are removed from the
survey statistics, the discovery frequency is similar to that found in
other studies.

A similar survey of six DA stars and one sdO star by \citet{Valetal06}
using the low-resolution prime focus UAGS spectropolarimeter at the
6-m BTA telescope at the Special Astrophysical Observatory yielded
marginal evidence for a field in WD\,1105$-$048. This field was still
not fully confirmed in spite of numerous further measurements (but see
Sect.~\ref{Sect_Discussion_Science} in this paper).

\citet{Lanetal12} presented the results of another small VLT
spectropolarimetric survey focussed on 10 relatively cool WDs, with a
typical \bz\ error bar between 1 and 1.5\,kG. They detected a field in
a WD which \citet{Koeetal98} had flagged as possibly magnetic,
but considered also the possibility that the H$\alpha$ line core is
broadened by rotation.  \citet{Lanetal12} revised the data reductions
of earlier FORS1 samples of 36 WDs with high-precision field
measurements, showing that some of the previous marginal detections
obtained with FORS1 were probably spurious.  \citet{Lanetal12}
concluded that WDs with very weak magnetic fields are not much more
common than WDs with strong and very strong magnetic fields.

Recent VLT spectropolarimetric field measurements of five DAZ stars, most
with \bz\ uncertainties \sz\  below 1\,kG, are reported by
\citet{Faretal18}, but fields were found only in the two WDs that
were already known to be magnetic from previous
measurements (see Sect.~\ref{Sect_High_Res}).

\subsection{High-resolution spectroscopic surveys}\label{Sect_High_Res}

In Sect.~\ref{Sect_Earliest} we noted that the first large survey for
MG magnetic fields, with negative results, was effectively carried out
by classification spectroscopy in the 1960s. More recently,
high-resolution spectroscopy has been found to be a tool that is
effective for searching for MWDs with fields of some tens of kG or
more.

\citet{Koeetal98} reported detection of three new MWDs with
\bs\ fields in the kG range from close examination of high-resolution
(flux) spectra of a sample of WDs, taken to find evidence of rotation
by searching for rotational broadening of the core of H$\alpha$. High
resolution spectroscopy of WDs was continued within the framework of
the ESO Supernova Ia Progenitor Survey (SPY) project, a search for
close SB2 white dwarf pairs (possible SN Ia progenitors) with the ESO
UVES spectrograph. SPY was the first project for which a substantial
sample of WDs was systematically observed with high spectral
resolving power ($R \sim 18\,000 - 80\,000$). The SPY survey confirmed
that such observations provide, as a side product, a powerful method
of detecting not only magnetic fields in the 1 -- 100\,MG range, but
that the sensitivity extends down to mean surface fields \bs\ of the
order of 50\,kG, which can be detected via Zeeman splitting in the
core of the H$\alpha$ line. Still weaker fields, down to about 20\,kG,
broaden the core of the H$\alpha$ line significantly, but to distinguish
this from rotational broadening, spectropolarimetry is required.

Based on SPY data, \citet{Koeetal01,
  Koeetal09} identified several further MWDs with \bs\ below 1\,MG.
Because fields as weak as $\bs \sim 50$\,kG will generally have
longitudinal fields \bz\ below 15\,kG, this is a field detection
method almost as powerful as sensitive spectro-polarimetric surveys of
Sect.~\ref{Sect_Weak}.

High-resolution spectroscopic studies of DAZ WDs by \citet{Zucetal11}
and \citet{Faretal11} revealed sub-MG fields in two cool stars via
detection of the Zeeman effect in the metal lines. Another cool DAZ
was found to have a sub-MG field by \citet{KawVen14}, who point out
that four of the 13 known cool ($T_{\rm eff} < 7000$\,K) DAZ stars are
not only magnetic, but have sub-MG fields. This is a very high
fraction compared to estimates of the weak-field magnetic WDs in the
general WD population of the order of 5\,\%.

\subsection{Discoveries of large magnetic field white dwarfs
  from the Sloan Digital Sky Survey}\label{Sect_SDSS}

Until early in this century, magnetic WDs were discovered at a rate of
at most a few per year. The rate of discovery of magnetic WDs
increased dramatically as a result of the Sloan Digital Sky Survey, a
huge project searching for and measuring the redshifts of millions of
nearby galaxies. As a byproduct, this project has revealed roughly
30\,000 new WDs, of which several hundred host magnetic fields
(\citeauthor{Schetal03} \citeyear{Schetal03},
\citeauthor{Kueletal09} \citeyear{Kueletal09},
\citeauthor{Kleetal13} \citeyear{Kepetal13},
\citeauthor{Kepetal15} \citeyear{Kepetal15}
\citeauthor{Kepetal16} \citeyear{Kepetal16};
see also the summary of \citeauthor{Feretal15} \citeyear{Feretal15}). 
However, the rate of discovery of MWDs with fields below about
1\,MG is still only one or two per year.

The data produced by the SDSS project are qualitatively different from
those resulting from earlier studies. The MWDs discovered in earlier
surveys are mostly among the brighter WDs, with magnitudes of $13 < V
< 16$. Many of these stars have been observed multiple times, often at
relatively high \snr\ and/or high spectral resolution. The fields
detected range over five orders of magnitude, from a few kG up to
nearly $10^3$\,MG. In contrast, the MWDs discovered in the SDSS are
mostly in the brightness range between $V=16$ and $V=20$. The spectra,
taken with the low resolving power of $R \sim 1800$, have low
\snr, usually $\la 20$. Because of the low \snr\ and $R$, the
threshold for detecting fields is roughly 2\,MG, and field strengths
can only be usefully estimated for the better spectra. In conclusion,
this enormous increase in the number of known magnetic WDs has mainly
increased our knowledge about the 2--80\,MG range of the \bs\ field strength
distribution, and the newly discoverd MWDs can mostly only be studied
further using the largest telescopes.

\subsection{The scope of our new surveys}

The total sample of known MWDs, single and in binary systems, was
recently analysed by \citet{Feretal15}. Among the approximately 250
best-characterised magnetic WDs, only about 30 are known with fields
(\bz\ or \bs) below 1\,MG, and a slightly smaller number with fields
$\bs \ga 80$\,MG; about 80\,\% of well-characterised magnetic WDs have fields
in the range of 1.5 to 80\,MG.

Because of the difficulty of identifying the largest fields in DC or
nearly DC spectra, and the lack of sensitivity to the weakest fields
in most general surveys of WDs, it is unlikely that this distribution
represents the true frequency of very low and very high fields.  The
considerably higher relative frequency of kG fields found in surveys
sensitive to them, as discussed above, implies that there should still
be a substantial number of such weak-field magnetic WDs to be
discovered even among relatively bright ($V$ less than 15 or 16)
WDs. Thus, {\it a survey to find more weak-field WDs has the potential
  to substantially improve our knowledge of the actual distribution of
  magnetic field strengths among WDs}, to provide more bright examples
of weak-field stars for detailed modelling and analysis, and to assist
us in understanding whether magnetic fields decay during white dwarf
cooling or whether some process(es) generate new magnetic flux.  Our
survey target list was originally based on a magnitude-limited sample, in
order to better understand the capabilities and limitations of FORS2
and (especially) ISIS, and on the desire to monitor some of the MWDs
that were discovered to be variable. The target lists of our future
surveys will be aimed mainly at surveying a volume-limited sample of
WDs such as the 20 or 25-pc samples described by \citet{Holetal16},
and will certainly benefit from the April 2018 release of new
high-precision astrometry from the Gaia mission.

\section{Magnetic field detection techniques: threshold and accuracy}\label{Sect_Techniques}
On the base of the large body of experiment and experience applied to
detect and measure fields in MWDs, we summarise here the strengths and
limitations of the various methods.

\subsection{Broad-band circular polarimetry}\label{Sect_BBCP}
Broad-band circular polarimetry is sensitive in practice only to
magnetic fields with typical strength \bz\ larger than roughly 10\,MG,
which are usually able to polarise the continuum at a detectable level
(above about 0.1\,\% polarisation) regardless of the spectral
type. These measurements may be interpreted in terms of longitudinal
field through the relationship between \bz\ in MG and circular
polarisation in \%\ \citep{Kemp70}
%%%%%%%%%%%%%%%%%%%%%%%%%%%%%%%%%%
\begin{equation}
  \bz(MG) \sim 10\,\frac{V}{I} (\,\%) \,.
\label{Eq_Kemp}
\end{equation}
%%%%%%%%%%%%%%%%%%%%%%%%%%%%%%%%%%
Equation~(\ref{Eq_Kemp}) is actually little more than an order of
magnitude estimate, and often underestimates the actual field as
determined by modelling by a factor of order 10 \citep[for example,
  compare the polarisation and \bs\ data in Table~2
  of][]{Landstreet92}.

\subsection{Circular spectropolarimetry}
Circular narrowband or spectro-polarimetry of spectral lines is
sensitive to the mean longitudinal field \bz, with a sensitivity that
is determined by the \snr\ that may be reached with current
telescopes: practically, with current mid- to large-size telescopes,
of a few hundred G in bright ($V \sim 13$) DA stars. The threshold
field sensitivity depends on the nature of the stellar spectrum, but
polarisation may be detected in H Balmer lines, He and/or metal lines
even when Zeeman splitting is negligible compared to the intrinsic
line broadening (see Sect.~\ref{Sect_Data_Reduction}). The detection
threshold of the mean longitudinal magnetic field also depends on the
instrument spectral resolution. There is no upper limit to the
magnetic field that may be detected with circular polarimetry, because
at the field regimes at which the magnetic fields washes out spectral
features, the continuum is certainly polarised (see
Sect.~\ref{Sect_BBCP}). We note that since spectro-polarimetry resolves
the WD intensity spectrum to a greater or lesser extent, depending on
the spectral resolution, the same data may be used to measure \bs\ if
the field is large enough for the Zeeman splitting to be significantly
larger than the spectral resolution element (see Sect.~\ref{Sect_I}
below).

\subsection{High-resolution spectroscopy}\label{Sect_I}
The mean magnetic field modulus \bs\ may be measured from the Zeeman
splitting of spectral lines observed in intensity. The sensitivity of
this technique increases with spectral resolution, but there is a
lower limit to the value of \bs\ that can be detected that is set by
the intrinsic broadening of the line cores of the lines used. It is a
little difficult to specify the typical threshhold field above which
high-resolution spectroscopy can reliably detect a field, as this
depends on the field geometry (whether the geometry leads to a
well-defined Zeeman triplet or not) and \snr, as well as the spectral
class. However, for the commonest case of spectra of DA stars, we may
take observation as a guide, and note that the (estimated) $\bs
\approx42$\,kG field of WD\,2105$-$820
\citep{Lanetal12} does not lead to a clear Zeeman pattern in the core
of H$\alpha$, while the 60\,kG field of WD\,2047+372 \citep{Lanetal17}
does. Based on this, we estimate that the available high-resolution
spectra of the WDs that show no clear Zeeman splitting provide upper
limit of $\bs \la 50$\,kG to possible fields. Spectroscopy may become
less useful in the presence of fields above 80\,MG, when spectral
lines sometimes nearly disappear, as they do for G195-19
\citep{Greetal71}. Obviously, Zeeman splitting cannot be measured in
DC stars, which have no spectral lines, and in which a magnetic field
may be detected with circular polarimetric techniques only if strong
enough to measurably polarise the continuum.

\subsection{Accuracy of the field measurements}
Regardless of the sensitivity that may be achieved in the measurement of
polarisation, and the accuracy quoted for derived measurements such as
\bz\ and \bs\ that are obtained with the various techniques, it is
essential to recall that the interpetation of the actual $I$ and $V/I$
spectra in terms of \bz\ and \bs\ relies on simplifying assumptions
that are not very accurate. The adopted transformations may lead to
precise values for the field strengths, but the precise
meaning of these values is rarely clear. The result is that
different methods of field strength measurement, and even measurements
with the same techniques in different wavelength regions, may lead to
field strength values that differ by much more than the nominal
uncertainties. This does not invalidate the usefulness of these field
values for estimating the strength of observed fields, but indicates
that caution is required in interpreting them. The issue has recently
been discussed in some detail with respect to field measurements of
main sequence stars using FORS1 by \citet{Lanetal14}, and we will
comment on it again in Sect.~\ref{Sect_Reference_Stars}.

\section{Instruments and instrument settings of our
survey}\label{Sect_Instruments}

Both \bs\ and \bz\ measurements are needed for simple dipole
modelling of the stellar magnetic structure, using the customary
modelling techniques historically applied to Ap and Bp stars
\citep[e.g.][]{LanMat00,Bagetal02b,Lanetal17}. However, even for
detection purpose, it is useful to have both high-resolution $I$ and
$V/I$ spectra, since there are examples of MWDs with large \bs\ values
with little or no signal of circular polarisation \citep[e.g.\
  WD\,2359$-$434,][]{Lanetal17}, or, viceversa, with no clear indication of
Zeeman splitting but with a measurable signal of circular polarisation
in spectral lines \citep[e.g.\ WD\,0446--789,][]{Aznetal04}.

Based on these considerations, it is clear that the ideal instrument
to detect magnetic fields in WDs would be a high-resolution
spectropolarimeter that can accurately measure both continuum and line
polarisation. Being fibre-fed, high-resolution spectropolarimeters
lack accuracy in the determination of polarisation in the continuum,
so the best viable option is a multi-instrument approach that makes
use of both low- and mid-resolution spectropolarimetry (with
capabilities in the continuum) and mid- to high-resolution
spectroscopy or spectro-polarimetry of spectral lines. Based on these considerations
we have decided to use the FORS2 instrument of the ESO VLT, the ISIS
instrument of the 4.2\,m William Herschel Telescope of the Issac
Newton Group (ING), and the high-resolution spectropolarimeter ESPaDOnS
at the Canada-France-Hawaii Telescope (CFHT).

The first two instruments are those used in this survey, and are
briefly described in the following paragraphs. In a forthcoming
paper we will report survey results obtained with ESPaDOnS.

Both FORS2 and ISIS have polarimetric optics arranged according to the
optical design described by \citet{Appenzeller67}. The polarimetric
module consists of an achromatic retarder waveplate ($\lambda/2$ for
observation of linear polarisation, or $\lambda/4$ for observation of
circular polarisation) which can be rotated to a series of fixed
positions, followed by a beam splitting device: a Wollaston prism in
the case of FORS2, and a Savart plate in case of ISIS. Essentially,
these devices split the incoming radiation into two beams polarised in
directions perpendicular to each other, one along the principal plane
of the plate (the parallel beam \fo), and one perpendicularly to that
plane (the perpendicular beam \fe). The beams split by the Wollaston
prism are tilted at an angle of about $20^\circ$, while the beams
split by a Savart plate propagate parallel to each other but
separated. A Wollaston mask
\citep{Scaretal83} or a special dekker prevents the superposition of
each beam split by the beam splitter with the light coming from the
other parts of the observed field of view.

%%%%%%%%%%%%%%%%%%%%%%%%%%%%%%%%%%%%%%%%%%%%%%%%%%%%%%%%%%%%%%%%%%%%%%%%%
\begin{figure}
\begin{center}
\includegraphics[width=9cm]{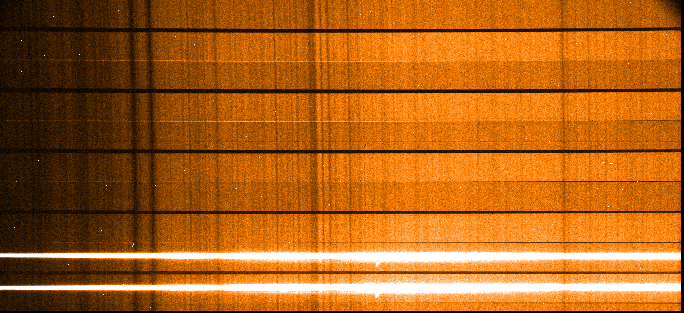}\\
\bigskip

\includegraphics[width=3cm,trim={0.0cm 10.0cm 0.0cm 10.0cm},clip]{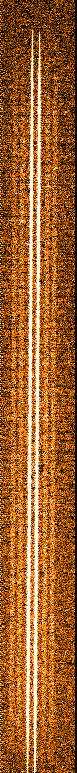}\ \ \ \
\includegraphics[width=3cm,trim={0.0cm 10.0cm 0.0cm 10.0cm},clip]{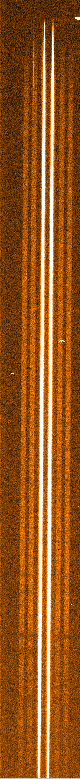}
\end{center}
\caption{\label{Fig_Raw} Raw image of spectropolarimetric data
  obtained with FORS2 (top panel), ISIS blue CCD (bottom left panel)
  and ISIS red CCD (bottom right panel). For all images, the dynamic
  range is set to show the sky background, which is at the level of a
  few hundred ADUs (while the spectra are at the level of several
  thousand ADUs). The FORS2 image refers to WD\,2039$-$202 observed on
  2015-06-02. The ISIS image refers to WD\,2111+498 observed on
  2015-08-30. The dispersion direction of the ISIS spectra has been
  heavily trimmed. Note in ISIS images the four strips illuminated by sky background.}
\end{figure}
%%%%%%%%%%%%%%%%%%%%%%%%%%%%%%%%%%%%%%%%%%%%%%%%%%%%%%%%%%%%%%%%%%%%%%%%%

\subsection{FORS2}
FORS2 (FOcal Reducer Spectrograph) is a multi-purpose instrument
capable of imaging and low-resolution spectroscopy in the optical,
equipped with polarimetric optics. It is attached at the Cassegrain
focus of Unit 1, Antu, of the ESO VLT of the Paranal Observatory. The
instrument is described in \citet{AppRup92} and \citet{Appetal98}. A
raw spectropolarimetric image obtained with FORS2 is shown in the
upper panel of Fig.~\ref{Fig_Raw}. In our survey we used
grism 1200B, which, with a dispersion of 0.71\,\AA\ per pixel (2x2
rebinning), covers the spectral range 3700-5200\,\AA. We used a
1\arcsec\ slit width for a spectral resolving power of 1400. A
discussion about the best choice for spectral range is presented in
Sect.~\ref{Sect_Red_vs_Blue}.

Currently, the FORS2 calibration plan includes regular monitoring
of standard stars for linear polarisation, but does not include a
regular check of the $\lambda/4$ waveplate. However, the polarimetric
optics are fixed in one of the instrument wheels, and need not be realigned
when they are used. It is probably safe to assume that the correct
alignment of the $\lambda/4$ retarder waveplate may be monitored from
occasional measurements of well known magnetic stars obtained within
a few months. For instance, the well known magnetic Ap stars HD\,94660
and HD\,188041 were observed in 2015 and 2016 and the measured values of
the field were in agreement with previous literature values
\citep{Bagetal17a}.

\subsection{ISIS}
The Intermediate dispersion Spectrograph and Imaging System (ISIS) is
mounted at the Cassegrain focus of the 4.2\,m William Herschel
Telescope. The instrument is equipped with polarimetric optics, and
the use of dichroic filters permits simultaneous observing
in two arms (blue and red), covering different spectral regions.

The observations in the blue arm were obtained with grating R600B,
centred at $\lambda = 4400$\,\AA, for a dispersion of 0.44\,\AA\ per
pixel. Most of the observations were carried out with a 1\arcsec\ slit
width, but for a few cases, under some less ideal seeing conditions,
we widened the slit to 1.2\arcsec, or even 2.0\arcsec\ (for four
observations). Spectral resolving power, as measured near the central
wavelength on the lines of the arc lamp, was about 2600, 2200 and 1250
for slit widths of 1.0\arcsec, 1.2\arcsec\ and 2.0\arcsec,
respectively.  In the red arm we employed grating R1200R centred at
$\lambda=6500$\,\AA, with 0.26\,\AA\ per pixel. Observations were
mostly made with a 1.0\arcsec\ slit width, and only in a few occasions
with a 1.2\arcsec\ slit width. Spectral resolving power measured at
the central wavelength with slit widths 1.0\arcsec\ and 1.2\arcsec\
was 8600 and 7200, respectively. Five measurements in the red arm were
obtained with grating R158R (dispersion = 1.8\,\AA\ per pixel) with a
1.0\arcsec\ slit width, for a spectral resolving power of 1100.

\subsubsection{Alignment of the polarimetric optics}

The alignment of the retarder waveplate of ISIS is checked during the
afternoon before the first night of observations by inserting a linear
polariser in the optical beam, and finding at which angles of the
retarder waveplate the contrast between the fluxes measured in the two
beams is minimum or maximum. This way, the instrument scientist
measures the offset $\alpha_0$ between the zero position of the
encoders, and the angle between the one of the axes of the retarder
waveplate and the ordinary beam of the Savart plate. It is then left
to the user to set the retarder waveplate at the correct positions
separated by $90^\circ$ to measure the reduced Stokes parameter $V/I$.

\subsubsection{The use of the dichroic in spectropolarimetric mode}\label{Sect_DIC}
Observations may be carried out simultaneously in the red and blue arm
by inserting a dichroic, or in one arm at a time, by inserting a
mirror (when observing in the blue arm)\footnote{The introduction of a
mirror in the optical path does not change the sign of the measured
circular polarisation, because the mirror is inserted {\it after} the
polarisation analyser.} or leaving the optical path free (in which
case only the red arm is fed). We note that in the two different arms
the positions of the ordinary and extra-ordinary beam are swapped
(due to the different direction of the readout of the two CCDs in the
two arms). Without taking this into account, the polarisation of the
same object observed in the two arms would be measured with opposite
sign.

ISIS documentation warns that ``using a dichroic is not recommended for
spectropolarimetric observations due to the reflected light from its
rear. The reflected light is displaced along the slit, partly into the
spectrum of the other polarisation, which may compromise the
polarimetry measurements.'' To investigate this problem, we carried
out some experiments in linear polarimetric mode, and observed some
discrepancy between observations obtained in the same arm with and and
without dichroic. We originally thought that these discprepancies
could be ascribed to scattered light due to the presence of the
dichroic, and we decided to use only one arm at a time (mainly the
blue arm).  Later, we discovered that these discrepancies had probably
an alternative explanation, linked to an inaccurate
positioning of the retarder waveplate, as discussed by
\citet{Bagetal17b} regarding FORS2 measurements. Therefore in
our second run we decided to experiment with the use of the dichroic in
circular polarimetric mode. Two magnetic stars, HD\,157751 and
$\gamma$\,Equ, were observed simultaneously in both arms with the
dichroic; then, immediately afterwards, in the blue arm and in the red
arm individually. The results of our experiments, presented in Sect.~\ref{Sect_Both_Arms},
showed us that the use of the dichroic does not have any impact
on our field measurements. Therefore in our August 2015 run we proceeded
with simultaneous observations in the blue and in the red arm.
Among the five available dichroics we used the standard one with
cut-off centred at 5300\,\AA.

\section{Observing strategy and tests}\label{Sect_Observations}

Our new observations were carried out with the FORS2 instrument of the
ESO VLT in service mode between March and September 2015 (during
semester P95) and between April and July 2016 (during semester P97),
and with the ISIS instrument of the 4.2\,m William Herschel Telescope
of the ING during two dedicated observing runs in visitor mode, in
February 2015 and in August 2015 (during semesters 15A and 15B,
respectively). Prior to these observing runs, four field measurements
of three WDs were obtained with ISIS in January 2014 as a pilot
experiment, and two stars were observed with FORS2 as a backup programme.
In total we have observed 48 WDs. Twelve of these stars
were observed more than once, and the total number of new spectropolarimetric
observations of WDs is 79. Science targets and observations are discusssed in later
Sections. Here we discuss more technical aspects of our campaign.

\subsection{Observations of well known magnetic stars
  to check ISIS performance}\label{Sect_Calib_Targets}

During our WHT observing runs we frequently checked the instrument
performance. A reliable way to do this is to observe bright stars with
well known magnetic fields. The most obvious candidates are the
magnetic Ap/Bp stars of the main sequence, many of which have been
well studied in the past. Ap/Bp stars usually have variable
magnetic fields, but their variability is often well known, and prediction
can be made for the expected field values \bz\ and sometimes \bs\  at
a given epoch. The
reference stars that we used in our survey are HD\,65339 (= 53\,Cam),
HD\,157751, HD\,201601 (= $\gamma$~Equ) and HD\,215441 (Babcock's
star). Below we describe the characteristics of the magnetic fields of
these stars; the comparison with our observations will be made in
Sect.~\ref{Sect_Reference_Stars}.

\subsubsection{HD\,65339}
HD\,65339 is a extremely well studied magnetic stars with a
longitudinal magnetic field that changes from $+4500$ to $-4600$ (as
measured using an H$\beta$ photoelectric polarimeter) with a rotation
period of $P=8.02681 \pm 0.00004$\,d. The zero point of the ephemeris
refers to the magnetic maximum HJD = $2448498.186 \pm 0.022$, and the
mean longitudinal field curve may be fit by
\begin{equation}
\bz\,(\phi) = B_0 + B_1 \cos (\phi)
\label{Eq_Bzero}
\end{equation}
with $B_0 = -53$\,G and $B_1 = 4572$\,G  \citep{Hiletal98}.

\subsubsection{HD\,157751}
HD\,157751 is a star discovered as magnetic by \citet{Hubetal06}. It is
not clear if its field is variable, but it was observed because it
has a strong field and could serve for the purpose of comparing field
measurements with and without dichroic (see Sect.~\ref{Sect_DIC}).

\subsubsection{HD\,201601}
HD\,201601=$\gamma$~Equ has a fairly strong magnetic field with an extremely
long rotation period, of the order of a century; \citet{Bycetal16} report
a rotation period of $P=35462.5 \pm 1150$\,d, refer the rotation phase
to the positive cross-over on HJD = 2425176.5, and provide for Eq.~(\ref{Eq_Bzero})
the coefficients $B_0 = -265$\,G and  $B_1 = 850$\,G. At the time of
our observations of this star, the nominal value of $\bz \approx -745$\,G.

\subsubsection{HD\,215441}
With a mean field modulus of about 35\,kG, Babcock's star HD\,215441
is the star with the strongest magnetic field known among Ap
stars. The field measurements available in the literature do not allow
us to calculate the rotation period: the mean field modulus is nearly
constant, while only a few longitudinal field measurements are
available.  Therefore we rely on the photometric ephemeris brightness
in the $B$ filter, $P=9.487574 \pm 0.000030$\,d, with a light maximum
at HJD=2448733.714 \citep{NorAde95}, and on a qualitative correlation
between field strength and star brightness: by comparing data of
\citet{BorLan78} with the light curve of \citet{Lec74} we
find that the magnetic maximum occurs close to maximum
brightness. From a fit to the H$\beta$ data by \citet{BorLan78} we
find $B_0 = 15700$\,G and $B_1=4800$\,G.
   
\subsection{Zero field measurements}
In a few experiments we also observed reference stars after setting the
retarder waveplate at position angles such that the polarisation signal
should be zero, that is, we set the retarder waveplate
at position angles of $0^\circ$ and $90^\circ$ instead of $\pm
45^\circ$ with respect to the principal axes of the beam splitter.
Any non--zero field resulting from this experiment would point either to
a misalignemnt of the retarder waveplate or to instrument flexures
\citep[for a discussion on how instrument flexures may lead to
  spurious detections see][]{Bagetal13}. We call these measurements
``Zero field measurements'' to distinguish them from the null
field, i.e., the field estimate that one would obtain by using
the null profiles instead of the $V/I$ profiles
(for definition and discussion of the use of the null profiles
and the null fields, see \citeauthor{Bagetal12}
\citeyear{Bagetal12} and \citeauthor{Bagetal13} \citeyear{Bagetal13}).

\subsection{The effect of a small offset of the retarder waveplate}
The possibility to set the retarder waveplate at an arbitrary position
angle allowed us to perform some experiments, namely: to
experimentally check if and how the measured  value of the longitudinal
magnetic field changes for small deviations of the position of the
retarder waveplate from the nominal values, and to check that after
offsetting the positions of the retarder waveplate by $45^\circ$, one
actually measures a null polarisation signal (and a magnetic field
consistent with zero) even in strongly magnetic stars.  We also
measured the magnetic field of two reference stars,
HD\,215441 and $\gamma$\,Equ, after systematically offsetting the
position angle of the retarder waveplate by $\pm 5\deg$, and compared
the results with the measurements obtained without this artificial
offset. The results of these experiments are described in
Sect.~\ref{Sect_QC}.

\section{Data reduction}\label{Sect_Data_Reduction}
Steps for data reduction are similar for both FORS2 and ISIS instruments.
After bias-subtraction, background subtraction, flux extraction and
wavelength calibration, we obtained the reduced Stokes $V$ profiles
($\pv=V/I$) by combining the various beams according to the well known
formulas
%%%%%%%%%%%%%%%%%%%%%%%%%%%%%%%%%%%%%%%%%%%%%%%%%%%%%%%%%%%%%%%%%%%%
\begin{equation}
\begin{array}{rcl}
\pv &=& \frac{1}{N}\sum_i^{N} \pv^{(i)} \\
 \pv^{(i)} &=& \frac{1}{2} 
\Bigg\{ \left(\frac{\fo^{(i)} - \fe^{(i)}}{\fo^{(i)} + \fe^{(i)}}\right)_{\alpha= \alpha_i} -
        \left(\frac{\fo^{(i)} - \fe^{(i)}}{\fo^{(i)} + \fe^{(i)}}\right)_{\alpha=\alpha_i+90^\circ}
\Bigg\} \ ,
\label{Eq_Diff}
\end{array}
\end{equation}
%%%%%%%%%%%%%%%%%%%%%%%%%%%%%%%%%%%%%%%%%%%%%%%%%%%%%%%%%%%%%%%%%%%%%
where \fo\ and \fe\ are the flux measured in the parallel and
perpendicular beam of the beam splitter device, respectively
\citep[e.g.][]{Bagetal09}, and $\alpha_i=315\degr$ or $135\degr$
(in the large majority of cases we had $N=4$ and $\alpha_i = 315\degr$).
The uncertainty of the \pv\ profile in a
certain spectral bin is approximately given by $\snr^{-1}$, where \snr\ is the
signal-to-noise ratio accumulated in that spectral bin adding up
the fluxes measured in both beams at all positions of the retarder
waveplate \citep[e.g.][]{Bagetal09}.

In the Zeeman regime, field measurements may be obtained by using the
technique described by \citet{Bagetal02}, i.e., by minimising the
expression:
\begin{equation}
\chi^2 = \sum_i \frac{(y_i - \bz\,x_i - b)^2}{\sigma^2_i}\; ,
\label{EqChiSquare}
\end{equation}
where, for each spectral point $i$, $y_i = V(\lambda_i)/I(\lambda_i)$, $x_i =
-g_\mathrm{eff} \cz \lambda^2_i \,(1/I_i\ \times
\mathrm{d}I/\mathrm{d}\lambda)_i$, and $b$ is a constant introduced to
account for possible spurious polarisation in the continuum,
$g_\mathrm{eff}$ is the effective land\'{e} factor, and
%%%%%%%%%%%%%%%%
\begin{equation}
\cz = \frac{e}{4 \pi m_\mathrm{e} c^2}
\ \ \ \ \ (\simeq 4.67 \times 10^{-13}\,\AA^{-1}\ \mathrm{G}^{-1})
\end{equation}
%%%%%%%%%%%%%%%%
where $e$ is the electron charge, $m_\mathrm{e}$ the electron mass,
and $c$ the speed of light. An extensive discussion of these
techniques is provided by \citet{Bagetal12}. For this survey we have
adopted the same $\sigma$ clipping algorithm used by
\citet{Lanetal12}, but also decided to clean the spectra from
cosmic-rays using the relevant option in the IRAF {\tt apall}
procedure. In some cases, especially those with the longest exposure
times, this has allowed us to decrease the uncertainties of our
measurements.  Our spectral analysis is shown in the many panels for
individual observations in the Appendix, which are organised as
follow.  In the upper panel, the black solid line shows the intensity
profile, the shape of which is heavily affected by the transmission
function of the atmosphere + telescope optics + instrument. The red
solid line is the $V/I$ profile (in \% units) and the blue solid line
is the null profile offset by $-2$\,\% for display
purpose. Photon-noise error bars are centred around $-2$\,\% and
appear as a light blue background.  Spectral regions highlighted by
green bars have been used to detemine the \bz\ value from H Balmer
lines. The two bottom panels show the best-fit obtained by minimising
the expression of Eq.~(\ref{EqChiSquare}) using the $V/I$ profiles
(left panels) and the \nnv\ profiles (right panels).

\subsection{Determining the sign of Stokes $V$}\label{Sect_Calib_Measurements}

Field measurements of the reference stars are crucial not only to
check instrument performance, but also, at a very basic level, to
establish the correct sign of the Stokes $V$ profiles (hence, of the
magnetic field).

The sign of $V$ can be obtained via analytical formulas once the
orientation of the polarimetric optics is known. However, getting the
\pv\ profile with the correct sign from these ``first principles'' is
a challenging task. One needs to identify which of the parallel and
perpendicular beams of the beam splitted device illuminates which image on the
CCD (remembering that, for ISIS, red and blue CCDs are read out in different
ways, and that therefore the position of the parallel and perpendicular beams
are swapped in the red and blue arms). Then one has to be sure of the convention used
by the software package of preference regarding the naming of the
aperture (i.e., whether aperture number increases towards the left or
towards the right).  Each of these issues may well be gotten under
control, but there is no doubt that a comparison of the field
measurement of a reference star with the expected value from previous
literature studies represents an attractive short-cut to determine the
correct sign of our field determination, and this is the method that
we have used. In summary, our determination of the sign of the $V$
profile was guided by the goal to make the sign of the longitudinal
magnetic field measured for our reference stars consistent with the
value expected from previous literature data.

\subsection{Measurements in the H$\alpha$ profile}
\begin{figure*}
\includegraphics[width=8.8cm,trim={1.2cm 0.8cm 1.1cm 1.0cm},clip]{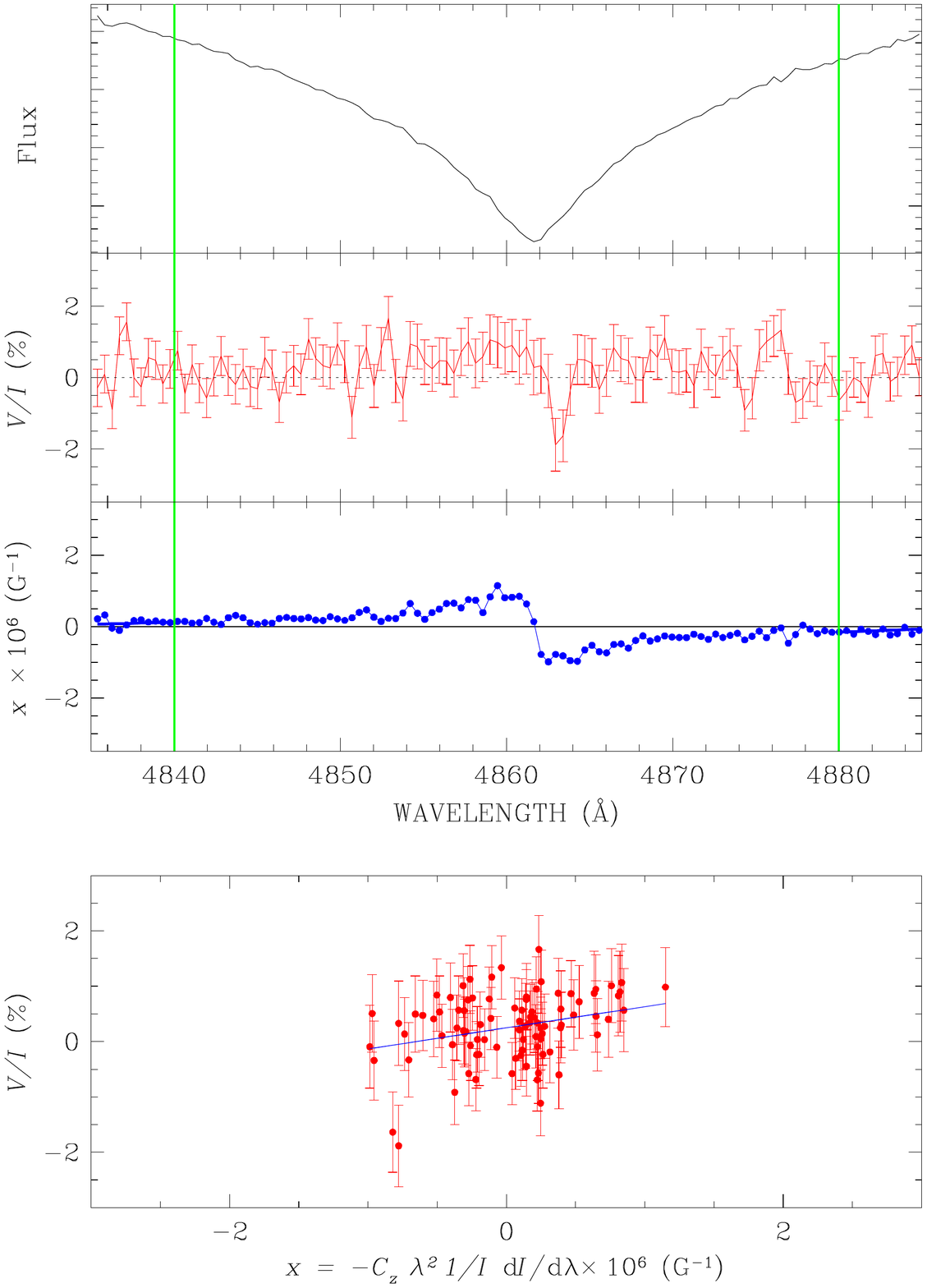}
\includegraphics[width=8.8cm,trim={1.2cm 0.8cm 1.1cm 1.0cm},clip]{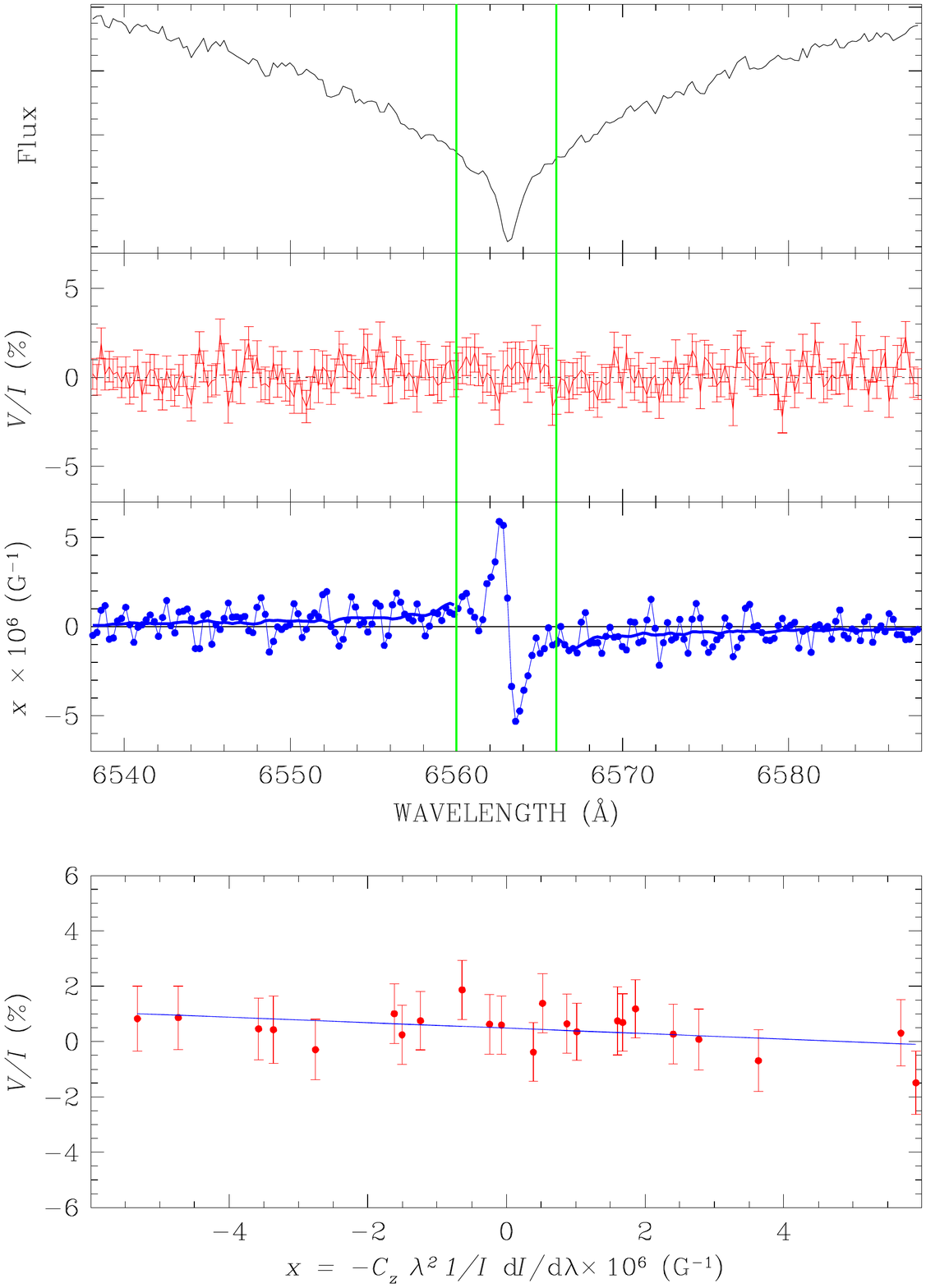}
\caption{\label{Fig_dI} H$\beta$ (left panel) and H$\alpha$ (right panel)
  lines of star WD\,1840$-$111 observed with ISIS in the blue and red arm,
  respectively.
    {\it Top panel:} the $I$ profile (in arbitrary units).
    {\it Second panel from top:} $V/I$ profile.
    {\it Third panel from top:} $x=\cz\,(1/I)\,{\rm d}I/{\rm d}\lambda$ {\it vs.}\ wavelength.
    The green vertical lines show the wavelength intervals in which the $x$
    values ($x=-\cz\,\lambda^2\,1/I\ {\rm d}I/{\rm d}\lambda$) clearly depart
    from zero, and which have been choosen to apply the least-square technique.
    {\it Bottom panel:} the relation between $V/I$ and $x$
    for the points within the solid green lines in the upper panels.
    Note that the $x$ and $V/I$ ranges of the H$\beta$ plots are half the size of 
    the corresponing ranges in the H$\alpha$ plots.
    }
\end{figure*}

It is important to note that the H$\alpha$ profile is often affected
by fringing issues (in the case of ISIS) and water vapour features, and, less often, by
features due to a WD or dMe companion, like the double degenerate
WD\,0135$-$052 \citep{Safetal88} and the WD+dMe system WD\,1213+528,
or non ideal background subtraction (the solar spectrum may pollute the target spectrum in
observations obtained during full moon nights or during twilight, if
the background is not accurately subtracted). These spurious signals
clearly have a negative impact on the field measurements, and in a
subtle way.  While photon noise scatters $V/I$ along the $y$
direction, fringing and water vapour features scatter the points along
the $x$ direction, invalidating the use of a least squares technique
under the approximation that the errors on $x$ are much smaller than
the errors on $y$.  Practically, these issues are effectively
indistinguishable from the contribution of spectral lines showing no
polarisation, and they dilute the field (if present) while still
formally adding precision to the measurements.  Figure~\ref{Fig_dI}
shows that this situation is mitigated by considering only the core of
H$\alpha$, where $I$ is sufficiently steep that fringing issues
becomes negligible with respect to photon-noise.

%%%%%%%%%%%%%%%%%%%%%%%%%%%%%%%%%%%%%%%%%%%%%%%%%%%%%%%%%%%%%%%%%%%%%%%%%
%%%%%%%%%%%%%%%%%%%%%%%%%%%%%%%%%%%%%%%%%%%%%%%%%%%%%%%%%%%%%%%%%%%%%%%%%

\section{Test results and quality checks}\label{Sect_QC}

\subsection{Magnetic field measurements of the reference stars}\label{Sect_Reference_Stars}
The values of the field measurements of the reference stars of
Sect.~\ref{Sect_Calib_Targets} allow us to perform a first basic
quality check of our observations and to find the sign of our Stokes
$V$ measurements.

HD\,65339 was observed during our January 2014 run at rotation phase
$\sim 0.9$ and during our February 2015 runs at various rotational
phases from $\sim 0.1$ to $\sim 0.5$. The remaining reference stars
were all observed during our August 2015 run. Table~\ref{Tab_Ap} shows
the observing log of the magnetic reference stars, and the measured
magnetic field.

%%%%%%%%%%%%%%%%%%%%%%%%%%%%%%%%%%%%%%%%%%%%%%%%%%%%%%%%%%%%%%%%%%%%%%%%%
\begin{figure}
\includegraphics[width=8.8cm,trim={0.2cm 5.8cm 1.0cm 3.0cm},clip]{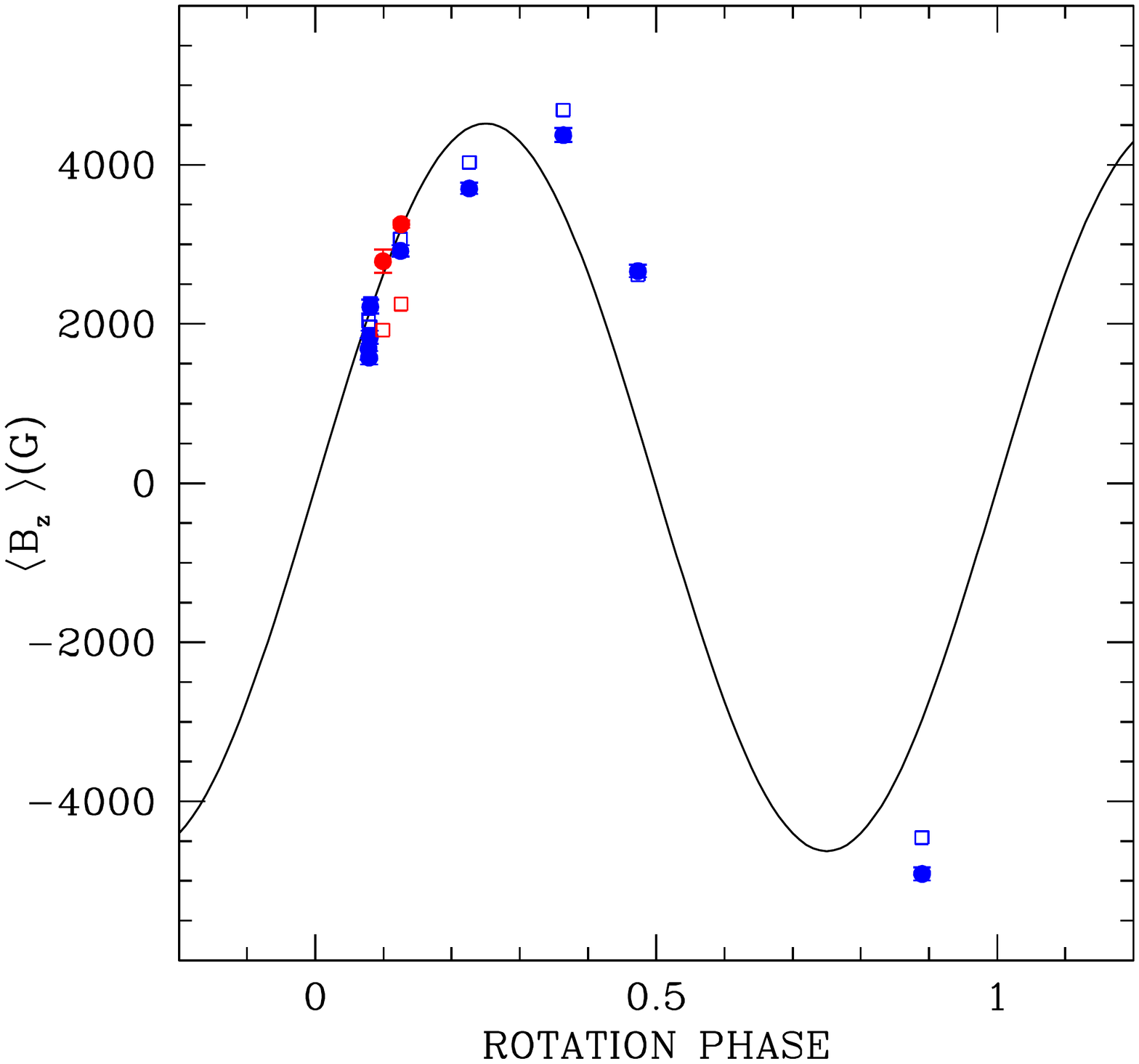}
  \caption{\label{Fig_Reference_Stars} Comparison of our ISIS measurements of reference star
    53\,Cam=HD\,65339 with the sinusoidal curve obtained by \citet{Hiletal98}. Filled circles refer to measurements
    obtained from metal lines, and empty squares to field measurements from H Balmer lines. Blue
    symbols refer to data obtained with grating R600B, and red symbols to data obtained
    with the R1200R grating. Error bars are shown only for the measurements obtained
  from H Balmer lines; their size is comparable to the symbol size.}
\end{figure}
%%%%%%%%%%%%%%%%%%%%%%%%%%%%%%%%%%%%%%%%%%%%%%%%%%%%%%%%%%%%%%%%%%%%%%%%%

We found that all our field measurements of magnetic Ap stars are
broadly consistent with the field value predicted by simple sinusoidal
curves based on their known rotation period,
(Figure~\ref{Fig_Reference_Stars} shows the case of 53\,Cam) but with
some discrepancies that have various explanations.

First of all, we note that high \snr\ measurements of $\gamma$~Equ and
53\,Cam that were obtained during a short interval of time (a few
minutes) with identical instrument settings are sometimes inconsistent
with their (small) errorbars. Similarly, Table~\ref{Tab_Zero} shows
the results of high \snr\ observations obtained with the retarder
waveplate at 0\degr\ and 90\degr, which depart from zero by much more
than can be explained by photon-noise. This suggests that, as in
the FORS spectrograph \citep{Bagetal12}, small flexures create
spurious signals that become important when photon-noise is very
low. This interpretation is confirmed by the fact that if we consider
the high \snr\ measurements of the Ap stars, the distribution of the
null field values normalised by their error bars departs from a
Gaussian distribution with $\sigma=1$.

Simultaneous or quasi-simultaneous observations of the same Ap star in
two different arms show remarkable discrepancies, and so do the field
measurements obtained from H Balmer lines and those obtained from He
and metal lines. This phenomenon is well known and discussed by
\citet{Lanetal14}. The quantity \bz\ is conceptually an average of the
line-of-sight component of the surface magnetic field over the visible
hemisphere, but which average? It is clear that the surface does not
emit the same specific intensity in the direction of the observer from
every surface element (for example, due to limb darkening), and also
that the spectral lines in which the polarisation is measured do not
have exactly the same shape and strength over the whole visible
hemisphere (lines tend to weaken mildly towards the limb). Thus the
hemispheric average of the longitudinal field is weighted somewhat
towards the centre of the visible disk. This weighting will vary with
wavelength, and from one line to another (due to the details of line
formation). The result is that we can confidently expect that values
of \bz\ determined in different wavelength regions, or with different
spectral lines, will have similar magnitudes but will frequently
differ from one another by considerably more than the nominal
uncertainties imply. This is the case for the measurements we report.
The discrepancies between the fit to previous measurements and our
new datapoints seen in Figure~\ref{Fig_Reference_Stars} are also
to be ascribed to the fact that the ephemeris of 53\,Cam
is not accurate after a time interval of 20 years.

Another important point to keep in mind is the following. The
technique we use here for deducing the value of \bz\ from a polarised
spectrum is valid in the ``weak-field'' limit, where the splitting of
the components of the Zeeman multiplet is small compared to the
natural width (as convolved with the spectrograph profile) of the
line. In general, this assumption is valid. An obvious exception are
the \bz\ measurements of HD\,215441 with ISIS using the R1200R
grating. In this case many line components are close to being resolved
by the instrument, and the peaks of the polarisation no longer
coincide with the global line wings, but with the centres of the
Zeeman $\sigma$ components. Thus the proportionality between large
$V/I$ and large line slope in $I$ is broken and the value of \bz\ is
underestimated.

Finally, another implicit assumption of this method is that the
spectral lines are mostly well separated from neighbours. In the cases
of metal lines of red spectra of HD\,215441 obtained with grating
R1200R, and of both H Balmer and metal lines of HD\,65339 obtained
with the R158R grating, the resolving power is so small that many
lines are blended together with near neighbours. Again the
proportionality of $V/I$ with d$I/$d$\lambda$ is broken, and the field
is underestimated.  The case of HD\,215441 was discussed in detail by
\citet{Lanetal15}.  For the remaining cases, the \bz\ values should be
realistic estimates of field on the observed stellar hemisphere. For a
more fundamental discussion of these points, see \citet{Landstreet82}
or \citet{Mathys89}.

Table~\ref{Tab_Ap} includes also the results of some observations
obtained with the retarder waveplate delibrately offset by $\pm
5\degr$, immediately before or after observations obtained with the
retarder waveplate set in the correct position. \citet{Bagetal09}
showed that deviations from the nominal values of the position angle
of the retarder waveplate are compensated to first order by the beam
swapping technique (in circular polarisation only); their Fig.~3 shows
that a systematic offset of 5\degr\ would introduced a relative error
on Stokes $V/I$ of about 1\,\%. In HD\,215441, we see that field
measurements are within error bars, confirming the expectation that a
misaligment as small as 5\degr\ does not significantly alter the field
measurements. More significant differences are seen on the field
measurements of HD\,201601, which are characterised by a much higher
\snr, and may be ascribed again to instrument flexures.

\subsection{The impact of using a dichoric on the circular polarisation
  measurements}\label{Sect_Both_Arms}
Table~\ref{Tab_Ap} includes also the results of the observations of reference stars obtained
simultaneously in the red and in the blue arms, and
(quasi-simultaneosuly) in the red and in the blue arm separately, presented
in Sect.~\ref{Sect_DIC}; the
differences in field measurements obtained when observing in one arm
at a time and with dichroic actually agree surprisingly well within photon-noise error bars. Furthermore, when
comparing the \pv\ profiles, we did not discover significant
discrepancies beyond those due to photon-noise. In conclusion, we
found that the dichroic could be used without affecting our
measurements.

\subsection{The spurious polarisation removed by the beam-swapping technique}
%%%%%%%%%%%%%%%%%%%%%%%%%%%%%%%%%%%%%%%%%%%%%%%%%%%%%%%%%%%%%%%%%%%%%%%%%
\begin{figure}
\includegraphics[width=8.8cm,trim={0.2cm 4.8cm 1.0cm 3.6cm},clip]{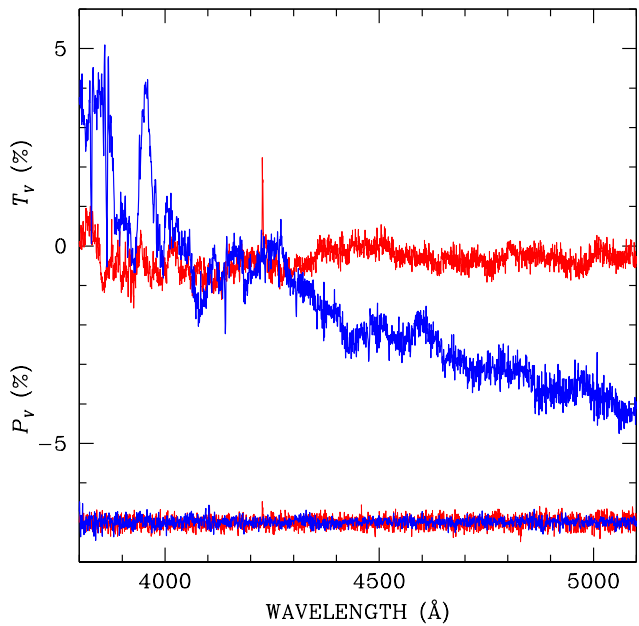}
\caption{\label{Fig_InsPol} First-order effects of imperfect flat-fielding that 
  are in fact canceled out by adopting the beam-swapping technique.
  The solid thick blue line refers to
  FORS2 data (with grism 1200B), the thin solid red line to ISIS data (grating R600B).
  Offset by -7\,\% (for display purpose) are the \pv\ profiles obtained with the
  beam swapping technique.}
\end{figure}
%%%%%%%%%%%%%%%%%%%%%%%%%%%%%%%%%%%%%%%%%%%%%%%%%%%%%%%%%%%%%%%%%%%%%%%%%
The beam swapping technique removes the instrumental polarisation
introduced by the polarimetric optics \citep[see][]{Bagetal09}.
It is of some interest to check how much is the contribution that
would be introduced mostly by imperfect flat-fielding
\citep[see Eq.~(34) of][]{Bagetal09}. This can be calculated
simply by replacing $\pv^{(i)}$ in Eq.~(\ref{Eq_Diff}) with the
expression
%%%%%%%%%%%%%%%%%%%%%%%%%%%%%%%%%%%%%%%%%%%%%%%%%%%%%%%%%%%%%%%%%%%%%%%%
\begin{equation}
  \label{Eq_TV}
   T_V^{(i)} = \frac{1}{2} 
\Bigg\{ \left(\frac{\fo^{(i)} - \fe^{(i)}}{\fo^{(i)} + \fe^{(i)}}\right)_{\alpha= \alpha_i} +
        \left(\frac{\fo^{(i)} - \fe^{(i)}}{\fo^{(i)} + \fe^{(i)}}\right)_{\alpha=\alpha_i+90^\circ}
\Bigg\} \ 
\end{equation}
%%%%%%%%%%%%%%%%%%%%%%%%%%%%%%%%%%%%%%%%%%%%%%%%%%%%%%%%%%%%%%%%%%%%%%%%
\citep[see also][]{Maund08,Ilyin12}. Figure~\ref{Fig_InsPol} compares
examples taken from FORS2 and ISIS observations, and shows that the
quantity $T_v =\frac{1}{N}\sum_i^{N} T_V^{(i)}$ is much higher (in
absolute value) and much more wavelength dependent in FORS2
observations than in ISIS observations. This exercise suggests that in
cases of ISIS, spectropolarimetric observations obtained with just at
one position of the retarder waveplate may still be useful, perhaps
after a normalisation to the continuum.  The impact of imperfect
flat-fielding in FORS2 data is such that data obtained at just one
position of the retarder waveplate are difficult to correct.

\subsection{Distribution of the null field measurements}

\begin{figure}
\includegraphics[width=8.8cm,trim={1.0cm 5.8cm 1.0cm 3.0cm},clip]{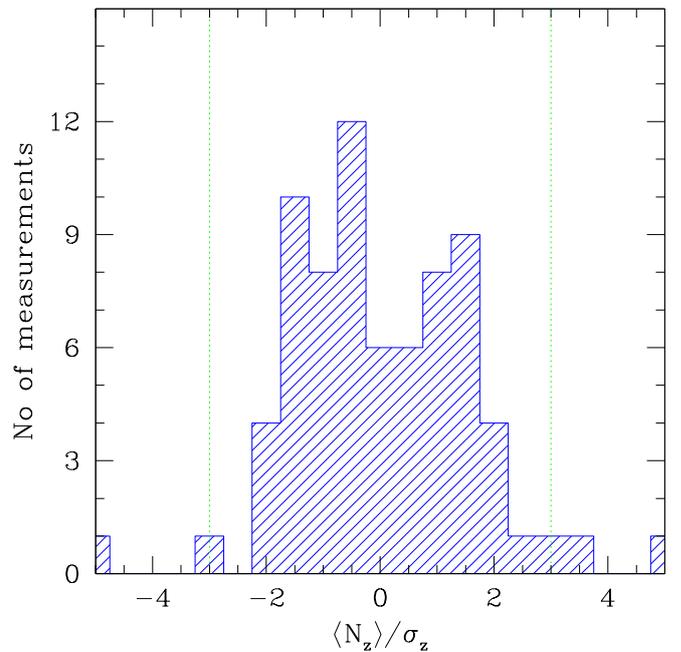}
\caption{\label{Fig_Histo_Nz} Distribution of the \nz\ measurements
normalised by their error bars for the WD observations.}
\end{figure}

Figure~\ref{Fig_Histo_Nz} shows the histogram of the ratio between
null field and its error bar (in the ideal case it should be a
Gaussian distribution with $\sigma=1$). This histogram is obtained
considering all FORS2 and ISIS measurements. When ISIS observations were
obtained simultaneously in the red and in the blue ISIS arm,
the \nz\ estimate is obtained by combining the two spectra.
The distribution appears generally well within 3\,$\sigma$ and has only
just a few outliers. It is not affected by instrument flexures because
spectral lines of WDs are broad and photon-noise error bars are relatively
high.

%%%%%%%%%%%%%%%%%%%%%%%%%%%%%%%%%%%%%%%%%%%%%%%%%%%%%%%%%%%%%%%%%%%%%%%%%
%%%%%%%%%%%%%%%%%%%%%%%%%%%%%%%%%%%%%%%%%%%%%%%%%%%%%%%%%%%%%%%%%%%%%%%%%

\section{Results}\label{Sect_Results}

We have obtained 79 \bz\ field measurements of 48 different WDs, 12 of
them observed twice or more. With FORS2, we obtained 27 observations
of a total of 13 stars (one of which is featureless, for which we
could only rule out the presence of a field strong enough to polarise
the continuum). With ISIS, we obtained 54 polarisation spectra of 38
stars (three targets were in common with the VLT sample). Twenty-four
ISIS observations were obtained simultaneously in the blue and red
arm, 25 observations were obtained only with in the blue arm, and five
in the red arm only (four of them on the same star, with the low
resolution grating R158R).

The log of the observations of WDs and the field measurements are
given in Table~\ref{Tab_Log} (spectra of an additional star, WD\,1900+705, will be
presented in a forthcoming paper. All our spectra (Stokes $I$ and
$V/I$ profiles) are made available at CDS.  We note that some of our
target WDs are also spectro-photometric standard stars with absolute
fluxes tabulated in literature, for example WD\,0501+527 = BD+52\,913 = G\,191-B2B
\citep{Oke90,Bohetal95} was observed with both ISIS arms; WD\,0549+158 =
GD\,71 \citep{Bohetal95,Moeetal14b}, was observed with both ISIS arms;
WD\,1134+300 = GD\,140 \citep{Masetal88,Masetal90} observed with ISIS
blue arm; WD\,2032+248 = HD\,340611 = Wolf\,1346 \citep{Masetal88,Masetal90},
was observed with both ISIS arms; WD\,0310$-$688 = GJ\,127.1 = EG\,21
\citep{Hametal92} was observed with FORS2 / grism 1200B.  The spectra of
these stars may be used for an approximate flux calibration of all our
targets, neglecting the effect of slit losses. Airmass extinction
coefficients for La Palma and for Paranal are given by \citet{King85}
and \citet{Patetal11}, respectively.

Table~\ref{Tab_Log}
includes distances obtained from the most recent Gaia release
(Gaia Collaboration \citeyear{gaia18}), and spectral types taken from Simbad database.  In
some cases we have adopted a more accurate spectral classification
(DAZ) than that listed in Simbad (DA); these cases are WD\,1116+026
\citep{Xuetal14}, WD\,1202-232 \citep{Zucetal03}, WD\,1337+705
\citep{Holetal97,Zucetal03}, WD\,1632+177 \citep{Zucetal03} and
WD\,2105$-$820 \citep{Koeetal05,Koeetal09}. We also note that Gaia
distances are generally known with much better accuracy than what is
shown in Table~\ref{Tab_Log}, but higher accuracy is not important in
the context of this work.

In addition, because of a typo in our target list, we obtained two
observations of an sdOp star with the blue arm of ISIS, and because of
a mistake from our side in the preparation of the observations, for two
stars observed with FORS2 we obtained only intensity spectra. The log of these
observations is given in Table~\ref{Tab_LogOne}.

\subsection{General discussion of our results}

\begin{figure}
\includegraphics[width=8.8cm,trim={1.0cm 5.8cm 1.0cm 3.0cm},clip]{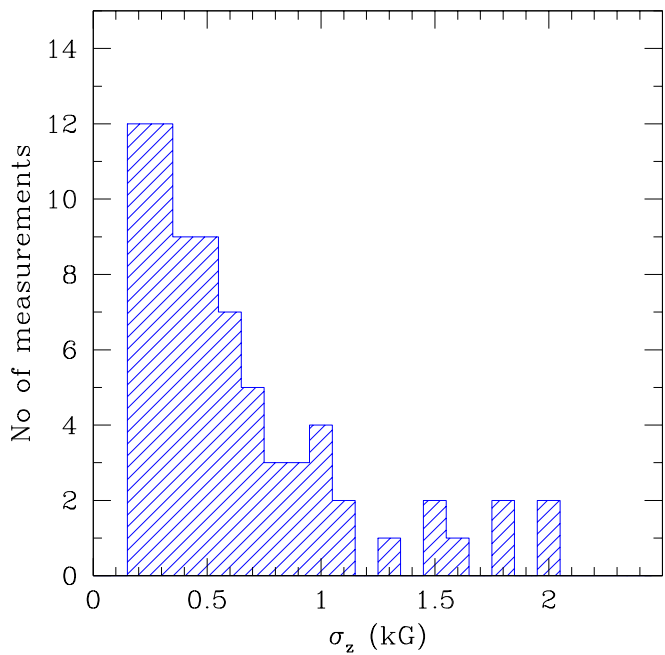}
\caption{\label{Fig_Histo_Sigma} Distribution of the uncertainties of the
\bz\ measurements presented in this survey.}
\end{figure}

Figure~\ref{Fig_Histo_Sigma} shows the distribution of the error bars
of all FORS2 and ISIS \bz\ measurments. As in the case of
Fig.~\ref{Fig_Histo_Nz}, observations obtained simultaneously
in the ISIS blue and red arm were combined together.
This Figure shows that the large majority of our new observations
have an uncertainty smaller than 1\,kG. The mean uncertainty of all
field measurements of our survey is 0.6\,kG.
%%%%%%%%%%%%%%%%%%%%%%%%%%%%%%%%%%%%%%%%%%%%%%%%%%%%%%%%%%%%%%%%%%%%%%%%%
%%%%%%%%%%%%%%%%%%%%%%%%%%%%%%%%%%%%%%%%%%%%%%%%%%%%%%%%%%%%%%%%%%%%%%%%%

In the course of the surveys presented in this paper we detected a
magnetic field in six WDs: WD\,2359$-$434 (observed with FORS2,
already discovered as magnetic by \citeauthor{Koeetal98} \citeyear{Koeetal98} and
by \citeauthor{Aznetal04} \citeyear{Koeetal98}),
WD\,0446$-$789 \citep[observed with FORS2, already discovered as
  magnetic by][]{Aznetal04}, WD\,1105$-$048 \citep[observed with FORS2
  and ISIS, already suspected as magnetic by][]{Aznetal04},
WD\,2047+372 \citep[observed with ISIS, detected as magnetic through
  Zeeman splitting in H$\alpha$, and discussed already
  by][]{Lanetal16,Lanetal17}, WD\,2051$-$208 \citep[observed with
  FORS2, already discovered as magnetic by][]{Koeetal09},
WD\,2105$-$820 (suggested as possibly magnetic from spectroscopy by
\citeauthor{Koeetal98} \citeyear{Koeetal98} and confirmed with FORS1
spectropolarimetry by \citeauthor{Lanetal12} \citeyear{Lanetal12}).
In addition, we have obtained a single marginal field detection at
about the $3\sigma$ level in two white dwarfs: WD\,1031$-$114, a DA1.9
WD, and WD\,2138$-$332, a DZ WD within the 20-pc volume around the
Sun.  Neither of these detections is at a sufficiently high level of
significance to convince us that we have detected kG-level fields in
these stars, so further observations of each will be undertaken.  In
particular, we note that some high-resolution (but low \snr) spectra
obtained with the FEROS instrument are available in the ESO
archive. None of the metal lines show sign of Zeeman splitting. FEROS
observations do not support our marginal detection, but are not
inconsistent with the presence of a very weak field (say $\bz \la
20$\,kG). We note that the detection in WD\,2138$-$332, if confirmed,
would further support the suggestion by \citet{Holetal15} that the
incidence of magnetic fields in cool DZ stars is higher than in WDs
of other spectral types.

A strong polarisation signal was detected with ISIS in the well
known magnetic star WD\,1900+705 \citep[the first WD discovered as
  magnetic,][]{Kemetal70}, which will be studied in a forthcoming
paper.

All the remaining stars, if magnetic, have a field that is not
sufficiently strong to be detected in our survey, or that at the time
of our observations was seen in a unfavourable geometrical
configuration. The results of our observations of WDs will be
discussed in more detail in Sect.~\ref{Sect_Discussion_Science}.

\section{The efficiency of the magnetic field measurements in WDs}\label{Sect_Efficiencies}
In this survey we have used two different instruments, FORS2 and ISIS,
and the ISIS instrument was used with two different settings, one with
the red arm and one with the blue arm. The moderate-resolution
spectropolarimeter ISIS is a rather different instrument than FORS2 in
two important ways. First, all of the Balmer lines in the visible
window, including H$\alpha$ can be observed simultaneously. Secondly,
the spectral resolving power in the blue arm is about 2500, almost
twice the value ($R \approx 1400$) used in FORS2 in the same region,
and about 8600 in the red arm. Measurements using two arms may be made
simultaneously, and therefore combined to improve the measurement
precision. In this Section we will make a comparison of the efficiency
of the results obtained adopting different instruments and instrument
settings, by analysing separately the results obtained with FORS2,
with the blue arm and with the the red arm of the ISIS instrument, and
we will also comment on the efficiency of the field measurements as a
function of stellar temperature. In our forthcoming papers, this
analysis will be continued by incorporating new magnetic field data
obtained with the ESPaDOnS instrument of the CFHT and with the grism
1200R of the FORS2 instrument.

\subsection{The efficiency of the instruments as photon counters}

%%%%%%%%%%%%%%%%%%%%%%%%%%%%%%%%%%%%%%%%%%%%%%%%%%%%%%%%%%%%%%%%%%%%%%%%%
\begin{figure}
\includegraphics[width=8.8cm,trim={1.7cm 6.3cm 11.0cm 12.8cm},clip]{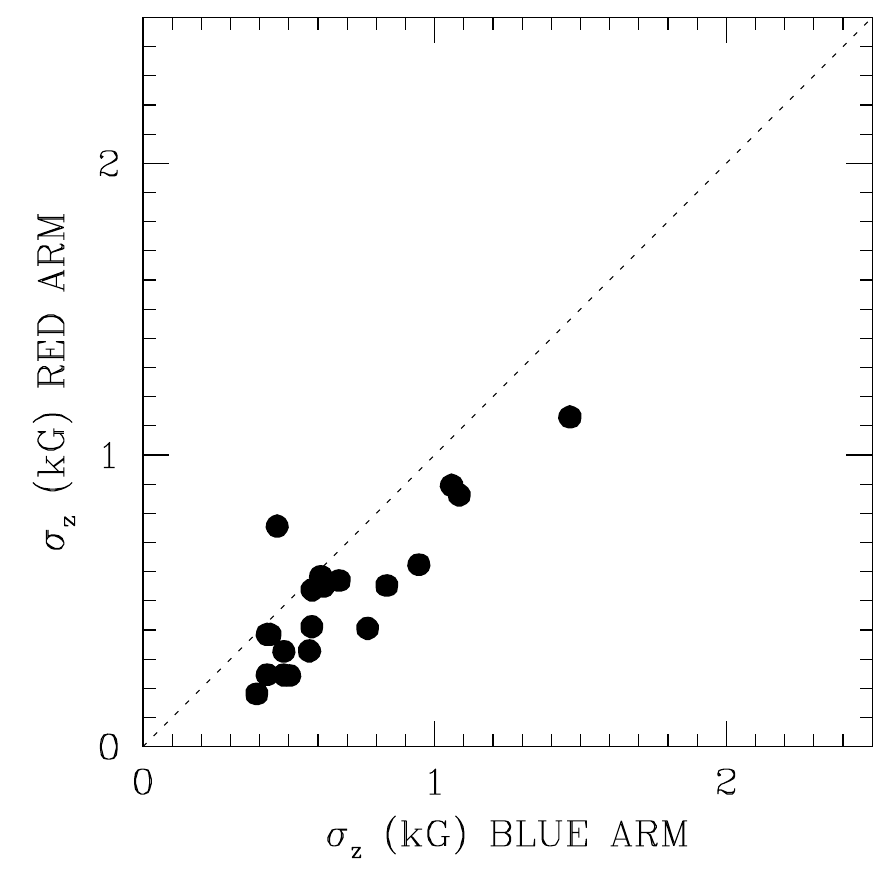}
\caption{\label{Fig_Blue_Red} Uncertainties of the observations obtained
  simultaneously in the red arm and in the blue arm of ISIS.
  The single star above the equality line is a DB star with only a
  rather weak line at 6678\,\AA\ and a low ratio red/blue flux because of
  $\teff \sim 25000$\,K. All the remaining points refer to DA WDs.
}
\end{figure}
%%%%%%%%%%%%%%%%%%%%%%%%%%%%%%%%%%%%%%%%%%%%%%%%%%%%%%%%%%%%%%%%%%%%%%%%%

The most basic comparison is to check the efficiency of the different
spectropolarimeters simply as photon-counters, independently of
telescope size. This comparison could be carried out with the Exposure
Time Calculators of the respective instruments, but the data obtained
from our survey allow us to make a more realistic comparison that
takes into account also the polarimetric optics of the instruments. An
obvious way to perform this exercise is to compare the measurements
obtained on the same stars. In our survey, three targets were observed
(at different epochs) with both FORS2 and the blue arm of ISIS:
WD\,1031$-$114, WD\,1105$-$048 and WD\,1327--083. Thin cirrus and
clouds were present on sky during WHT observations, and during the
observation of WD\,1105$-$048 obtained on 2016-07-02 with FORS2, so
this dataset cannot be used to properly measure the photon collection
efficiency. However, since the magnitudes of almost all of our targets
are known from previous studies with a good accuracy, it is possible
to calculate the quantity
%%%%%%%%%%%%%%%%%%%%%%%%%%%%%%%%%%%%%%%%%%%%%%%%%%%%%%%%%%%%%%%%%%%%
\begin{equation}
E = \frac{(\snr)_{\rm max}^2}{A \, 10^{-0.4(V-kX)}\ t_{\rm exp}}
\end{equation}
%%%%%%%%%%%%%%%%%%%%%%%%%%%%%%%%%%%%%%%%%%%%%%%%%%%%%%%%%%%%%%%%%%%%
where $t_{\rm exp}$ is the exposure time, $V$ the star magnitude, $A$ the
telescope primary mirror area, $(\snr)_{\rm max}$ the peak \snr\ per \AA, $k$ an average
extinction coefficient in the observed spectral range, and $X$
the airmass. By considering only the observations obtained during clear nights 
and with good seeing, we compared statistically the $E$ values for
the three instruments, and we found that for our target list of WDs
the efficiency of FORS2 in the blue regions and that of ISIS in the blue
arm are roughly similar.

\subsection{Comparison of the precision of the field measurements obtained with the two
  arms of the ISIS instrument, and with the FORS2
instrument}\label{Sect_Red_vs_Blue}

The uncertainty of field measurements decreases with the number of the
observed spectral lines, their strength and slope, and their
wavelength (the Zeeman  $\pi - \sigma$ splitting is proportional
to $\lambda^2$).  In the case of our ISIS observations of DA WDs, the
spectral range observed in the blue arm includes Balmer lines from
H$\beta$ to H9, while the red arm includes H$\alpha$
only. Roughly speaking, in terms of measurement precision, the fact
that H$\alpha$ is at longer wavelength than the remaining Balmer lines
does not fully balance the fact that the blue arm includes up to six times
more Balmer lines than the red arm (as $(6500/4400)^2 \sim
2.1$). Furthermore, for equal exposure time, a higher \snr\ is reached
in the blue arm than in the red arm. On the other hand, the red arm
has about three times larger spectral resolution than the blue
arm. Figure~\ref{Fig_Blue_Red} shows the uncertainties of the fields
measurements obtained in the red and in the blue arm for DA WDs
(observations in the two arms were obtained simultaneously with the
use of the dichroic).

\begin{figure}
\includegraphics[width=8.8cm,trim={0.9cm 5.8cm 1.1cm 3.0cm},clip]{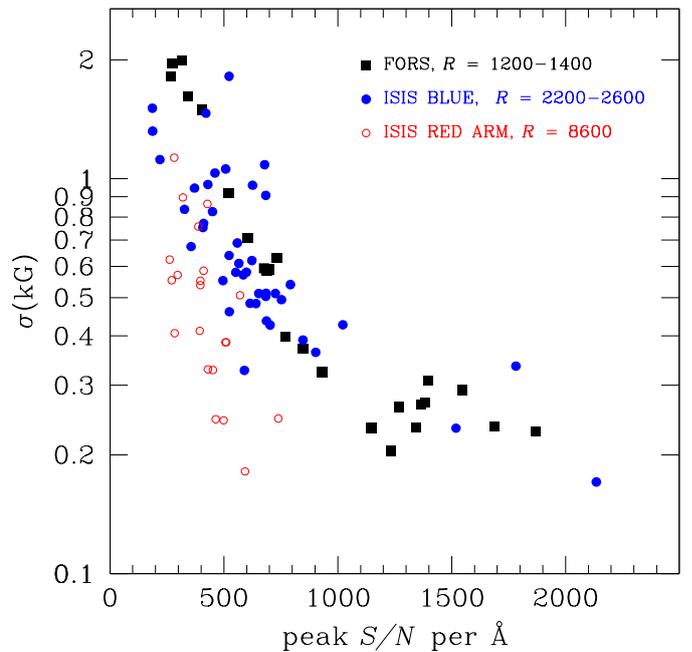}
\caption{\label{Fig_SIGMAvsSNR} Uncertainties versus \snr\ per \AA\ for the
  field measurements obtained with the FORS2 instrument (black solid squares),
  and with the ISIS instrument in the blue arm (blue filled
  circles) and in the red arm (red empty circles).}
\end{figure}

Another way to analyse the efficiencies of the field measurements
is by plotting error bars agains the \snr\ for the various instrument
modes and setting, as is done in Fig.~\ref{Fig_SIGMAvsSNR}.

%%%%%%%%%%%%%%%%%%%%%%%%%%%%%%%%%%%%%%%%%%%%%%%%%%%%%%%%%%%%%%%%%%%%%%%%%
%%%%%%%%%%%%%%%%%%%%%%%%%%%%%%%%%%%%%%%%%%%%%%%%%%%%%%%%%%%%%%%%%%%%%%%%%

It appears that the relationship between the uncertainty \sz\ of the
\bz\ measurements and the \snr\ per \AA\ as obtained with the blue arm
of ISIS is very similar to the relationship that describes the FORS2
data. Thus for blue arm field measurements, the standard errors are
about twice as large using ISIS as using FORS2 for the same shutter
time. In contrast, the red arm provides uncertainties that range from
roughly 25\,\% smaller than those from the blue arm to almost two times
smaller, even though the \snr\ per \AA\ is always smaller in the red
arm measurement than in the blue arm data (with the exceptions of the
observations obtained in the red arm with the low resolution grating
R158R). In spite of lower continuum \snr, the red arm provides
substantially more accurate measurements than the blue. This is partly
because of the larger Zeeman splitting at H$\alpha$ compared to the
higher Balmer lines (the splitting varies as $\lambda^2$), but mainly
because the higher resolving power allows us to exploit the large
slope and polarisation signal near the deep and sharp core of
H$\alpha$ to obtain a more tightly constrained slope in the
correlation diagram. In the future we will investigate whether the use
of grism 1200R (and $R \approx 2800$) with FORS2 would bring to a
higher sensitivity in our measurements.

\subsection{Efficiency of the field measurements as a function of spectral type}

\begin{figure}
\includegraphics[width=8.8cm,trim={0.4cm 5.3cm 1.1cm 3.5cm},clip]{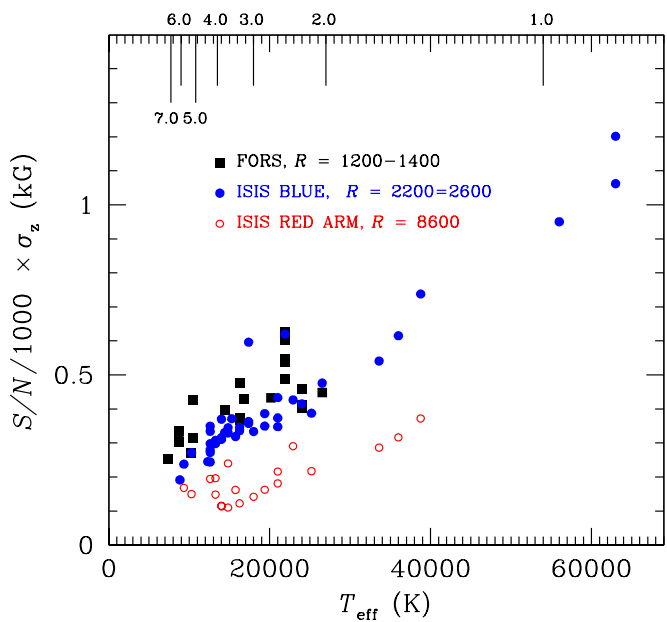}
  \caption{\label{Fig_SpT} Efficiency of the field measurements
    versus spectral classes for DA WDs.  Spectral class (shown at the
    top of the diagram) is linked to the
    stellar temperature through the relation class=50400/\teff. The
    meaning of the symbols is the same as in Figs.~\ref{Fig_SIGMAvsSNR}.}
\end{figure}

It was pointed out in Sect.~\ref{Sect_Red_vs_Blue} that the precision
of field measurements depend on the specific features of spectral
lines. These in turn depend on stellar temperature. In general we can
expect that in two DA WDs of similar magnitude but different temperature,
field measurements will be more precise in cooler than in hotter
stars, because cooler WDs have deeper and steeper Balmer lines, at
least down to $\teff \approx 7000$\,K. To
express this concept in a more quantitative way, we may consider the
product $\snr\,\times\,\sz$ as a function of the spectral type (or
stellar temperature) as a proxy for the efficiency of the field
measurement. In order to get figures close to unity, it is convenient
to divide the \snr\ by 1000, and expresss \sz\ in kG.
Figure~\ref{Fig_SpT} shows the product of the errorbar \sz\ and the
peak \snr\ per \AA\ versus the effective temperature of all DA WDs
observed in this survey. This plot tells us for instance that a field
measurements obtained in the blue arm of the ISIS instrument with a
\snr\ per \AA\ = 1000 in a DA1.0 WD has a 1\,kG uncertainty. A
measurement with same peak \snr\ on a DA4.0 WD would have an
uncertainty of about 0.3\,kG.  More generally, measurements with the
same peak \snr\ in DA1.0 stars have error bars 3-4 four times higher
than in the coolest WDs. From the practical point of view, one should
also remember that cooler WDs are fainter than hotter stars; therefore,
in terms of shutter time the comparison may be more favourable to
hotter stars.

%%%%%%%%%%%%%%%%%%%%%%%%%%%%%%%%%%%%%%%%%%%%%%%%%%%%%%%%%%%%%%%%%%%%%%%%%
%%%%%%%%%%%%%%%%%%%%%%%%%%%%%%%%%%%%%%%%%%%%%%%%%%%%%%%%%%%%%%%%%%%%%%%%%

\subsection{Precision versus exposure time}

\begin{figure}
\includegraphics[width=8.8cm,trim={0.9cm 5.8cm 1.1cm 3.0cm},clip]{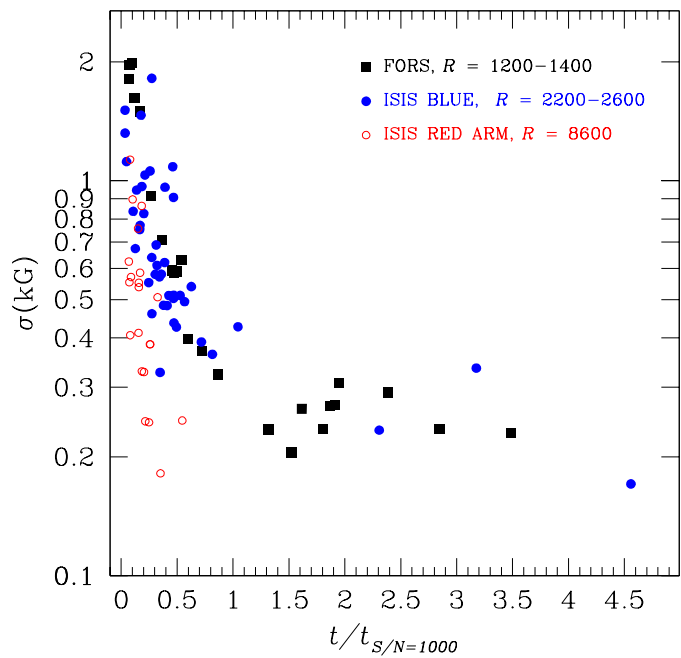}
\caption{\label{Fig_texp} Uncertainty reached as a function of the exposure
  time
      for the field measured obtained with the FORS2 instrument
      (black solid circles), and with the ISIS instrument in the red and in
      the blue arm (blue filled circles and red
      empty circles, respectively). Exposure time is normalised to the
      time needed to obtained $\snr=1000$ per \AA. The
      meaning of the symbols is the same as in Figs.~\ref{Fig_SIGMAvsSNR}
      and \ref{Fig_SpT}.}
\end{figure}

During observation planning it is clearly useful to have an idea of
the precision than may be achieved as a function of exposure
time. While it is easy to anticipate that $\sz \propto t^{-1/2}$, it
is less obvious how to express the resulting precision in real field
strength units. Figure~\ref{Fig_texp}, combined with the use of the
instrument Exposure Time Calculator, helps to associate the precision
that may be achived as a function of exposure time. Clearly, the final
numbers depend on the WD spectral class, but as a first approximation
one can see that in order to reach a precision of 3-400\,G, one has to
reach a \snr\ ratio of 1000, but that with half the exposure time
needed to reach a \snr=1000 per \AA\ one can still obtain an
uncertainty better than 0.5\,kG, while to go below 0.2\,kG one needs a
four times longer exposure time.

%%%%%%%%%%%%%%%%%%%%%%%%%%%%%%%%%%%%%%%%%%%%%%%%%%%%%%%%%%%%%%%%%%%%%%%%%
%%%%%%%%%%%%%%%%%%%%%%%%%%%%%%%%%%%%%%%%%%%%%%%%%%%%%%%%%%%%%%%%%%%%%%%%%

\section{Discussion}\label{Sect_Discussion_Science}
In our surveys we had two main science objectives.

\noindent
(1) Increasing substantially the number of WDs that have been 
searched for kG-level longitudinal magnetic fields, in order to
clarify the frequency of occurrence of the weakest detectable fields
in various types of WDs. A basic goal was to determine whether fields
occur at the $\bz \sim 1$\,kG level (frequently, sometimes, never),
and particularly to discover whether we have reached a lower limit of
the field distribution yet.

\noindent
(2) Improve the monitoring of the (currently small) number of MWDs
  of any type known to have kG-level fields, in order to develop our
  modelling techniques and obtain models of as many such stars as
  possible.

Our FORS2 spectropolarimetry has focussed on both aspects.
Two-thirds of the FORS2 polarised spectra (19 spectra out of 27)
have been obtained for detailed studies of previously identified
MWDs. The remaining eight FORS2 spectra are observations of relatively
bright WDs, mostly DA stars, not previously checked for weak
longitudinal magnetic field.

All ISIS spectropolarimetry, except for one star (40\,Eri\,B, see
Sect.~\ref{Sect_Eri}) were obtained only for survey purpose, and not
to monitor any individual star.

In this section we will discuss separately the survey component of
the FORS2 observing runs (Sect.~\ref{Sect_FORS_Survey}), the ISIS
survey (Sect.~\ref{Sect_ISIS_Survey}), and the monitoring of
individual stars (Sect.~\ref{Sect_Individual}). Finally, we will discuss
some preliminary statistical conclusions
(Sect.~\ref{Sect_Statistical}).

\subsection{FORS2 spectropolarimetric survey}\label{Sect_FORS_Survey}

Of the ten newly observed FORS2 stars, two observations failed because
of incorrect wave-plate settings, and one of the stars, WD\,1917$-$077,
was found to be effectively a DC star (it is classified as a DBQA5,
but all absorption features are broad and shallow, so no field
measurement was attempted).

Of the seven remaining survey stars observed with FORS2, six are DA
white dwarfs with $V$ magnitudes in range of 11.4 to 13.1, while one
is a fainter DZ star. All have been previously observed for magnetic
fields, but with higher field detection thresholds. All have been
studied using low-resolution optical spectra
\citep[e.g.][]{Giaetal11}.  In such data, fields of about $\bs > 1$\,MG
are readily visible, so normal parameter fitting identifies and
filters for such fields. Furthermore, the DA stars have been observed
with high-resolution optical spectroscopy,
\citep[mostly in the context of the SPY project, e.g.][]{Koeetal09}, 
providing upper limits on \bs\ of roughly 50\,kG.
 
Low S/N ratio spectra of the DZ star WD\,2138-332 taken with FEROS and
obtained from the ESO Archive show deep and sharp metal
lines. Modelling of these lines would probably allow one to set upper
limits on \bs\ of the order of 20\,kG. A field upper limit of this
level would be consistent with the marginal detection of non-zero \bz,
but does nothing to support the possible field detection.

For all DA WDs, the computed value of \sz\ is between 220 and
310\,G. Only one DA star appears to show a marginal detection:
WD\,1031$-$114, with $\bz = -870 \pm 290$\,G. This detection is
significant at just barely 3\,\sz. The single DZ star, WD\,2138$-$332,
also shows a marginally significant field detection, at the 3.3\,\sz\
level. These two stars are discussed individually below. 

Our uncertainties for the DA star \bz\ measurements are about a factor
of two smaller than those obtained for previous measurements of some
of the same stars. This improvement is due to the longer integration
times that we have used and the higher-resolution grism that we have
adopted (1200B instead of 600B). We have reached a level of
uncertainty that should reveal fields of order $\bz \approx 1-1.5$\,kG
if they are present in any of the stars of this small sample, and it
appears that we may have detected one such field.

\subsection{ISIS spectropolarimetric survey}\label{Sect_ISIS_Survey}

Using ISIS we made 54 \bz\ measurements on 38 WDs. Most of the stars
observed have $V$ magnitudes between 12.5 and 14. Using total shutter
times of somewhat more than one hour, the data have uncertainties
\sz\ below 300\,G for one observation of each of 11 stars; for the
remainder, \sz\ is mostly in the range between 300 and 700\,G. It
appears that the high measurement efficiency of ISIS in the red arm is
a sufficiently important factor to compensate for the relative mirror
area of about 4:1 in favour of FORS2; the two spectropolarimeters
provide data with roughly similar uncertainties for observations with
similar shutter times on stars with similar magnitude. A larger \bz\ survey of DA and related WDs with
ISIS would also be capable of revealing fields of about 1\,kG in a
statistically interesting sample of some tens of WDs.

Of the 36 WDs for which useful \bz\ measurements were obtained with
ISIS (two stars had featurless spectra), all had been previosuly studied at low resolution, and were known to be
non-magnetic at the 1\,MG or higher level. About 40\,\% have been
observed at high resolution in unpolarised light, and thus probably do
not have fields exceeding $\bs \sim 50$\,kG. Three stars observed with
ISIS (WD\,1031$-$114, WD\,1105$-$048, WD\,1327$-$083) have also been
observed  with FORS2.  WD\,1105$-$048 is strongly suspected of
hosting a variable field near the detection threshhold, and will be
discussed in Sect~\ref{Sect_Individual}. For the other two stars,
measurements with the two instruments provide concordant
non-detections.

As discussed above, the values of \sz\ for the ISIS \bz\ measurements
mostly lie in the range of 180 to 700\,G, and are thus sensitive to
fields larger than roughly 1 to 2\,kG. One new MWD was
discovered in this sample: WD\,2047+372 \citep{Lanetal16}. This WD has
a constant field modulus of $\bs \approx 60$\,kG, and a longitudinal
field which varies between $-12$ and $+15$\,kG with a period of
0.243\,d \citep{Lanetal17}. No other stars of the ISIS survey sample
have significant fields. This discovery increased the number of
confirmed MWDs having fields below 200\,kG from six to seven
\citep[see Table 6 of ][]{Lanetal17}. Even a single discovery is
capable of having a substantial effect on the statistics of weak-field
frequency.

\subsection{Comments on individual stars}\label{Sect_Individual}

\subsubsection{WD\,2359$-$434}
With FORS2, we have obtained four new \bz\ measurements of the known
MWD WD\,2359$-$434 \citep{Koeetal98,Aznetal04}. These data have already
been reported by us in a previous paper \citep{Lanetal17}, and were
analysed together with a large data set obtained with ESPaDOnS on the
CFHT. The full data set reveal a field whose \bz\ values always lie
between roughly 0 and 10\,kG, while the field modulus varies between
about 50 and 100\,kG. These data have been used in the modelling of
the strongly non-dipolar magnetic field geometry of this star
\citep{Lanetal17}.

\subsubsection{40 Eri\,B = WD\,0413$-$077}\label{Sect_Eri}
This star has been monitored both with ISIS and with the ESPaDOnS
instrument \citep{Lanetal15} to attempt to confirm the magnetic field
of $\bz \sim 4$\,kG reported by \citet{Fabetal03}. \citet{Lanetal15}
did not find any evidence of the presence of a magnetic field above
the level of about 250\,G.  In this survey we also report previously
unpublished measurements obtained using the red arm of the ISIS
instrument with the low resolution grating R158R. All are consistent with
no field detection.

\subsubsection{WD\,0446$-$789}\label{Sect_WDZero}
This star was originally discovered to be a MWD with FORS1 by
\citet{Aznetal04}. Two of their non-zero \bz\ measurements hinted at
possible variability. Our four new data points, while all having the
same sign as the discovery observations, further confirm at higher
precision the probability that the star is a magnetic variable. The
data are not numerous enough to strongly constrain the star's period,
but if we assume that the \bz\ variations are roughly sinusoidal in
form, then the longest period consistent with the \bz\ data is about
10\,d, the mean field $B_0$ is between $-5500$ and $-4000$\,G, and
the amplitude of the \bz\ curve is $\sim 1400 - 2400$\,G.

The line core of H$\alpha$ is not broadened enough to provide a very
strong constraint on \bs, apart from establishing $20-25$\,kG as an
upper limit \citep[see][Fig.~3]{Lanetal12}.

Combining all these constraints, it appears possible that that the
global field of WD\,0446$-$789 may be roughly dipolar, with a polar
field of the order of 30\,kG. If this model is basically correct, both
the angle between the line of sight and the stellar rotation axis,
and the angle between the rotation and magnetic axes, are small, with
one of order $20^\circ - 30^\circ$, and the other of the order of
$5^\circ$ to $10^\circ$.

\subsubsection{WD\,1031$-$114}

Our survey data from FORS2 and ISIS reported in this paper has
identified as a possible very weak field MWD the DA1.9 star
WD\,1031$-$114. One of our FORS2 measurement was $\bz = -870 \pm
290$\,G, barely significant at the $3\,\sigma$ level, while another ISIS
measurement (with uncertainty more than three times larger than the
measurement obtained with FORS2) did not show a significant field. This star is not yet
confirmed as magnetic, but is well worth following up with further
measurements achieving uncertainties of the order of 300\,G or
better. Assuming a roughly dipolar configuration for the magentic
field, our positive field detection also places a constraint on the mean
field modulus \bs, namely $\bs\ \ga 3$\,kG.  Zeeman
broadening or splitting of the Stokes $I$ profile of H$\alpha$ may be
detected only if $\bs \ga 25$\,kG, i.e., several times larger than the
lower limit of 3\,kG.

\subsubsection{WD\,1105$-$048} 

WD\,1105$-$048 has a possible magnetic field detected as $\bz = 3340
\pm 655$\,G, significant at the $5\,\sigma$ level, in one of two
previously published field measurements with FORS1; the other
measurement of about the same precision showed no significant field
\citep{Aznetal04,Lanetal12}. A barely significant detection of a --8\,kG field was also reported by \citet{Valetal06}. On the basis of this evidence we have
considered WD\,1105$-$048 to be probable but unconfirmed MWD. 
We have obtained three new FORS2 field measurements and one new ISIS
observation. Two of the FORS2 measurements have uncertainties of $\sim
270$\,G; the other two measurements have substantially larger
uncertainties. One of the two new high-precision FORS2 measurements
shows a field of $\bz = 2145 \pm 270$\,G, significant at the
$7.9\,\sigma$ level. The other three measurements are all consistent
with zero field. However, with now two high-significance field
detections of this star, {\it we now consider WD\,1105-048 to be a confirmed MWD.}

The field of the star appears to vary between a
\bz\ value of roughly zero and about $+3$\,kG in such a way that \bz\
is usually closer to the minimum than to the maximum value. This
suggests a field \bs\ of the order of 10\,kG or larger. The intensity
profile of H$\alpha$ does not show any obvious broadening of the
non-LTE line core \citep[Fig. 3]{Lanetal12}, suggesting that \bs\ is
less than about 20--25\,kG.

\subsubsection{WD\,2047$-$372}

Our single ISIS field measurement of WD\,2047$+$372 showed a non-zero
field of $\bz = 1005 \pm 410$\,G, a marginally significant value, but
confirmed the presence of a field through an almost resolved Zeeman
triplet in H$\alpha$. This star was then observed extensively with
ESPaDOnS, revealing a longitudinal field \bz\ varying sinusoidally
between $-12$ and $+15$\,kG, and a \bs\ value that is virtually
constant at 60\,kG. The resulting data set has been described and
modelled with a simple dipolar field structure \citep{Lanetal17}.

\subsubsection{WD\,2105$-$820}

This star was previously suggested as possibly magnetic by
\citet{Koeetal98}, who estimated that a field of $\bs \approx 42$\,kG
might be present. \citet{Lanetal12} obtaind five spectropolarimetric
observations from FORS2, which showed a longitudinal field ranging
in \bz\ values from $+8.2$ to $+11.4$\,G, fully confirming the presence of a
magnetic field. This range is about as large as would be
expected if \bz\ were actually constant at $+9500$\,G. We have three
new FORS2 measurements of this WD, all with \bz\ values that are positive
but smaller than the five previous measurements. One of the
measurements, of $\bz = 3545 \pm 685$\,G, is different enough from the
earlier data to establish with high probability that the field of
this star is in fact mildly variable. It is clear from the
precision of these measurements that a FORS2 time series with the
attainable precision would probably reveal a weakly variable field from
which the rotation period could be derived.  However, the data are
sufficiently widely spaced to provide no useful constraints on the
rotation period. 

A single measurement of $\bz = +5.3 \pm 0.3$\,kG was obtained by
\citet{Faretal18}. This measurement confirms the presence of a field
in this star, but because this measurement was made using only
H$\alpha$, in contrast to our measurements using only higher Balmer
lines, it is unlikely that this measurement is on exactly the same
scale as ours.

\citet{Lanetal12} modelled the star's magnetic field as a simple
dipole having either the magnetic axis nearly aligned with the
rotation axis, or a viewing geometry in which the star is viewed with
the rotation axis nearly aligned with the line of sight. The new data
do not discredit this model, but suggest that the inclination of the
rotation axis to the line of sight is slightly larger than previously
estimated, or that the angle between the rotation axis and the dipole
axis is slightly larger than estimated. However, the smaller of these
two angles is still probably less than about $20^\circ$.

\subsubsection{WD\,2138$-$332}
Our single observation of the DZ star WD\,2138-332 yields a field of
$\bz = 3300 \pm 990$\,G, significant at the $3.3 \sigma$ level. The
larger value of \sz\ compared to the DA stars is due to a combination
of fainter magnitude, shorter integration time, and the weakness of
the Mg\,{\sc i}, Ca\,{\sc i} and {\sc ii}, and Fe\,{\sc i} lines used
to measure the field.  This single detection is not significant enough
for us to consider that this star definitely has a weak magnetic
field, but it certainly calls for further observations, which have
been proposed. Note that this star is within the sphere of radius
20\,pc centred on the Sun.

\subsection{Preliminary statistical considerations}\label{Sect_Statistical}

The surveys of \citet{Aznetal04}, \citet{Joretal07} and \citet{Lanetal12}
looked in total at 30 stars for which no previous study had pointed to
a magnetic field of any strength, plus two stars (WD\,2105$-$820 and
WD\,2359$-$434) for which \citet{Koeetal98} had already suggested the
presence of weak fields, both of which were confirmed by
spectropolarimetry to be MWDs. Among the remaining 30 stars, two new
very weak field stars were found (WD\,0446$-$789 and
WD\,1105$-$048). This suggests a frequency among what is essentially a
magnitude limited survey of about $6 \pm 4$\% for fields in the range
of $1 \leq \bz\ \leq 10$\,kG.

We have surveyed 14 WDs with measurements obtaining $\sz \leq 300$\,G
and a total of 20 with $\sz \leq 500$\,G. Both these samples contain
the one suspected magnetic star (WD\,1031$-$114), for which the largest
measured \bz\ value is actually slightly below 1\,kG. Considering this
star to be really magnetic, with a \bz\ values in the 1 to 10\,kG
range, we have a frequency of occurence of fields in this range of
roughly $5 \pm 5$\%, in reasonable agreement with the older
result. These two frequency estimates both have rather large relative
uncetainties, but both suggest that very weak WD fields occur, in the
range of $1 \leq \bz\ \leq 10$\,kG, roughly as frequently as larger
fields, which appear to have a frequency of occurence of the order of
$3-4$\% per dex (per factor of ten in field strength) as found for
example by \citet{Kawetal07}.

To summarise, our data do not suggest that we have reached a minimum
field strength at $\bz \sim 1$\,kG below which WD fields become
extremely rare, nor have we reached a field strength at this level at
which WD fields become ubiquitous. Instead, the situation prevailing
for stronger field MWDs, that fields are uncommon but not rare, seems
to continue.

We shall postpone a more detailed discussion of the statistics of
detection of very weak fields to the next paper in this series, which
will add to the data here the results of a parallel survey using
ESPaDOnS at the CFHT, and examine these data in a larger context.

\section{Conclusions}\label{Sect_Conclusions}

We have reported several significant results, both of instrumental
nature and concerning the observation of very weak magnetic fields in
WDs.

\subsection{Instrumental advances}
We have shown that with the ESO's FORS2 instrument in
spectropolarimetric mode it is practical to measure the mean
longitudinal field with uncertainties of the order of 250 -- 300\,G in
DA and DB stars having $V \la 13$.

Similarly, we have established that the spectropolarimeter ISIS on the
WHT is an effective instrument for searching for and measuring very
weak fields in WDs. First we have shown that field measurements with ISIS
on known magnetic Ap/Bp stars are consistent with those obtained with other
spectropolarimeters. We have
further established that the use of a dichroic does not introduce
spurious effect when we measure the magnetic fields, and therefore the
red and blue arm of the instrument can be used simultaneously. This
significantly improves the power of ISIS for measuring very weak WD
fields. Like FORS2, ISIS is capable of achieving \sz\ values of around
300\,G on bright WDs.

We have compared the efficiency of ISIS and FORS2 in several ways,
finding that, in terms of photon detection efficiency, in the blue
spectral region FORS2 with grism 1200B and the blue arm of ISIS with
grating R600B have a comparable efficiency (per unit of telescope
collecting area). When we compare field measurements in the blue
spectral region obtained with similar \snr\ (or photon count) per \AA,
the two instruments yield essentially similar field
uncertainties. Thus their effectiveness as faint object,
high-precision spectropolarimeters is similar, apart from the
difference in efficiency and telescope aperture.

A very important asset of ISIS is that the red arm is equipped with a
grating that allows one to observe with $R \approx 8600$. It is found
that this is an extremely valuable tool for weak field measurements of
stars with a substantial H$\alpha$ line. For a given \snr\ level, the
field uncertainty derived from H$\alpha$ alone is of the order of two
times {\it smaller} than that obtained from using five Balmer lines in
the blue window. This high detection efficiency is possible because
the resolving power is high enough to almost resolve the sharp, deep
non-LTE core of H$\alpha$, which substantially reduces \sz\ compared
to a meaurement which does not resolve this core \citep[see the
discussion in][]{Lanetal15}. Using the red and blue arms of ISIS
together, we are able to achieve roughly the same uncertainties \sz\
in a given integration time as is achieved with FORS2 + grism 1200B on a
telescope with four times larger area. 

The conclusion is that both FORS2 and ISIS can quite practically carry
out a large survey searching for fields of $\bz \sim 1$\,kG in the
more than 150 DA stars of $V \la 14$\,mag.

\subsection{Astrophysical results}
In the course of the FORS2 survey presented in this paper we have
obtained several \bz\ measurements of each of four previously known
MWDs: WD\,0446$-$789, WD\,2051$-$208, WD\,2105$-$820, and
WD\,2359$-$434. The first three of these MWDs all have somewhat
variable \bz\ values, and we have used the observed range of these
values, together with very rough estimates of \bs, to propose very
approximate dipole-like field models \citep[for WD\,0446$-$789, see
  Sect.~\ref{Sect_WDZero}; for WD\,2051$-$208 and WD\,2105$-$820,
  see][]{Lanetal12,Lanetal17}. The FORS2 \bz\ data for WD\,2359$-$434
have already been published, and have been used in the construction of
a reasonably constrained field geometry model of the MWD, which quite
clearly departs from a simple dipole \citep{Lanetal17}. With the ISIS
and the ESPaDOnS instruments we have also discovered and monitored the
very weak-field star WD\,2047+372, for which we now have enough
\bz\ and \bs\ data to establish the 0.24\,d rotation period and to
construct a well-constrained dipole model \citep{Lanetal16,Lanetal17}.

Another significant result of our survey is to confirm the magnetic
nature of WD\,1105-048, which was previously in doubt because only one
really convincing field detection had previously been achieved. This
star is the MWD with the weakest confirmed magnetic field so far
found. The field cleaerly varies between $\bz \approx 0$ and 2100\,G,
with a still unknown period.

About 20 bright WDs of our sample have been surveyed with
uncertainties $\la 500$\,G, so that we are sensitive to fields of
1--2\,kG. For most of the stars of our target list, we have strong
upper limits on \bz\ of at most about 2\,kG. In the course of the
survey, one DA and one DZ star, WD\,1031--114 and WD\,2138--332, have
yielded one \bz\ measurement each that is significantly different from
zero at about the $3\,\sz$ level. These could be new examples of
extremely weak fields like that of WD\,1105-048, or they could be
spurious detections. We plan to re-observe both these stars as soon as
practical.

From this modest survey, it already seems clear that even at the 1\,kG
level, magnetic fields are still present in WDs, but not common. We
have not found a floor beneath which fields die away, nor a level at
which fields appear to be ubiquitous.

\begin{acknowledgements}
  This survey is based on observations collected at the William
  Herschel Telescope operated on the island of La Palma by the Isaac
  Newton Group, programmes P28 during semester 15A and P17 during
  semester 15B, and at the European Organisation for Astronomical
  Research in the Southern Hemisphere under ESO programmes 095.C-0855
  and 096.C-0159. A smaller number of observations were obtained at
  the ING WHT in the context of ING programme P11 during semester 13B,
  and at the ESO VLT as backup during programme ID 093.D-0680. This
  project has made extensive use of the Montreal White Dwarf Database
  (www.MontrealWhiteDwarfDatabase.org). Work by one of us (JDL) has
  been supported by the Natural Sciences and Engineering Research Council of
  Canada. We thanks the referee Dr.\ Adela Kawka for very useful comments which
  contributed to improve the manuscript.
\end{acknowledgements}
\bibliography{sbabib}

%-------------------------------------------------------------------

\newpage

%%%%%%%%%%%%%%%%%%%%%%%%%%%%%%%%%%%%%%%%%%%%%%%%%
\input{Table_Ap.tex}
%%%%%%%%%%%%%%%%%%%%%%%%%%%%%%%%%%%%%%%%%%%%%%%%

%%%%%%%%%%%%%%%%%%%%%%%%%%%%%%%%%%%%%%%%%%%%%%%%
\input{Table_Zero.tex}

%%%%%%%%%%%%%
\newpage
\input{Table_Log.tex}

\clearpage
\input{Fig_Atlas}

\end{document}

%% file: Table_Ap.tex
\begin{table*}
\caption{\label{Tab_Ap} Log of the observations of well known magnetic Ap/Bp stars obtained with ISIS of the WHT.}
%               12345678        90        12        3
\begin{tabular}{lcccrlrr@{$\pm$}lr@{$\pm$}lr@{$\pm$}l}
\hline\hline
  &                                      %1
  &                                      %2
DATE&                                      %3
UT&                                      %4
EXP&                                     %5
&                                        %6
s.w.&                                    %7
\multicolumn{6}{c}{\bz\ (G)}\\           %8,9,10,11,12,13
STAR &                                   %1
&                                        %2
yyyy-mm-dd&                                        %3
hh:mm&                                        %4
(s)&                                     %5
GRATING &                                  %6
(\arcsec)&                               %7
\multicolumn{2}{c}{H Balmer}&            %8,9
\multicolumn{2}{c}{metal}&               %10,11
\multicolumn{2}{c}{H + metal}       \\   %12,13
\hline
\multicolumn{13}{c}{}\\
%-----------------------------------------------------------------------------------------------------------------
\multicolumn{13}{c}{}\\
53\,Cam            &m&  2014-01-19 & 02:55 &  960  &  R600B     & 1.0 &$-4455$& 80&$-4910$& 55&$-4870$& 45\\ [2mm] 
%-----------------------------------------------------------------------------------------------------------------
\hline
\multicolumn{13}{c}{}\\
53\,Cam            &m&  2015-02-01 & 21:26 &   40  &  R600B     & 1.0 &$ 2035$&135&$ 1685$& 70&$ 1760$& 65\\ [1mm]
                   &m&  2015-02-01 & 21:34 &  120  &  R600B     & 1.0 &$ 1855$& 85&$ 1575$& 50&$ 1625$& 40\\ [1mm]
                   &m&  2015-02-01 & 21:46 &  120  &  R600B     & 1.0 &$ 1965$& 80&$ 1835$& 45&$ 1840$& 40\\ [1mm]
                   &m&  2015-02-01 & 21:55 &  120  &  R600B     & 1.0 &$ 2260$& 90&$ 2220$& 50&$ 2245$& 45\\ [1mm]     
                   &-&  2015-02-02 & 01:29 &  240  &  R1200R    & 1.0 &$ 1925$&145&$ 2790$& 50&$ 2705$& 45\\ [1mm]
                   &m&  2015-02-02 & 06:25 &  240  &  R600B     & 1.0 &$ 3070$& 70&$ 2920$& 45&$ 2950$& 40\\ [1mm]
                   &-&  2015-02-02 & 06:39 &  240  &  R1200R    & 1.0 &$ 2250$&120&$ 3255$& 45&$ 3165$& 45\\ [1mm]
                   &m&  2015-02-03 & 01:52 &  480  &  R600B     & 1.0 &$ 4030$& 70&$ 3705$& 55&$ 3780$& 45\\ [1mm]
                   &-&  2015-02-03 & 02:08 &  120  &  R158R     & 1.0 &$ 6030$&510&$ 4320$&525&$ 5040$&395\\ [1mm]
                   &m&  2015-02-04 & 04:27 &  100  &  R600B     & 1.0 &$ 4690$& 90&$ 4375$& 55&$ 4445$& 50\\ [1mm]
                   &-&  2015-02-04 & 04:35 &   12  &  R158R     & 1.0 &$ 3640$&635&$ 3740$&470&$ 4015$&375\\ [1mm]
                   &m&  2015-02-05 & 01:31 &   88  &  R600B     & 1.0 &$ 2620$& 80&$ 2665$& 55&$ 2650$& 45\\ [1mm]
                   &-&  2015-02-05 & 01:38 &   12  &  R158R     & 1.0 &$ 1495$&470&$ 1945$&495&$ 1750$&470\\ [2mm]
%-----------------------------------------------------------------------------------------------------------------
\hline
\multicolumn{13}{c}{}\\
HD\,157751         &d&  2015-08-28 & 20:49 &  960  &  R600B     & 1.0 &$ 4350$& 60&$ 3595$& 60&$ 3905$& 45\\ 
                   &d&             &       &       &  R1200R    & 1.0 &$ 3560$&120&$ 3465$& 40&$ 3440$& 40\\ [1mm]     
                   &m&  2015-08-28 & 21:12 &  960  &  R600B     & 1.0 &$ 4250$& 50&$ 3525$& 50&$ 3800$& 50\\ [1mm]
                   &-&  2015-08-28 & 21:34 &  960  &  R1200R    & 1.0 &$ 3085$&115&$ 3390$& 40&$ 3360$& 40\\ [1mm]
                   &d&  2015-08-29 & 20:32 &  480  &  R600B     & 1.0 &$ 4115$& 60&$ 3340$& 40&$ 3645$& 40\\
                   &d&             &       &       &  R1200R    & 1.0 &$ 2815$&110&$ 3295$& 45&$ 3230$& 40\\ [2mm]
HD\,215441         &d&  2015-08-29 & 05:56 &  960  &  R600B     & 1.0 &$17613$&110&$12535$&120&$14395$&105\\ 
                   &d&             &       &       &  R1200R    & 1.0 &$15426$&800&$ 5010$&165&$ 5685$&165\\ [1mm]
                   &d&  2015-08-29 & 05:56 &  960  &  R600B     & 1.0 &$17613$&110&$12535$&120&$14395$&105\\ 
                   &d&             &       &       &  R1200R    & 1.0 &$15426$&800&$ 5010$&165&$ 5685$&165\\ [1mm]
                   &d&  2015-08-31 & 03:06 &  480  &  R600B     & 1.0 &$15250$&125&$15040$&165&$15135$&115\\ 
                   &d&             &       &       &  R1200R    & 1.0 &$11760$&420&$ 5365$&220&$ 6140$&210\\ [1mm]
\ \ \ \ \ [$-$5\degr]&d&2015-08-31 & 03:15 &  480  &  R600B     & 1.0 &$14925$&115&$14740$&110&$14900$&110\\ 
                   &d&             &       &       &  R1200R    & 1.0 &$12505$&410&$ 5145$&225&$ 6500$&200\\ [1mm] 
\ \ \ \ \ [+5\degr]&d&  2015-08-31 & 03:24 &  480  &  R600B     & 1.0 &$14245$&100&$14245$&100&$14245$&100\\ 
                   &d&             &       &       &  R1200R    & 1.0 &$11000$&340&$ 5015$&210&$ 5870$&200\\ [2mm]
$\gamma$\,Equ      &d&  2015-08-29 & 00:13 &  120  &  R600B     & 1.0 &$ -980$& 35&$-1070$& 35&$-1035$& 25\\ 
                   &d&             &       &       &  R1200R    & 1.0 &$ -860$& 75&$-1110$& 25&$-1090$& 25\\ [1mm]
                   &m&  2015-08-29 & 00:27 &  120  &  R600B     & 1.0 &$ -895$& 35&$ -985$& 35&$ -955$& 25\\ [1mm]
                   &-&  2015-08-29 & 00:34 &   80  &  R1200R    & 1.0 &$ -840$& 85&$-1110$& 25&$-1085$& 25\\ [1mm]
                   &d&  2015-08-31 & 03:36 &  120  &  R600B     & 1.0 &$ -700$& 35&$-1045$& 30&$ -970$& 25\\ 
                   &d&             &       &       &  R1200R    & 1.0 &$ -640$& 75&$-1025$& 25&$ -980$& 25\\ [1mm]
\ \ \ \ \ [$-$5\degr]&d&2015-08-31 & 03:39 &  120  &  R600B     & 1.0 &$-1005$& 25&$-1005$& 25&$-1005$& 25\\ 
                   &d&             &       &       &  R1200R    & 1.0 &$ -615$& 75&$ -980$& 25&$-1020$& 25\\ [1mm]
\ \ \ \ \ [+5\degr]&d&  2015-08-31 & 03:43 &  120  &  R600B     & 1.0 &$ -645$& 40&$-1010$& 35&$ -925$& 25\\ 
                   &d&             &       &       &  R1200R    & 1.0 &$ -670$& 70&$ -950$& 25&$ -910$& 25\\ 
%-----------------------------------------------------------------------------------------------------------------
\hline
\end{tabular}
\tablefoot{The symbol in Col.~2 refers to the instrument setting: `m' means that a mirror was
  insered in the optical path and that therefore only the blue arm was fed; `-' means that
  the mirror was removed and the beam would feed the red arm only; `d' means that a dichroic
  was inserted, and that the red arm and the blue arm would be fed simulataneously.\\
The horizontal lines define the observations obtained during different observing runs.}
\end{table*}

%% file: Table_Zero.tex
\begin{table*}
  \caption{\label{Tab_Zero} Observations obtained with the retarder waveplate at position angles 0\degr\ and 90\degr.
  Using this setting, the observed polarisation (hence the magnetic field) should be consistent with zero.}
%               12345678       90        12        3
\begin{tabular}{lccrlcr@{$\pm$}lr@{$\pm$}lr@{$\pm$}l}
\hline\hline
                 &             &       &  EXP  &        &s.w.   &\multicolumn{6}{c}{\bz\ (G)} \\
STAR             & DATE        &  UT   & (s)   & GRATING   &('')&\multicolumn{2}{c}{H Balmer}&\multicolumn{2}{c}{metal}&\multicolumn{2}{c}{H + metal}    \\
\hline
                 &             &       &       &             &\multicolumn{2}{c}{}\\
HD\,65339        &  2014-01-19 & 03:17 &  960  &  R600B      & 1.0 &$ 130$&  20&$   80$&  10&$  90$& 10\\ [2mm]
\hline
                 &             &       &       &             &\multicolumn{2}{c}{} \\
HD\,65339        &  2015-02-01 & 21:26 &   40  &  R600B      & 1.0 &$ -90$& 130&$  195$&  60&$ 130$& 55\\ 
HD\,65339        &  2015-02-01 & 21:35 &  120  &  R600B      & 1.0 &$-260$&  80&$ -250$&  40&$-250$& 35\\ 
HD\,65339        &  2015-02-01 & 21:48 &  120  &  R600B      & 1.0 &$ 220$&  70&$  125$&  35&$ 140$& 30\\ 
HD\,65339        &  2015-02-01 & 21:56 &  120  &  R600B      & 1.0 &$ -25$&  70&$ -160$&  35&$-120$& 30\\ [1mm]
HD\,65339        &  2015-02-02 & 01:29 &  240  &  R1200R     & 1.0 &$   5$&  65&$  -90$&  20&$ -80$& 20\\ 
HD\,65339        &  2015-02-02 & 06:25 &  240  &  R600B      & 1.0 &$ 465$&  40&$  140$&  20&$ 185$& 15\\ 
HD\,65339        &  2015-02-02 & 06:39 &  240  &  R1200R     & 1.0 &$ -85$&  65&$   35$&  20&$ -25$& 20\\ [1mm]
HD\,65339        &  2015-02-03 & 01:52 &  480  &  R600B      & 1.0 &$   0$&  40&$  -25$&  20&$ -20$& 20\\ 
HD\,65339        &  2015-02-03 & 02:08 &  120  &  R158R      & 1.0 &$ 270$& 635&$ -315$& 460&$  -5$&385\\ [1mm] 
HD\,65339        &  2015-02-04 & 04:27 &  100  &  R600B      & 1.0 &$ 135$&  50&$   25$&  25&$  40$& 20\\
HD\,65339        &  2015-02-04 & 04:35 &   12  &  R158R      & 1.0 &$-455$& 600&$-1230$& 480&$-1340$&365\\ 
HD\,65339        &  2015-02-05 & 01:31 &   88  &  R600B      & 1.0 &$-115$&  45&$ -190$&  20&$-175$& 20\\ 
HD\,65339        &  2015-02-05 & 01:38 &   12  &  R158R      & 1.0 &$-455$& 495&$  495$& 445&$ 400$&330\\ [2mm]
\hline
                 &             &       &       &             &\multicolumn{2}{c}{}\\
$\gamma$\,Equ    &  2015-08-29 & 00:20 &  120  &  R600B      & 1.0 &$-125$&  40&$ -215$&  20&$-200$& 20\\ 
$\gamma$\,Equ    &             &       &       &  R1200R     & 1.0 &$  10$&  45&$   85$&  15&$  80$& 15\\ 
                 &             &       &       &             &\multicolumn{2}{c}{}\\
\hline
\end{tabular}
\end{table*}

%% file: Table_Log.tex
\begin{tiny}
%                 123456789012            3
\begin{longtable}{llrrlccrrllr@{\,$\pm$\,}l}
\caption{Observing log} \label{Tab_Log} \\
\hline\hline
%   1       2       3         4          5            6           7          8       9                     
                            & %1
                            & %2
\multicolumn{1}{c}{$d$}     & %3
                            & %4  
\multicolumn{1}{c}{Spec.}   & %5
                            & %6
                            & %7
\multicolumn{1}{c}{EXP}     & %8
\multicolumn{1}{c}{\it S/N} & %9 
                            & %0
\multicolumn{1}{l}{s.w.}    & %11
\multicolumn{2}{c}{\bz}  \\ %12,13
\multicolumn{2}{c}{STAR}        & %1,2
\multicolumn{1}{c}{(pc)}        & %3
\multicolumn{1}{c}{$V$}         & %4
\multicolumn{1}{c}{type}        & %5
\multicolumn{1}{c}{DATE}        & %6
\multicolumn{1}{c}{UT}          & %7
\multicolumn{1}{c}{(s)}         & %8 
\multicolumn{1}{c}{\AA$^{-1}$}   & %9
\multicolumn{1}{l}{SETTING}     & %10
\multicolumn{1}{l}{(\arcsec)}        & %11
\multicolumn{2}{c}{(G)}        \\ %12,13
\hline
\endfirsthead
\hline\hline
                            & %1
                            & %2
\multicolumn{1}{c}{$d$}     & %3
                            & %4  
\multicolumn{1}{c}{Spec.}   & %5
                            & %6
                            & %7
\multicolumn{1}{c}{EXP}     & %8
\multicolumn{1}{c}{\it S/N} & %9 
                            & %0
\multicolumn{1}{l}{s.w.}    & %11
\multicolumn{2}{c}{\bz}  \\ %12,13
\multicolumn{2}{c}{STAR}        & %1,2
\multicolumn{1}{c}{(pc)}        & %3
\multicolumn{1}{c}{$V$}         & %4
\multicolumn{1}{c}{type}        & %5
\multicolumn{1}{c}{DATE}        & %6
\multicolumn{1}{c}{UT}          & %7
\multicolumn{1}{c}{(s)}         & %8 
\multicolumn{1}{c}{\AA$^{-1}$}   & %9
\multicolumn{1}{l}{SETTING}     & %10
\multicolumn{1}{l}{(\arcsec)}        & %11
\multicolumn{2}{c}{(G)}        \\ %12,13
\hline
\endhead
WD\,2359$-$434 & LTT\,9857  &$  8.3            $& 13.05 & DAP5.8 & 2015-07-31 & 07:10 & 4296 &1270 & 1200B & 1.2 &$  2750 $&  265 \\
               &            &$                 $&       &        & 2015-09-08 & 06:03 & 4296 &1340 & 1200B & 1.2 &$  3030 $&  235 \\
               &            &$                 $&       &        & 2016-06-05 & 08:59 & 2496 & 770 & 1200B & 1.0 &$  2895 $&  395 \\
               &            &$                 $&       &        & 2016-06-28 & 09:34 & 2496 & 930 & 1200B & 1.0 &$  2360 $&  325 \\ [1mm]
WD\,0046$+$051 & Wolf\,28   &$  4.3            $& 12.37 & DZ7.5  & 2015-08-29 & 03:54 & 4521 & 625 & R600B & 1.0 &$ -1685 $&  965  \\ 
               &            &$                 $&       &        &            &       &      & 555 & R1200R&&\multicolumn{2}{c}{n.m.}\\ [1mm]% +30sec in red
WD\,0050$-$332 & SB\,360    &$ 59.8            $& 13.36 & DA1.4  & 2015-08-31 & 04:26 & 3840 & 420 & R600B & 1.0 &$  -545 $& 1465  \\
               &            &$                 $&       &        &            &       &      & 280 & R1200R&     &$  -670 $& 1130  \\  
               &            &$                 $&       &        &            &       &      & 420 &R600B+R1200R&&$  -620 $&  940  \\ [1mm]  
WD\,0135$-$052 & NLTT\,5460 &$ 12.6            $& 12.83 & DA6.9  & 2015-07-31 & 08:39 & 4296 &1235 & 1200B & 1.2 &$    10 $&  205  \\ [1mm] 
WD\,0134$+$833 & GD\,419    &$ 27.0            $& 13.1: & DA2.6  & 2015-02-03 & 23:24 & 5600 & 560 & R600B & 1.0 &$   255 $&  690  \\ [1mm] 
WD\,0148$+$467 & GJ\,3121   &$ 16.6            $& 12.46 & DA3.6  & 2015-02-04 & 21:21 & 5600 & 725 & R600B & 1.0 &$  -220 $&  510  \\
               &            &$                 $&       &        & 2015-08-30 & 04:49 & 2880 & 640 & R600B & 1.0 &$  -810 $&  495  \\ 
               &            &$                 $&       &        &            &       &      & 465 & R1200R&     &$   210 $&  245  \\ 
               &            &$                 $&       &        &            &       &      & 665 &R600B+R1200R&&$   -40 $&  235  \\ 
               &            &$                 $&       &        & 2015-08-31 & 06:12 &  960 & 410 & R600B & 1.0 &$ -1230 $&  865  \\ 
               &            &$                 $&       &        &            &       &      & 285 & R1200R&     &$   690 $&  405  \\ 
               &            &$                 $&       &        &            &       &      & 410 &R600B+R1200R&&$   -80 $&  435  \\ [1mm]
WD\,0205$+$250 & LTT\,10723 &$ 39.1            $& 13.22 & DA2.4  & 2015-08-30 & 03:43 & 3840 & 600 & R600B & 1.0 &$  1100 $&  580  \\
               &            &$                 $&       &        &            &       &      & 400 & R1200R&     &$   890 $&  540  \\ 
               &            &$                 $&       &        &            &       &      & 600 &R600B+R1200R&&$   980 $&  380  \\  [1mm] 
WD\,0310$-$688 & GJ\,127.1  &$ 10.4            $& 11.39 & DA3.0  & 2015-09-02 & 08:50 &  800 &1865 & 1200B & 1.2 &$  -195 $&  230  \\ [1mm] 
WD\,0413$-$077 & 40\,Eri \,B&$ 199.5           $&9.50   & DA2.9  & 2015-02-02 & 21:27 & 4000 &1780 & R600B & 1.0 &$    50 $&  335  \\
               &            &$                 $&       &        & 2015-02-02 & 22:36 & 3200 &1590 & R158R & 1.0 &$  -715 $&  525  \\
               &            &$                 $&       &        & 2015-02-03 & 21:00 & 1600 &1525 & R158R & 1.0 &$ -1495 $&  755  \\
               &            &$                 $&       &        & 2015-02-03 & 21:24 &  800 &1070 & R158R & 1.0 &$  -760 $&  730  \\
               &            &$                 $&       &        & 2015-02-03 & 22:04 & 2800 &2135 & R600B & 1.0 &$    95 $&  170  \\
               &            &$                 $&       &        & 2015-02-04 & 22:45 & 3200 &1520 & R600B & 1.0 &$   640 $&  235  \\
               &            &$                 $&       &        & 2015-02-04 & 23:34 & 2000 &1250 & R158R & 1.0 &$  -840 $&  530  \\ [1mm] 
WD\,0426$+$588 & G\,175-34B &$  5.5            $& 12.43 & DC7.1& 2015-08-29 & 05:25 & 1200 & 375 & R600B & 1.0 &\multicolumn{2}{c}{$< 10^6$}\\  
               &            &$                 $&       &        &            &       &      & 350 & R1200R& 1.0 &\multicolumn{2}{c}{$< 10^6$}\\  [1mm]
WD\,0446$-$789 & BPM\,3523  &$ 43.9            $&13.47  & DA2.1  &2016-07-21  & 07:55 & 2496 & 730 &1200B  & 1.0 &$ -4400 $&  630 \\
               &            &$                 $&       &        & 2016-07-22 & 09:50 & 2496 & 690 &1200B  & 1.0 &$ -3310 $&  585 \\
               &            &$                 $&       &        & 2016-07-24 & 08:36 & 2496 & 700 &1200B  & 1.0 &$ -5515 $&  590 \\
               &            &$                 $&       &        & 2016-07-26 & 09:39 & 2496 & 675 &1200B  & 1.0 &$ -6350 $&  595 \\ [1mm] 
WD\,0501$+$527 & G\,191-B2B &$ 52.9            $& 11.69 & DA0.8  & 2015-08-30 & 05:40 & 2560 & 940 & R600B & 1.0 &$  1280 $& 1130 \\ 
               &            &$                 $&       &        &            &       &      & 565 & R1200R&     &\multicolumn{2}{c}{$< 10^6$}\\ 
               &            &$                 $&       &        & 2015-09-01 & 04:27 & 3840 &1245 & R600B & 1.2 &$  -280 $&  965 \\  
               &            &$                 $&       &        &            &       &      & 750 & R1200R&     &\multicolumn{2}{c}{$< 10^6$}\\ [1mm]
WD\,0549$+$158 & GD\.71     &$ 52.0            $& 13.03 & DA1.5  & 2015-09-01 & 05:31 & 2880 & 510 & R600B & 1.0 &$  -795 $& 1060 \\
               &            &$                 $&       &        &            &       &      & 320 & R1200R&     &$  -715 $&  895 \\ 
               &            &$                 $&       &        &            &       &      & 510 &R600B+R1200R&&$  -620 $&  660 \\ [1mm]
WD\,0644$+$375 & LFT\,487   &$ 17.1            $& 12.08 & DA2.3  & 2015-02-03 & 00:14 & 5600 & 685 & R600B & 1.0 &$   745 $&  905 \\ [1mm] 
WD\,0859$-$039 &RE\,J0902-04&$ 37.9            $& 12.4  & DA2.1  & 2015-02-04 & 03:12 & 5600 & 430 & R600B & 1.0 &$    -5 $&  965 \\ [1mm] 
WD\,0943$+$441 & SA\,29-130 &$ 32.0            $& 13.29 & DA3.8  & 2015-02-03 & 04:46 & 4800 & 410 & R600B & 1.0 &$   110 $&  750 \\ [1mm] 
WD\,1031$-$114 & LTT\,3870  &$ 35.8            $& 13.01 & DA1.9  & 2015-02-02 & 00:31 & 4800 & 460 & R600B & 1.0 &$ -1435 $& 1035 \\
               &            &$                 $&       &        & 2015-05-03 & 02:07 & 4296 &1545 & 1200B & 1.2 &$  -870 $&  290 \\ [1mm] 
WD\,1105$-$048 & NLTT\,26379&$ 24.8            $& 13.05 & DA3.5  & 2015-02-03 & 03:11 & 5600 & 525 & R600B & 1.0 &$   640 $&  640 \\  
               &            &$                 $&       &        & 2015-05-03 & 00:43 & 3712 &1380 & 1200B & 1.2 &$  -235 $&  270 \\
               &            &$                 $&       &        & 2015-05-20 & 01:51 & 3712 &1365 & 1200B & 1.2 &$  2145 $&  270 \\
               &            &$                 $&       &        & 2016-07-02 & 01:24 & 1248 & 520 & 1200B & 1.0 &$  -395 $&  875 \\ [1mm]
WD\,1116$+$026 & GD\,133    &$ 38.2            $&14.57  & DAZ4.0 & 2014-01-19 & 04:14 & 4800 & 185 & R600B & 2.0 &$   865 $& 1320 \\
               &            &$                 $&       &        & 2014-01-20 & 02:41 & 4800 & 185 & R600B & 2.0 &$  -200 $& 1510 \\ [1mm] 
WD\,1134$+$300 & GC\,140    &$ 15.7            $& 12.47 & DA2.2  & 2015-02-02 & 02:21 & 4800 & 790 & R600B & 1.0 &$  1110 $&  540 \\
               &            &$                 $&       &        & 2015-02-02 & 03:50 & 4800 & 570 & R1200R& 1.0 &$  -475 $&  510 \\ [1mm] 
WD\,1202$-$232 &EC\,12028-2316&$10.4           $& 12.80 & DAZ5.7 & 2015-02-05 & 02:34 & 5600 & 590 & R600B & 1.0 &$   425 $&  325 \\ [1mm] 
WD\,1213$+$528 & GJ\,459.1  &$ 28.7            $& 13.23 & DA3.3  & 2015-02-04 & 05:33 & 5600 & 450 & R600B & 1.0 &$   955 $&  825 \\ [1mm] 
WD\,1327$-$083 & G\,14-58   &$ 16.1            $& 12.33 & DA3.5  & 2015-02-05 & 05:50 & 4800 & 905 & R600B & 1.0 &$  -900 $&  365 \\
               &            &$                 $&       &        & 2015-05-27 & 01:06 & 3200 &1685 & 1200B & 1.2 &$   275 $&  235 \\ [1mm]
WD\,1337$+$705 & LAWD 52    &$ 26.5            $&12.77  & DAZ2.4 & 2015-02-02 & 05:25 & 5600 & 755 & R600B & 1.0 &$   710 $&  495 \\ [1mm]
WD\,1422$+$095 & GD\,165    &$ 33.4            $& 14.32 & DA4.1  & 2014-01-20 & 04:25 & 4800 & 220 & R600B & 2.0 &$  -275 $& 1115 \\ [1mm] 
WD\,1531$-$022 & BPM\,77964 &$ 41.5            $& 14.03 & DA2.6  & 2015-08-29 & 21:22 & 3840 & 370 & R600B & 1.0 &$   160 $&  945 \\
               &            &$                 $&       &        &            &       &      & 260 & R1200R&     &$   370 $&  625 \\  [1mm]  
               &            &$                 $&       &        &            &       &      & 370 &R600B+R1200R&&$  -115 $&  570 \\  [1mm]  
WD\,1632$+$177 &PG\,1632+177&$ 25.6            $&13.08& DAZ4.9   & 2015-04-14 & 08:13 & 4296 &1150 & 1200B & 1.2 &$   215 $&  235 \\ [1mm] 
WD\,1645$+$325 & GD\,358    &$ 43.1            $& 13.65 & DB2  & 2015-08-30 & 21:23 & 3360 & 525 &R600B  & 1.0   &$ 780 $&  385 \\
               &            &$                 $&       &        &            &       &      & 340 &R1200R &     &$  -435 $&  605 \\ 
               &            &$                 $&       &        &            &       &      & 525 &R600B+R1200R&&$   181 $&  135 \\ 
WD\,1647$+$591 & GJ\,1206   &$ 10.9            $& 12.24 & DA4.0  & 2014-01-19 & 05:51 & 5600 & 495 &R600B  & 2.0 &$  -545 $&  550 \\
               &            &$                 $&       &        & 2015-02-03 & 06:15 & 4400 & 655 &R600B  & 1.0 &$   675 $&  510 \\
               &            &$                 $&       &        & 2015-02-05 & 04:10 & 4800 & 685 &R600B  & 1.0 &$   350 $&  510 \\ [1mm] 
WD\,1655$+$215 & G\,169-34  &$ 21.0            $& 14.13 & DA5.4& 2015-08-31 & 20:58 & 4320 & 355 &R600B  & 1.0 &$   440 $&  670 \\ 
               &            &$                 $&       &        &            &       &      & 295 &R1200R &     &$    25 $&  570 \\ 
               &            &$                 $&       &        &            &       &      & 355 &R600B+R1200R&&$   170 $&  390 \\ [1mm] 
WD\,1713$+$695 & LTT\,18455 &$ 26.3            $& 13.2  & DA3.2& 2015-08-30 & 22:29 & 3840 & 550 &R600B  & 1.0 &$   550 $&  580 \\  
               &            &$                 $&       &        &            &       &      & 395 &R1200R &     &$   -80 $&  410 \\ 
               &            &$                 $&       &        &            &       &      & 550 &R600B+R1200R&&$    70 $&  285 \\ [1mm]
WD\,1840$-$111 & LTT\,7421  &$ 24.1            $&14.18& DA4.9& 2015-08-31 & 22:21 & 4320 & 325 &R600B  & 1.2 &$  1965 $&  835 \\ 
               &            &$                 $&       &        &            &       &      & 270 &R1200R &     &$    60 $&  555 \\ 
               &            &$                 $&       &        &            &       &      & 325 &R600B+R1200R&&$   890 $&  540 \\ [1mm] 
%WD\,1900$+$705 &Grw+70\,8247&$ 13.2^{+0.7}_{-0.7}$& 13.25& DAP4.3& 2015-08-28 & 22:10 & 1920 & 375 &R600B & 1.0  &\multicolumn{2}{c}{$< 10^6$}\\   
%               &            &$                $&       &        &            &       &      & 300 &R1200R&      &\multicolumn{2}{c}{$< 10^6$}\\   
%               &            &$                $&       &        & 2015-08-31 & 23:30 & 2400 & 365 &R600B & 1.0  &\multicolumn{2}{c}{$< 10^6$}\\   
%               &            &$                $&       &        &            &       &      & 305 &R1200R&      &\multicolumn{2}{c}{$< 10^6$}\\  [1mm] 
WD\,1917$-$077 & LTT\,7658  &$ 10.5            $& 12.29 & DBQA5& 2015-05-31 & 06:16 & 2592 &1420 & 1200B & 1.2 &\multicolumn{2}{c}{$< 10^6$}\\ [1mm] 
WD\,1935$+$276 & G\,185-32  &$ 18.3            $& 12.98 & DA4.0& 2015-08-28 & 23:18 & 4800 & 685 &R600B & 1.0  &$  -705  $& 435  \\
               &            &$                 $&       &        &            &       &      & 505 &R1200R&      &$    10  $& 385  \\        
               &            &$                 $&       &        &            &       &      & 685 &R600B+R1200R&&$  -265  $& 270  \\  [1mm] 
WD\,2028$+$390 & GD\,391    &$ 40.1            $&13.38& DA2.0& 2015-08-29 & 22:45 & 4800 & 625 &R600B & 1.0  &$   610  $& 620  \\
               &            &$                 $&       &        &            &       &      & 395 &R1200R&      &$   355  $& 550  \\ 
               &            &$                 $&       &        &            &       &      & 625 &R600B+R1200R&&$   465  $& 400  \\ [1mm] 
WD\,2032$+$248 & HD\,340611 &$ 14.8            $& 11.52 & DA2.4& 2015-08-29 & 01:38 & 5040 &1020 &R600B & 1.0  &$  -545  $& 425  \\
               &            &$                 $&       &        &            &       &      & 740 &R1200R&      &$   260  $& 245  \\  
               &            &$                 $&       &        &            &       &      &1020 &R600B+R1200R&&$    60  $& 220  \\ [1mm] 
WD\,2039$-$202 & LTT\,8189  &$ 21.7            $& 12.33 & DA2.5  & 2015-06-02 & 09:39 & 2528 &1395 &1200B & 1.2  &$   115  $& 310  \\ [1mm]
WD\,2047$+$372 & LTT\,16093 &$ 17.6            $& 12.93 & DA3.4  & 2015-09-01 & 01:58 & 3360 & 565 &R600B & 1.2  &$   605  $& 610  \\  
               &            &$                 $&       &        &            &       &      & 410 &R1200R&      &$  1355  $& 585  \\ 
               &            &$                 $&       &        &            &       &      & 565 &R600B+R1200R&&$  1005  $& 410  \\ [1mm]
WD\,2051$-$208 & HK\,22880  &$ 31.2            $&15.06  & DAH2.3 & 2016-04-27 & 09:24 & 2496 & 315 & 1200B & 1.0 &$ -2990  $& 1990 \\
               &            &$                 $&       &        & 2016-05-23 & 09:08 & 2496 & 405 & 1200B & 1.0 &$ 15355  $& 1490 \\
               &            &$                 $&       &        & 2016-06-05 & 06:46 & 2496 & 340 & 1200B & 1.0 &$ 18215  $& 1615 \\
               &            &$                 $&       &        & 2016-06-09 & 09:39 & 2496 & 275 & 1200B & 1.0 &$-10065  $& 1965 \\
               &            &$                 $&       &        & 2016-06-11 & 04:40 & 2496 & 270 & 1200B & 1.0 &$  3560  $& 1810 \\ [1mm]
WD\,2111$+$498 & GD\,394    &$ 50.4            $&13.09  & DA1.3  & 2015-08-31 & 02:18 & 4320 & 680 & R600B & 1.0 &$ -1135  $& 1085 \\  
               &            &$                 $&       &        &            &       &      & 430 & R1200R&     &$  1400  $&  865 \\ 
               &            &$                 $&       &        &            &       &      & 680 &R600B+R1200R&&$    15  $&  685 \\ [1mm]
WD\,2105$-$820 & LTT\,8381  &$ 16.2            $& 13.50 & DAZ4.8 & 2014-06-04 & 05:48 & 3200 & 355 & 1200B & 1.0 &$  3545  $&  685 \\
               &            &$                 $&       &        & 2015-08-23 & 02:55 & 4296 & 850 & 1200B & 1.2 &$  7090  $&  370 \\
               &            &$                 $&       &        & 2016-05-24 & 04:24 & 2496 & 605 & 1200B & 1.0 &$  8065  $&  705 \\ [1mm] 
WD\,2117$+$539 & G\,231-40  &$ 17.3            $&12.33  & DA3.4  & 2015-08-31 & 01:04 & 3840 & 845 & R600B & 1.0 &$   265  $&  390 \\
               &            &$                 $&       &        &            &       &      & 595 & R1200R&     &$   -50  $&  180 \\  
               &            &$                 $&       &        &            &       &      & 845 &R600B+R1200R&&$    20  $&  185 \\   [1mm] 
WD\,2126$+$734 & LTT\,18524 &$ 22.2            $& 12.82 & DA3.1  & 2015-08-30 & 00:12 & 4800 & 685 & R600B & 1.0 &$   850  $&  505 \\
               &            &$                 $&       &        &            &       &      & 500 & R1200R&     &$    15  $&  245 \\
               &            &$                 $&       &        &            &       &      & 685&R600B+R1200R& &$   190  $&  245 \\  [2mm]
WD\,2136$+$828 & LFT\,1649  &$ 26.4            $& 13.02 & DA2.8& 2015-08-30 & 23:47 & 4160 & 585 & R600B & 1.0 &$   795  $&  570 \\
               &            &$                 $&       &        &            &       &      & 430 & R1200R&     &$   145  $&  330 \\ 
               &            &$                 $&       &        &            &       &      & 585 &R600B+R1200R&&$   285  $&  265 \\   [1mm] 
WD\,2138$-$332 & NLTT\,51844&$ 16.1            $& 14.47 & DZ7 C  & 2014-06-04&  06:51 & 2400 & 230 &1200B  & 1.0 &$ 3300   $&  990 \\   [2mm]
WD\,2140$+$207 & LHS\,3703  &$ 11.0            $& 13.24 & DQ6.1  & 2015-09-01 & 01:08 & 1680 & 300 & R600B & 1.2 &\multicolumn{2}{c}{$< 10^6$}\\
               &            &$                 $&       &        &            &       &      & 260 & R1200R&     &\multicolumn{2}{c}{$< 10^6$}\\  [1mm]  
WD\,2309$+$105 & BPM\,97895 &$ 76.4            $& 13.89 & DA0.9  & 2015-08-30 & 01:23 & 2880 & 525 & R600B & 1.0 &$  1640  $& 1815  \\
               &            &$                 $&       &        &            &       &      & 320 & R1200R&     &\multicolumn{2}{c}{$< 10^6$}\\ [1mm]
WD\,2341$+$322 & LTT\,16991 &$ 18.6            $& 12.92 & DA3.8  & 2015-08-30 & 02:27 & 3840 & 615 & R600B & 1.0 &$  1110  $&  485 \\
               &            &$                 $&       &        &            &       &      & 450 & R1200R&     &$   290  $&  330 \\ 
               &            &$                 $&       &        &            &       &      & 615 &R600B+R1200R&&$   565  $&  280 \\ 
               &            &$                 $&       &        & 2015-09-01 & 03:10 & 4320 & 705 & R600B & 1.2 &$  1040  $&  425 \\ 
               &            &$                 $&       &        &            &       &      & 510 & R1200R&     &$   415  $&  385 \\ 
               &            &$                 $&       &        &            &       &      & 705 &R600B+R1200R&&$   605  $&  245 \\ [5mm]
\hline
\end{longtable}
\end{tiny}

\begin{tiny}
%                 123456789012            3
\begin{longtable}{llrrlccrrllr@{\,$\pm$\,}l}
\caption{Additional observations} \label{Tab_LogOne} \\
\hline\hline
%   1       2       3         4          5            6           7          8       9                     
                            & %1
                            & %2
\multicolumn{1}{c}{$d$}     & %3
                            & %4  
\multicolumn{1}{c}{Spec.}   & %5
                            & %6
                            & %7
\multicolumn{1}{c}{EXP}     & %8
\multicolumn{1}{c}{\it S/N} & %9 
                            & %0
\multicolumn{1}{l}{s.w.}    & %11
\multicolumn{2}{c}{\bz}  \\ %12,13
\multicolumn{2}{c}{STAR}        & %1,2
\multicolumn{1}{c}{(pc)}        & %3
\multicolumn{1}{c}{$V$}         & %4
\multicolumn{1}{c}{type}        & %5
\multicolumn{1}{c}{DATE}        & %6
\multicolumn{1}{c}{UT}          & %7
\multicolumn{1}{c}{(s)}         & %8 
\multicolumn{1}{c}{\AA$^{-1}$}   & %9
\multicolumn{1}{l}{SETTING}     & %10
\multicolumn{1}{l}{(\arcsec)}        & %11
\multicolumn{2}{c}{(G)}        \\ %12,13
\hline
\endfirsthead
\hline\hline
                            & %1
                            & %2
\multicolumn{1}{c}{$d$}     & %3
                            & %4  
\multicolumn{1}{c}{Spec.}   & %5
                            & %6
                            & %7
\multicolumn{1}{c}{EXP}     & %8
\multicolumn{1}{c}{\it S/N} & %9 
                            & %0
\multicolumn{1}{l}{s.w.}    & %11
\multicolumn{2}{c}{\bz}  \\ %12,13
\multicolumn{2}{c}{STAR}        & %1,2
\multicolumn{1}{c}{(pc)}        & %3
\multicolumn{1}{c}{$V$}         & %4
\multicolumn{1}{c}{type}        & %5
\multicolumn{1}{c}{DATE}        & %6
\multicolumn{1}{c}{UT}          & %7
\multicolumn{1}{c}{(s)}         & %8 
\multicolumn{1}{c}{\AA$^{-1}$}   & %9
\multicolumn{1}{l}{SETTING}     & %10
\multicolumn{1}{l}{(\arcsec)}        & %11
\multicolumn{2}{c}{(G)}        \\ %12,13
\hline
\endhead
WD\,0447$+$176 & HIP\,22485 &$322.9                $& 12.66 & sdOp & 2015-02-01 & 22:53 & 4800 & 515 & R600B & 1.0 &$   475  $&  265 \\ %!!!!!!
               &            &$                     $&       &      & 2015-02-04 & 01:12 & 5600 & 575 & R600B & 1.0 &$   355  $&  250 \\ [5mm] 
WD\,1544$-$377 &CD$-$37\,657&$ 15.2                $&12.78  &DA4.8 & 2015-04-05 & 09:08 & 4296 & 1245& 1200B & 1.2 &\multicolumn{2}{c}{---}\\ [1mm] 
WD\,1615$-$154 & LTT\,6497  &$ 47.1                $& 13.43 &DA1.7 & 2015-04-03 & 08:54 & 4296 & 1420& 1200B & 1.2 &\multicolumn{2}{c}{---}\\ 
\hline
\end{longtable}
\end{tiny}

%% file: Fig_Atlas.tex
\noindent
\includegraphics*[angle=270,width=8.0cm,trim={0.90cm 0.0cm 0.1cm 1.0cm},clip]{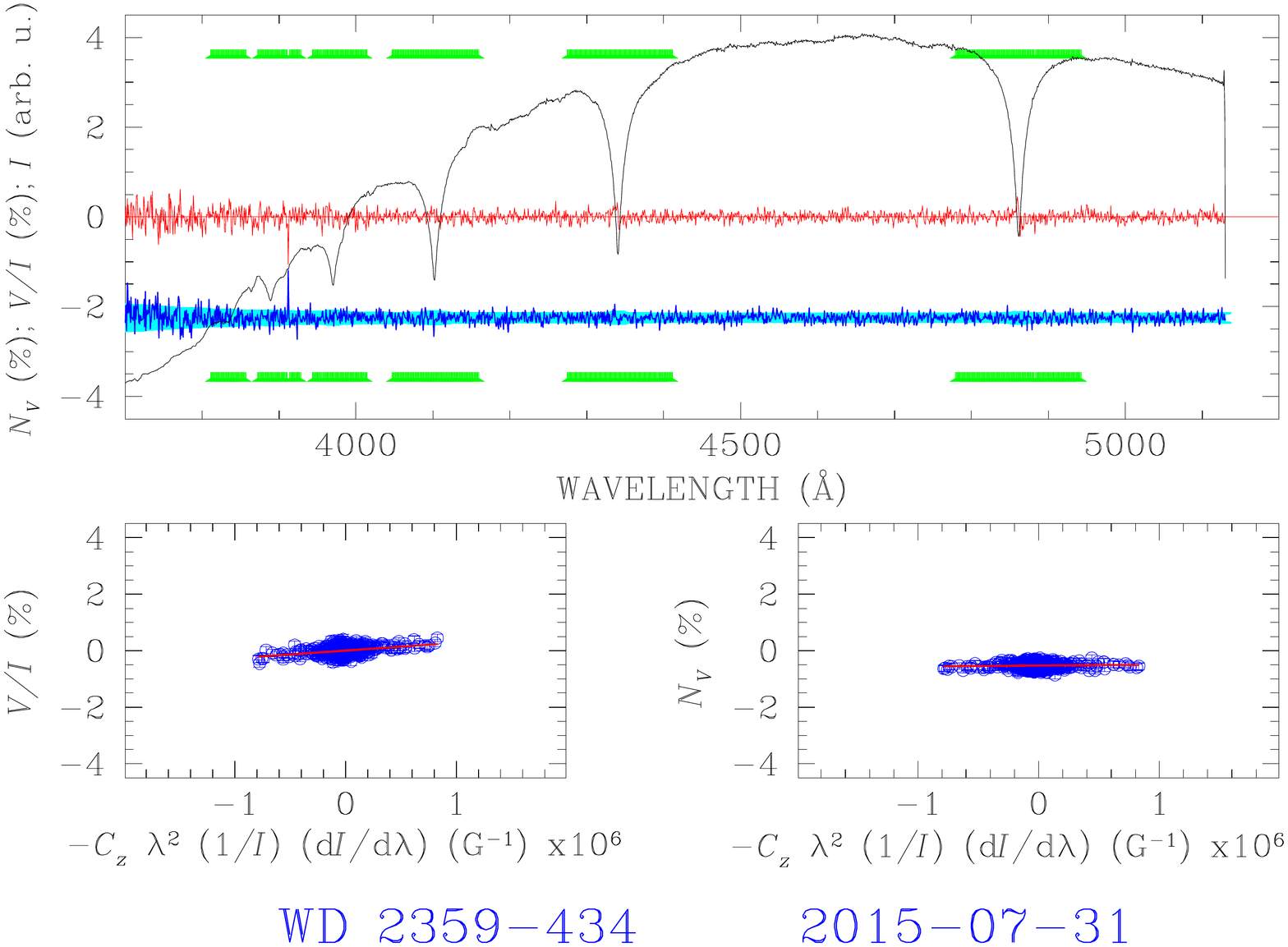} \\
\includegraphics*[angle=270,width=8.0cm,trim={0.90cm 0.0cm 0.1cm 1.0cm},clip]{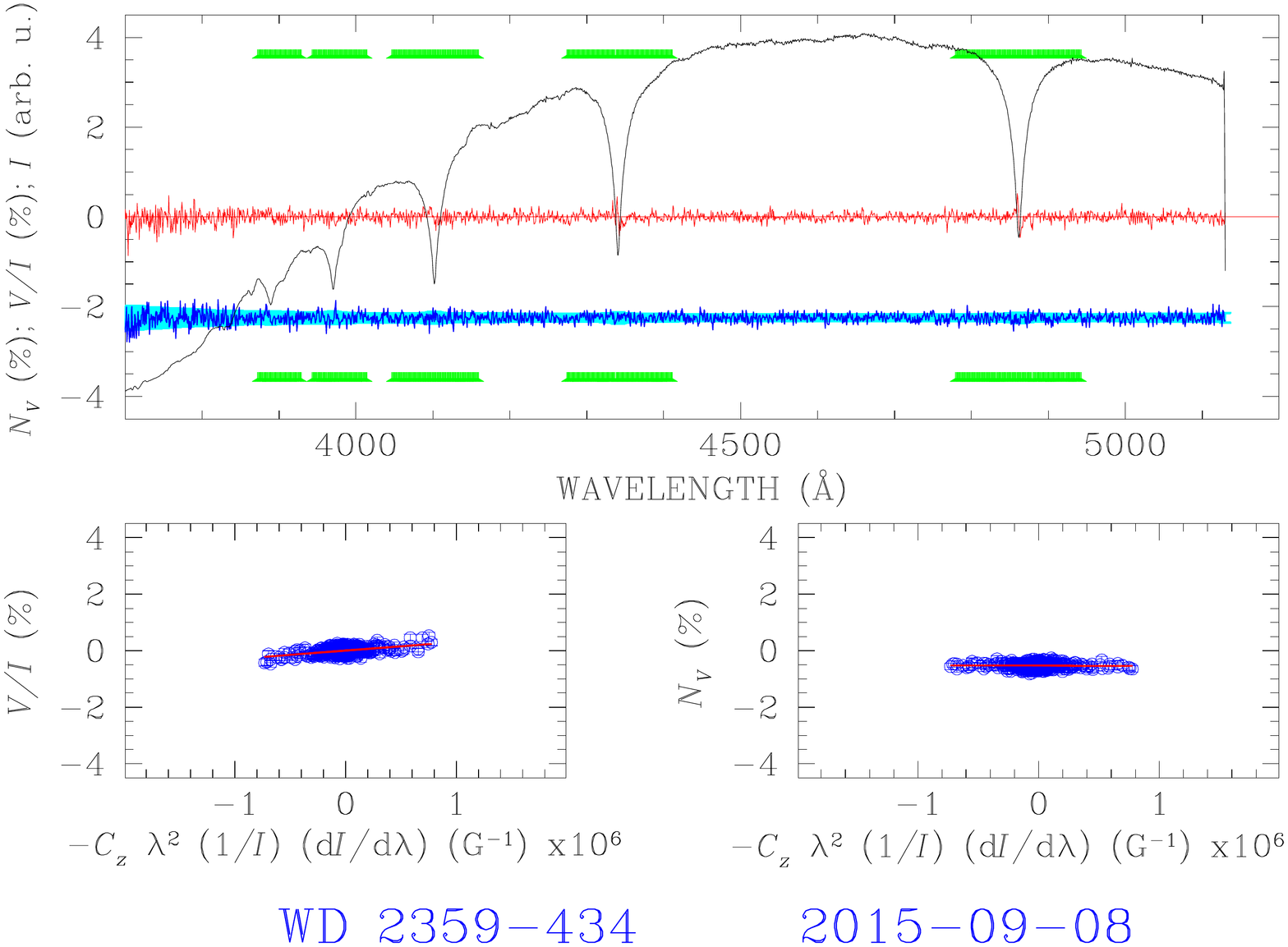} \\
\includegraphics*[angle=270,width=8.0cm,trim={0.90cm 0.0cm 0.1cm 1.0cm},clip]{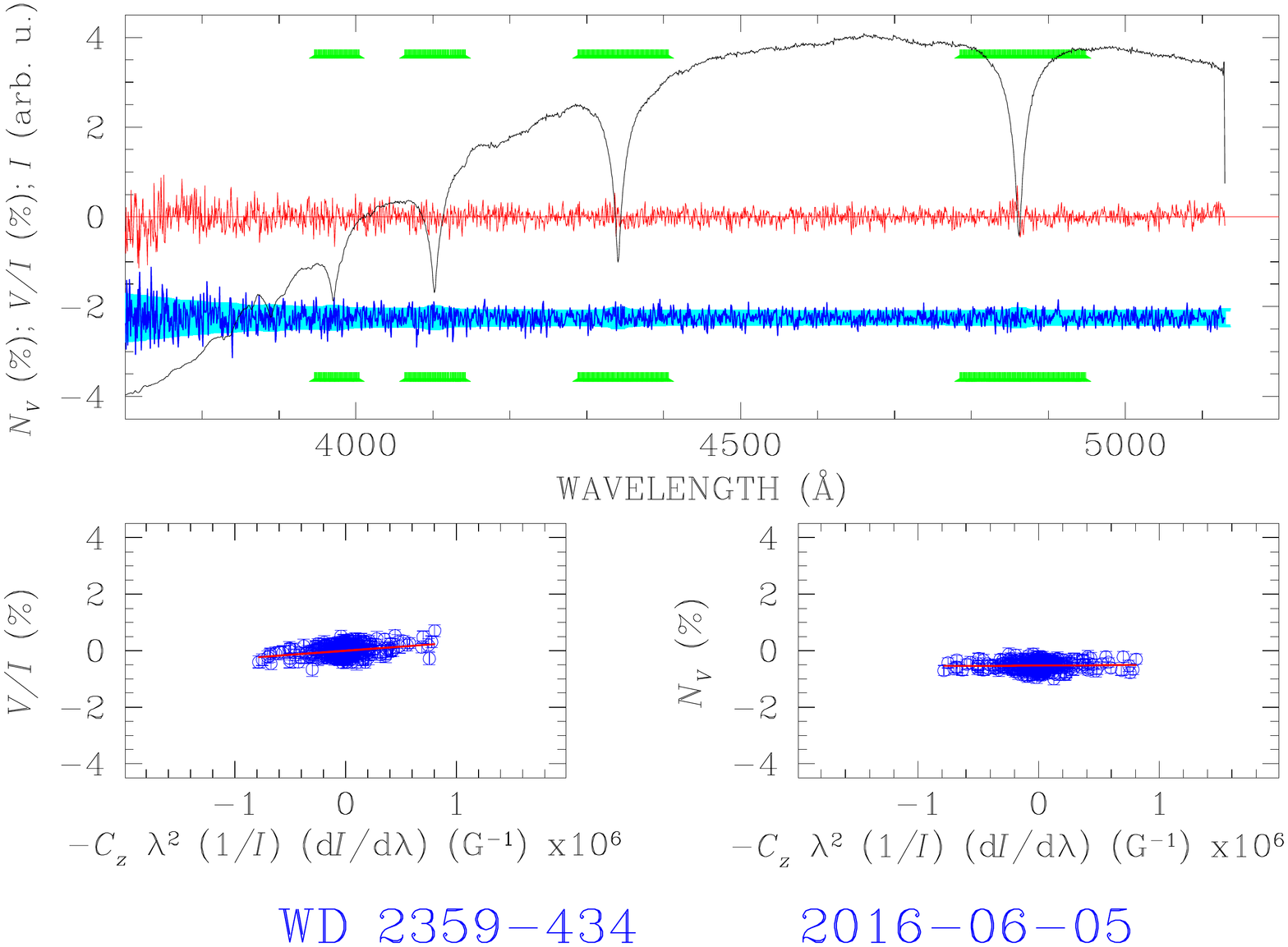} \\
\includegraphics*[angle=270,width=8.0cm,trim={0.90cm 0.0cm 0.1cm 1.0cm},clip]{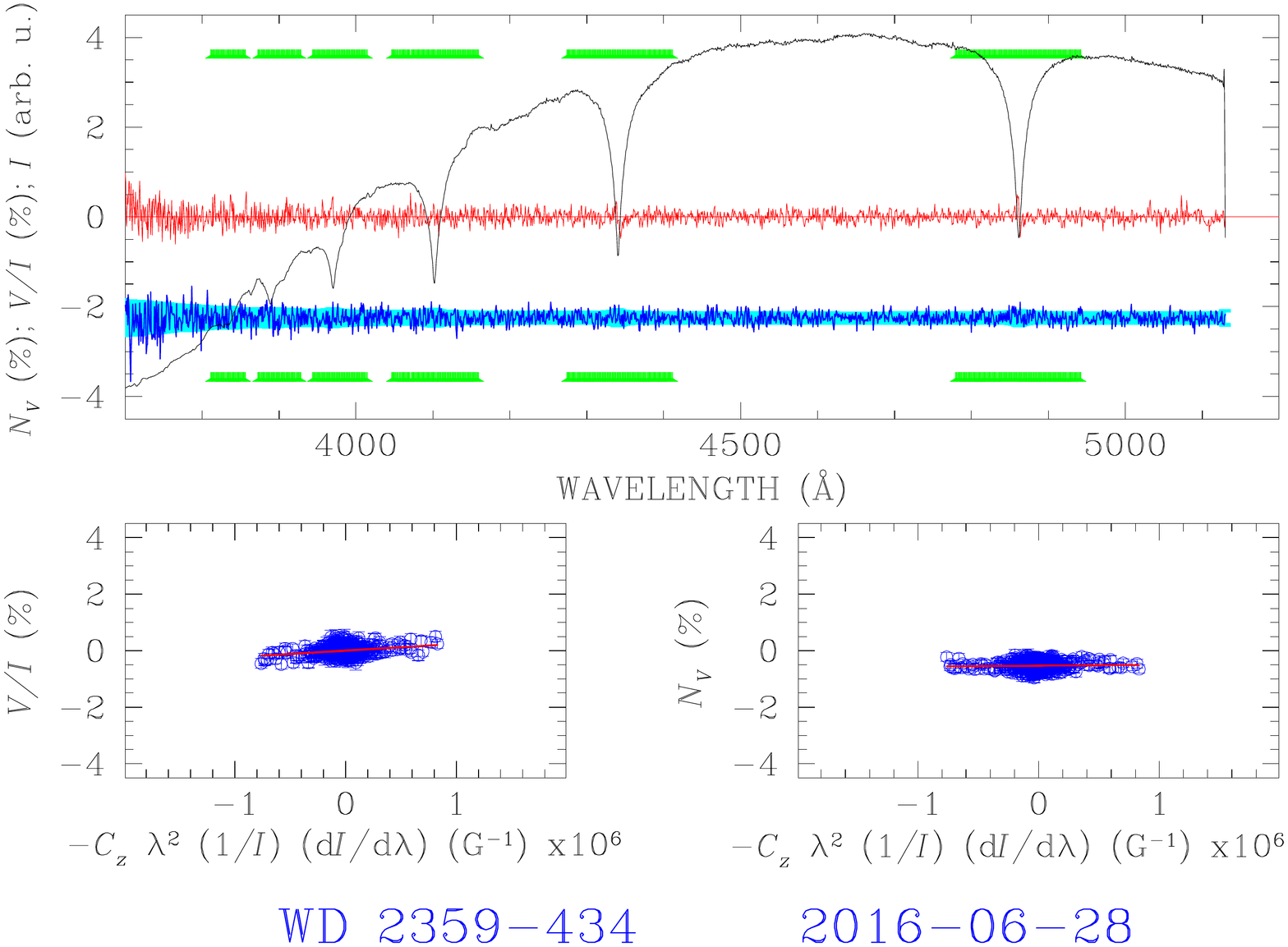} \\
\includegraphics*[angle=270,width=8.0cm,trim={0.90cm 0.0cm 0.1cm 1.0cm},clip]{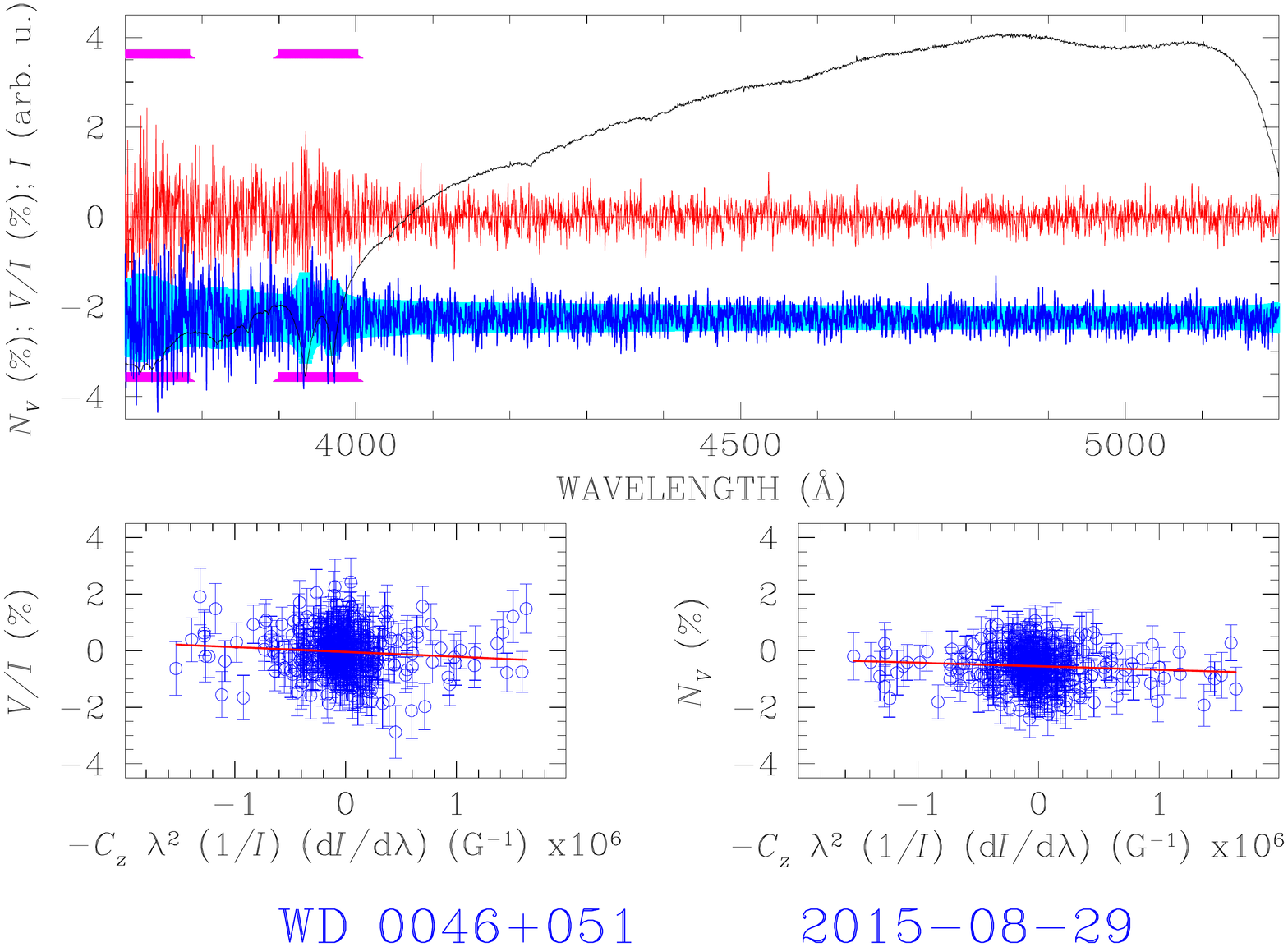}
\includegraphics*[angle=270,width=8.0cm,trim={0.90cm 0.0cm 0.1cm 1.0cm},clip]{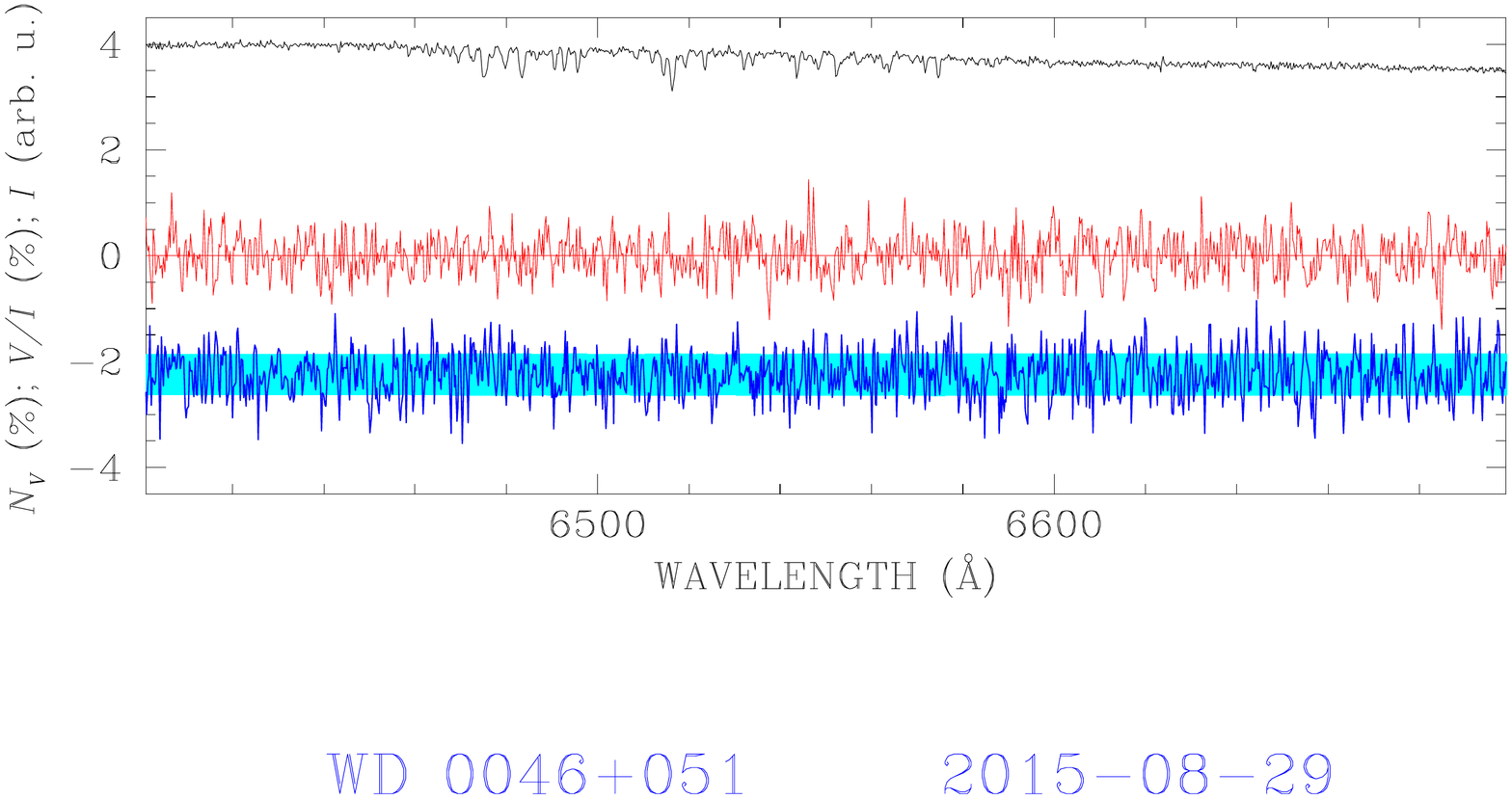}\\
\includegraphics*[angle=270,width=8.0cm,trim={0.90cm 0.0cm 0.1cm 1.0cm},clip]{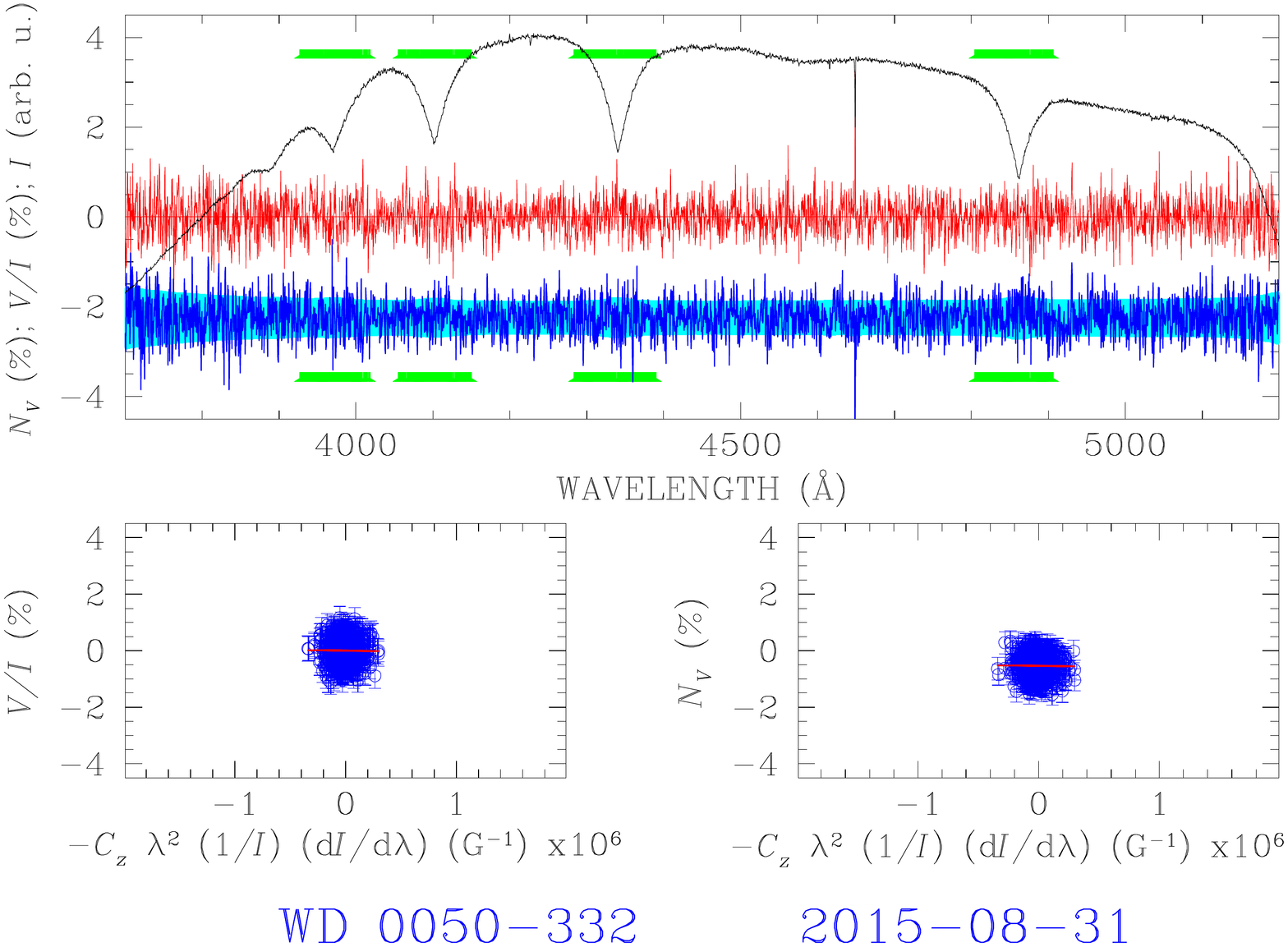}
\includegraphics*[angle=270,width=8.0cm,trim={0.90cm 0.0cm 0.1cm 1.0cm},clip]{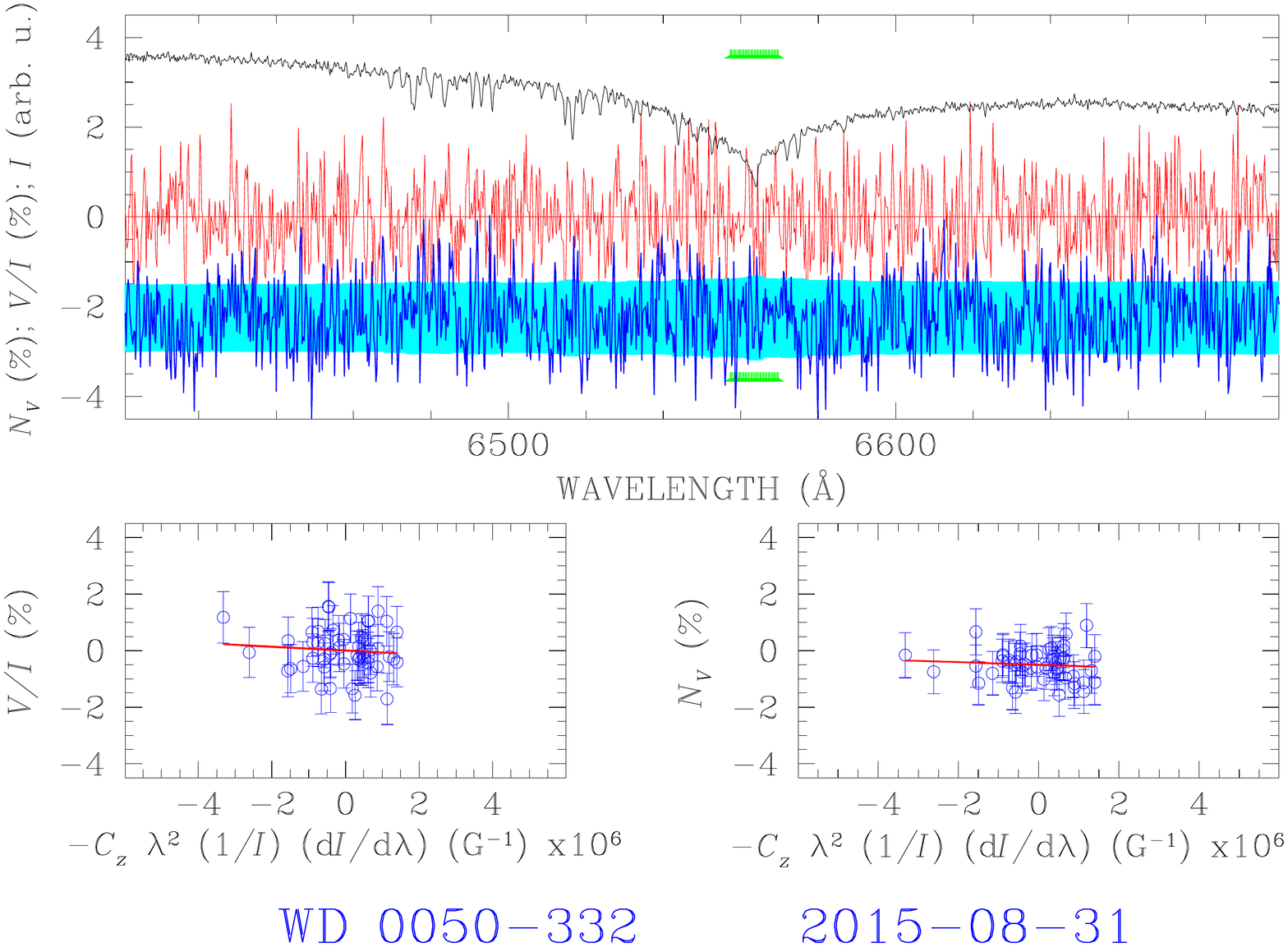}\\
\includegraphics*[angle=270,width=8.0cm,trim={0.90cm 0.0cm 0.1cm 1.0cm},clip]{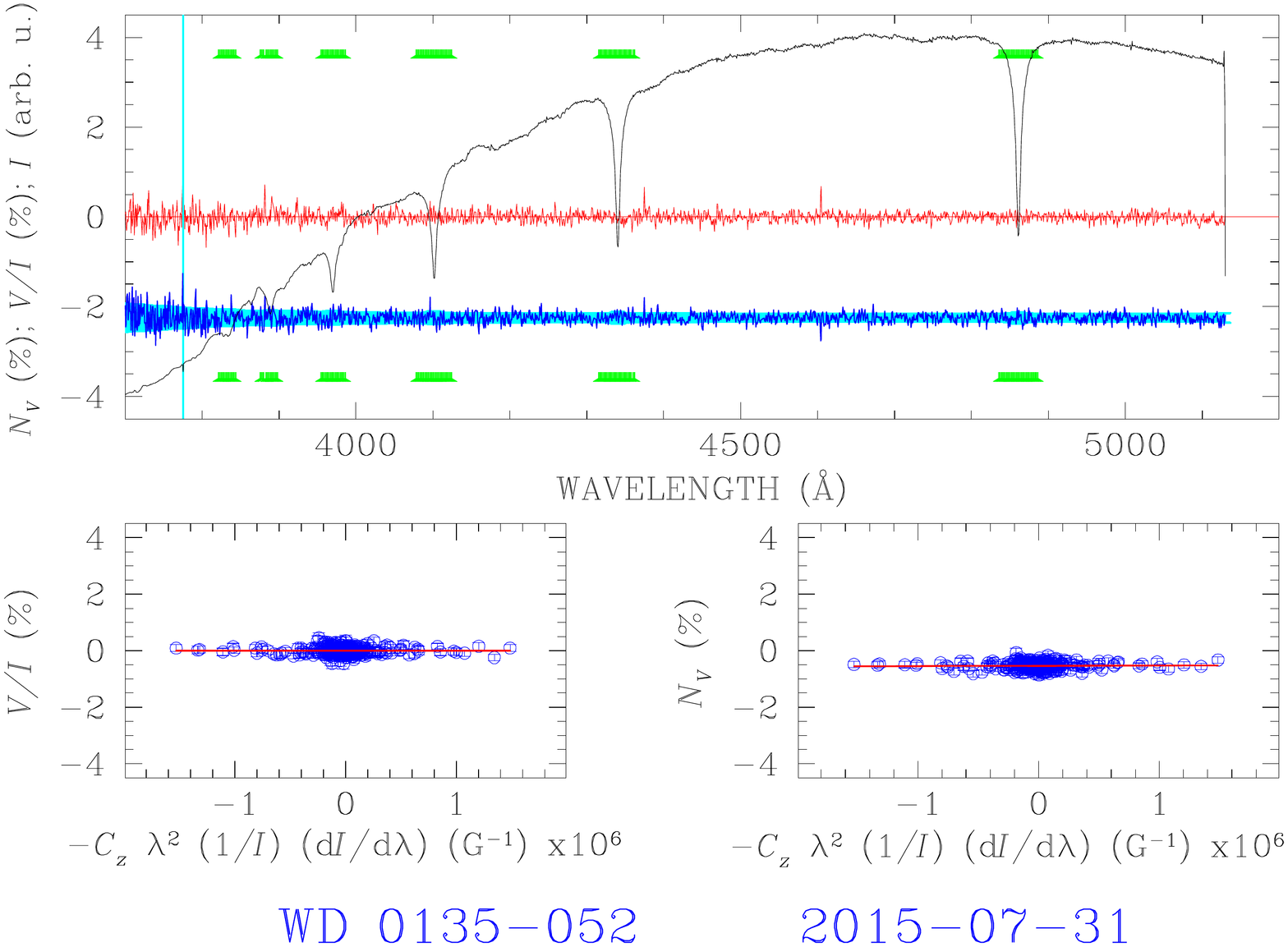}\\
\includegraphics*[angle=270,width=8.0cm,trim={0.90cm 0.0cm 0.1cm 1.0cm},clip]{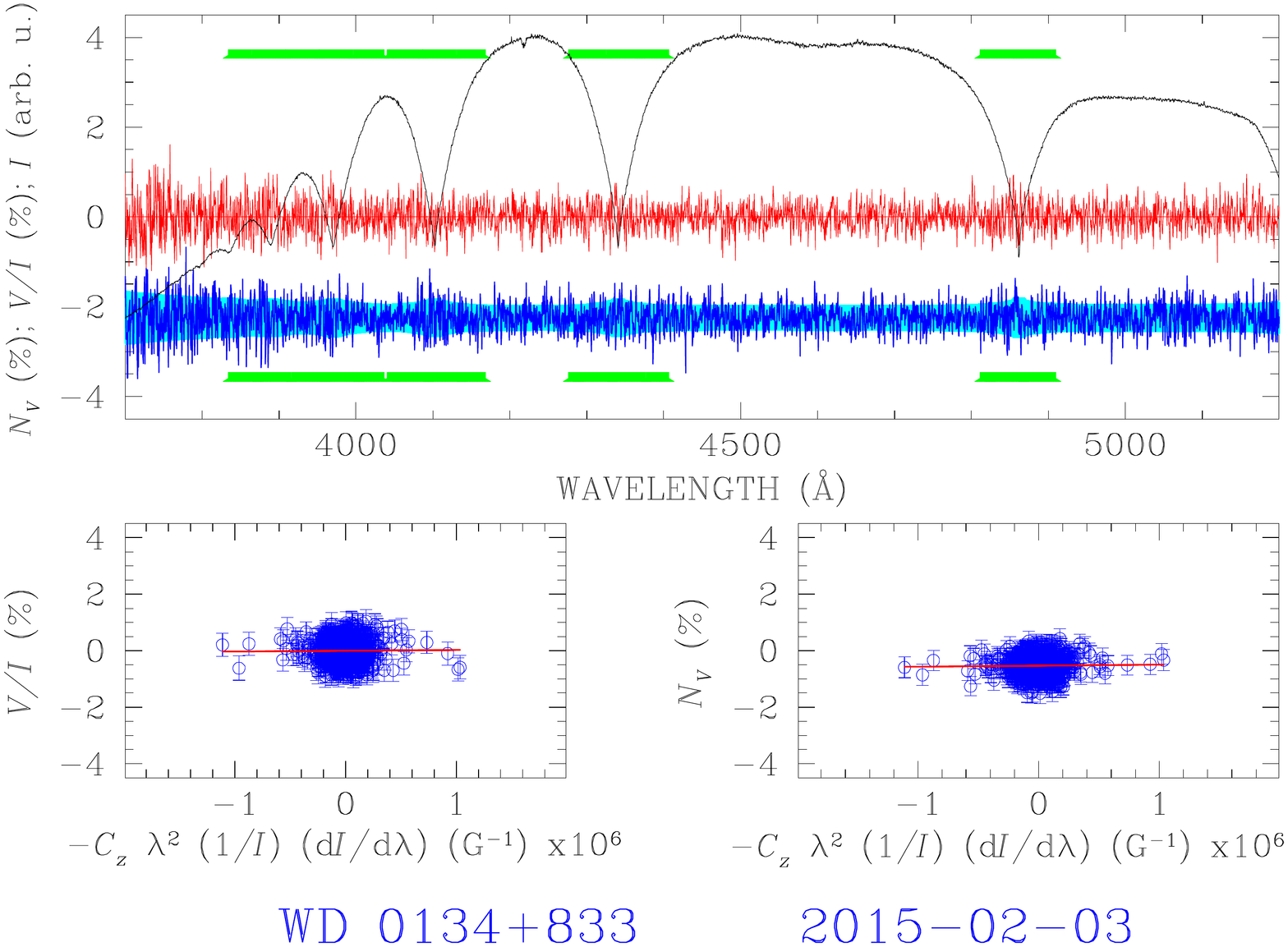}\\
\includegraphics*[angle=270,width=8.0cm,trim={0.90cm 0.0cm 0.1cm 1.0cm},clip]{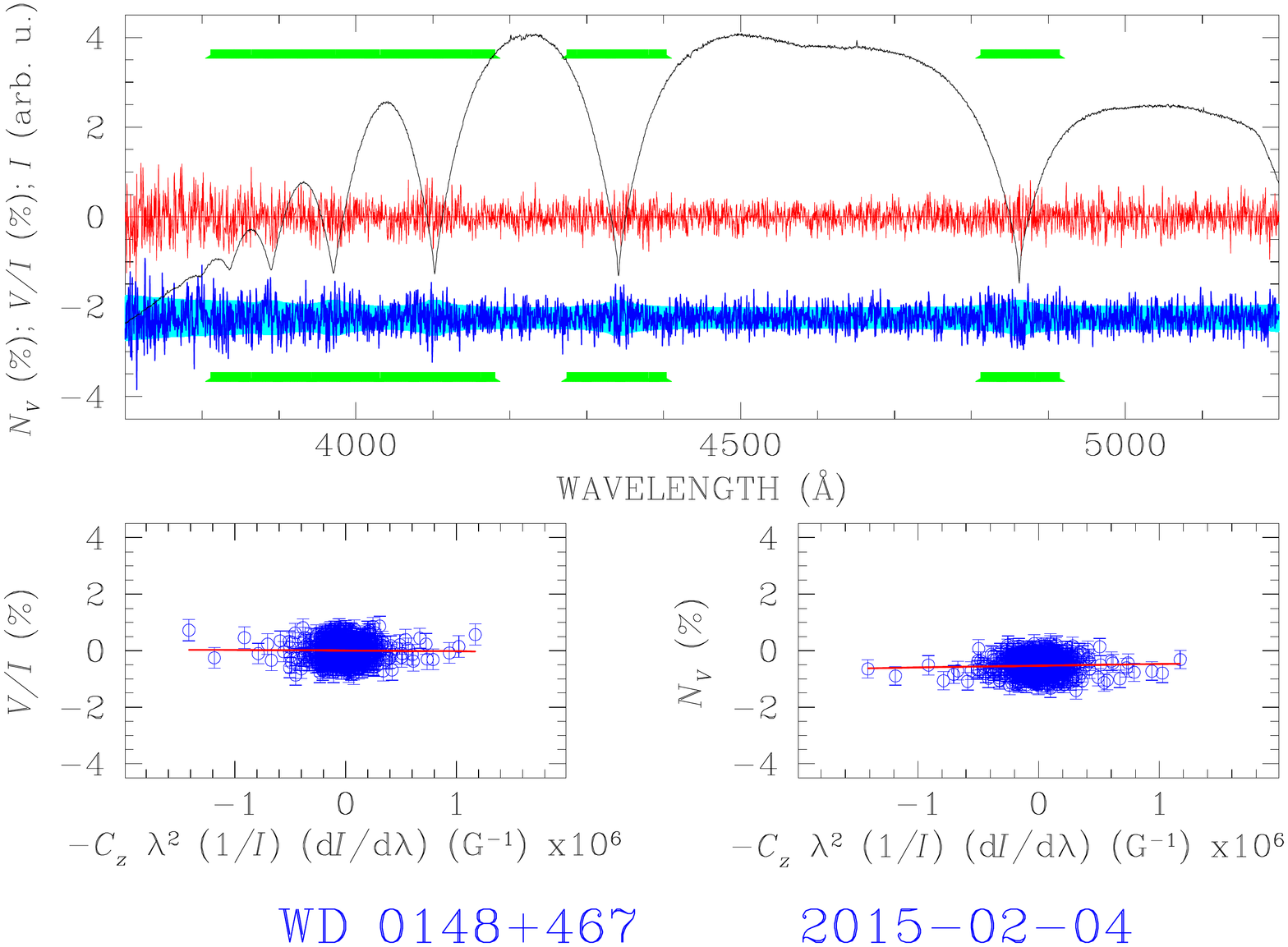}\\
\includegraphics*[angle=270,width=8.0cm,trim={0.90cm 0.0cm 0.1cm 1.0cm},clip]{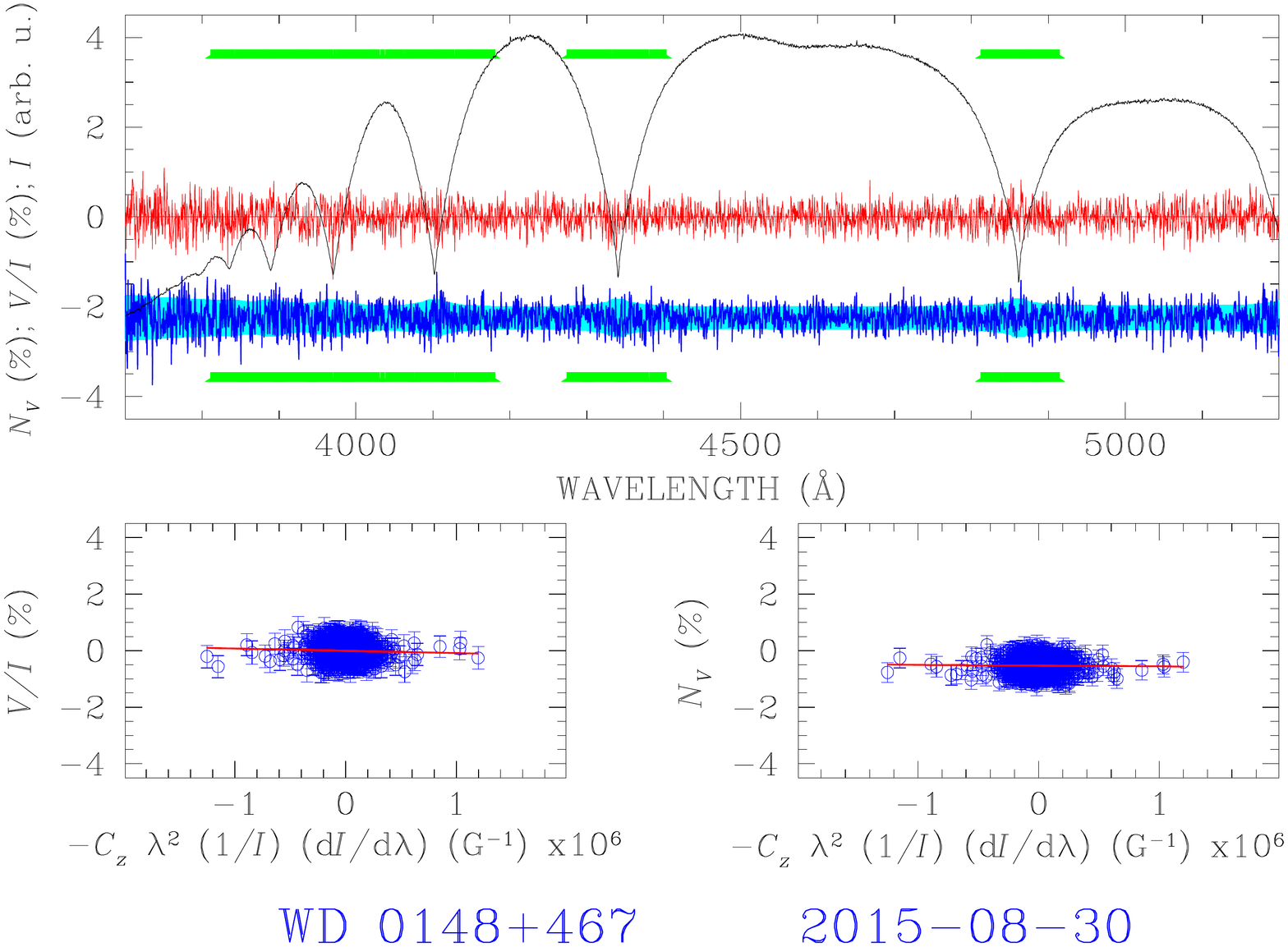}
\includegraphics*[angle=270,width=8.0cm,trim={0.90cm 0.0cm 0.1cm 1.0cm},clip]{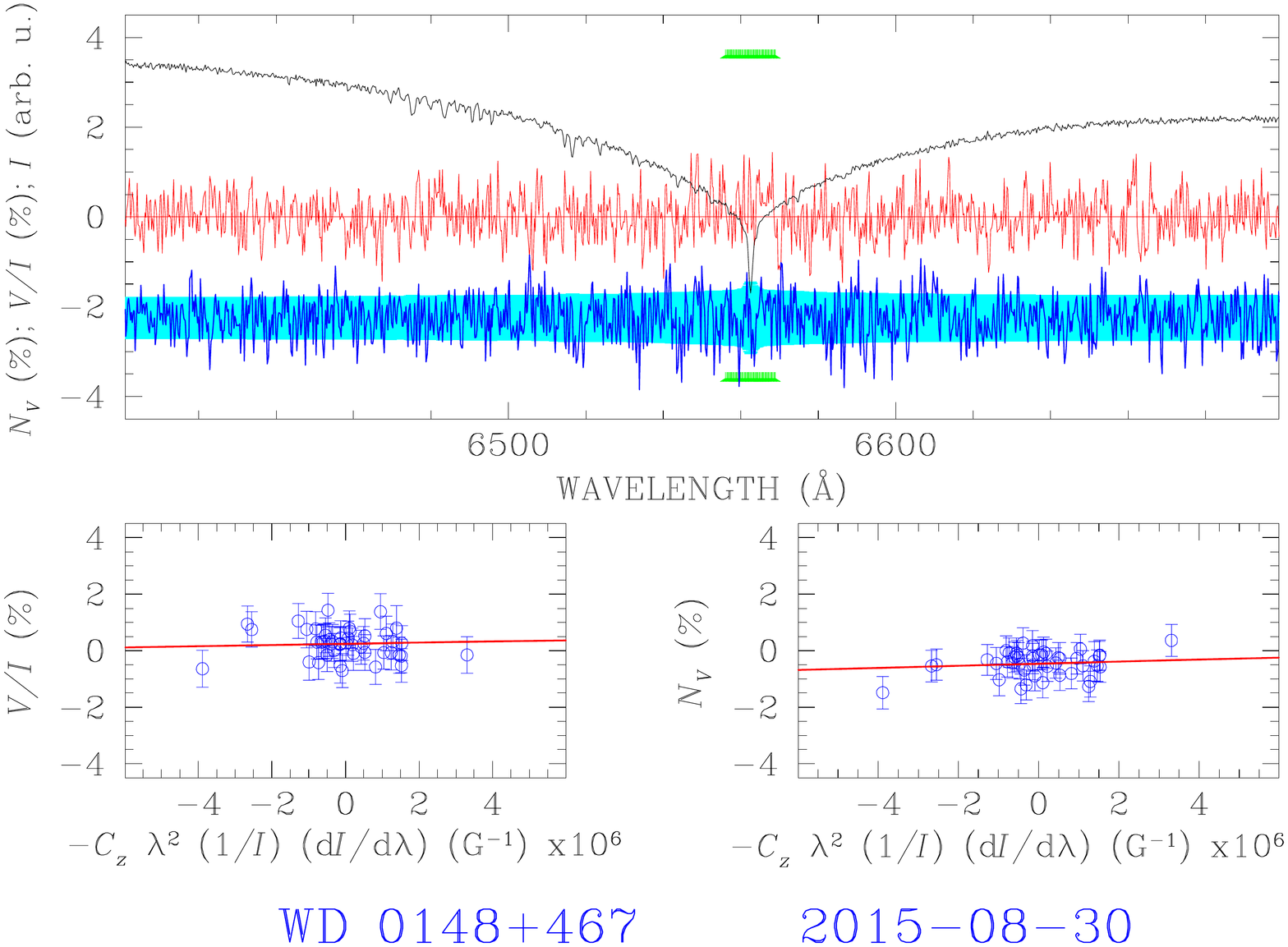}\\
\includegraphics*[angle=270,width=8.0cm,trim={0.90cm 0.0cm 0.1cm 1.0cm},clip]{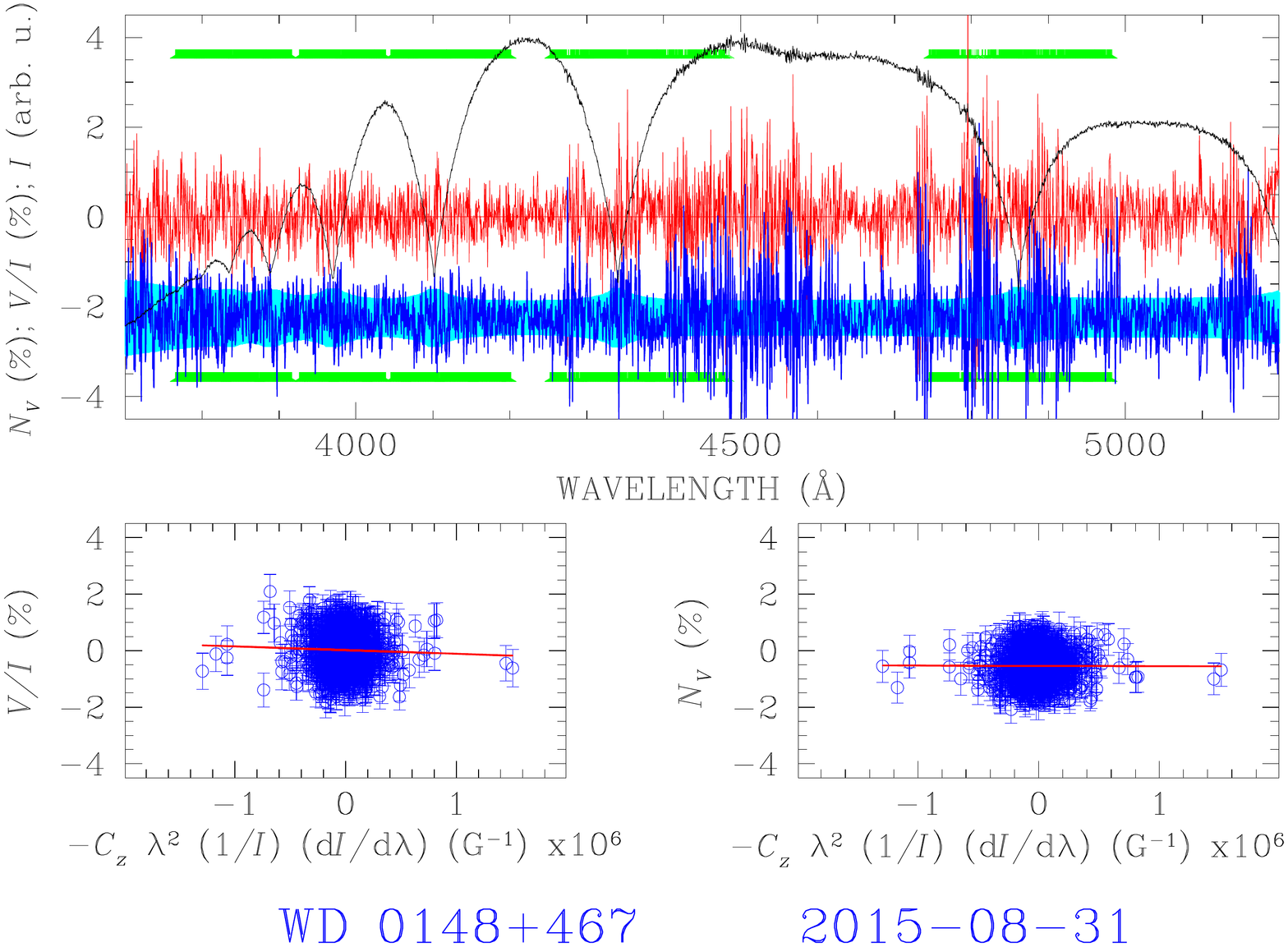}
\includegraphics*[angle=270,width=8.0cm,trim={0.90cm 0.0cm 0.1cm 1.0cm},clip]{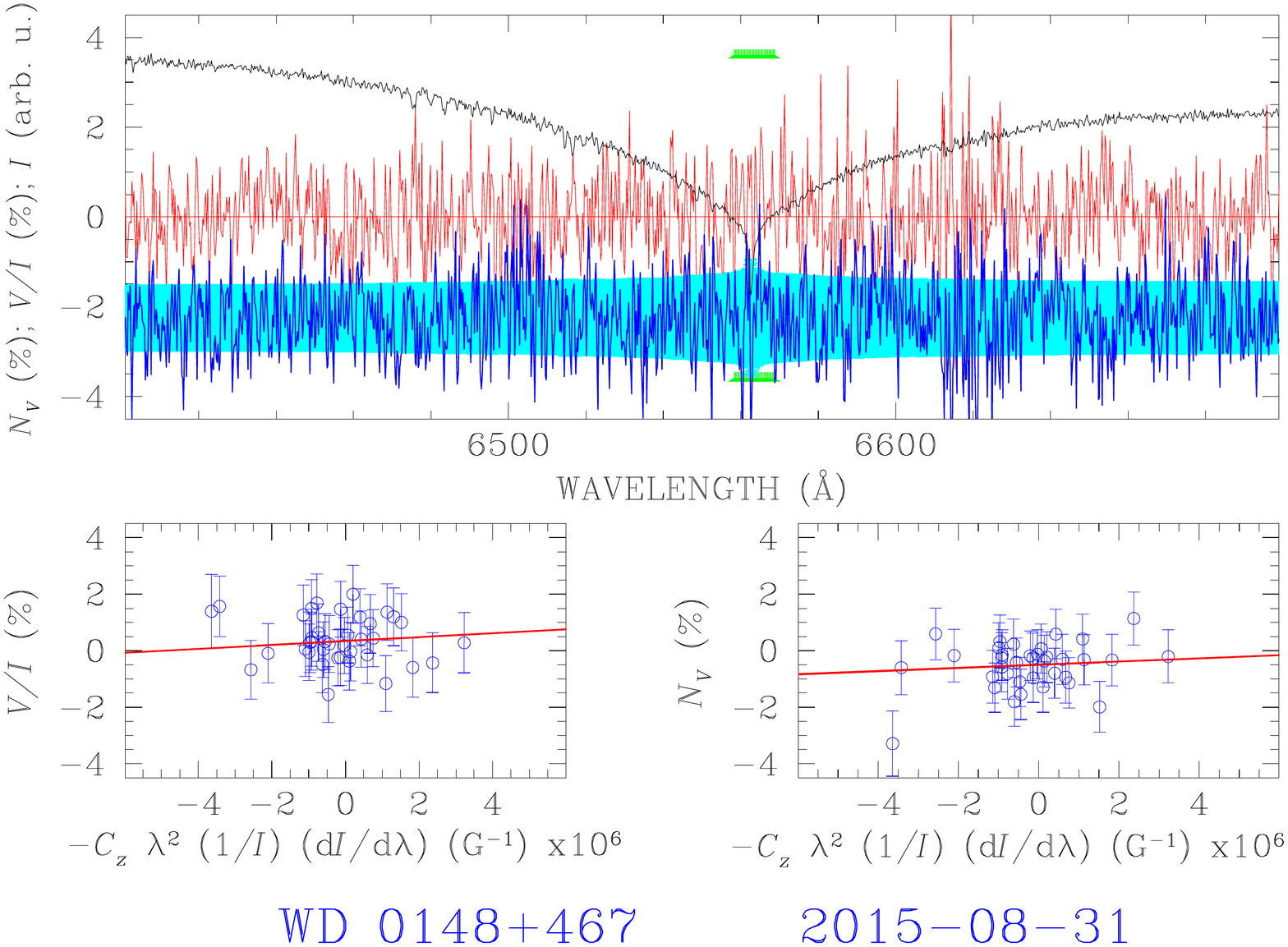}\\
\includegraphics*[angle=270,width=8.0cm,trim={0.90cm 0.0cm 0.1cm 1.0cm},clip]{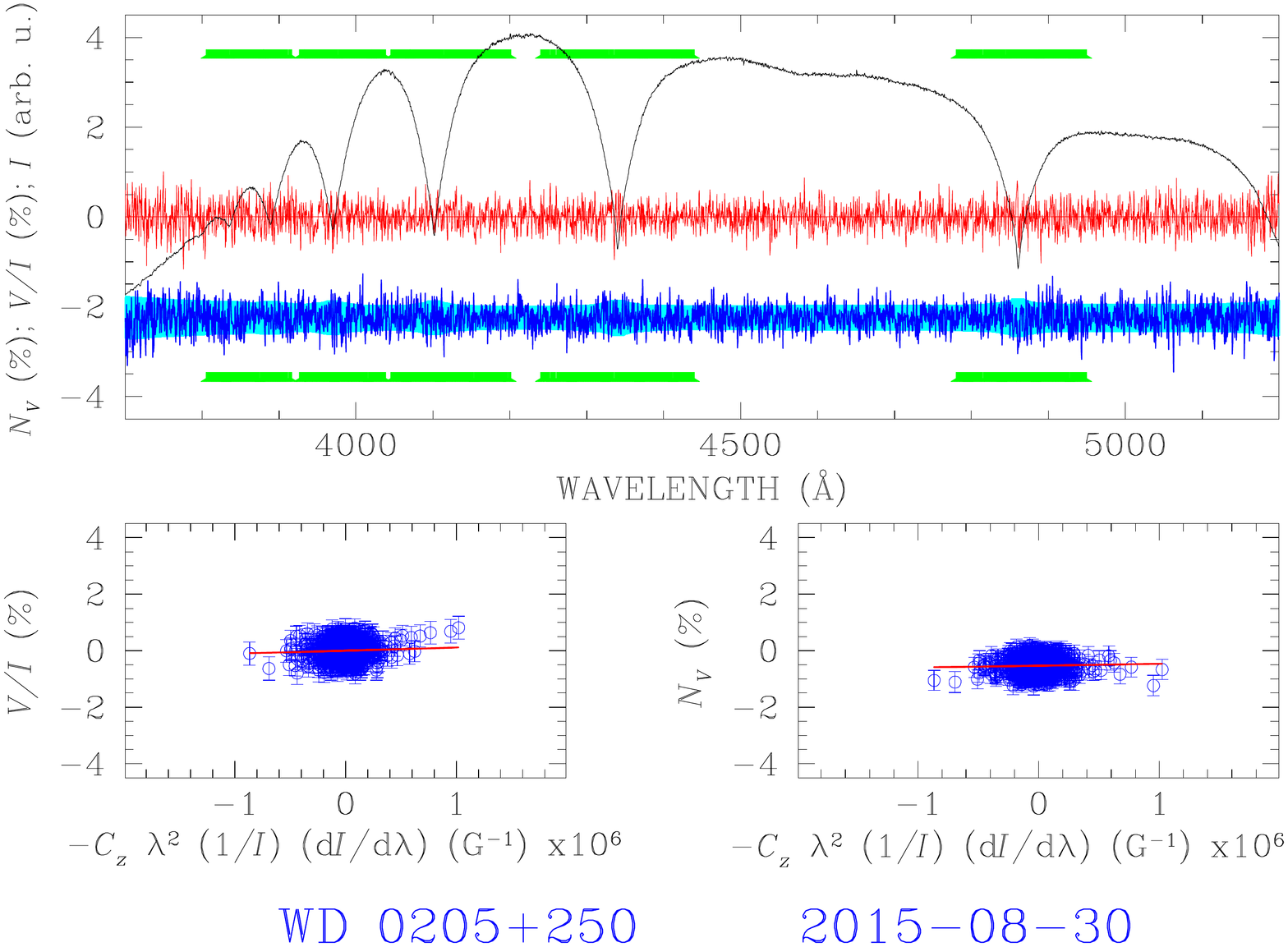}
\includegraphics*[angle=270,width=8.0cm,trim={0.90cm 0.0cm 0.1cm 1.0cm},clip]{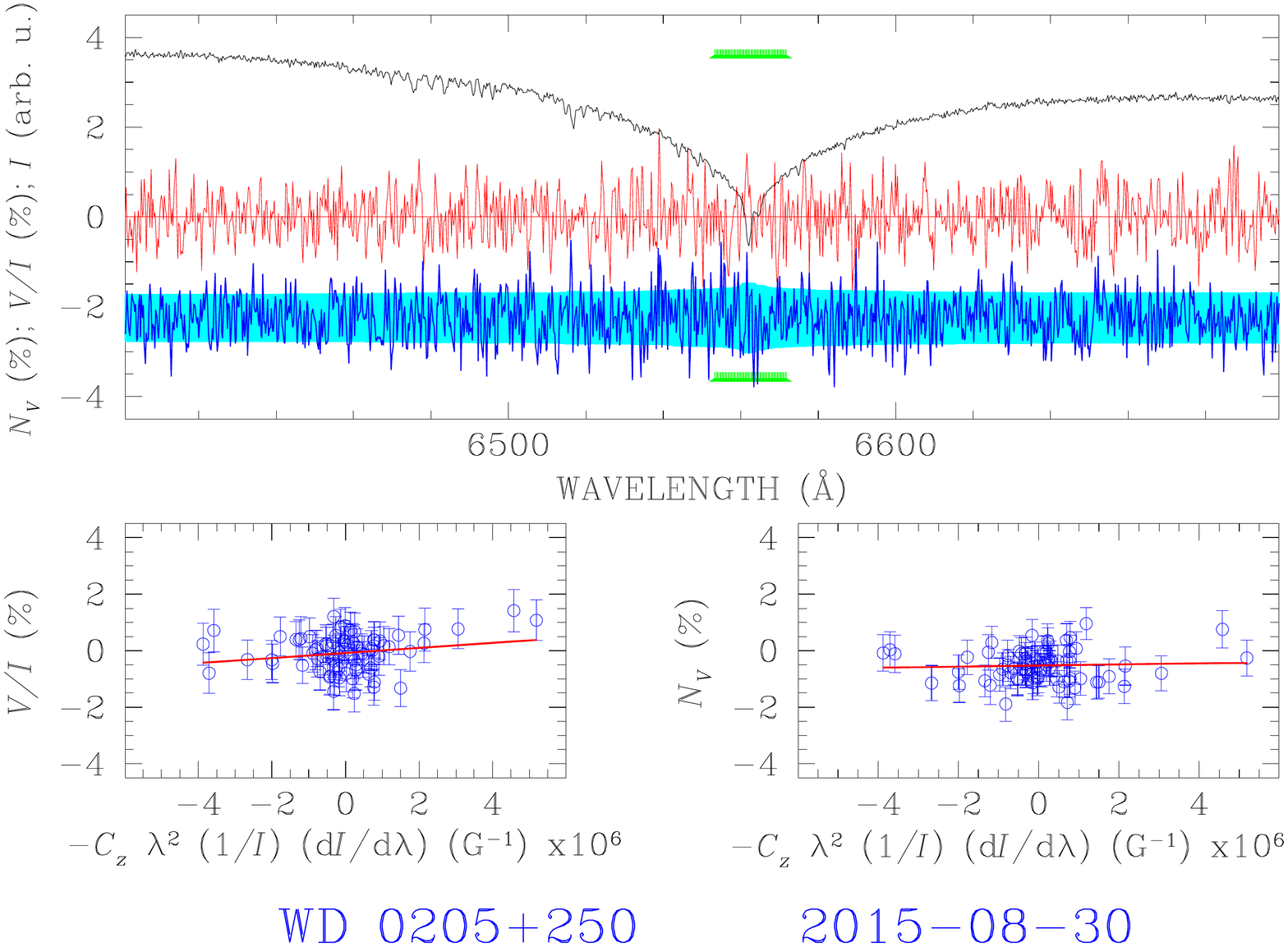}\\
\includegraphics*[angle=270,width=8.0cm,trim={0.90cm 0.0cm 0.1cm 1.0cm},clip]{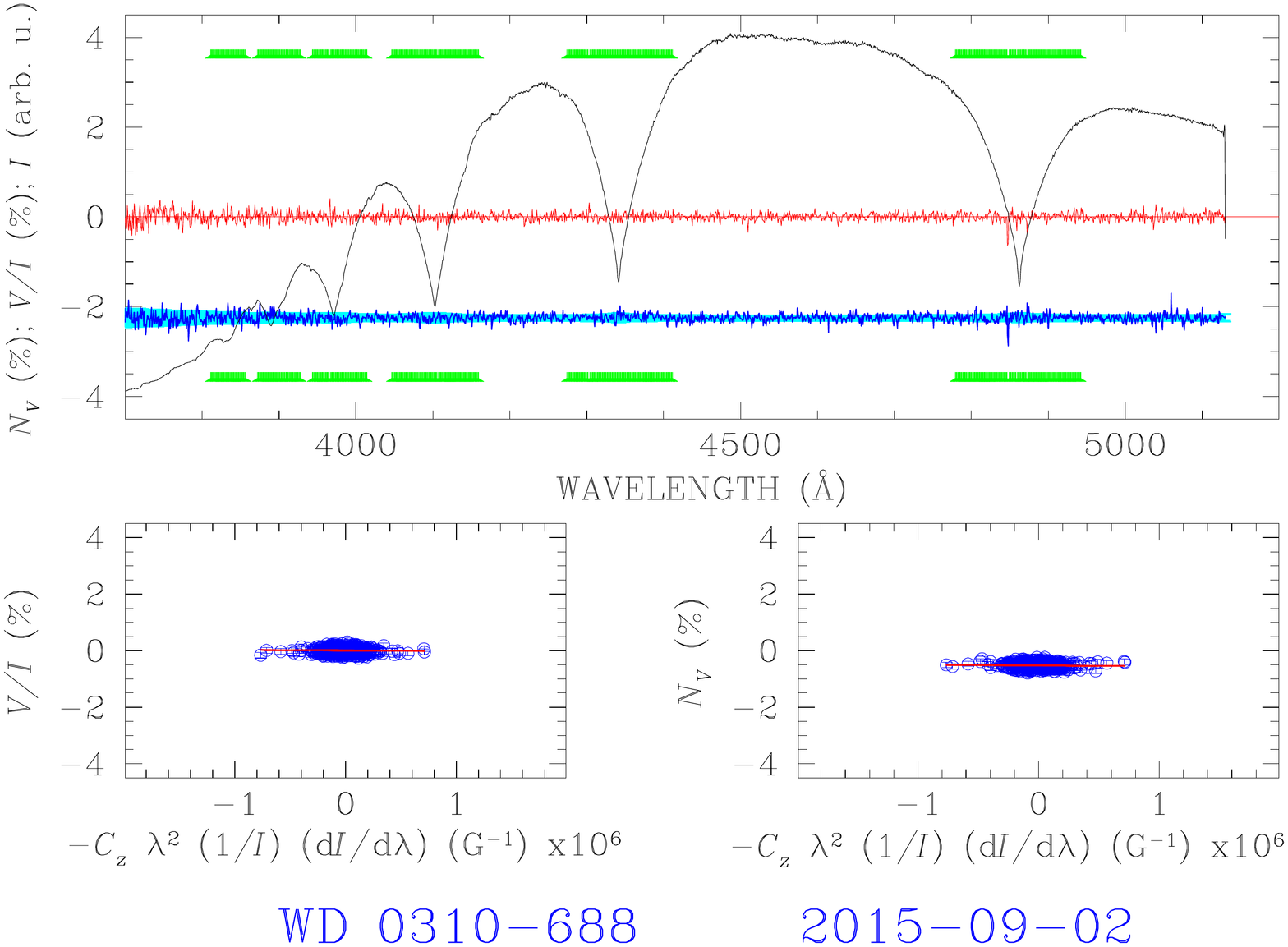}\\
\includegraphics*[angle=270,width=8.0cm,trim={0.90cm 0.0cm 0.1cm 1.0cm},clip]{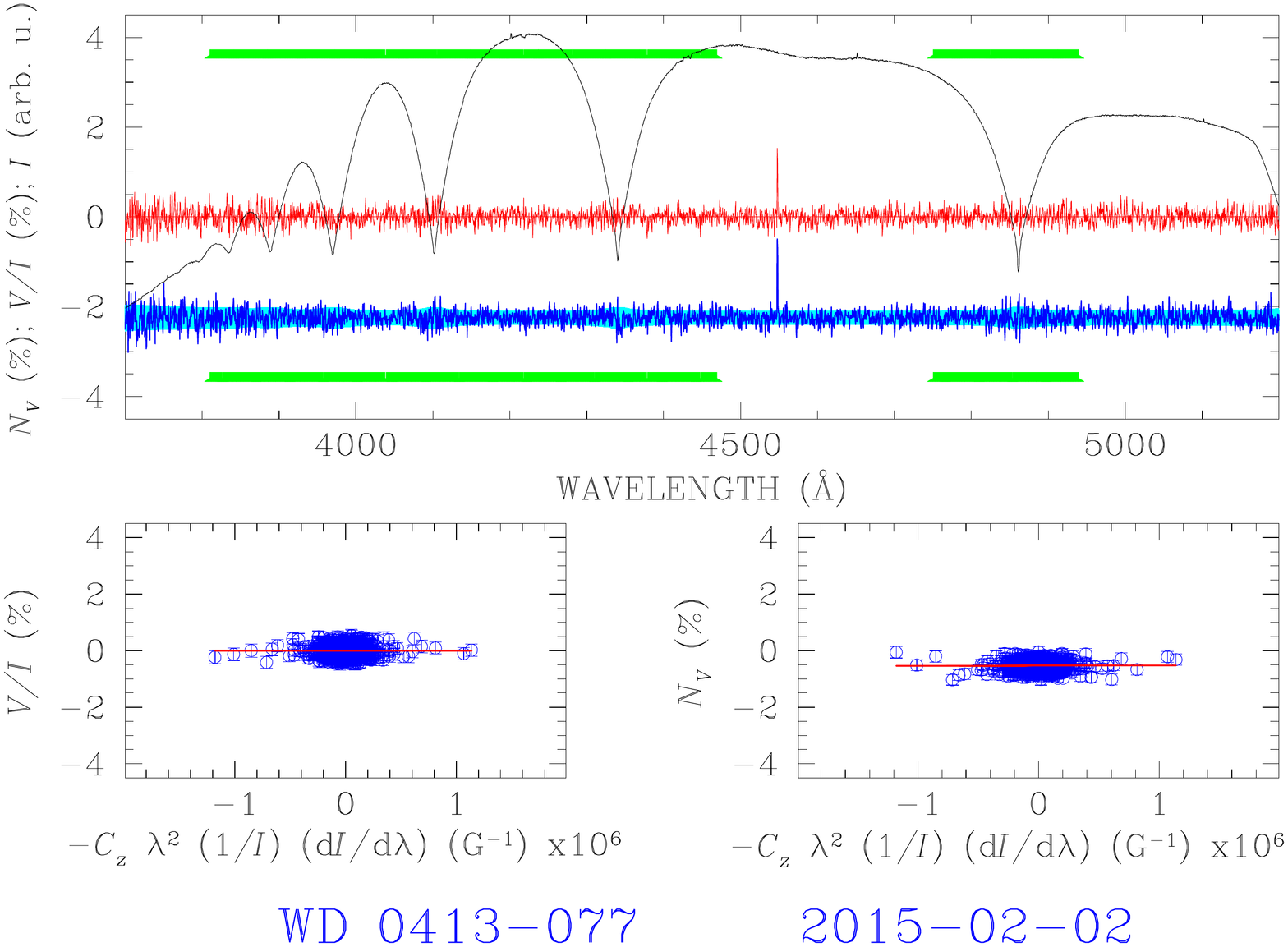}
\includegraphics*[angle=270,width=8.0cm,trim={0.90cm 0.0cm 0.1cm 1.0cm},clip]{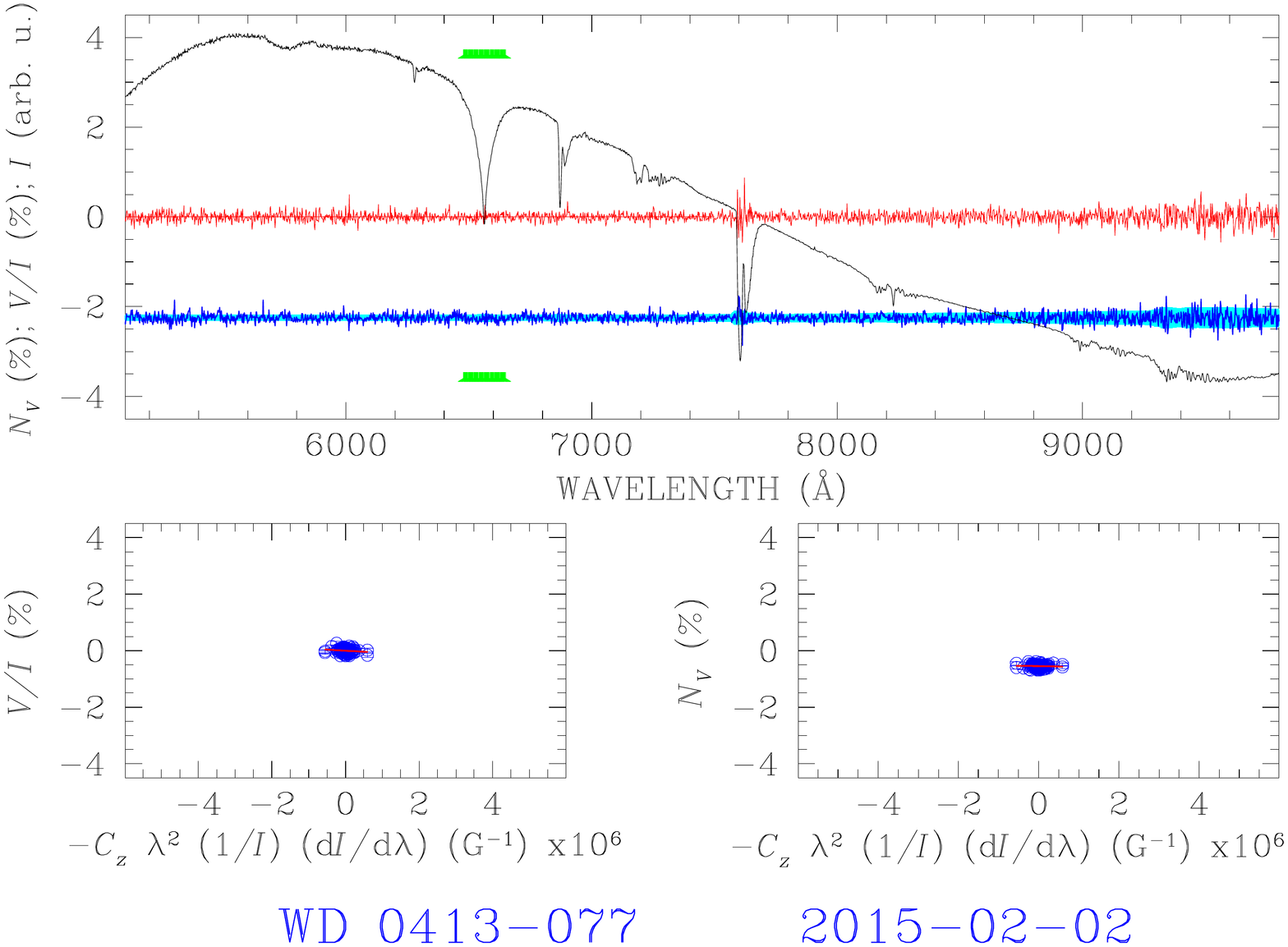}\\
\hspace*{8.8cm}
\includegraphics*[angle=270,width=8.0cm,trim={0.90cm 0.0cm 0.1cm 1.0cm},clip]{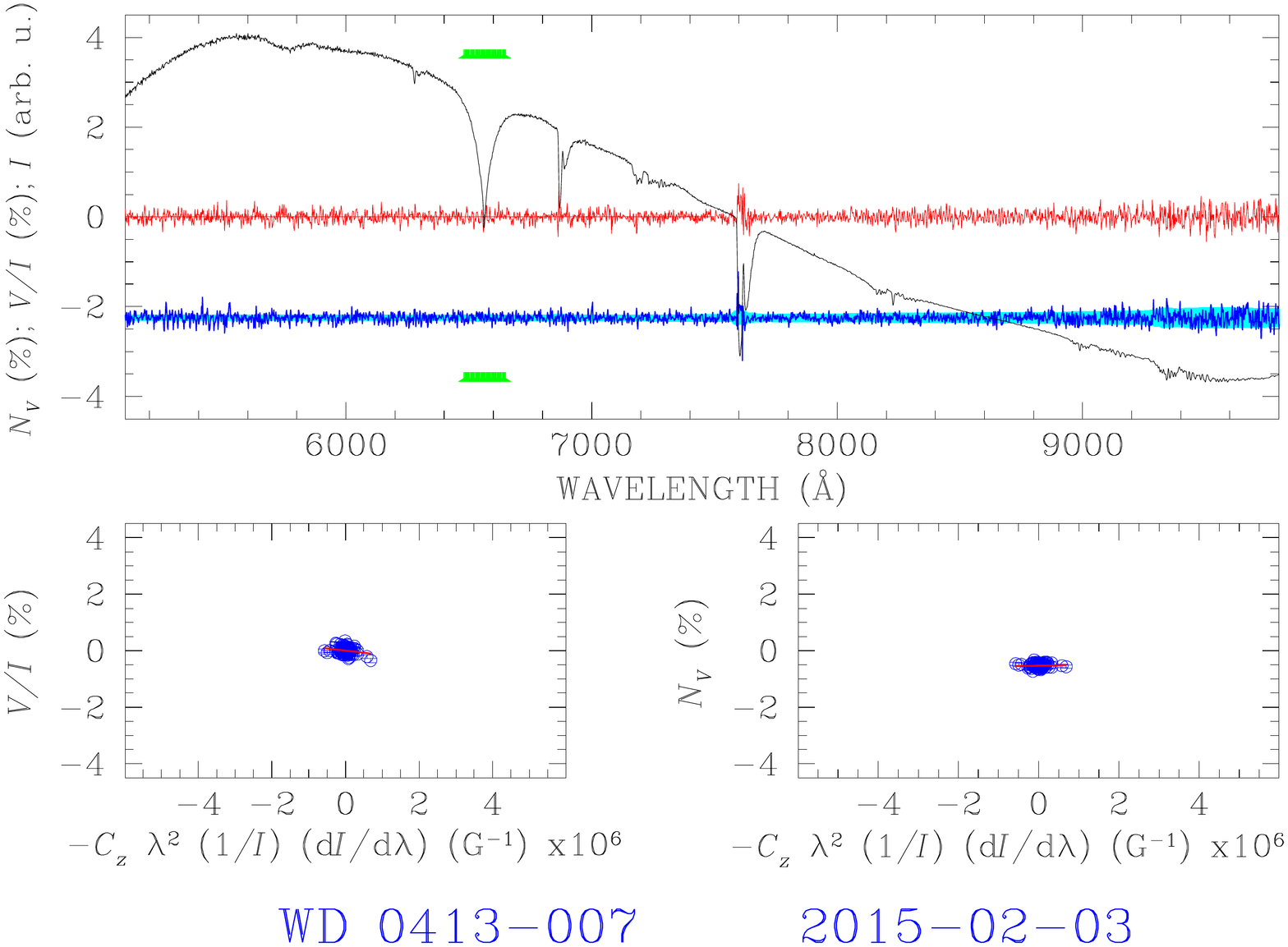} \\
\includegraphics*[angle=270,width=8.0cm,trim={0.90cm 0.0cm 0.1cm 1.0cm},clip]{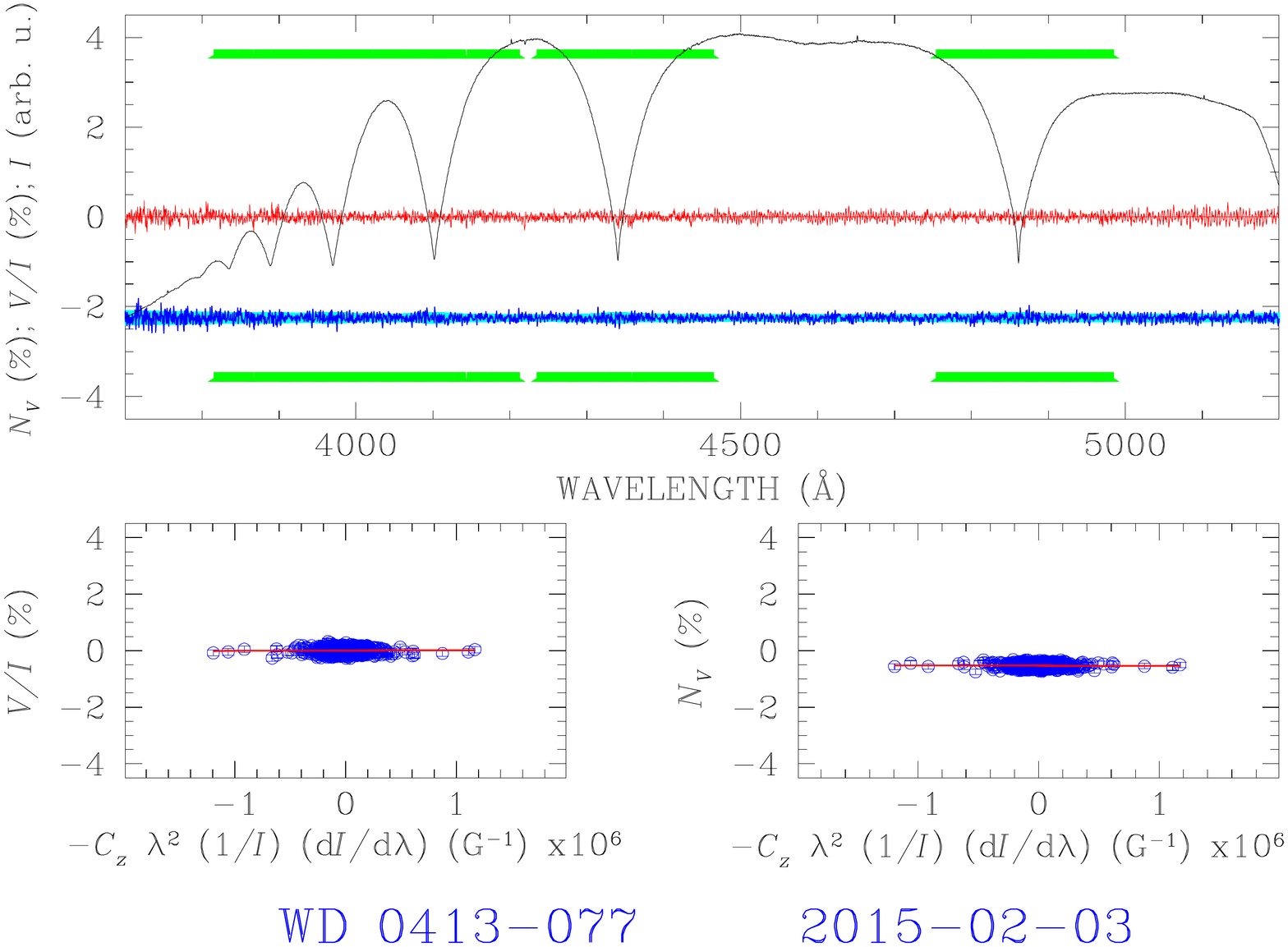}
\includegraphics*[angle=270,width=8.0cm,trim={0.90cm 0.0cm 0.1cm 1.0cm},clip]{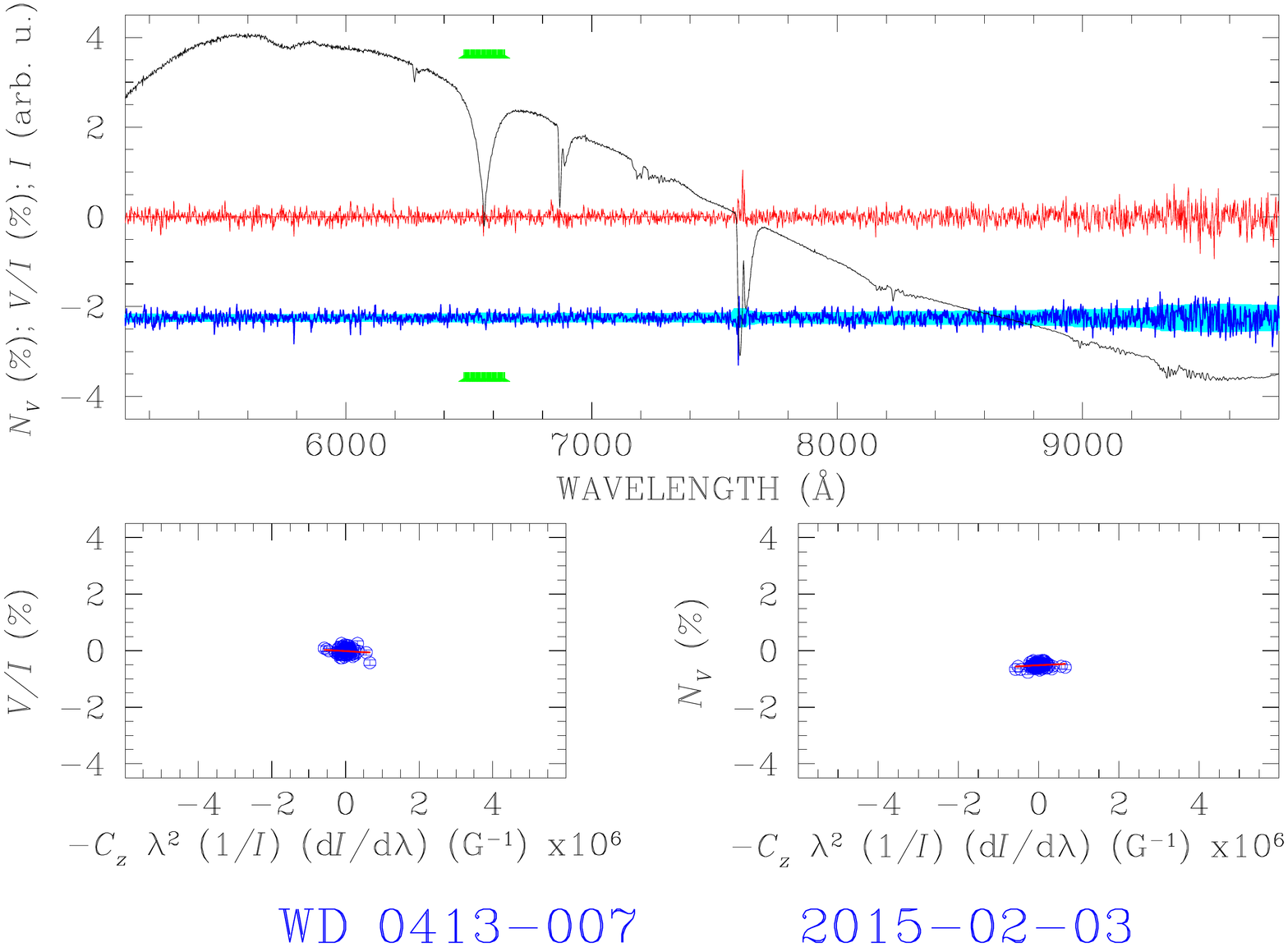} \\
\includegraphics*[angle=270,width=8.0cm,trim={0.90cm 0.0cm 0.1cm 1.0cm},clip]{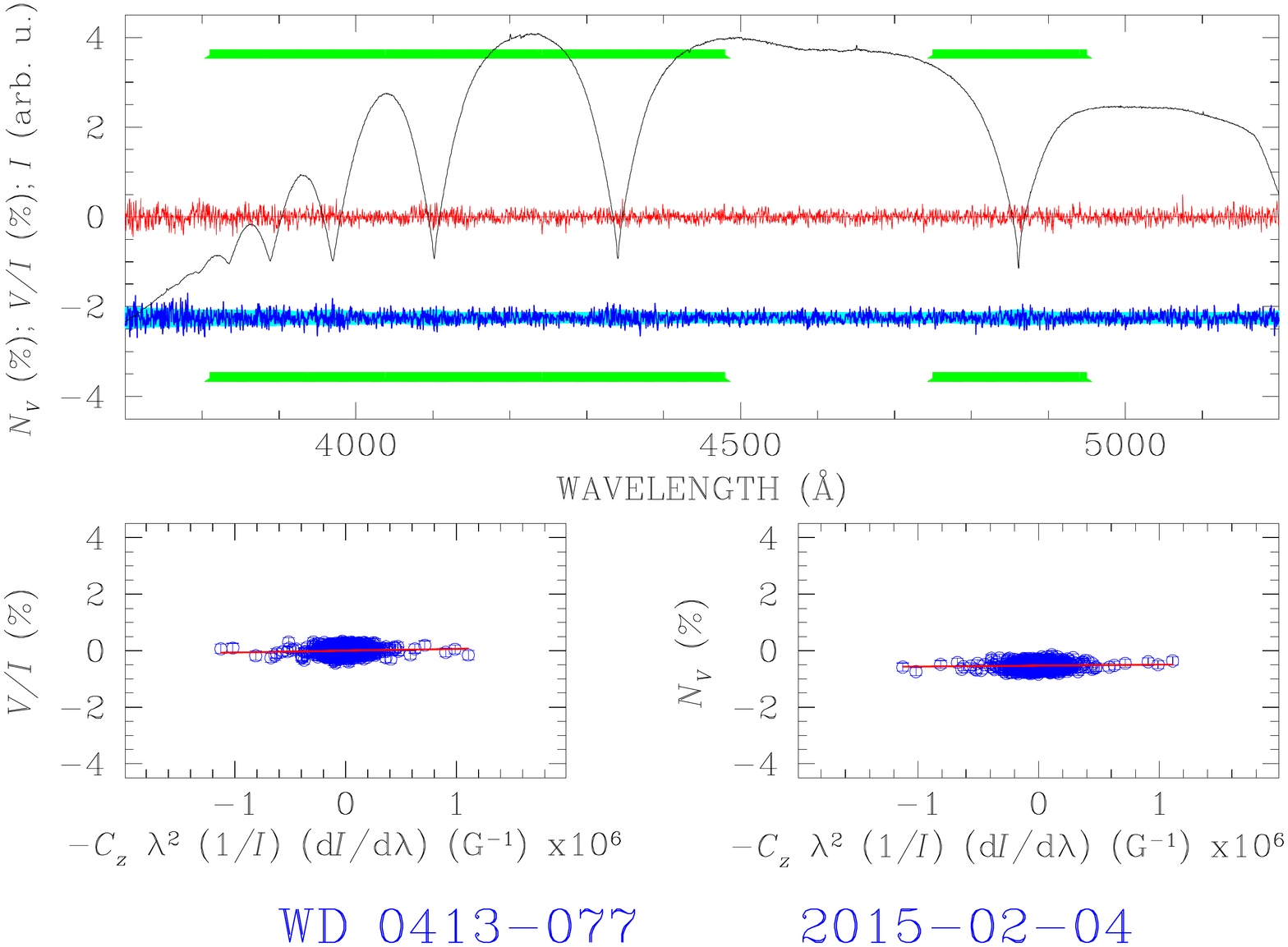}
\includegraphics*[angle=270,width=8.0cm,trim={0.90cm 0.0cm 0.1cm 1.0cm},clip]{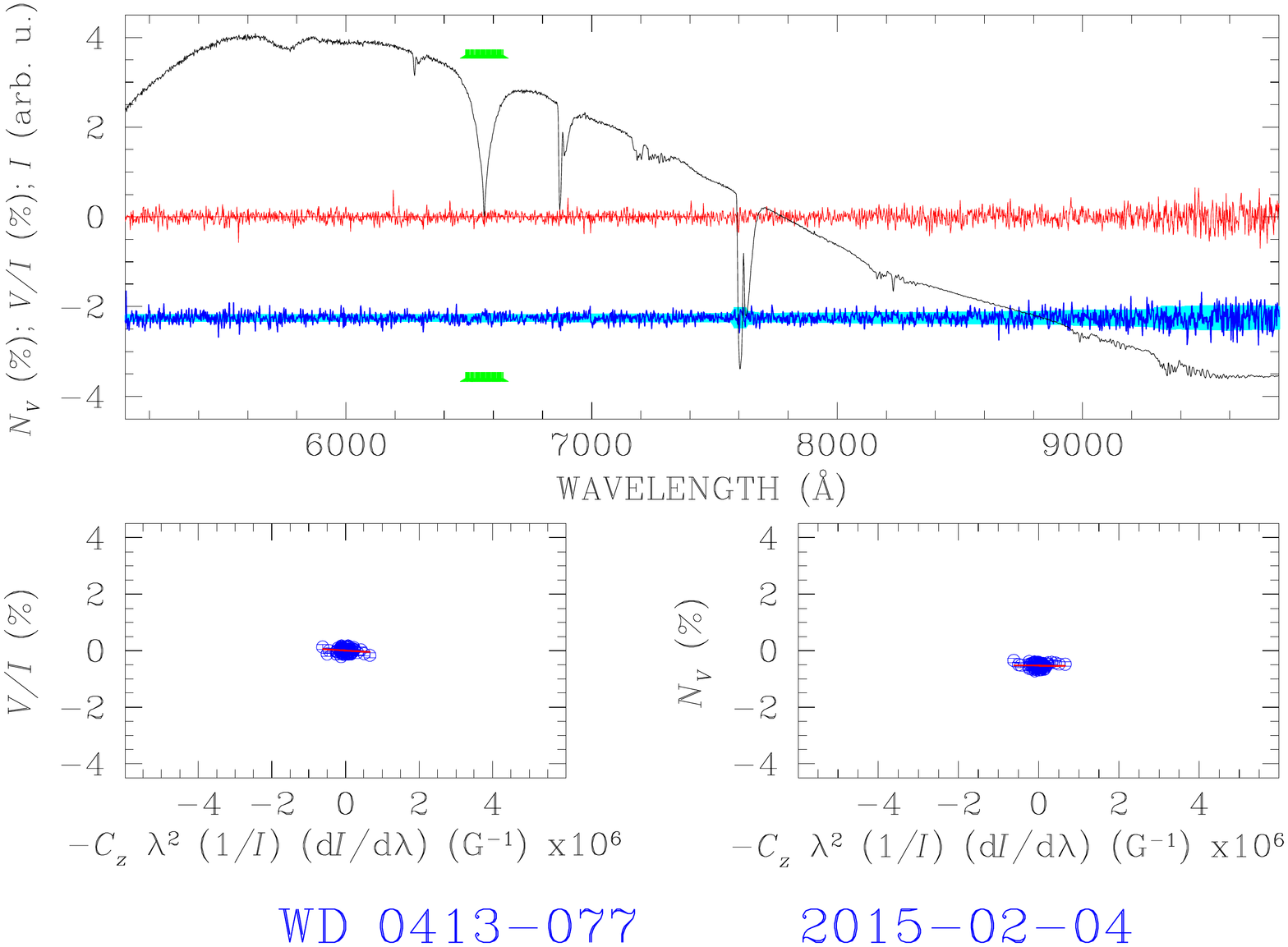} \\
\includegraphics*[angle=270,width=8.0cm,trim={0.90cm 0.0cm 0.1cm 1.0cm},clip]{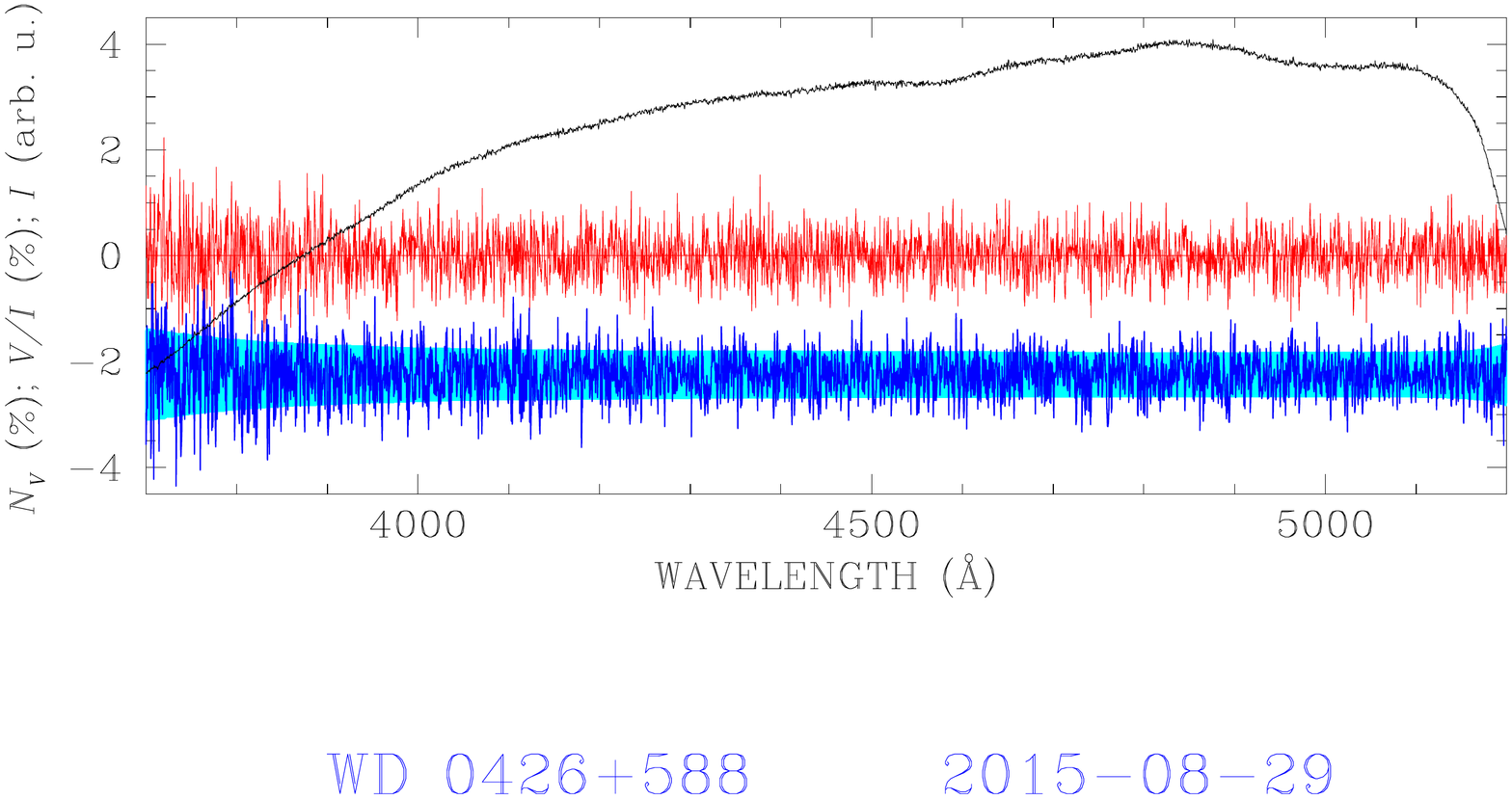}
\includegraphics*[angle=270,width=8.0cm,trim={0.90cm 0.0cm 0.1cm 1.0cm},clip]{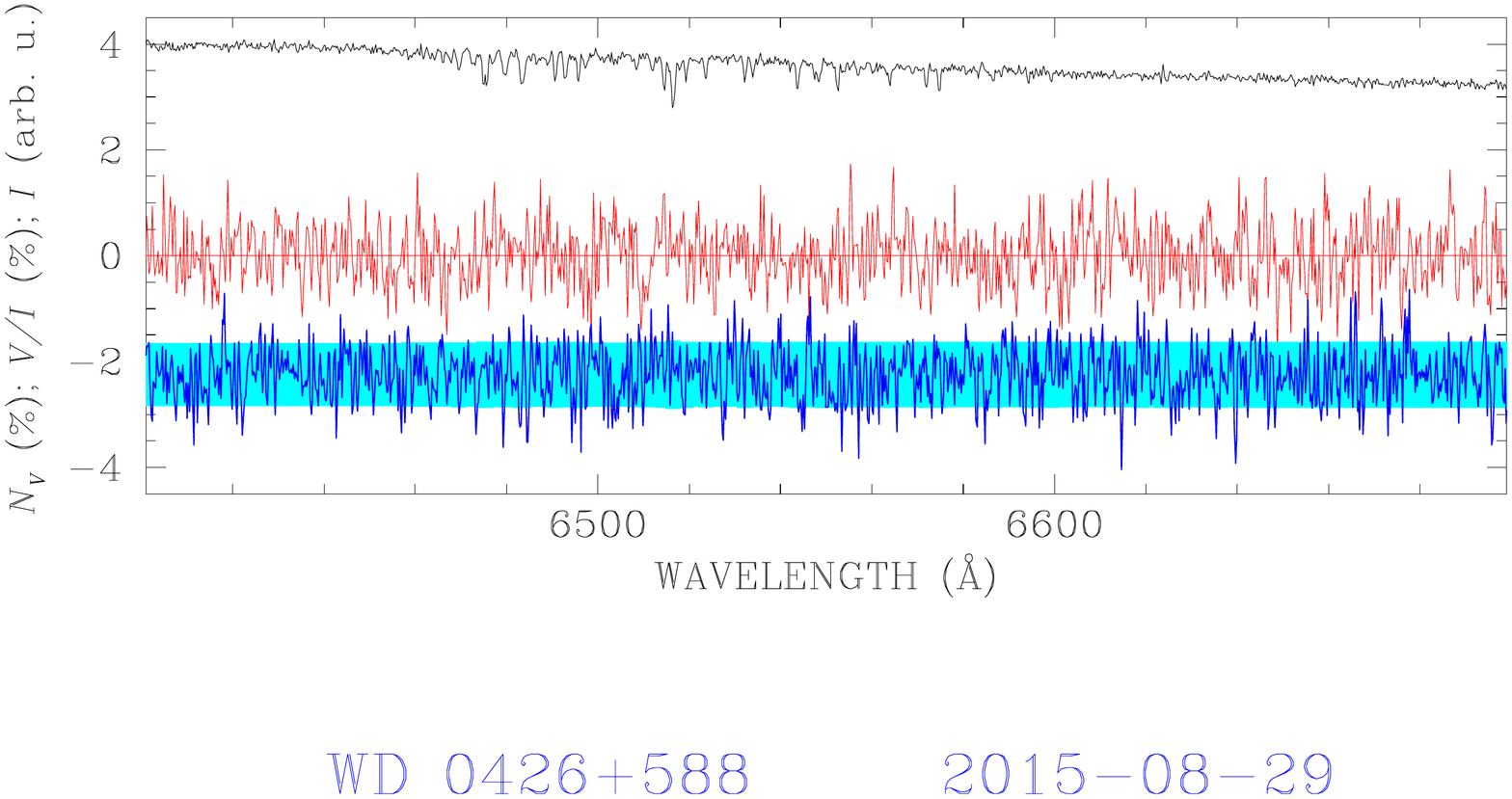} \\
\includegraphics*[angle=270,width=8.0cm,trim={0.90cm 0.0cm 0.1cm 1.0cm},clip]{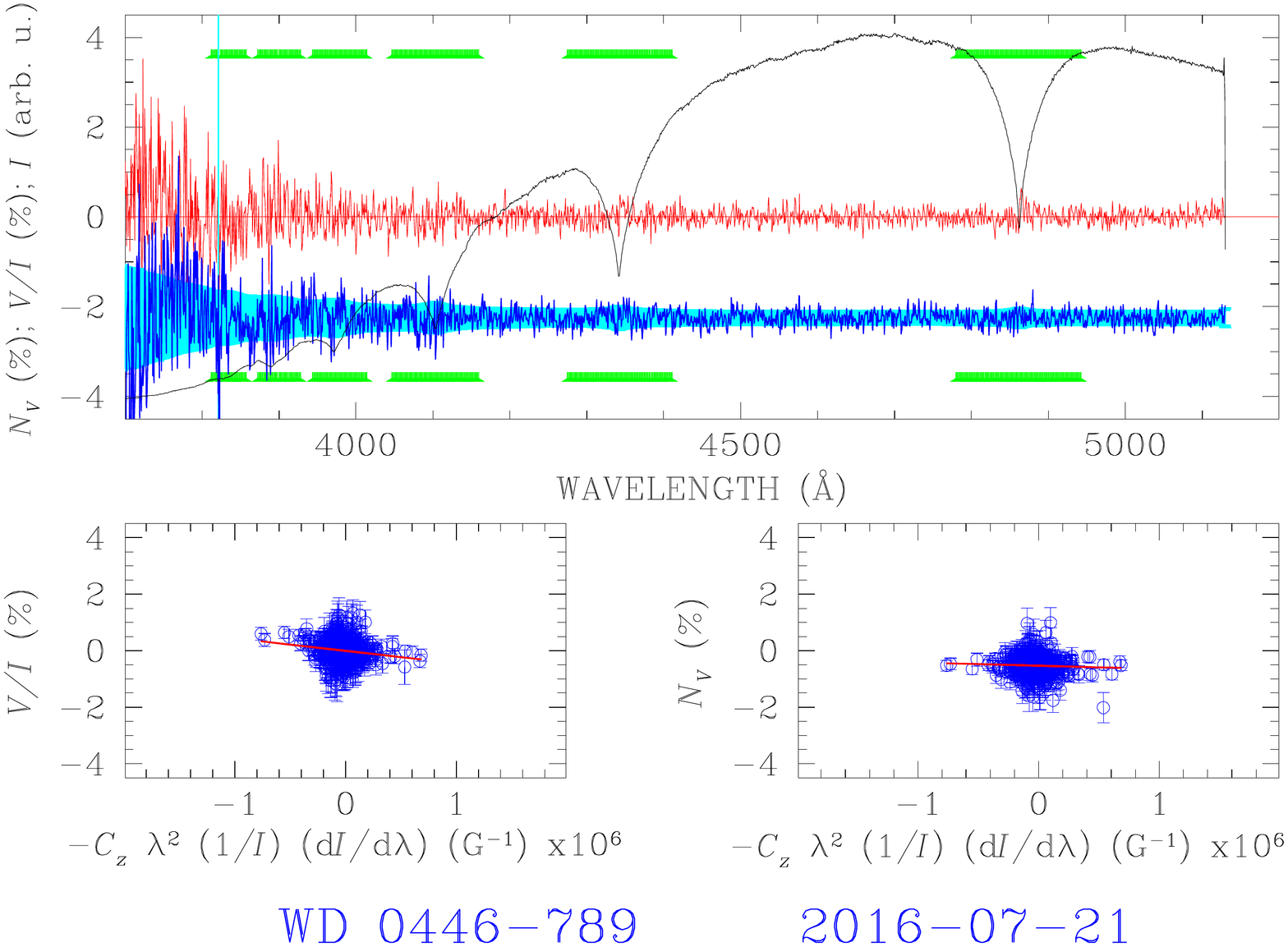}\\
\includegraphics*[angle=270,width=8.0cm,trim={0.90cm 0.0cm 0.1cm 1.0cm},clip]{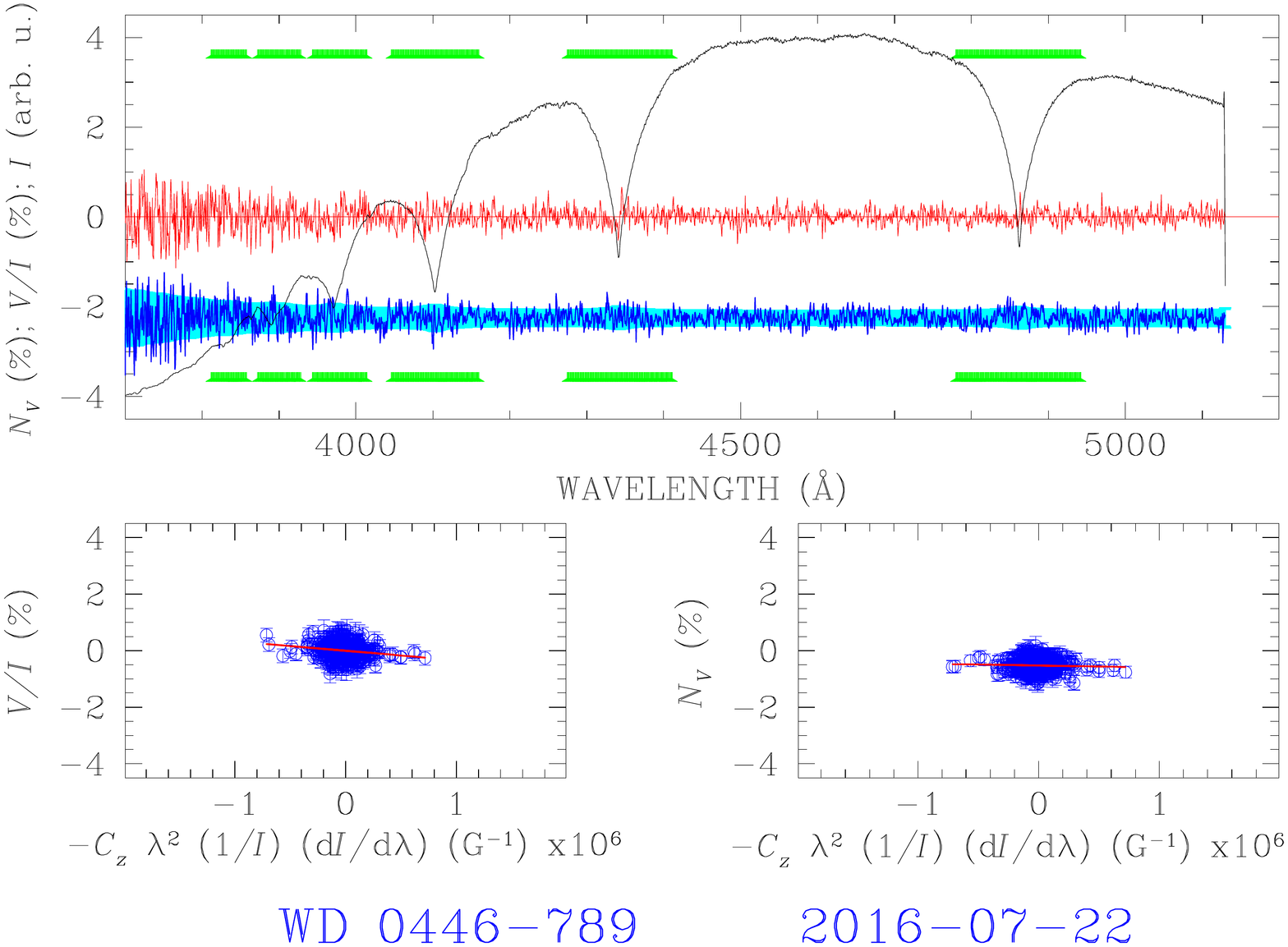}\\
\includegraphics*[angle=270,width=8.0cm,trim={0.90cm 0.0cm 0.1cm 1.0cm},clip]{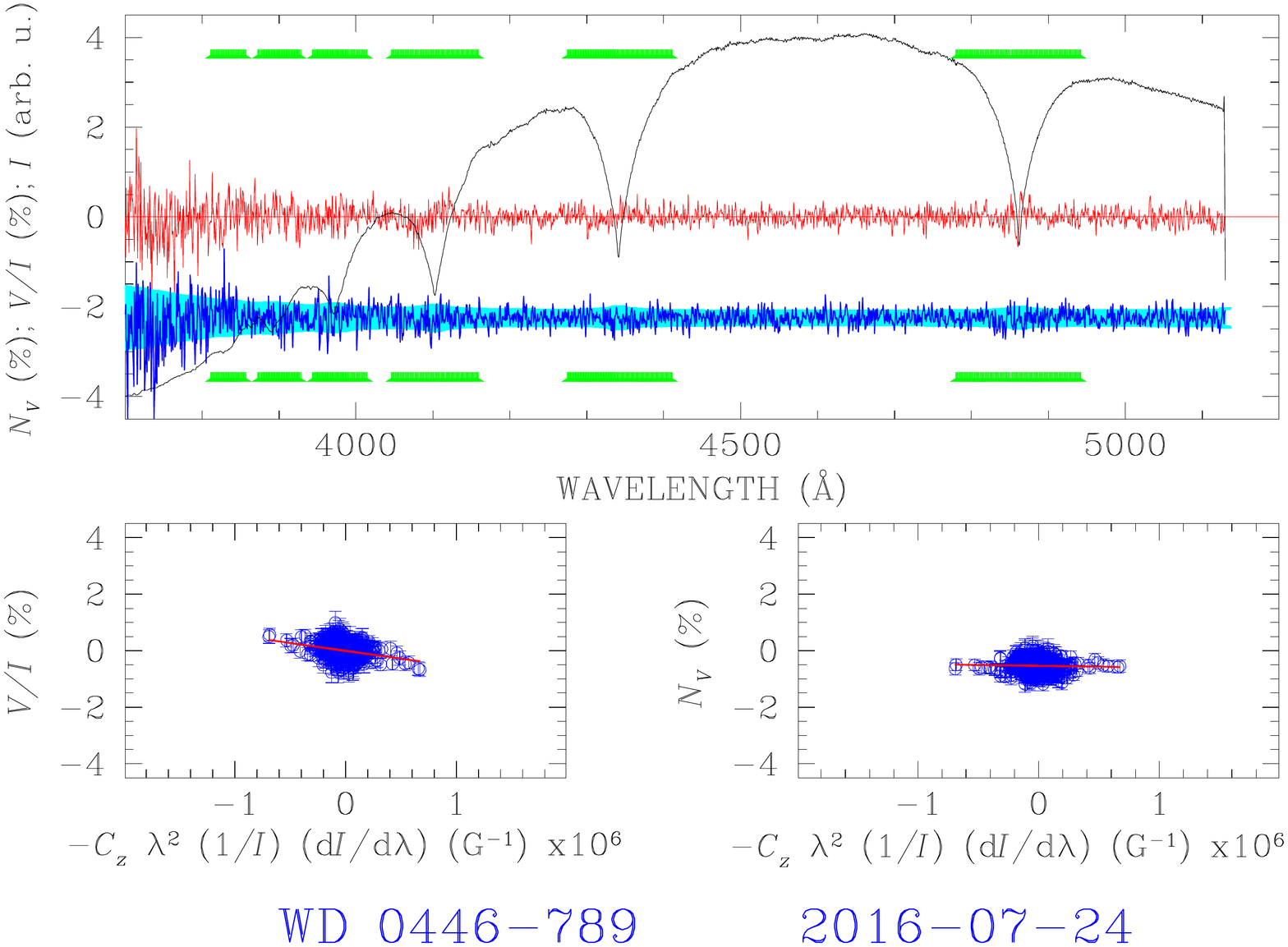}\\
\includegraphics*[angle=270,width=8.0cm,trim={0.90cm 0.0cm 0.1cm 1.0cm},clip]{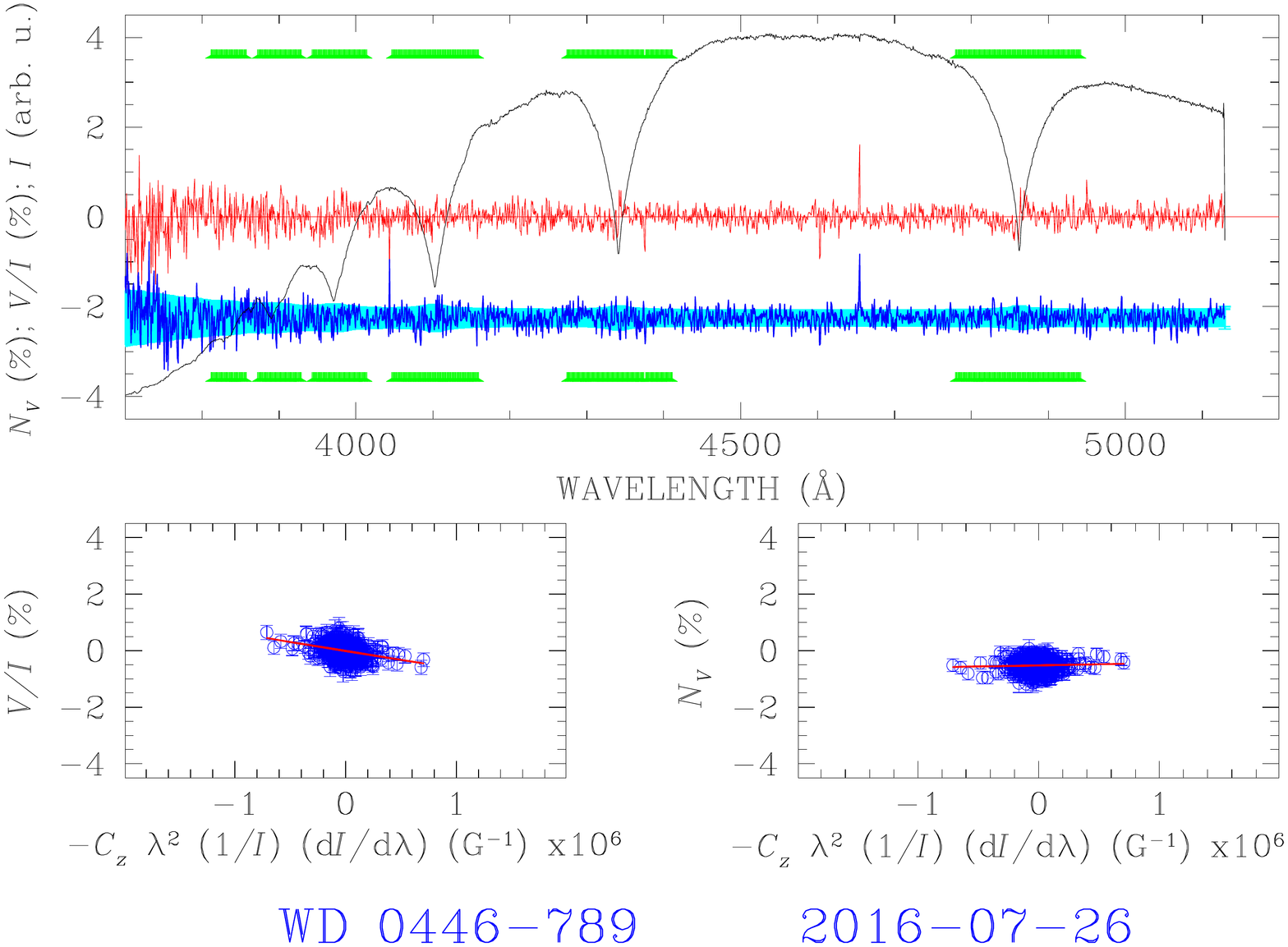}\\
\includegraphics*[angle=270,width=8.0cm,trim={0.90cm 0.0cm 0.1cm 1.0cm},clip]{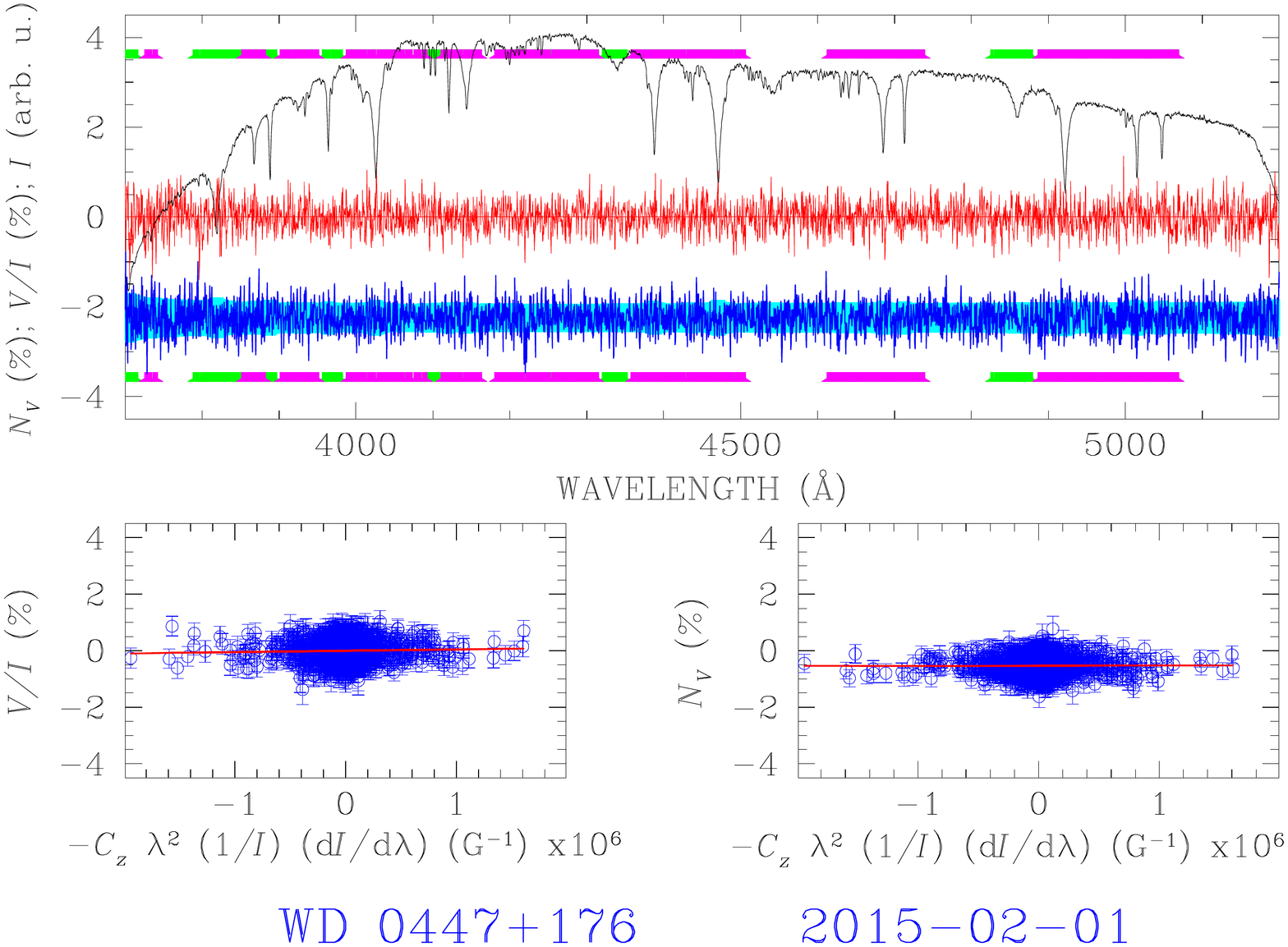}\\
\includegraphics*[angle=270,width=8.0cm,trim={0.90cm 0.0cm 0.1cm 1.0cm},clip]{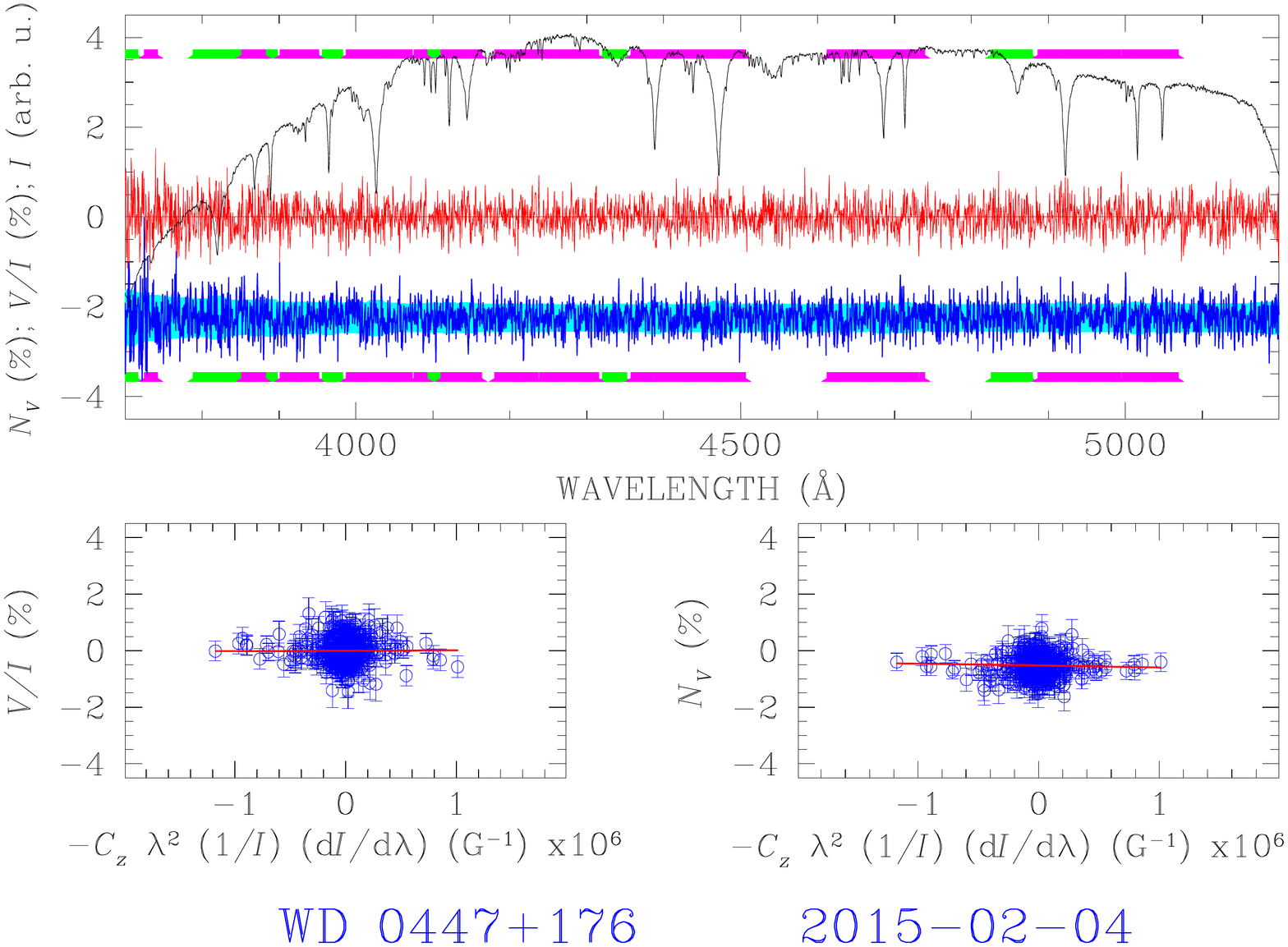}\\
\includegraphics*[angle=270,width=8.0cm,trim={0.90cm 0.0cm 0.1cm 1.0cm},clip]{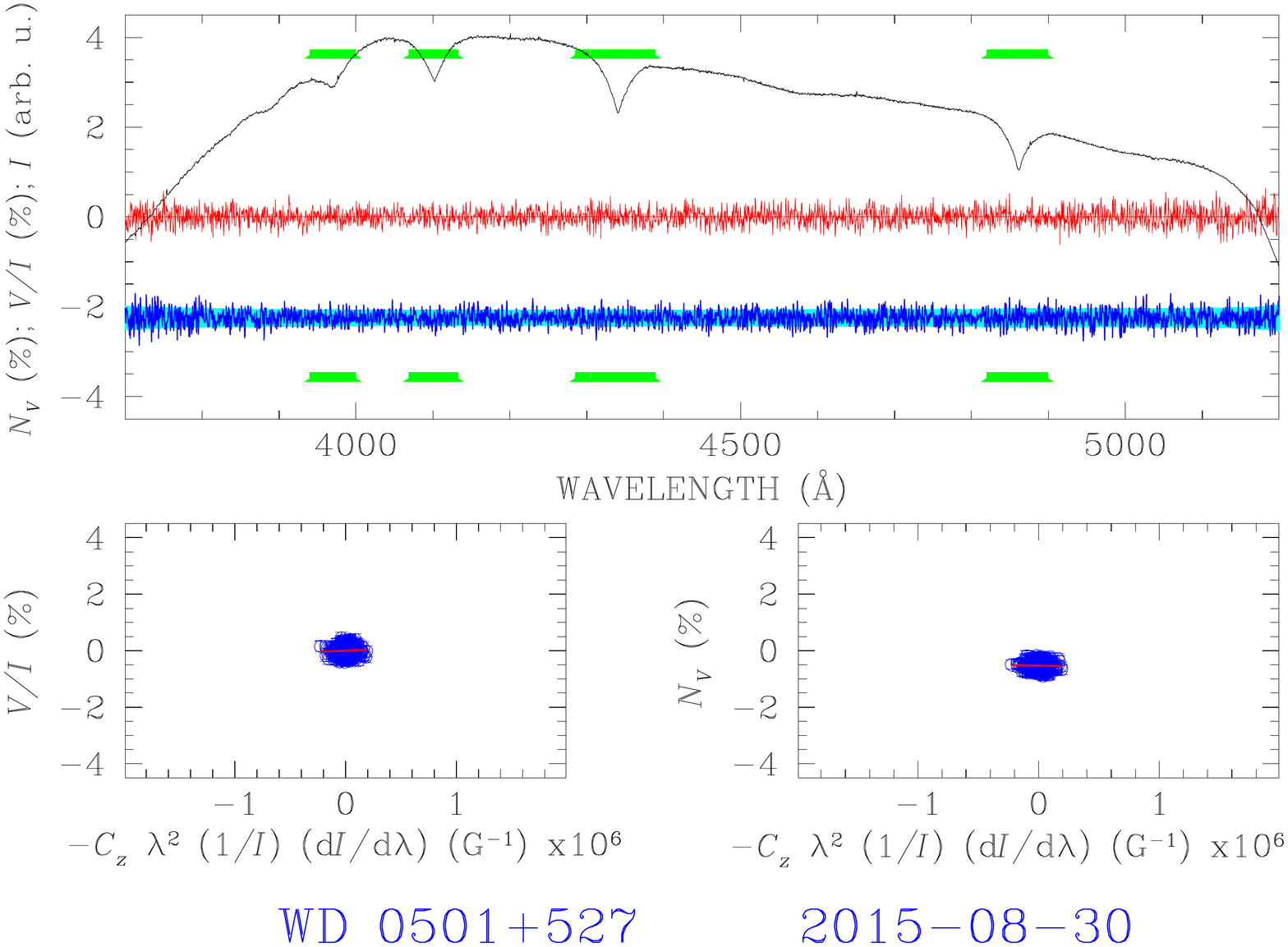}
\includegraphics*[angle=270,width=8.0cm,trim={0.90cm 0.0cm 0.1cm 1.0cm},clip]{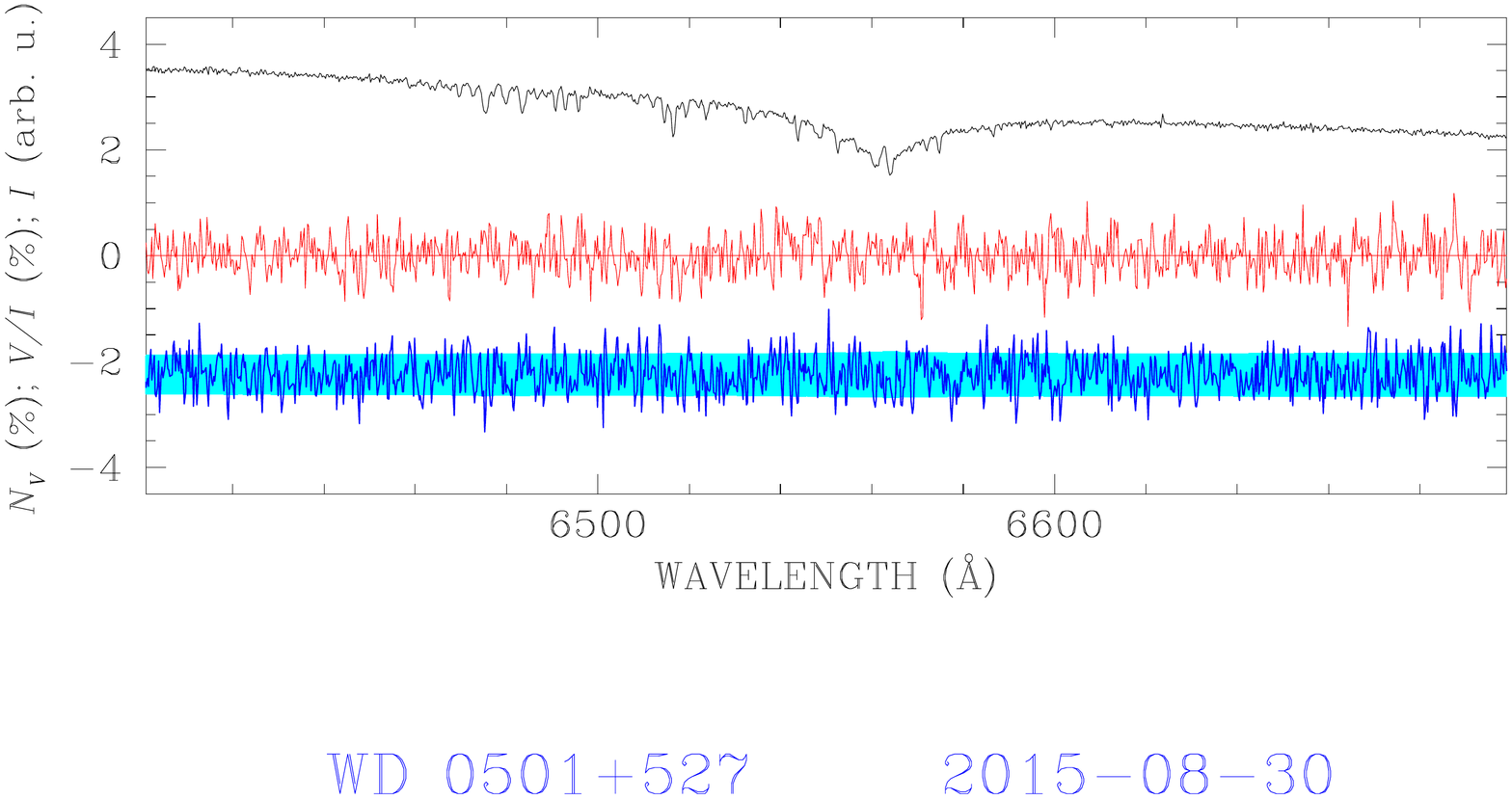} \\
\includegraphics*[angle=270,width=8.0cm,trim={0.90cm 0.0cm 0.1cm 1.0cm},clip]{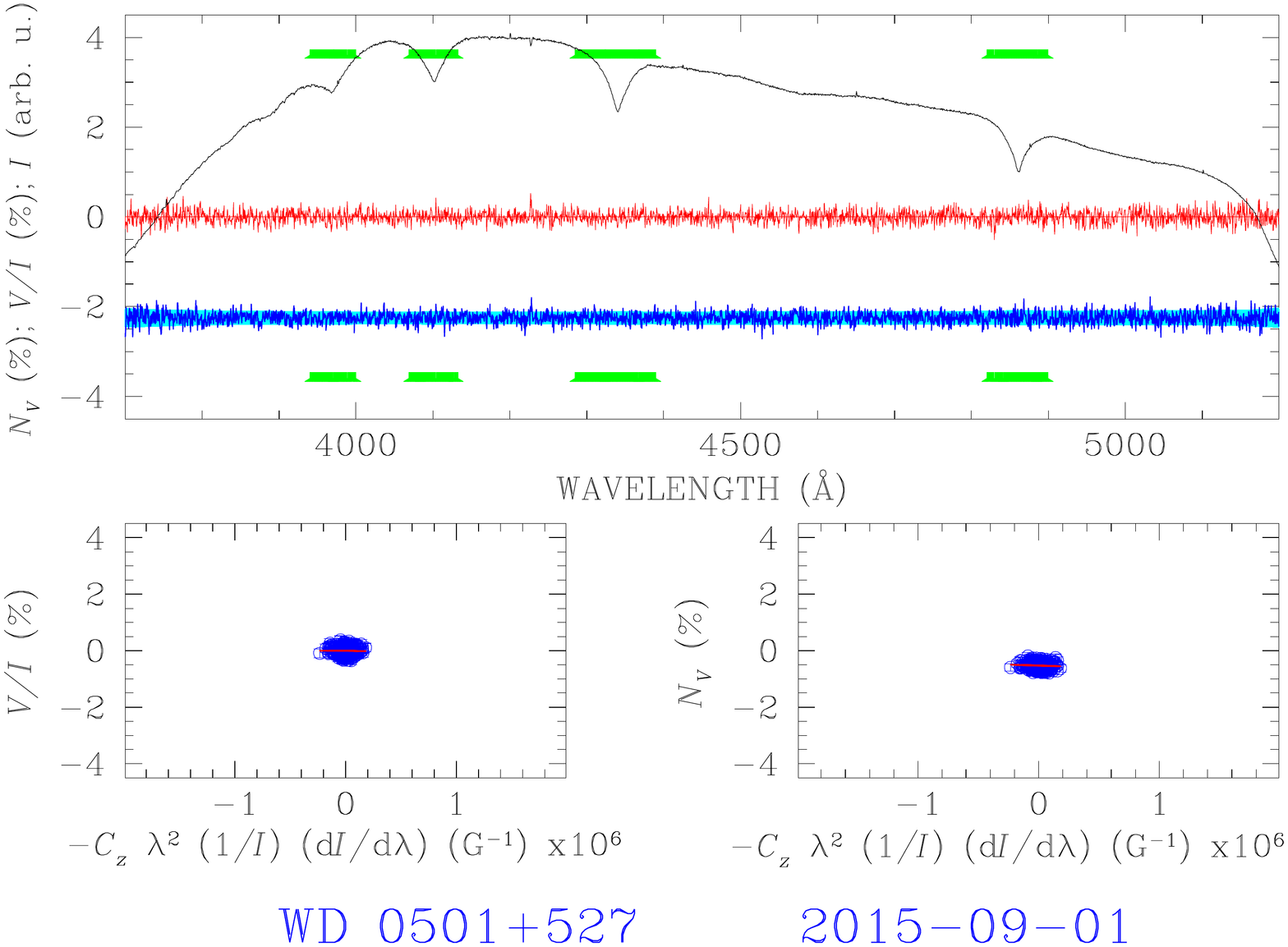}
\includegraphics*[angle=270,width=8.0cm,trim={0.90cm 0.0cm 0.1cm 1.0cm},clip]{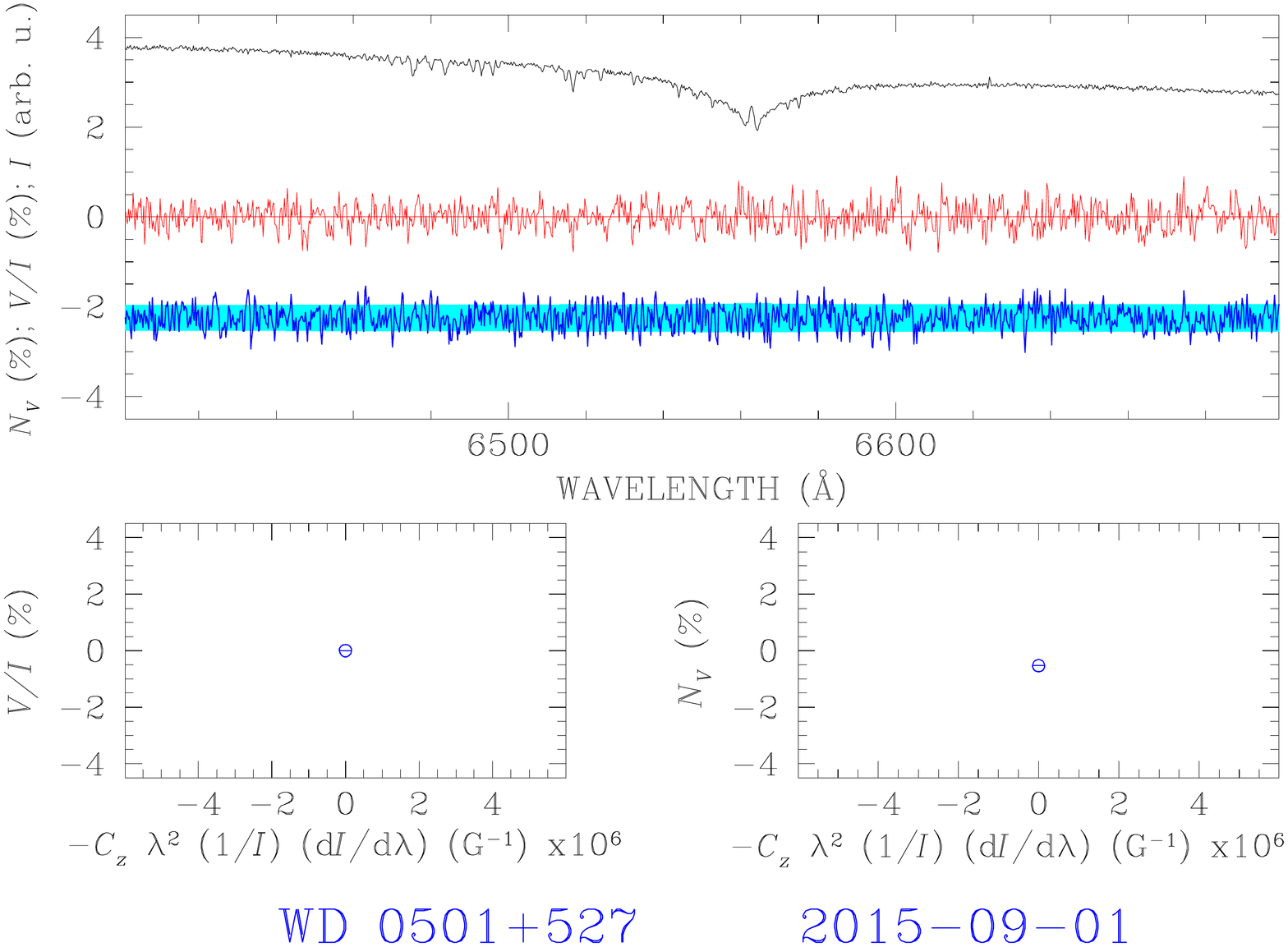} \\
\includegraphics*[angle=270,width=8.0cm,trim={0.90cm 0.0cm 0.1cm 1.0cm},clip]{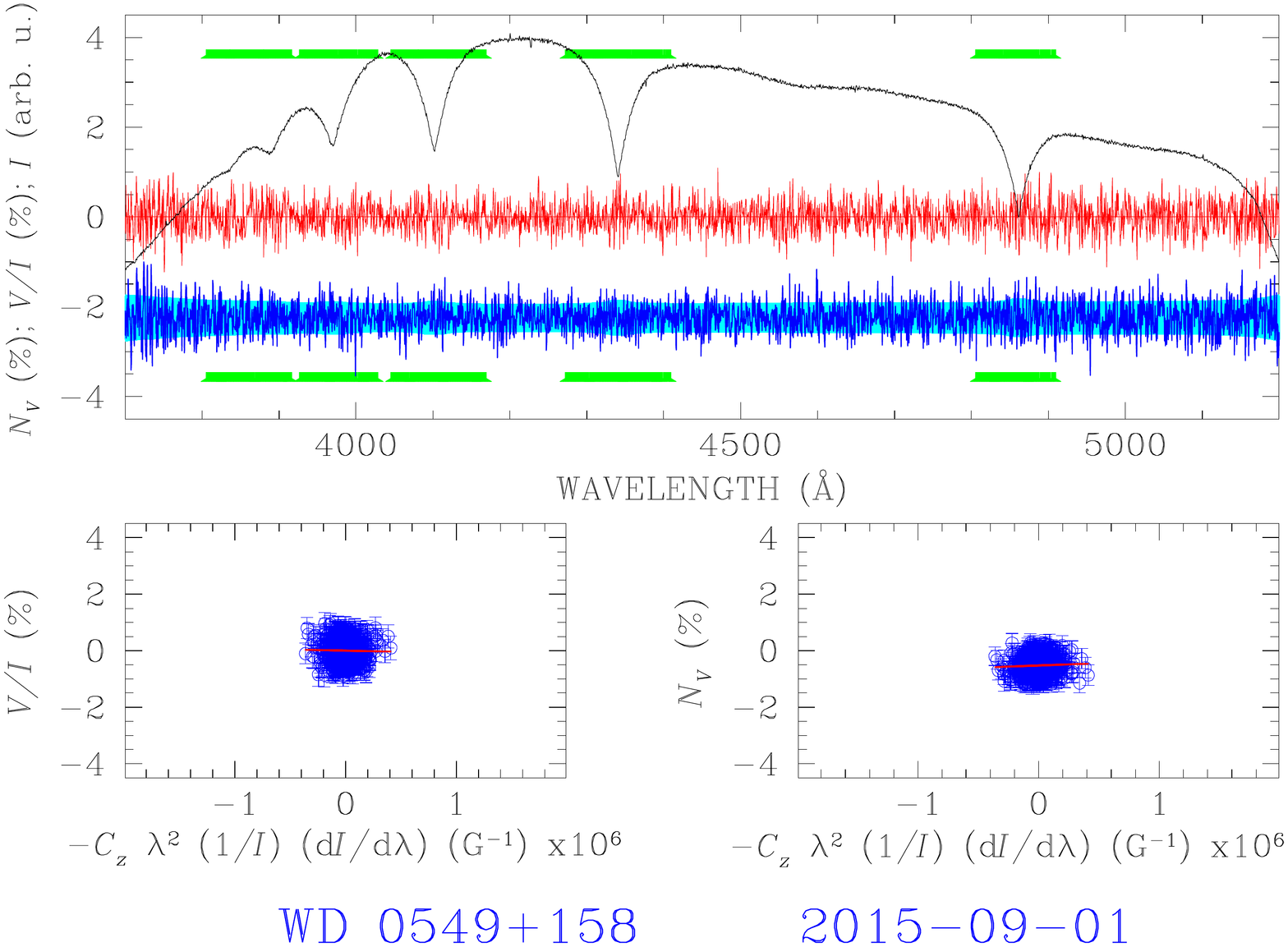}
\includegraphics*[angle=270,width=8.0cm,trim={0.90cm 0.0cm 0.1cm 1.0cm},clip]{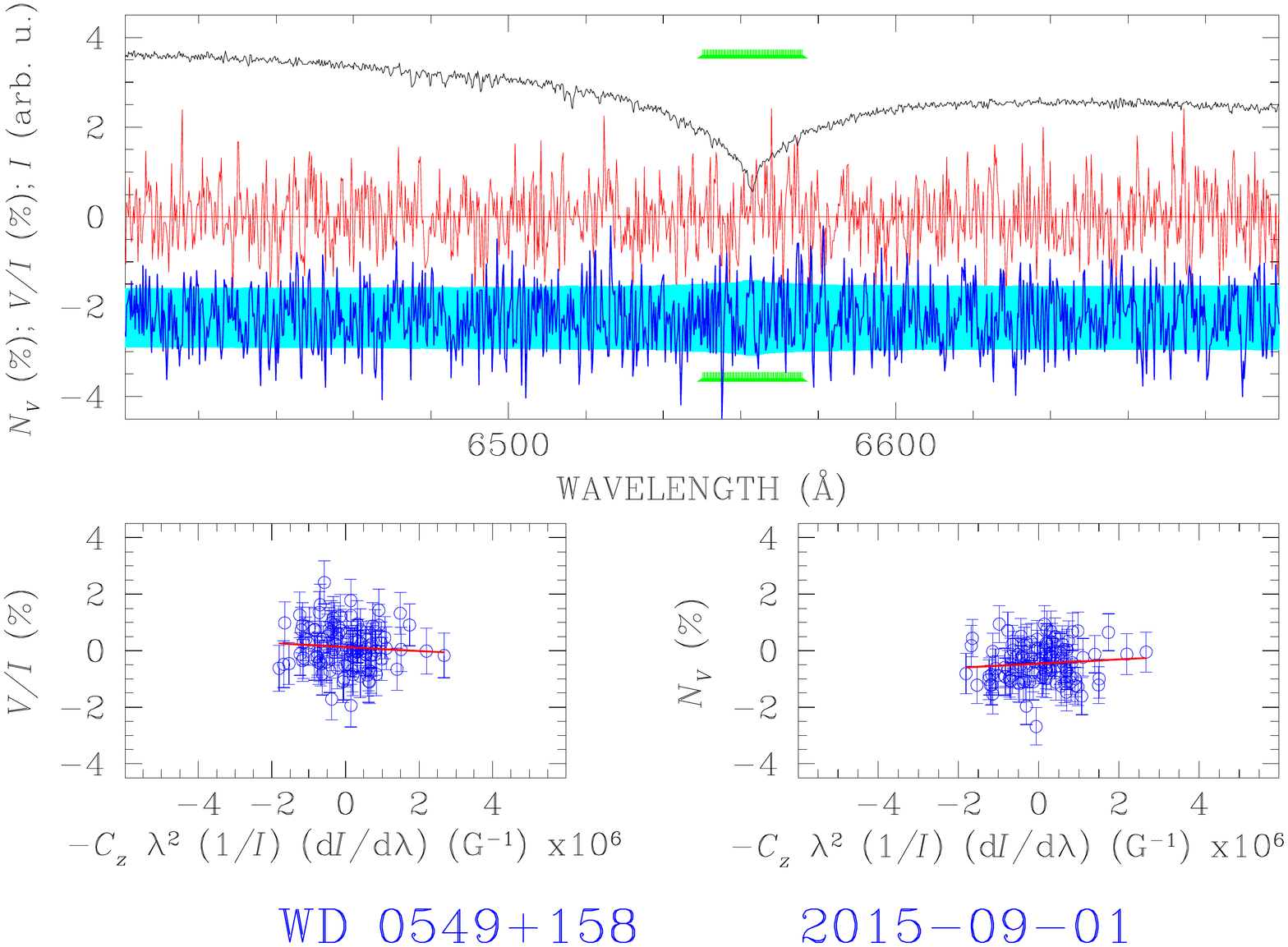} \\
\includegraphics*[angle=270,width=8.0cm,trim={0.90cm 0.0cm 0.1cm 1.0cm},clip]{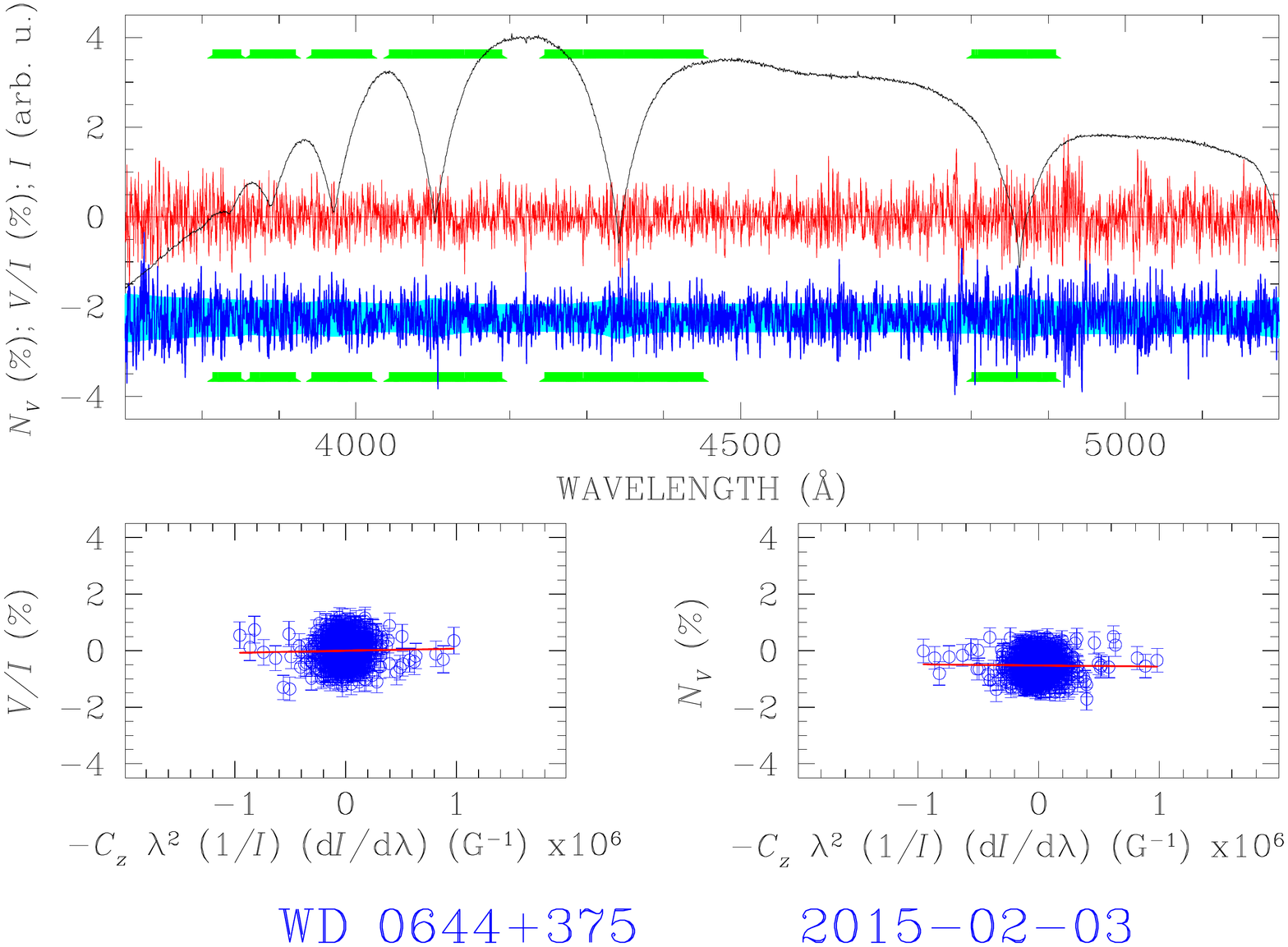} \\
\includegraphics*[angle=270,width=8.0cm,trim={0.90cm 0.0cm 0.1cm 1.0cm},clip]{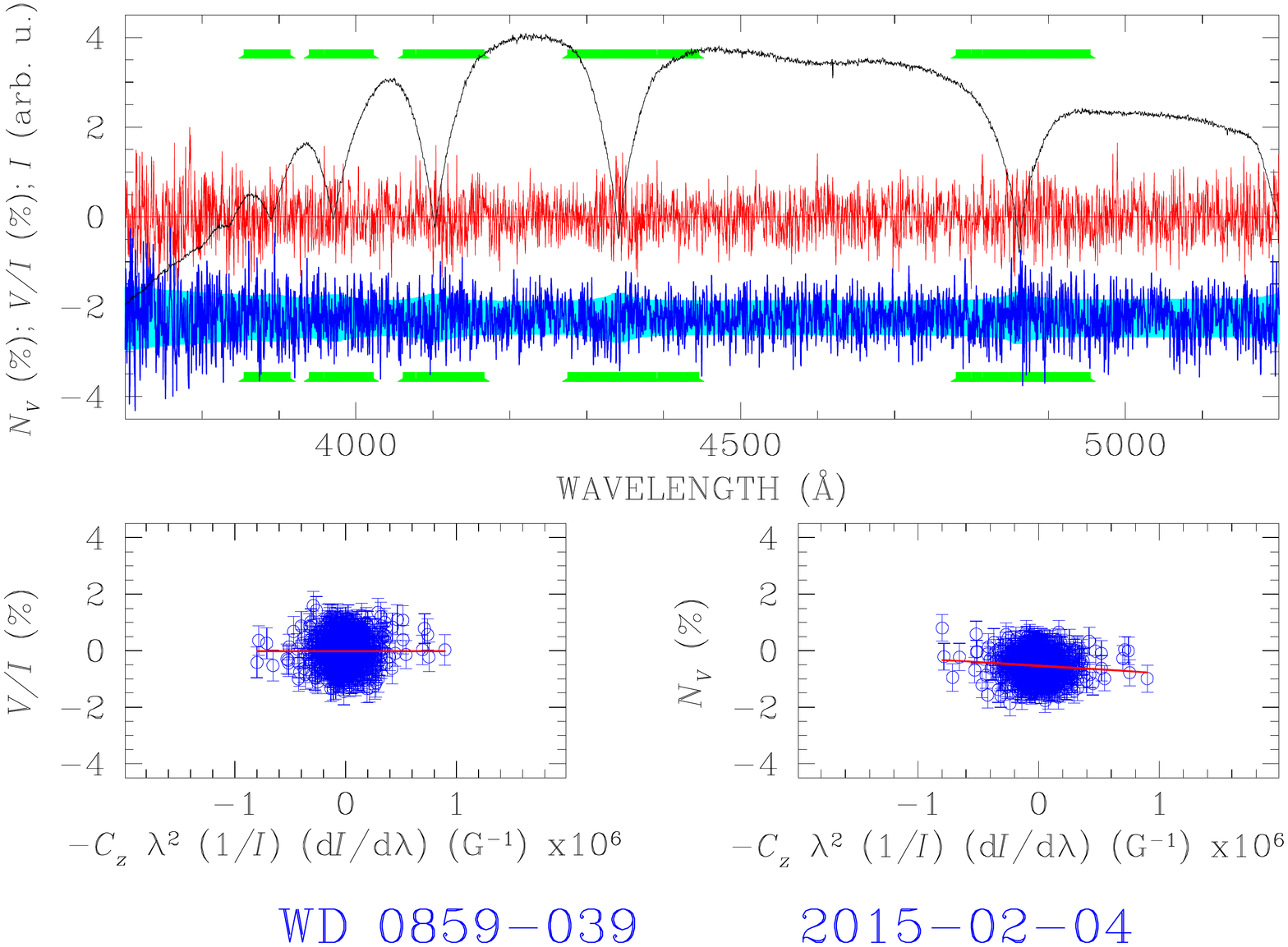} \\
\includegraphics*[angle=270,width=8.0cm,trim={0.90cm 0.0cm 0.1cm 1.0cm},clip]{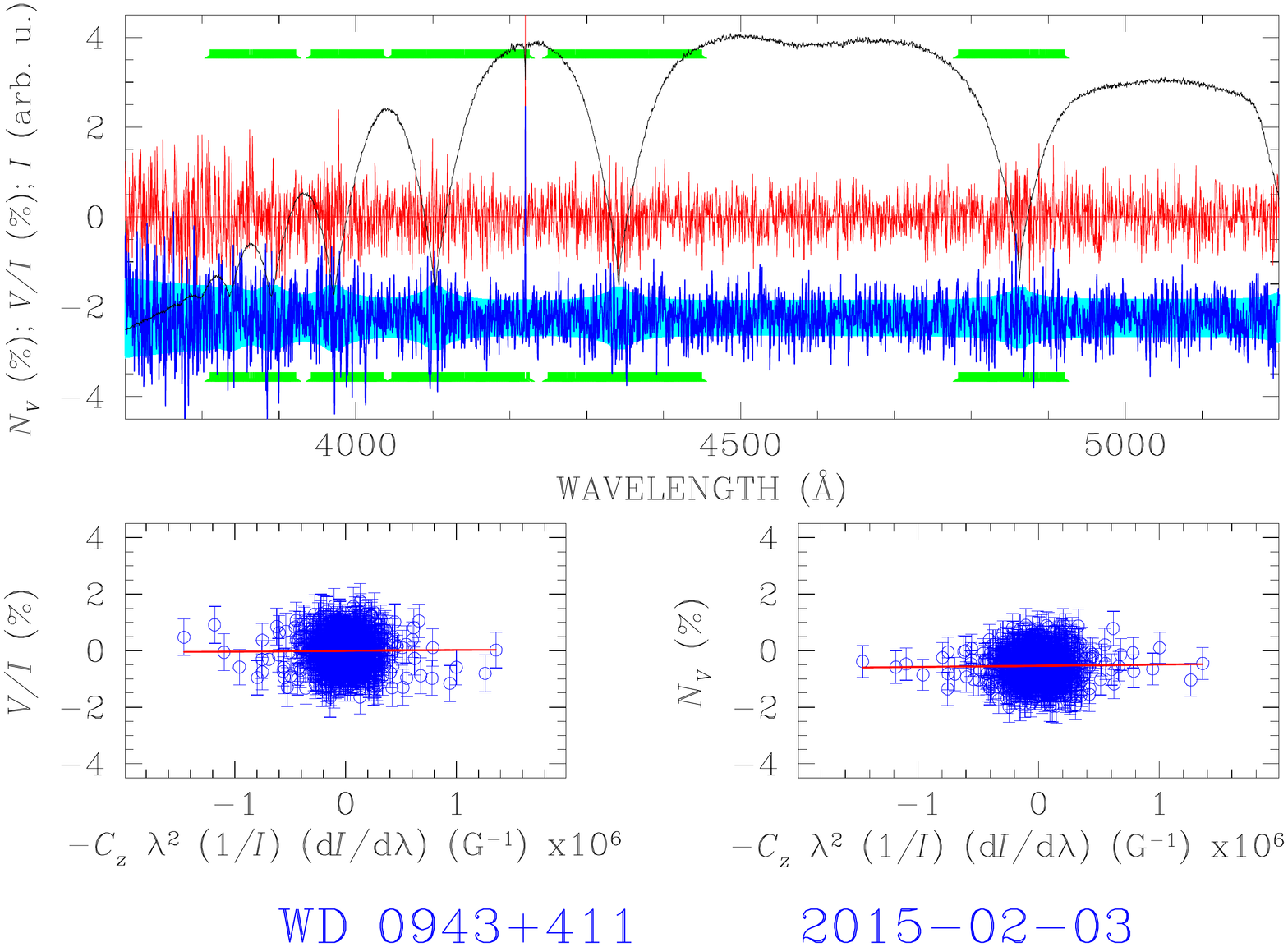} \\
\includegraphics*[angle=270,width=8.0cm,trim={0.90cm 0.0cm 0.1cm 1.0cm},clip]{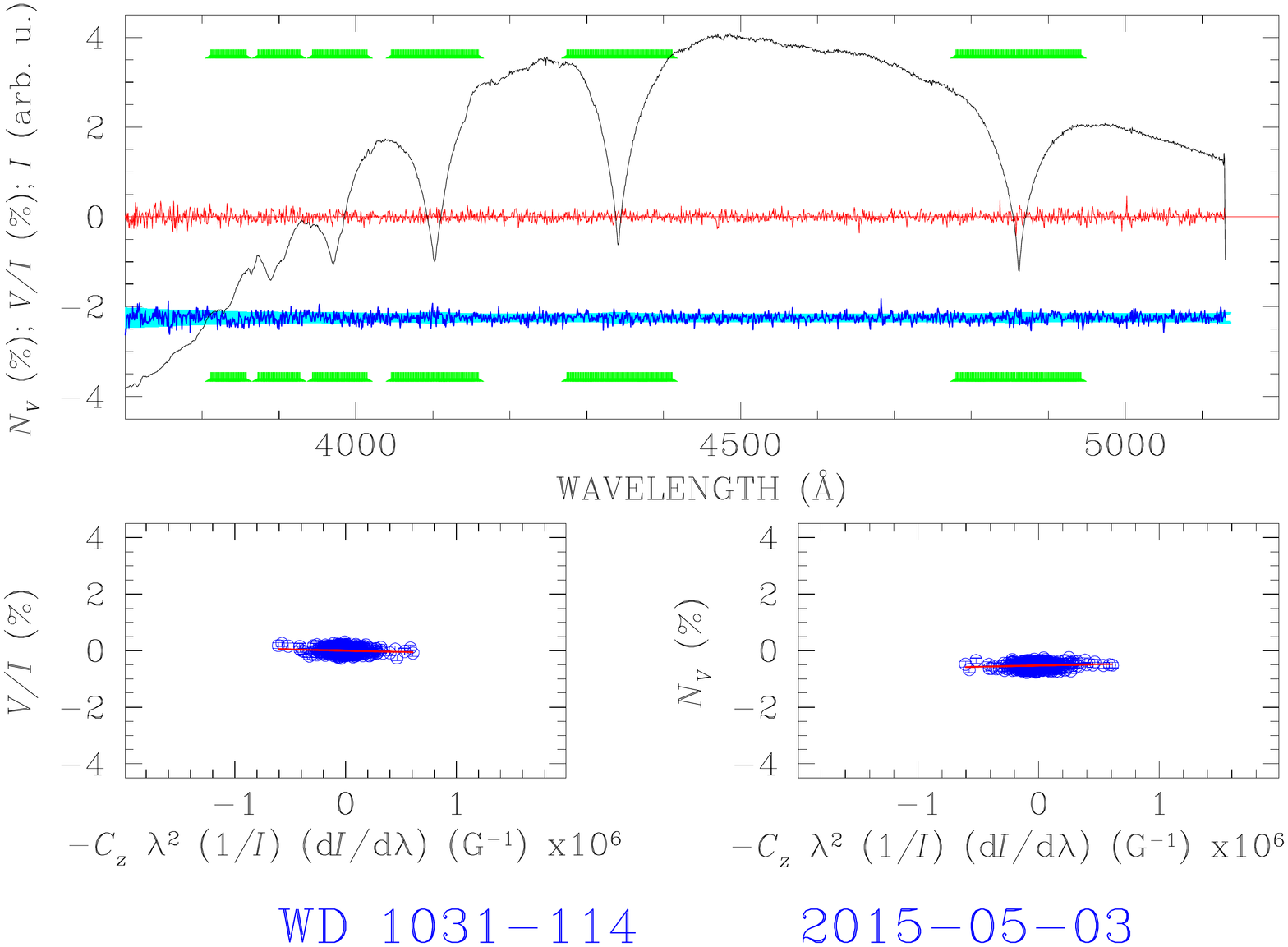} \\
\includegraphics*[angle=270,width=8.0cm,trim={0.90cm 0.0cm 0.1cm 1.0cm},clip]{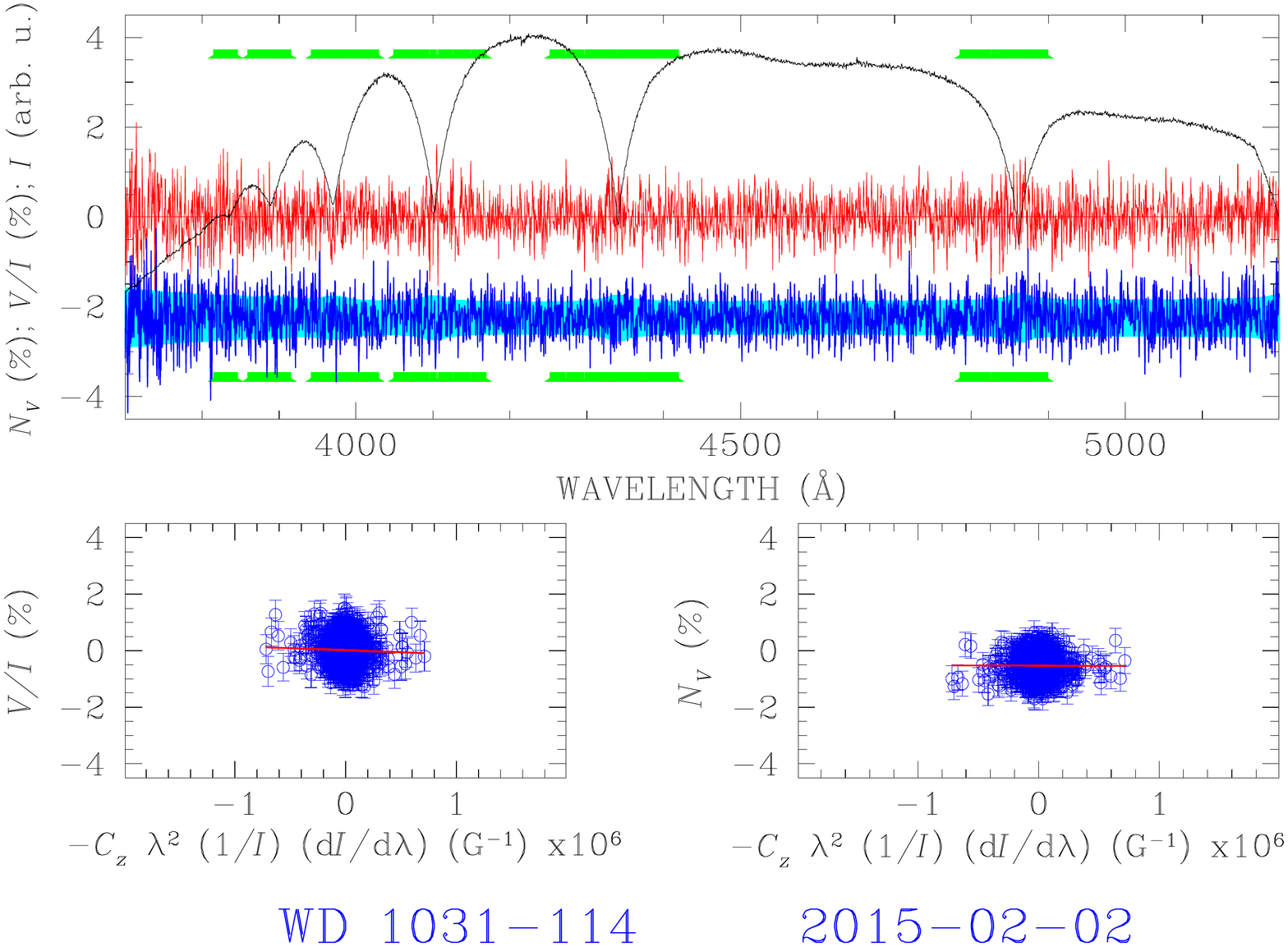} \\
\includegraphics*[angle=270,width=8.0cm,trim={0.90cm 0.0cm 0.1cm 1.0cm},clip]{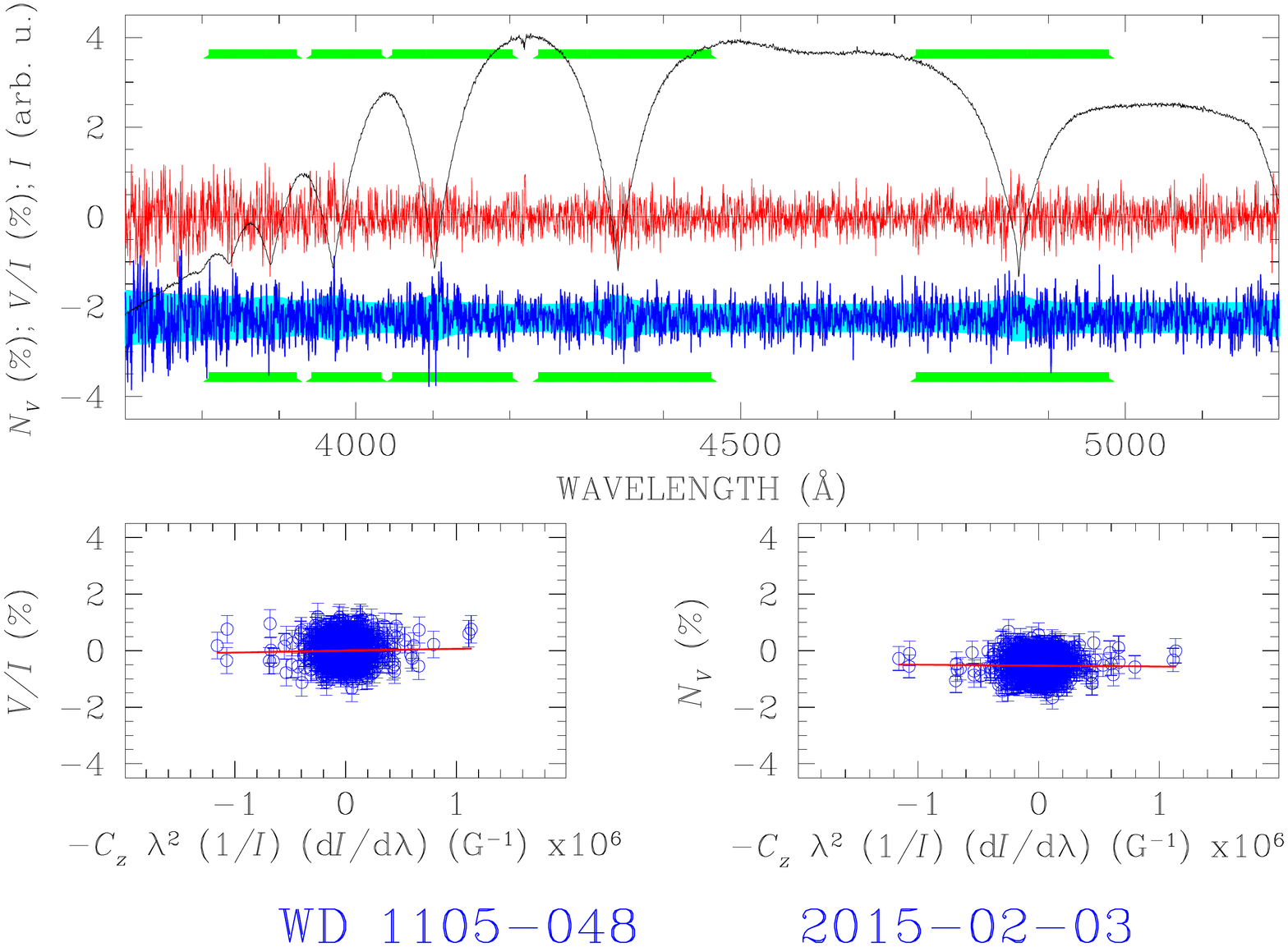} \\
\includegraphics*[angle=270,width=8.0cm,trim={0.90cm 0.0cm 0.1cm 1.0cm},clip]{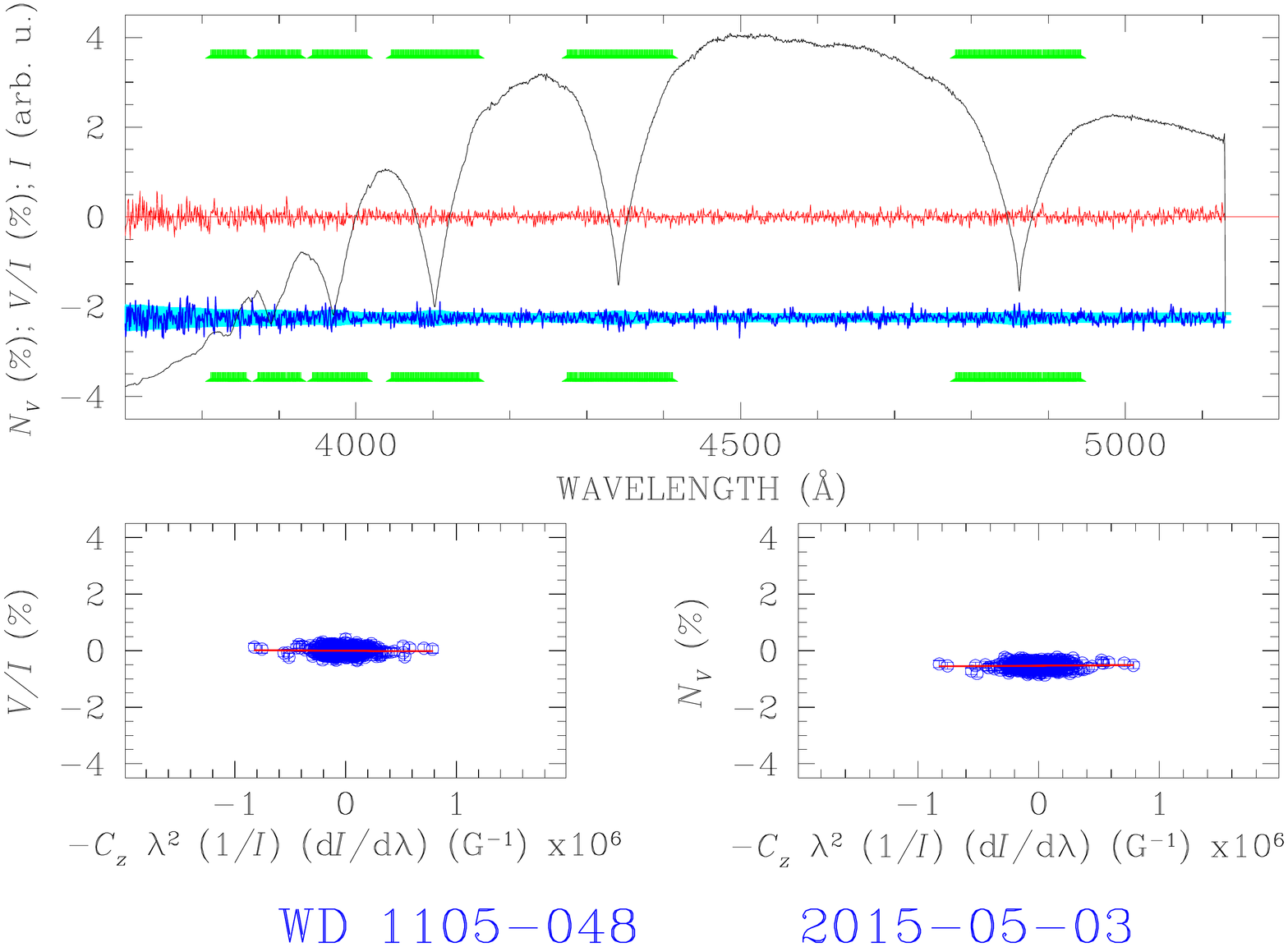} \\
\includegraphics*[angle=270,width=8.0cm,trim={0.90cm 0.0cm 0.1cm 1.0cm},clip]{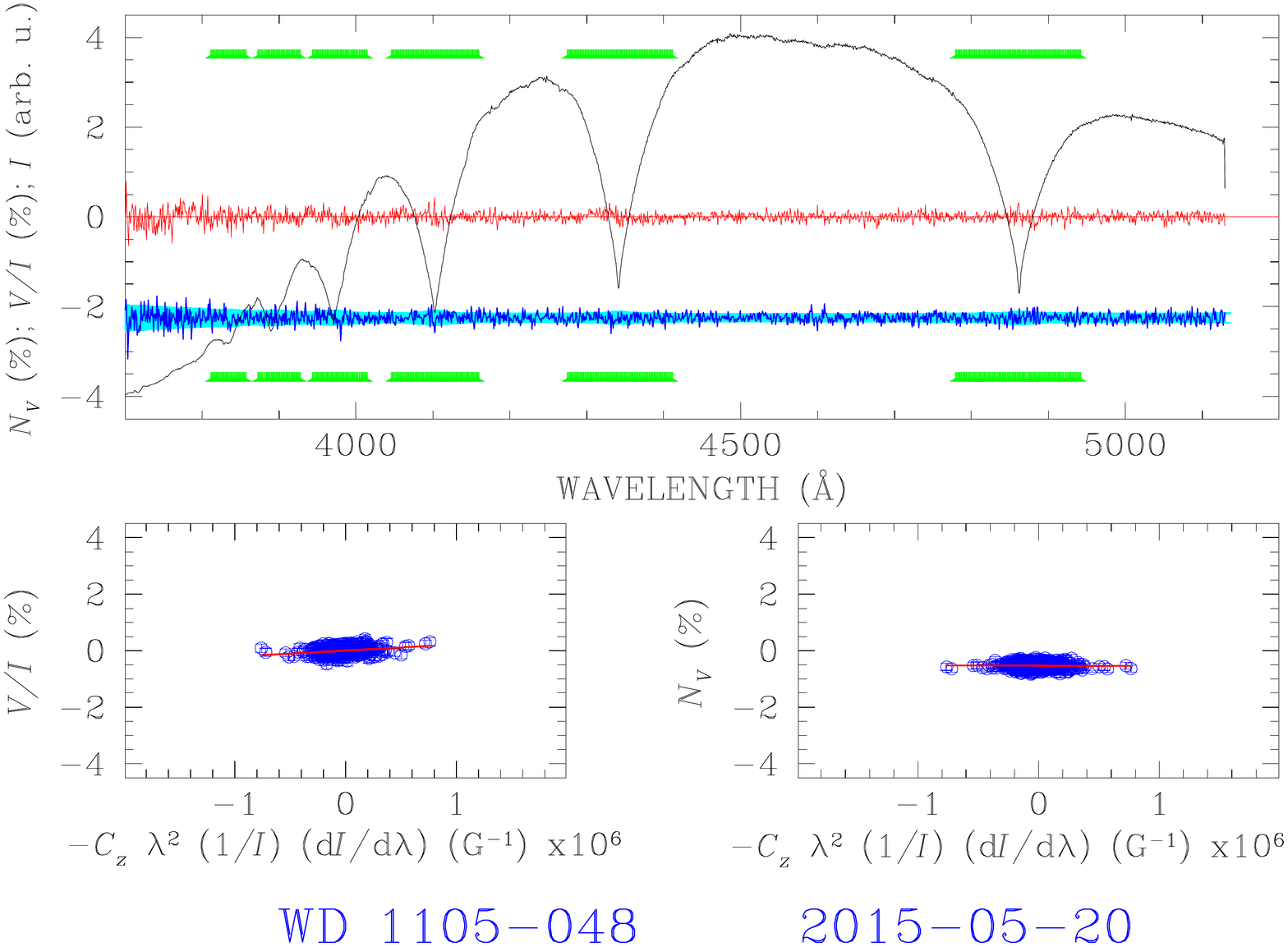} \\
\includegraphics*[angle=270,width=8.0cm,trim={0.90cm 0.0cm 0.1cm 1.0cm},clip]{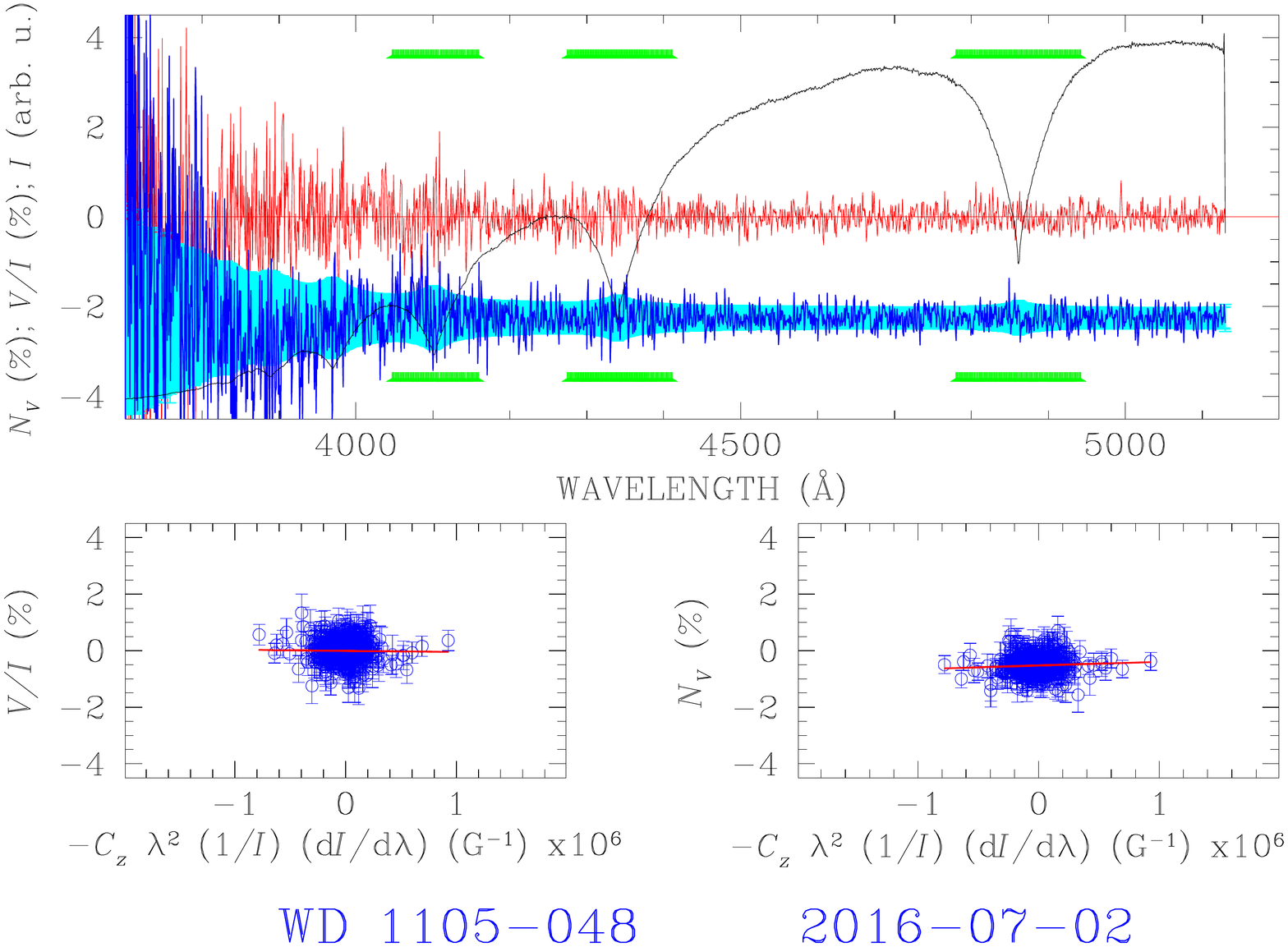} \\
\includegraphics*[angle=270,width=8.0cm,trim={0.90cm 0.0cm 0.1cm 1.0cm},clip]{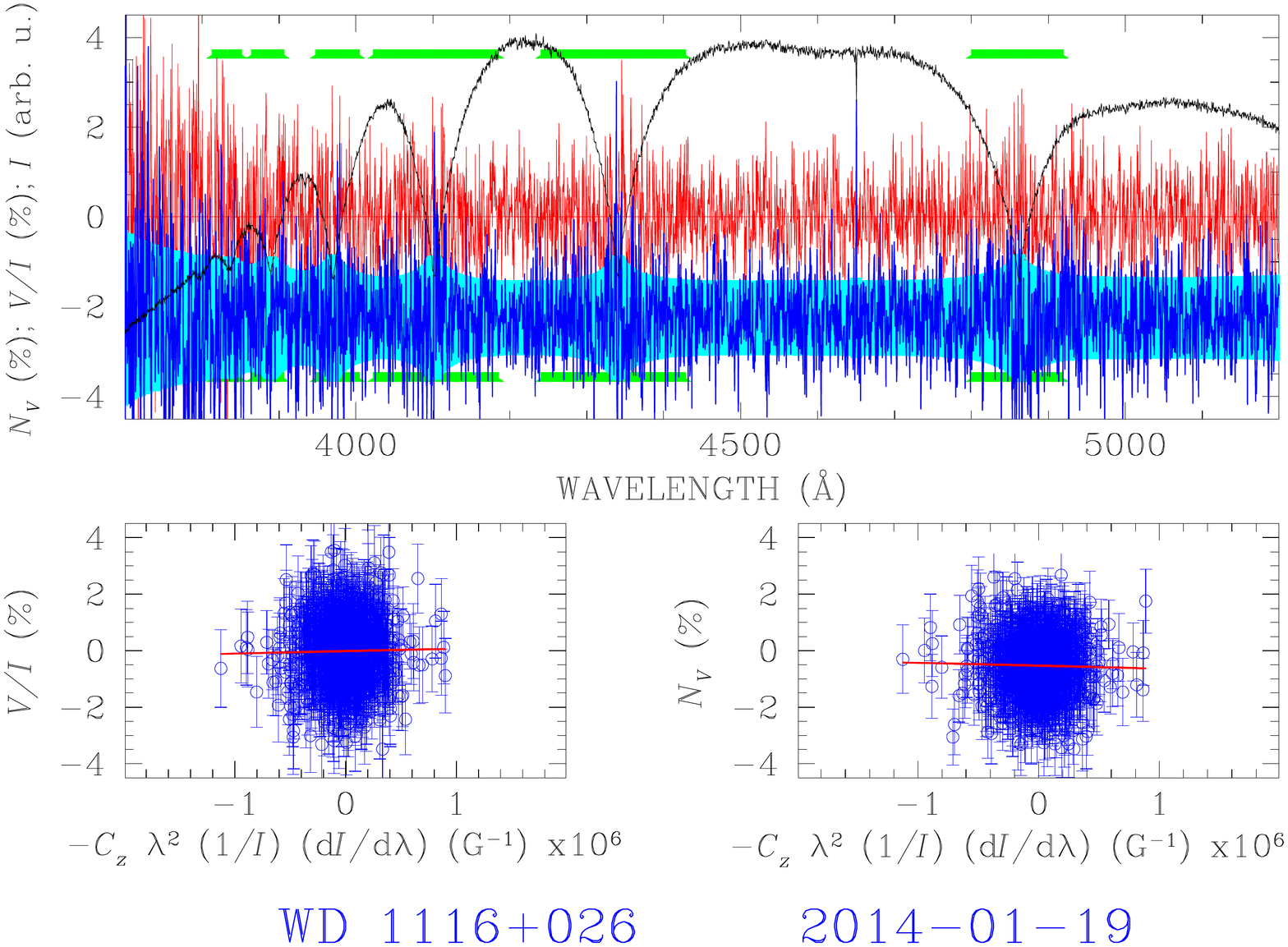} \\
\includegraphics*[angle=270,width=8.0cm,trim={0.90cm 0.0cm 0.1cm 1.0cm},clip]{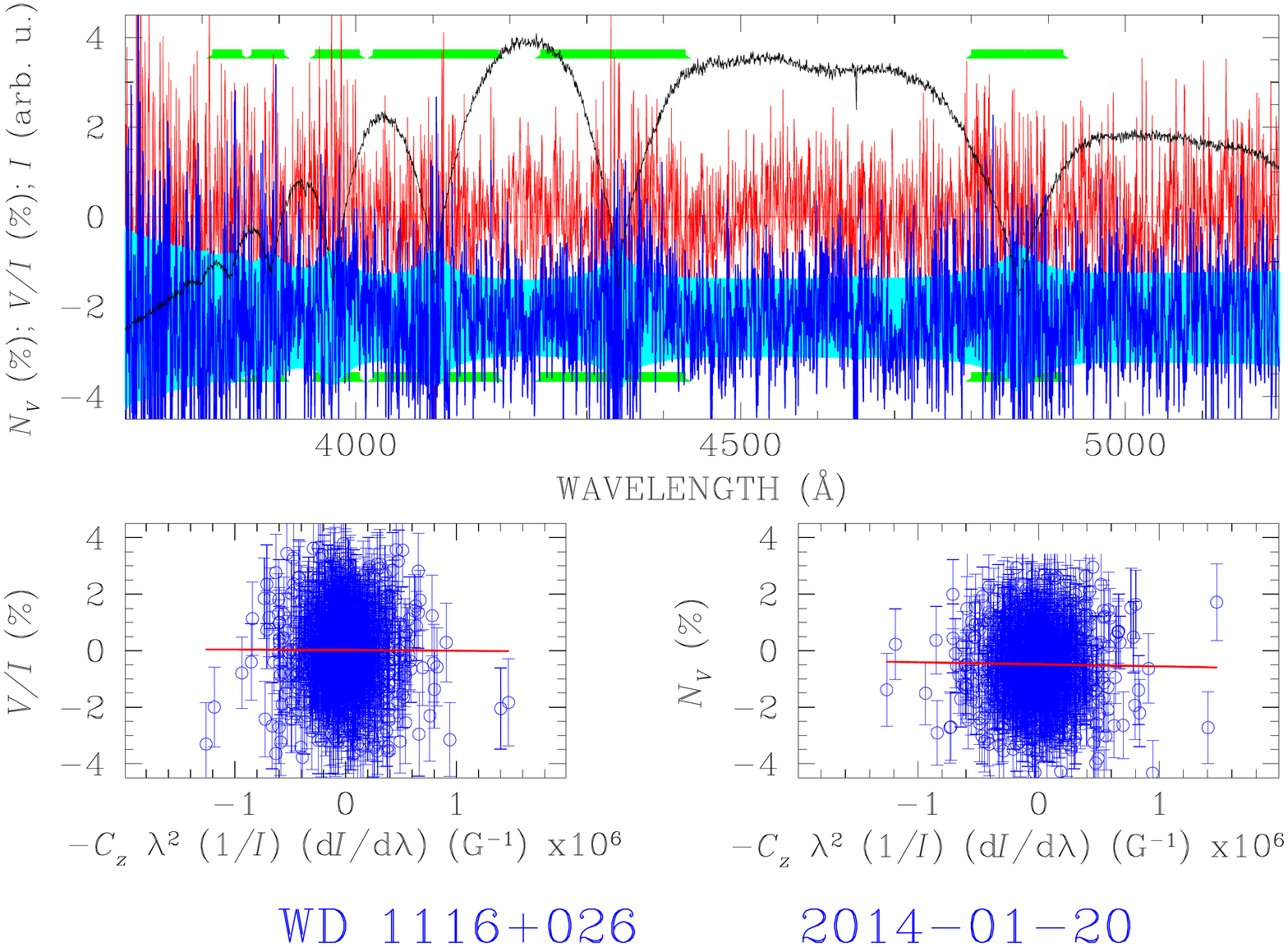} \\
\includegraphics*[angle=270,width=8.0cm,trim={0.90cm 0.0cm 0.1cm 1.0cm},clip]{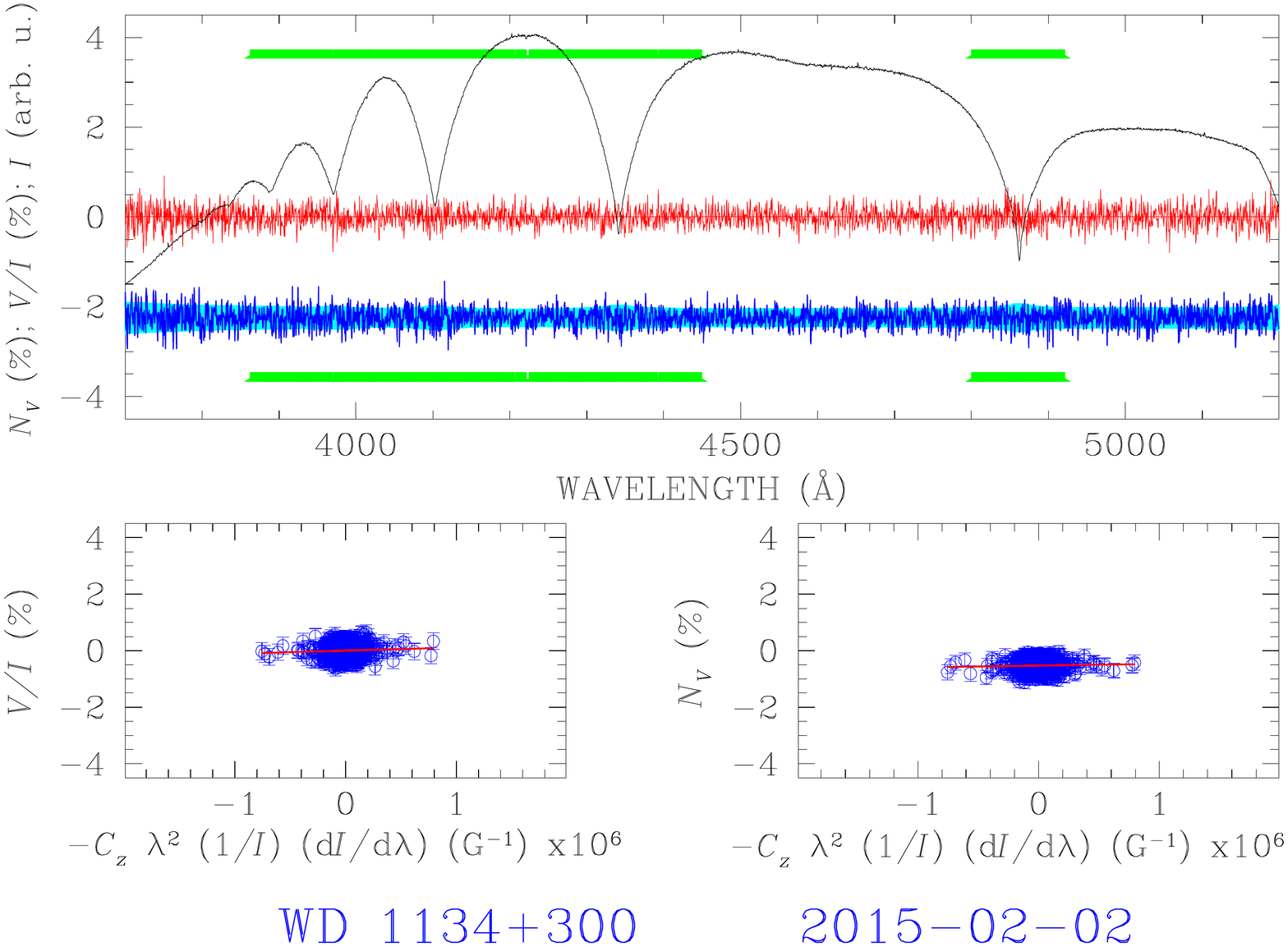}
\includegraphics*[angle=270,width=8.0cm,trim={0.90cm 0.0cm 0.1cm 1.0cm},clip]{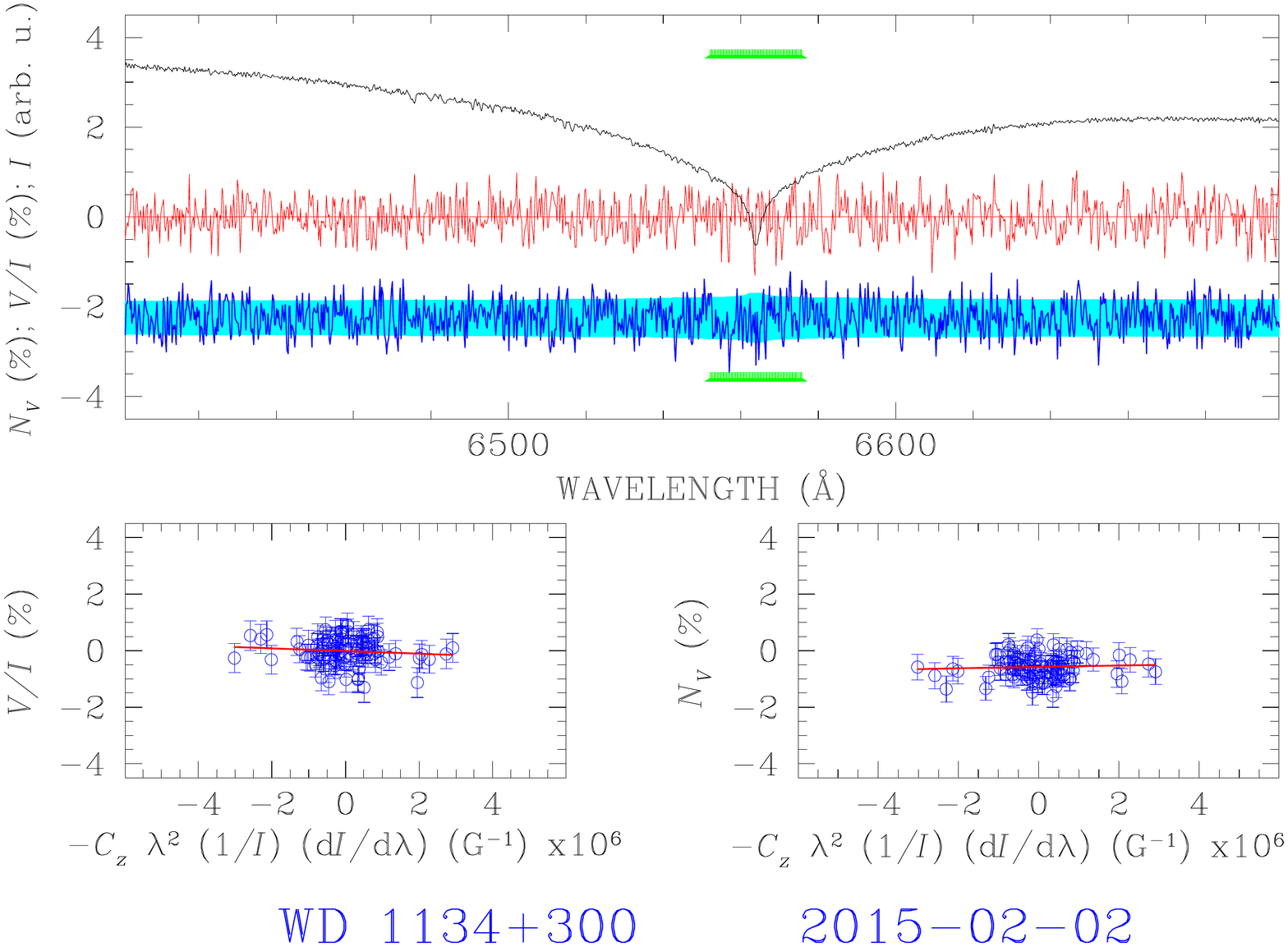} \\
\includegraphics*[angle=270,width=8.0cm,trim={0.90cm 0.0cm 0.1cm 1.0cm},clip]{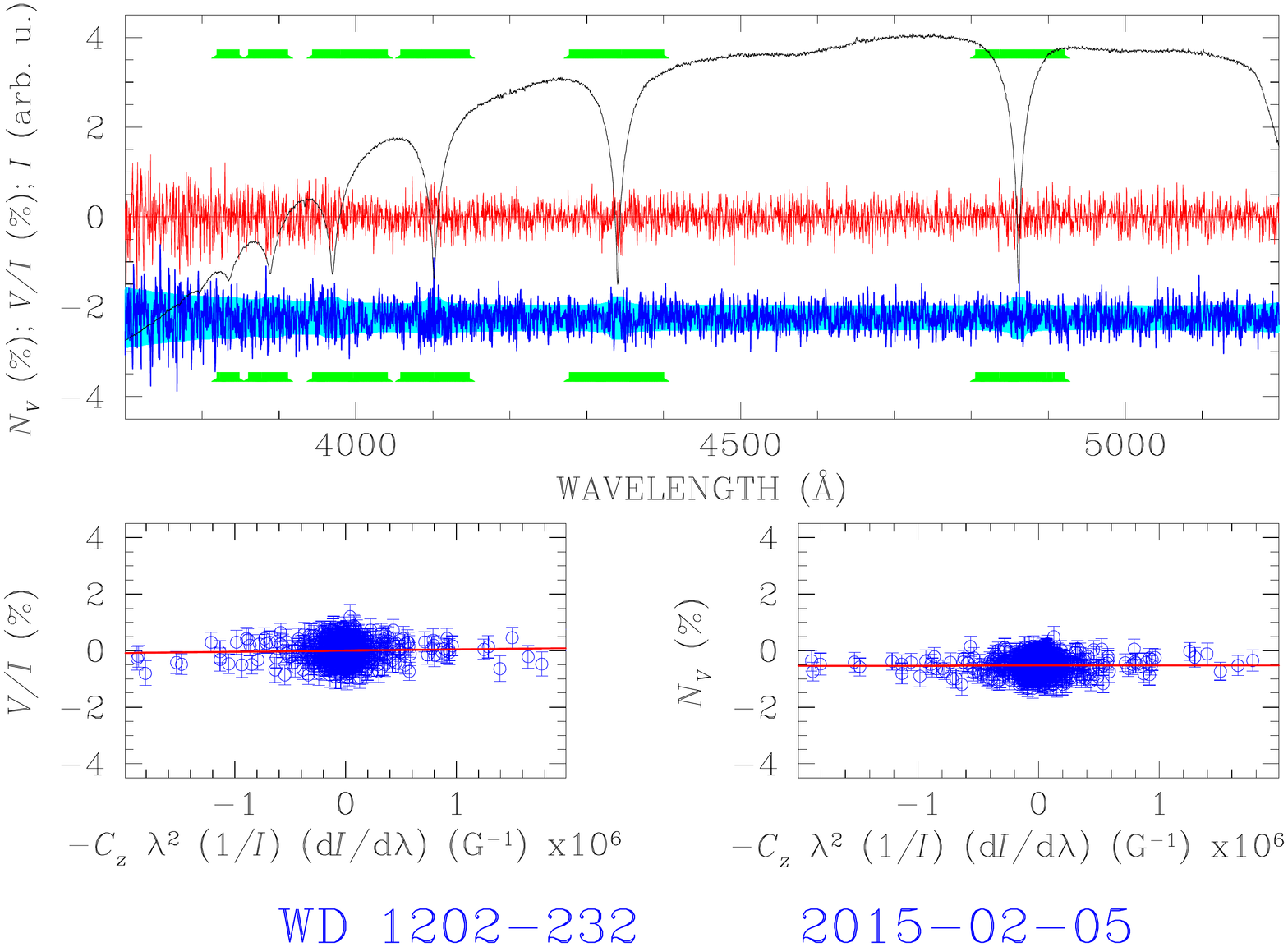} \\
\includegraphics*[angle=270,width=8.0cm,trim={0.90cm 0.0cm 0.1cm 1.0cm},clip]{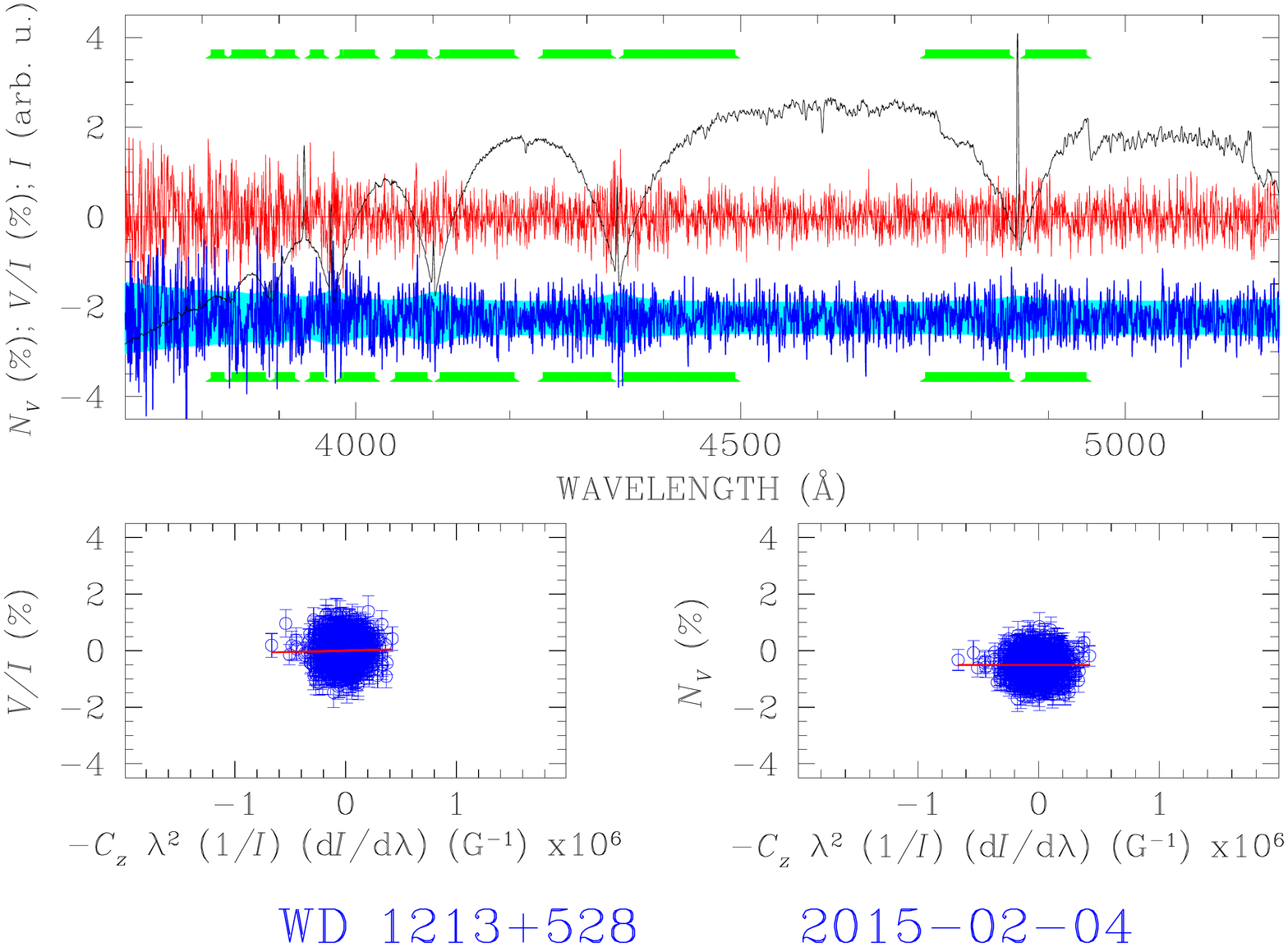} \\
\includegraphics*[angle=270,width=8.0cm,trim={0.90cm 0.0cm 0.1cm 1.0cm},clip]{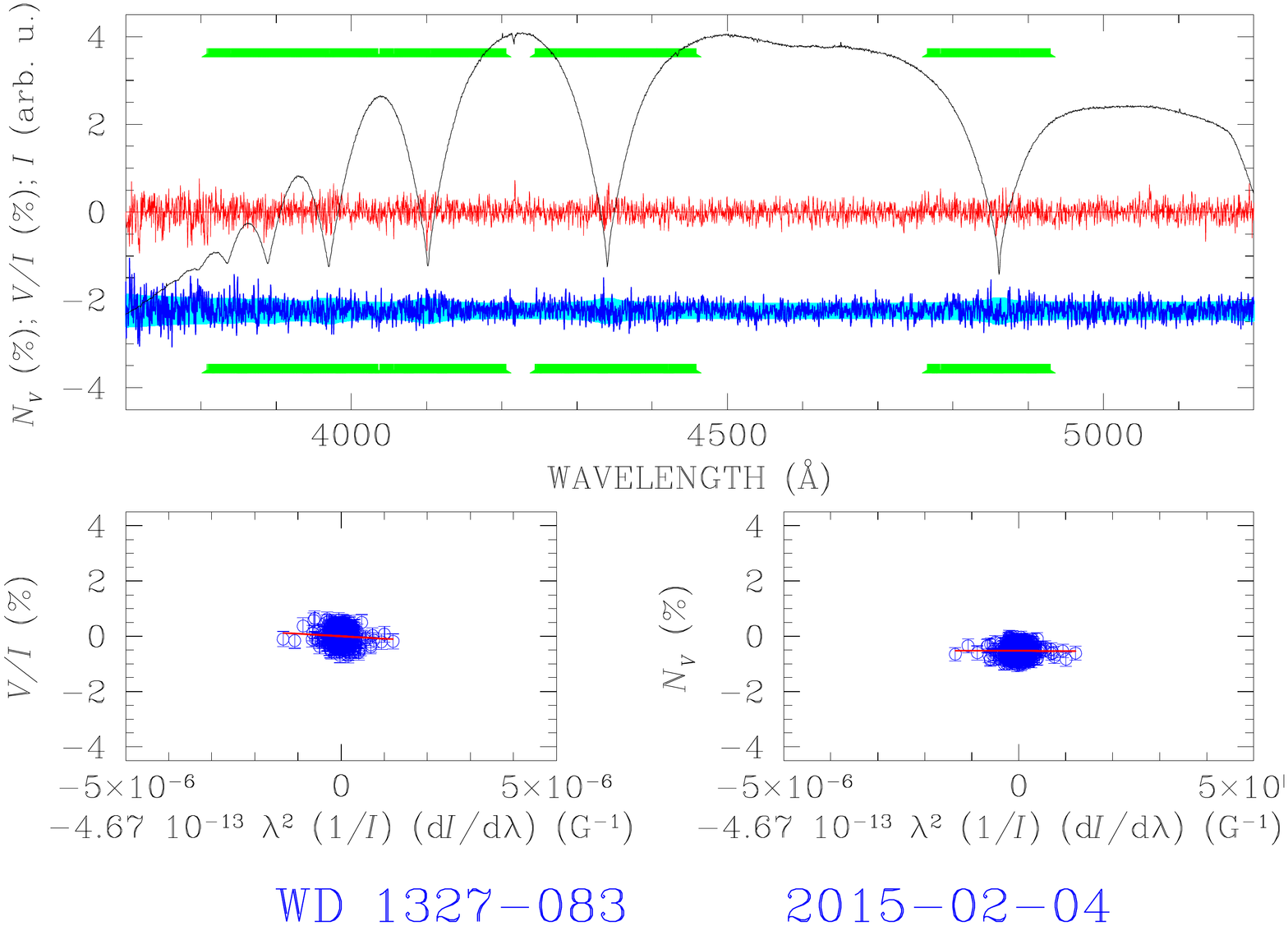} \\
\includegraphics*[angle=270,width=8.0cm,trim={0.90cm 0.0cm 0.1cm 1.0cm},clip]{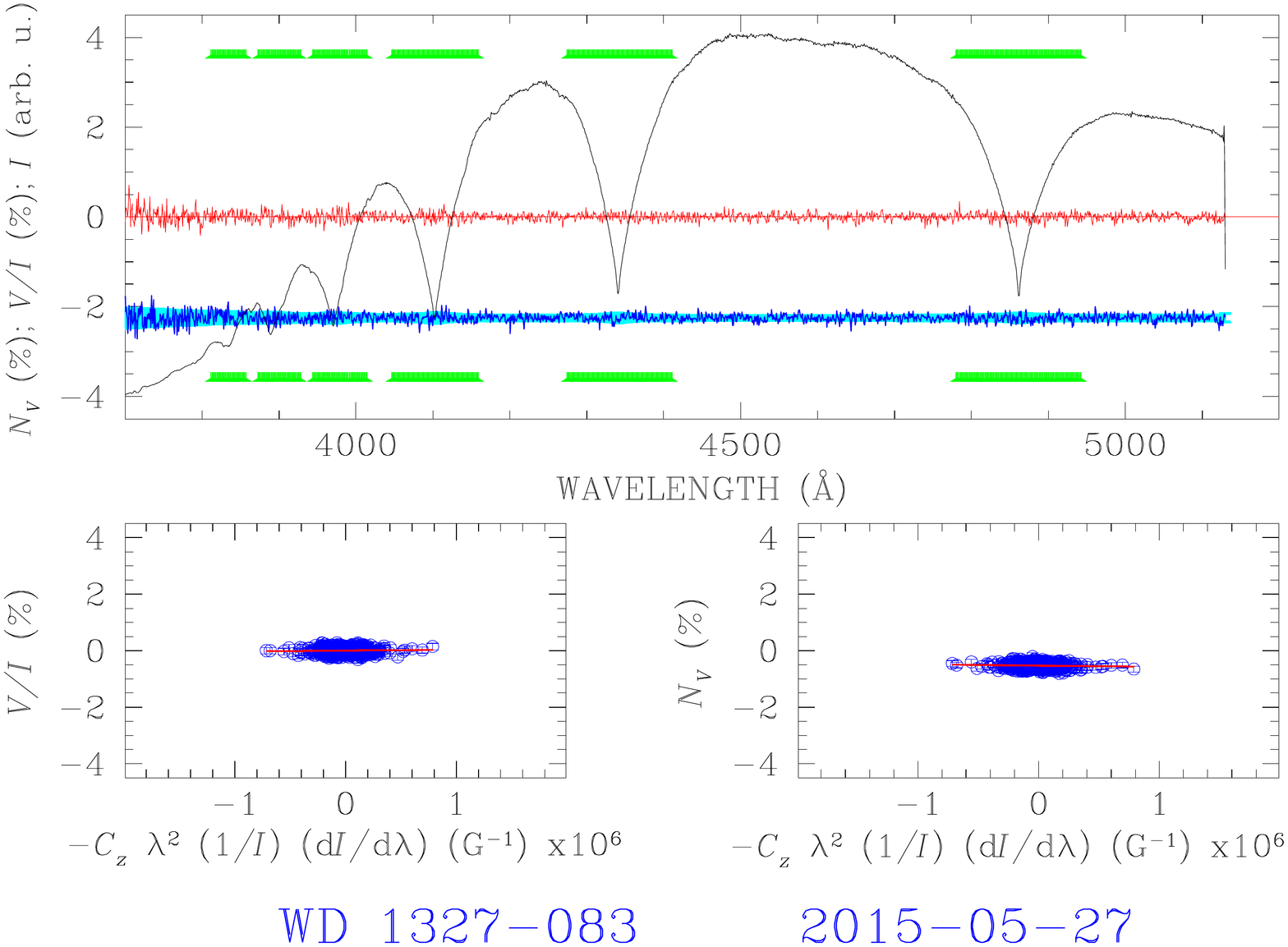} \\
\includegraphics*[angle=270,width=8.0cm,trim={0.90cm 0.0cm 0.1cm 1.0cm},clip]{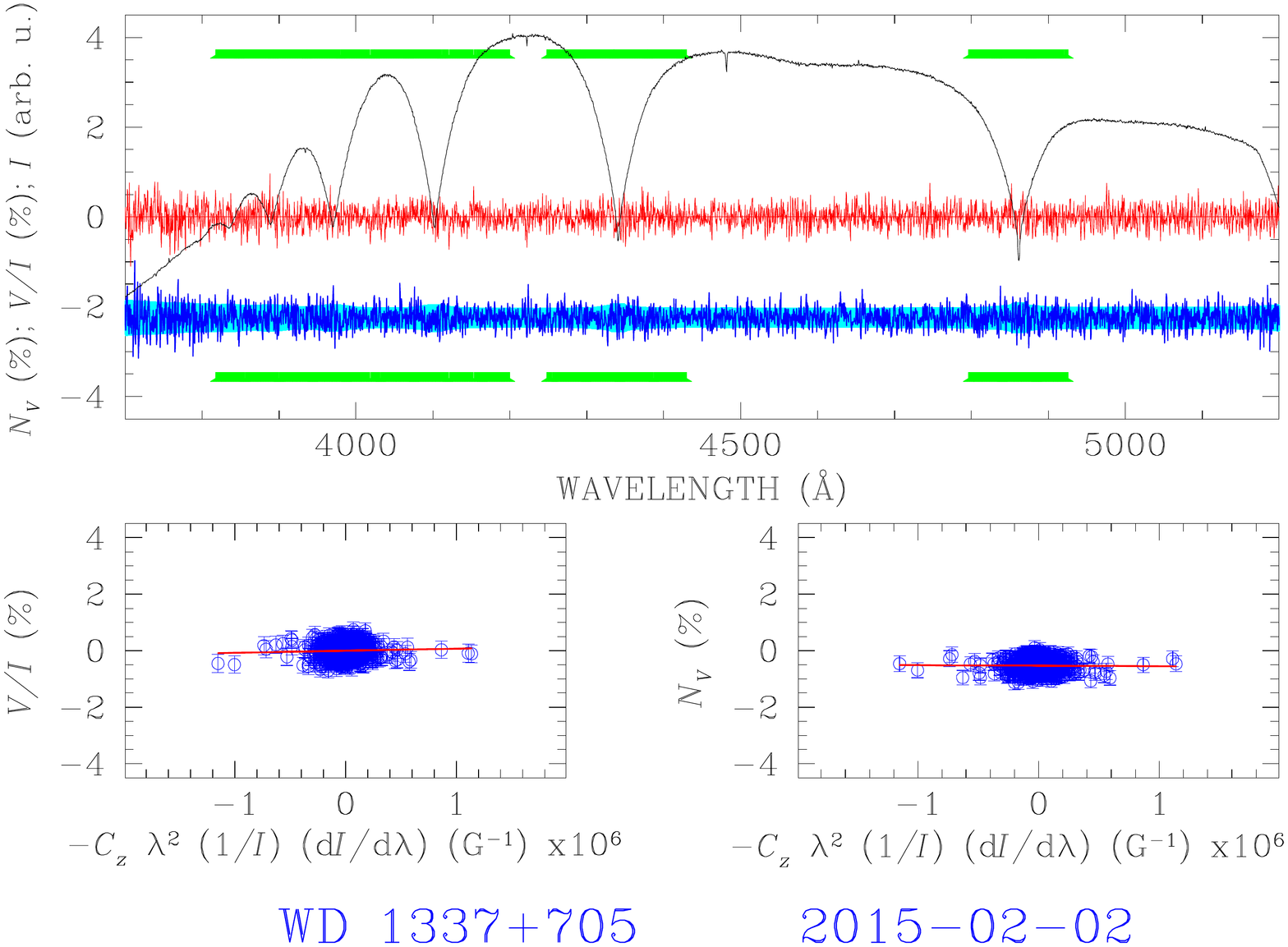} \\ 
\includegraphics*[angle=270,width=8.0cm,trim={0.90cm 0.0cm 0.1cm 1.0cm},clip]{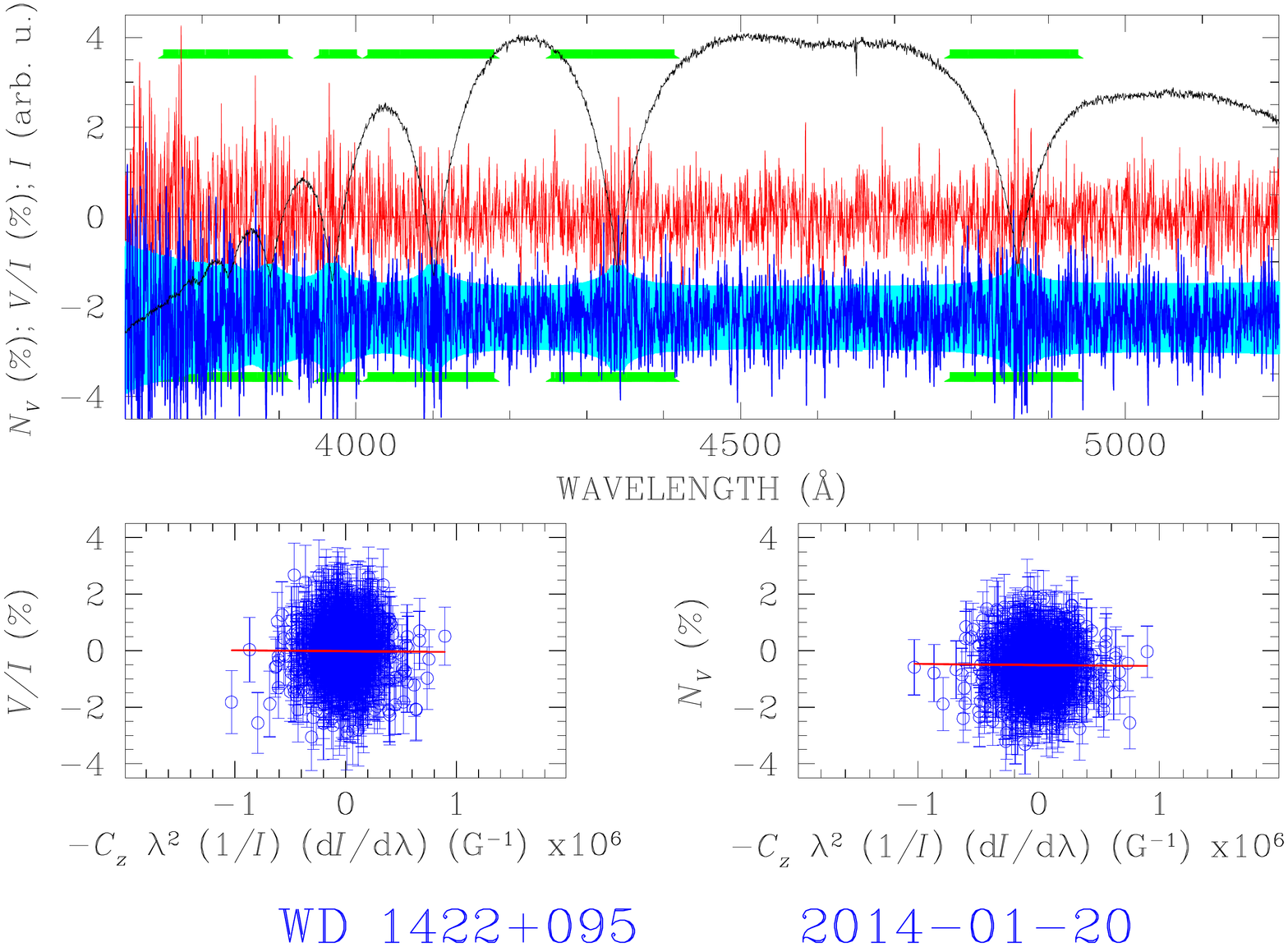} \\
\includegraphics*[angle=270,width=8.0cm,trim={0.90cm 0.0cm 0.1cm 1.0cm},clip]{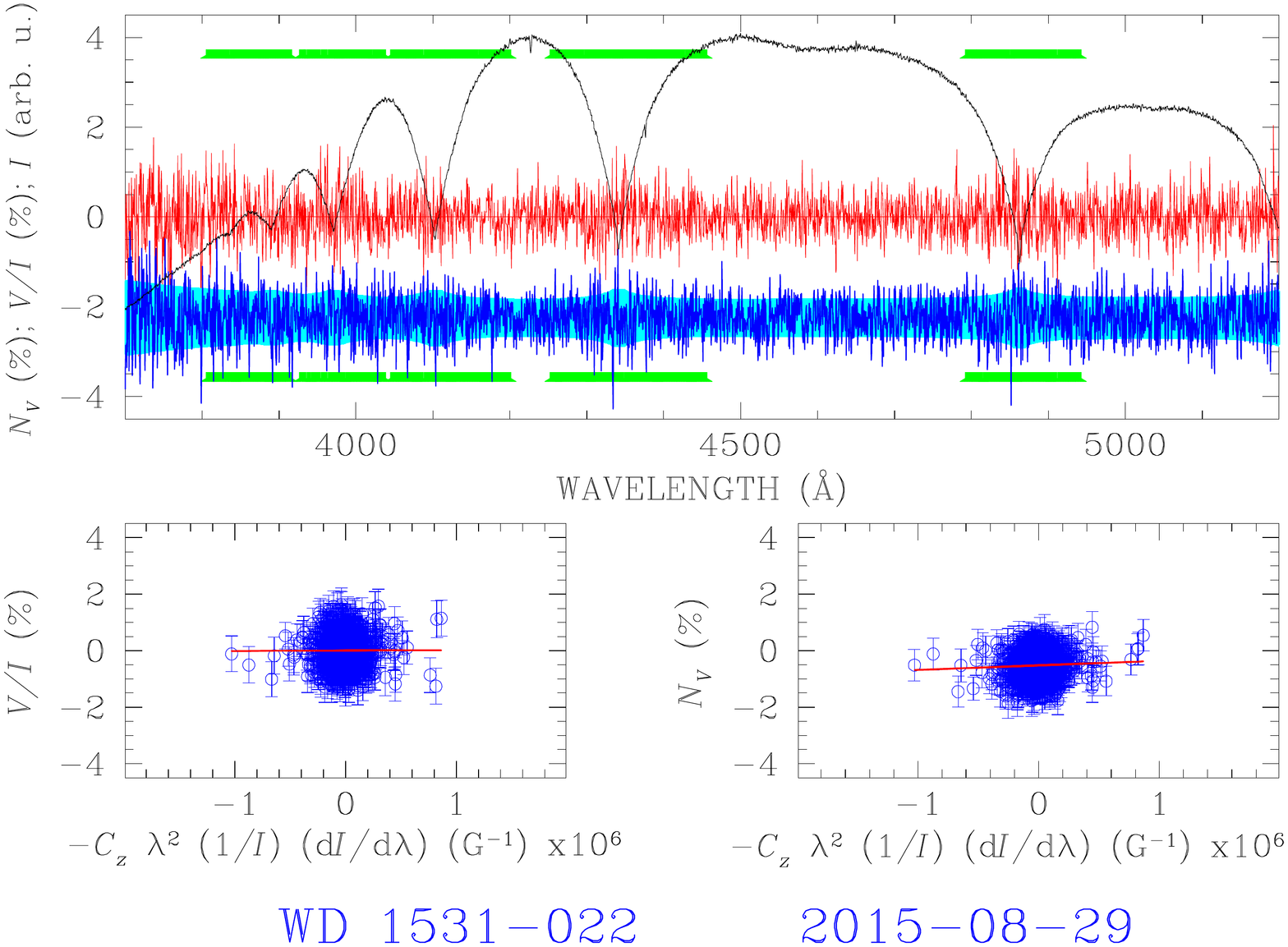}
\includegraphics*[angle=270,width=8.0cm,trim={0.90cm 0.0cm 0.1cm 1.0cm},clip]{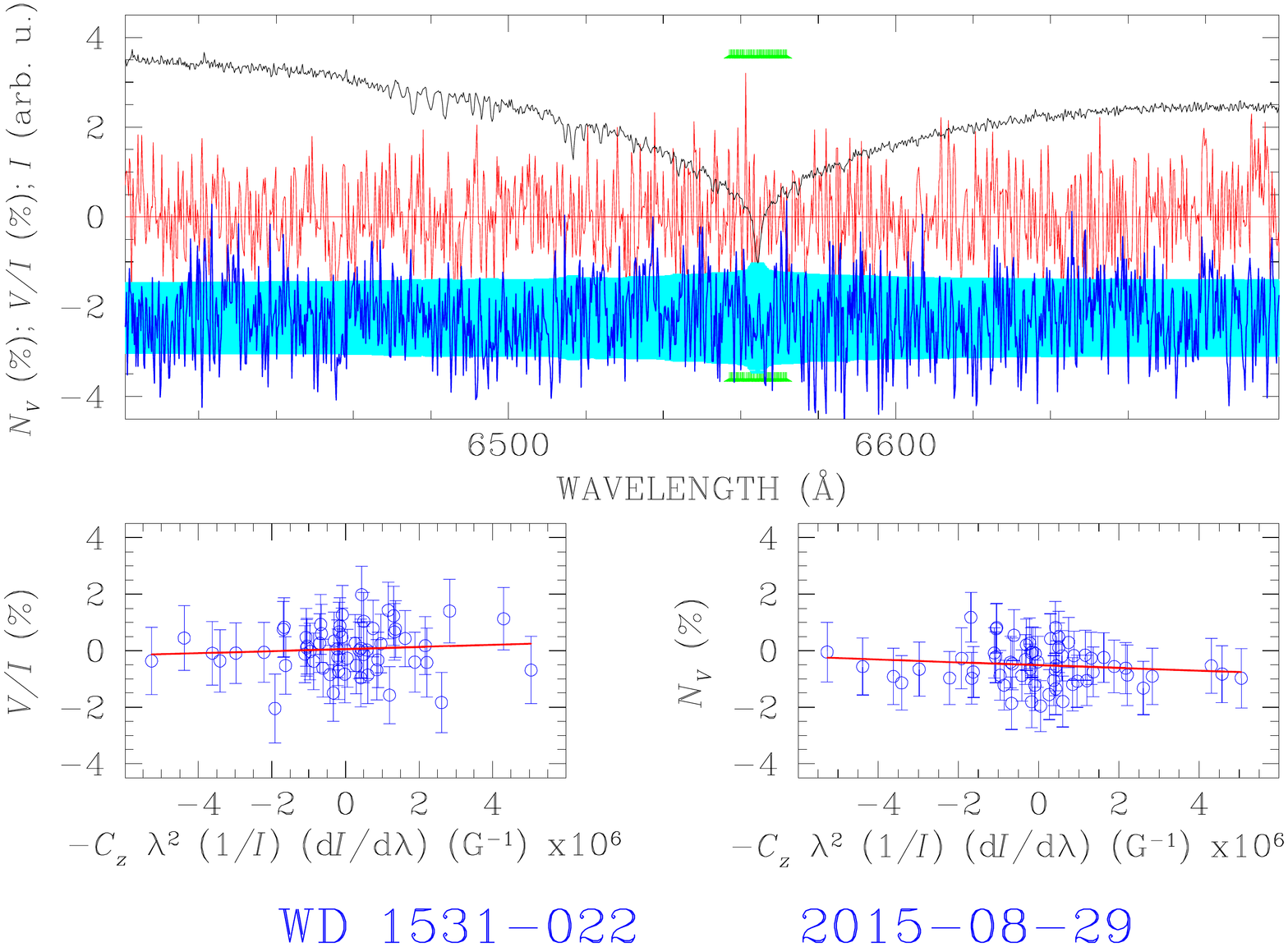}  \\
\includegraphics*[angle=270,width=8.0cm,trim={0.90cm 0.0cm 0.1cm 1.0cm},clip]{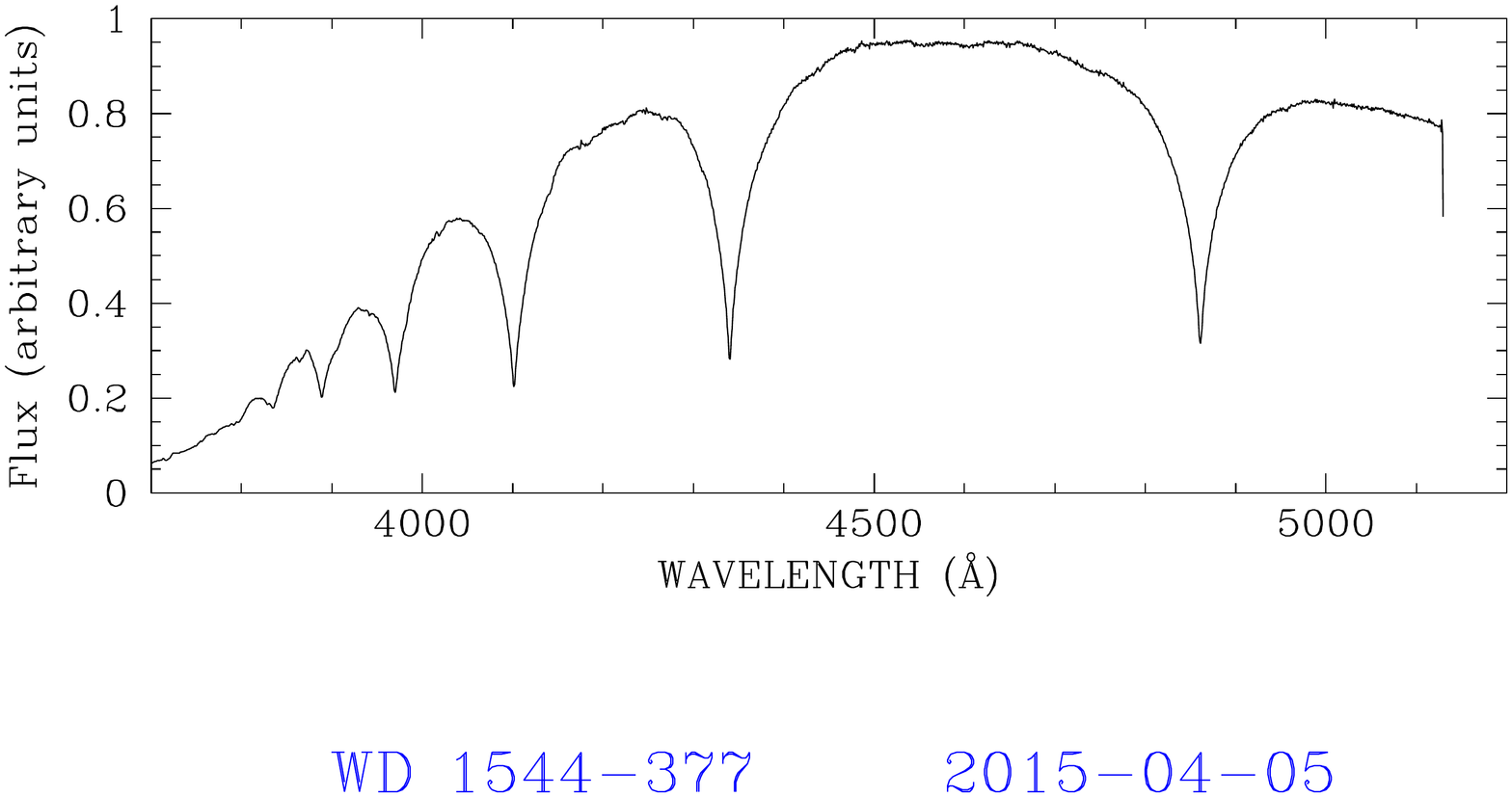} \\
\includegraphics*[angle=270,width=8.0cm,trim={0.90cm 0.0cm 0.1cm 1.0cm},clip]{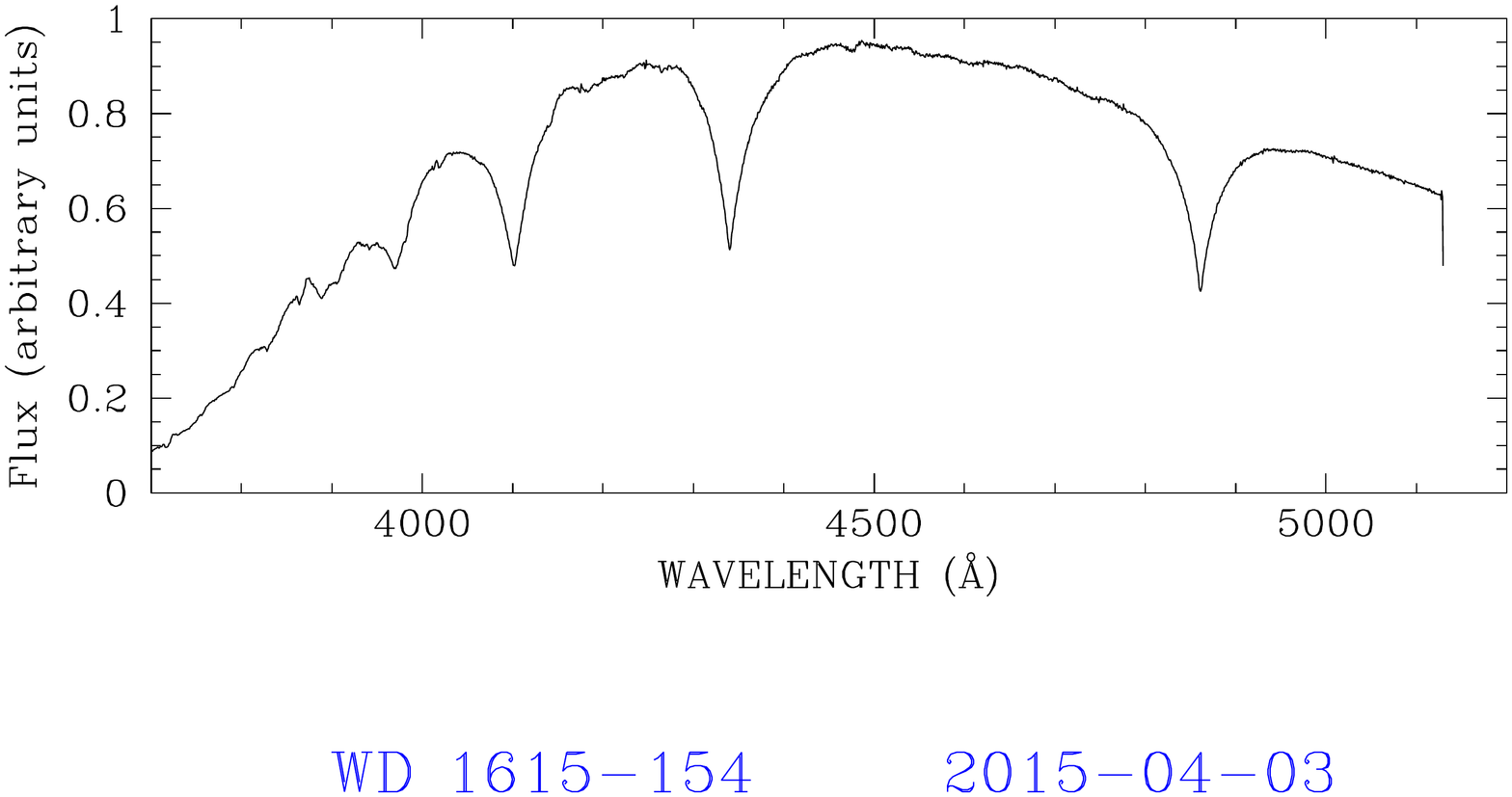} \\
\includegraphics*[angle=270,width=8.0cm,trim={0.90cm 0.0cm 0.1cm 1.0cm},clip]{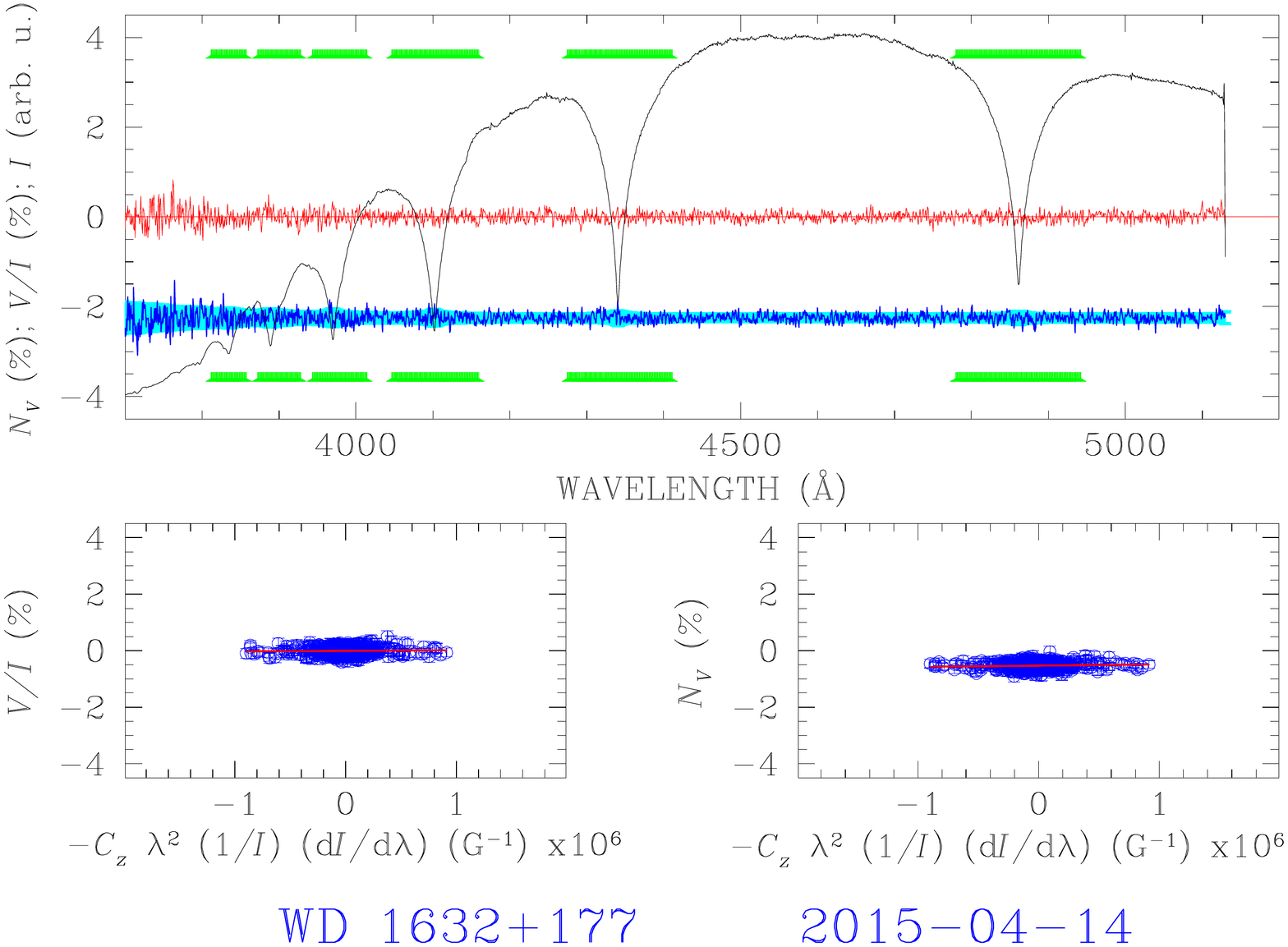} \\
\includegraphics*[angle=270,width=8.0cm,trim={0.90cm 0.0cm 0.1cm 1.0cm},clip]{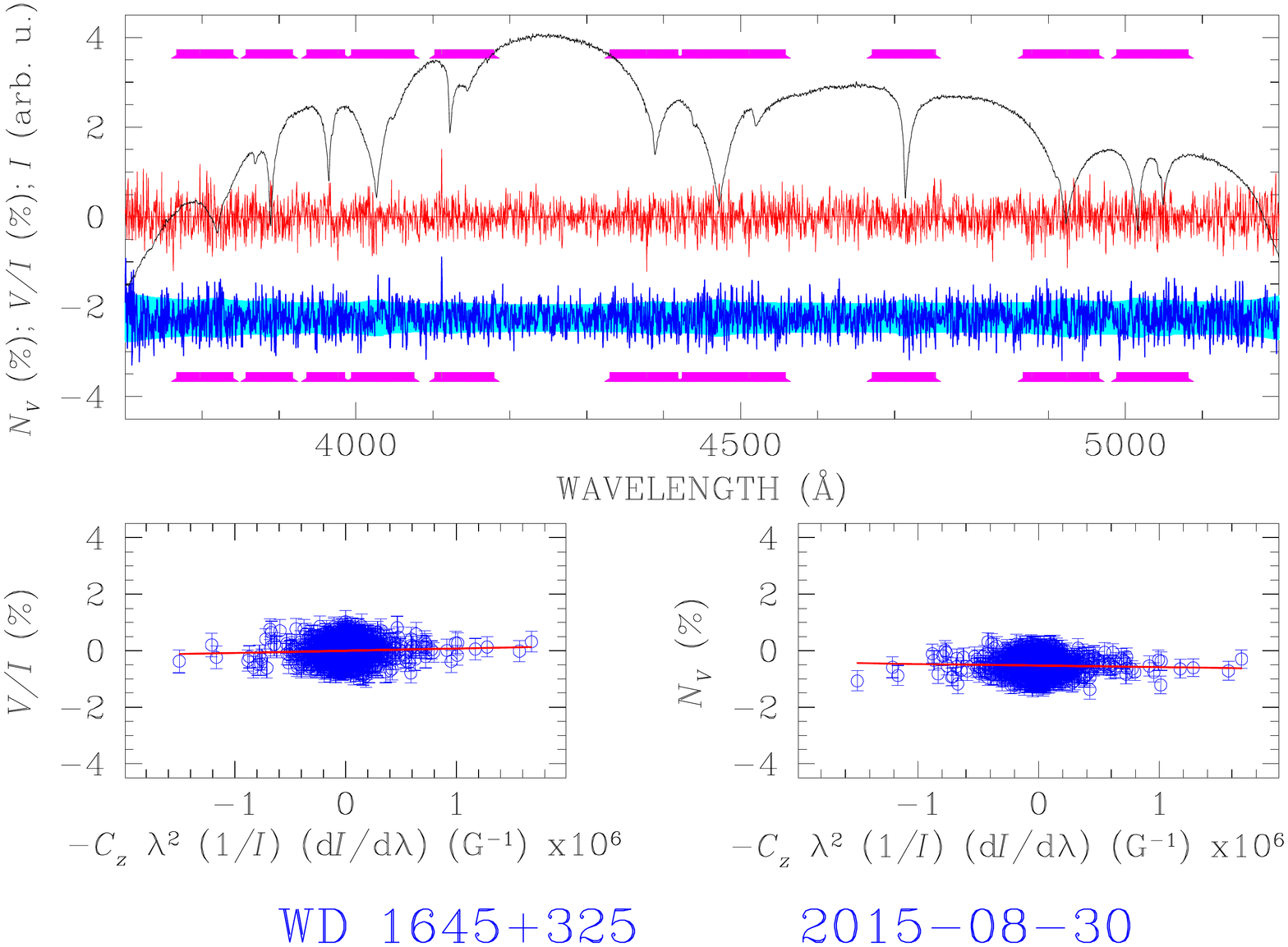}
\includegraphics*[angle=270,width=8.0cm,trim={0.90cm 0.0cm 0.1cm 1.0cm},clip]{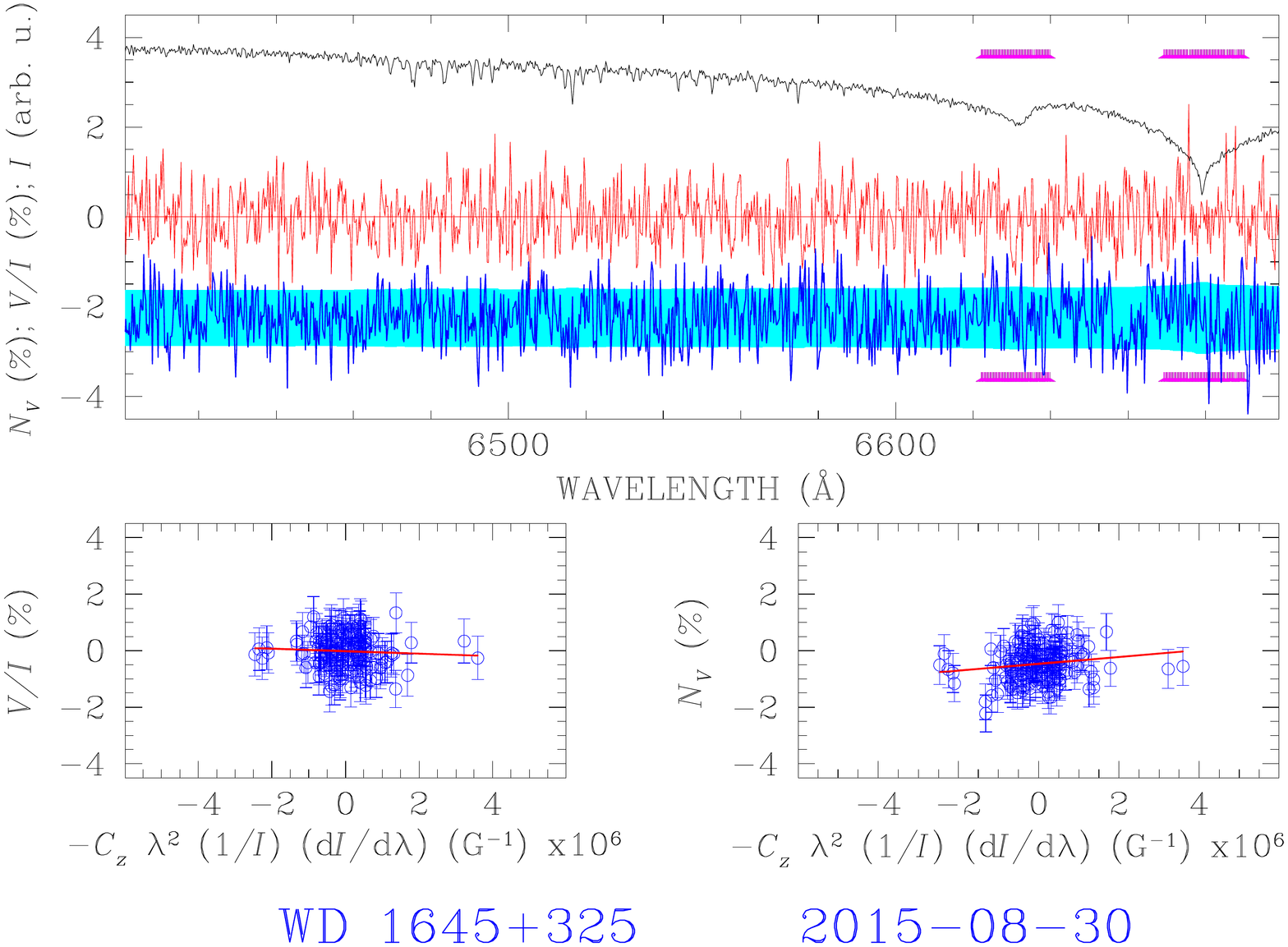} \\
\includegraphics*[angle=270,width=8.0cm,trim={0.90cm 0.0cm 0.1cm 1.0cm},clip]{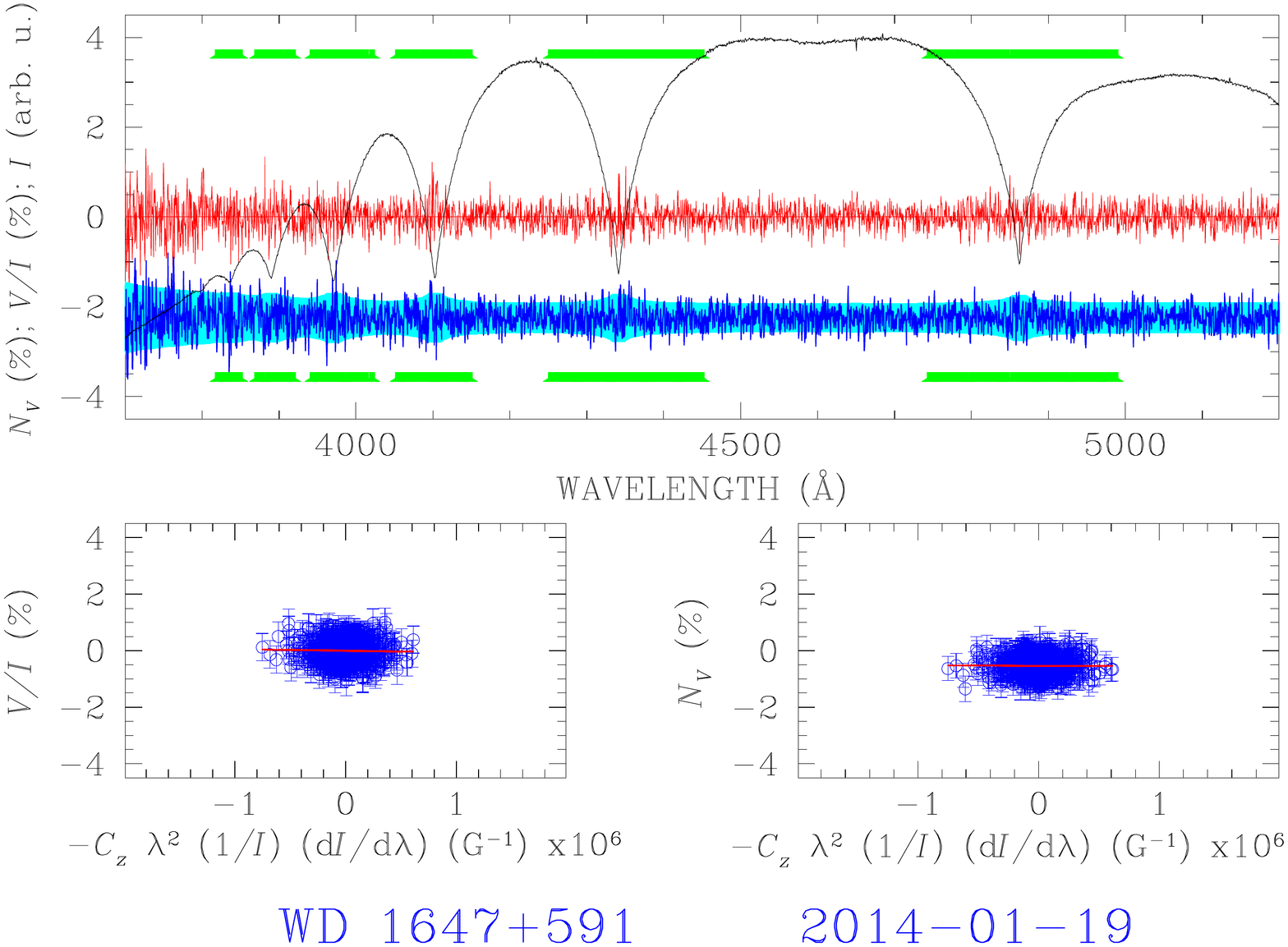} \\
\includegraphics*[angle=270,width=8.0cm,trim={0.90cm 0.0cm 0.1cm 1.0cm},clip]{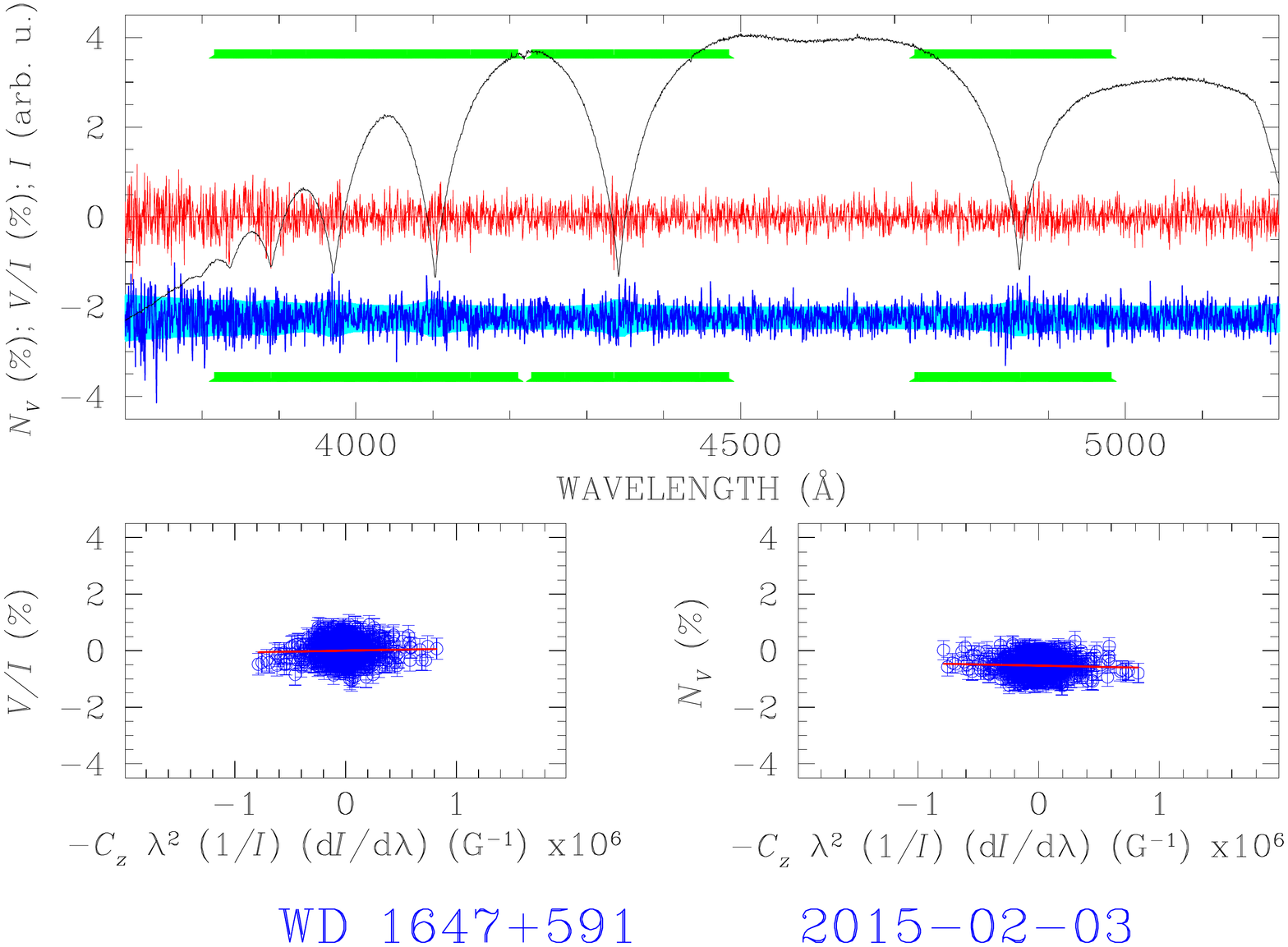} \\
\includegraphics*[angle=270,width=8.0cm,trim={0.90cm 0.0cm 0.1cm 1.0cm},clip]{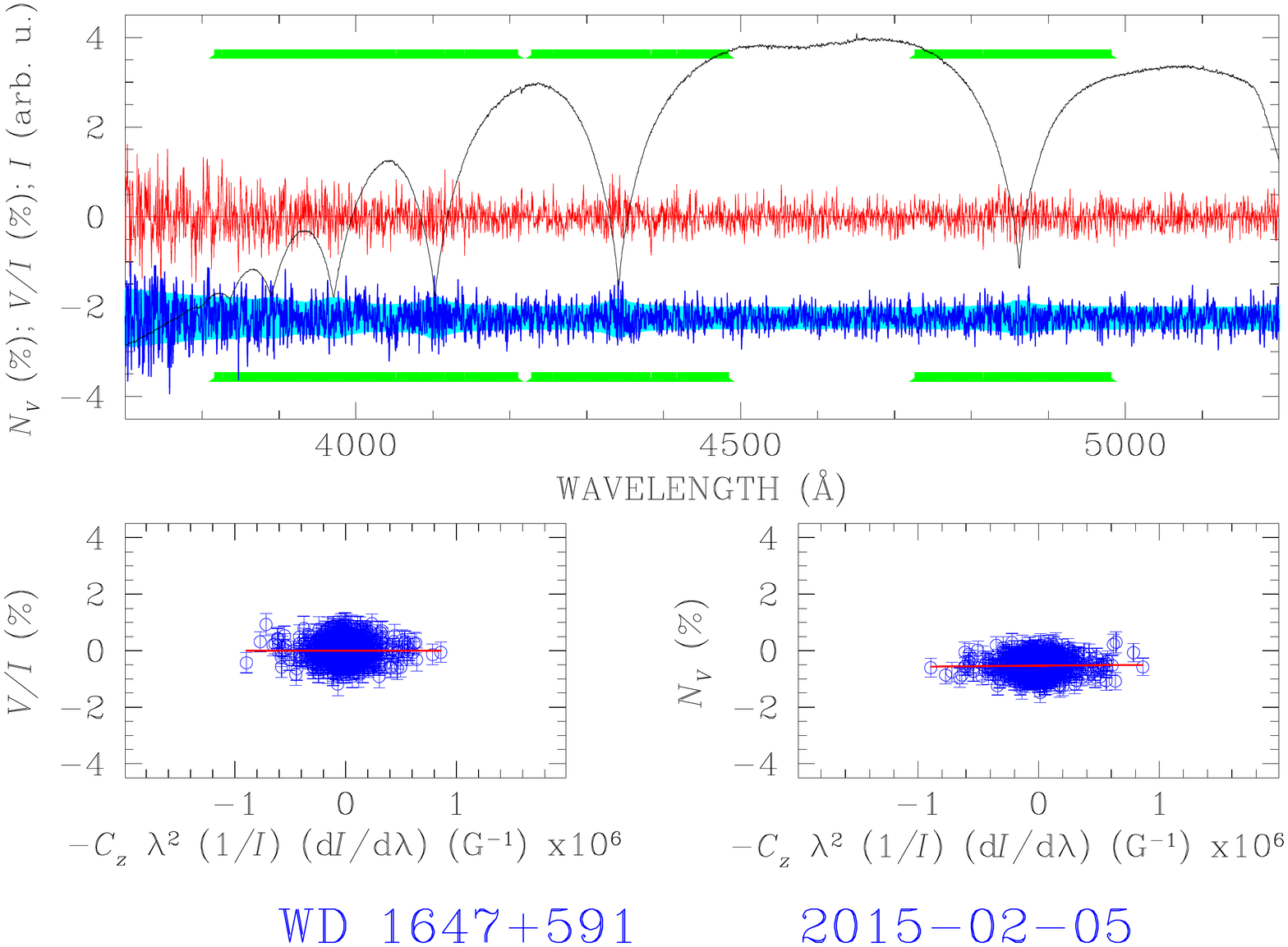} \\
\includegraphics*[angle=270,width=8.0cm,trim={0.90cm 0.0cm 0.1cm 1.0cm},clip]{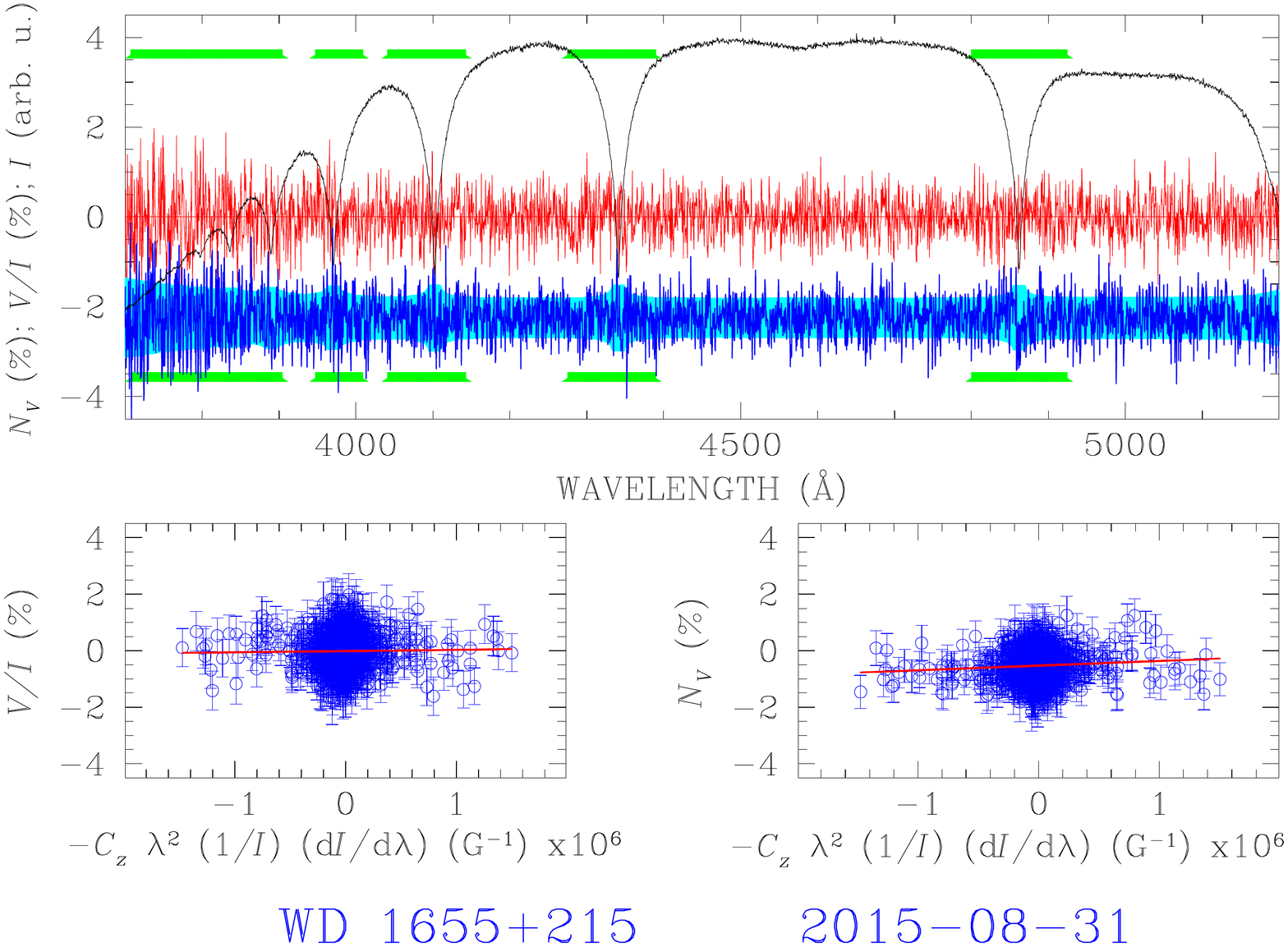}
\includegraphics*[angle=270,width=8.0cm,trim={0.90cm 0.0cm 0.1cm 1.0cm},clip]{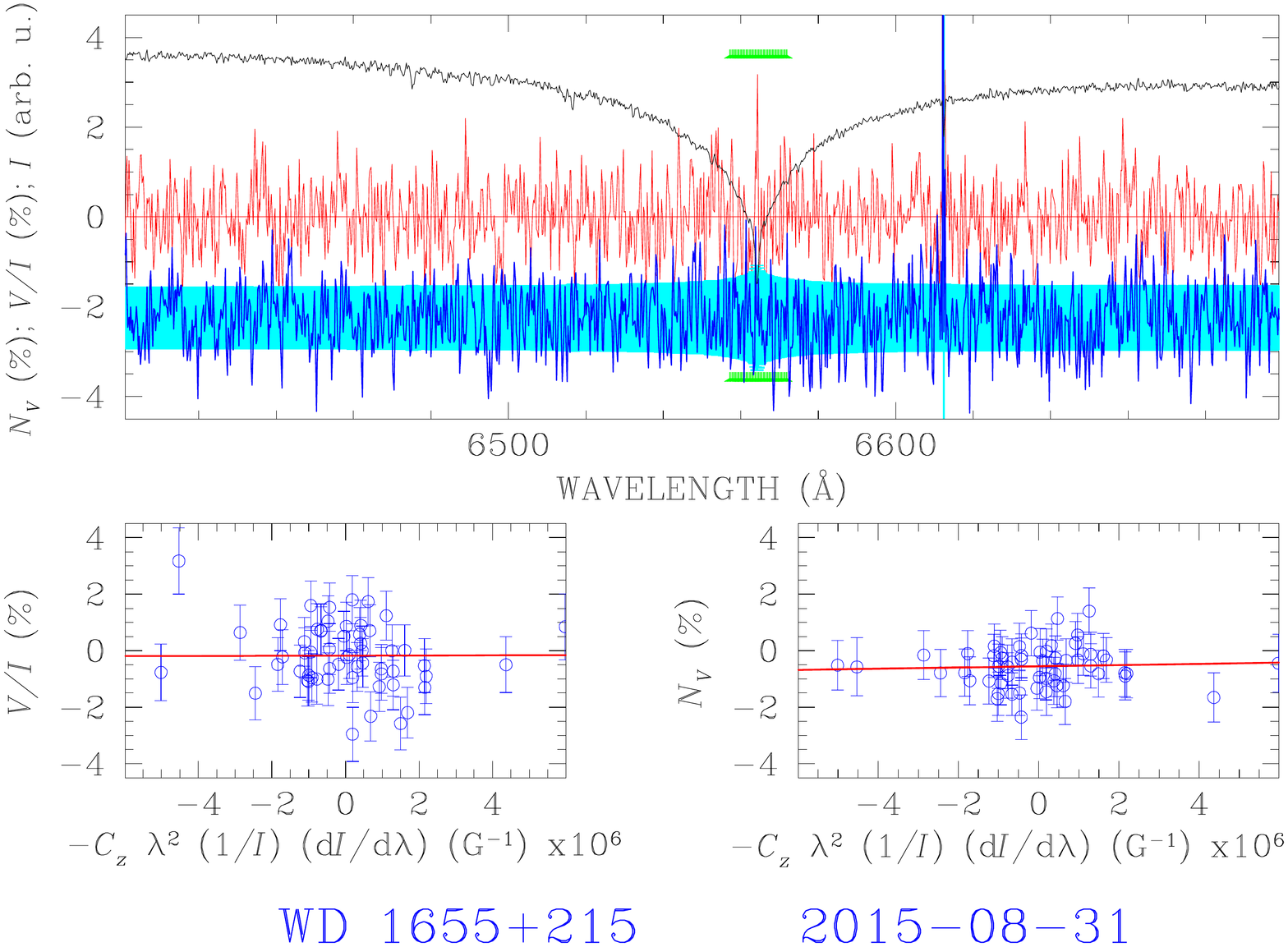} \\
\includegraphics*[angle=270,width=8.0cm,trim={0.90cm 0.0cm 0.1cm 1.0cm},clip]{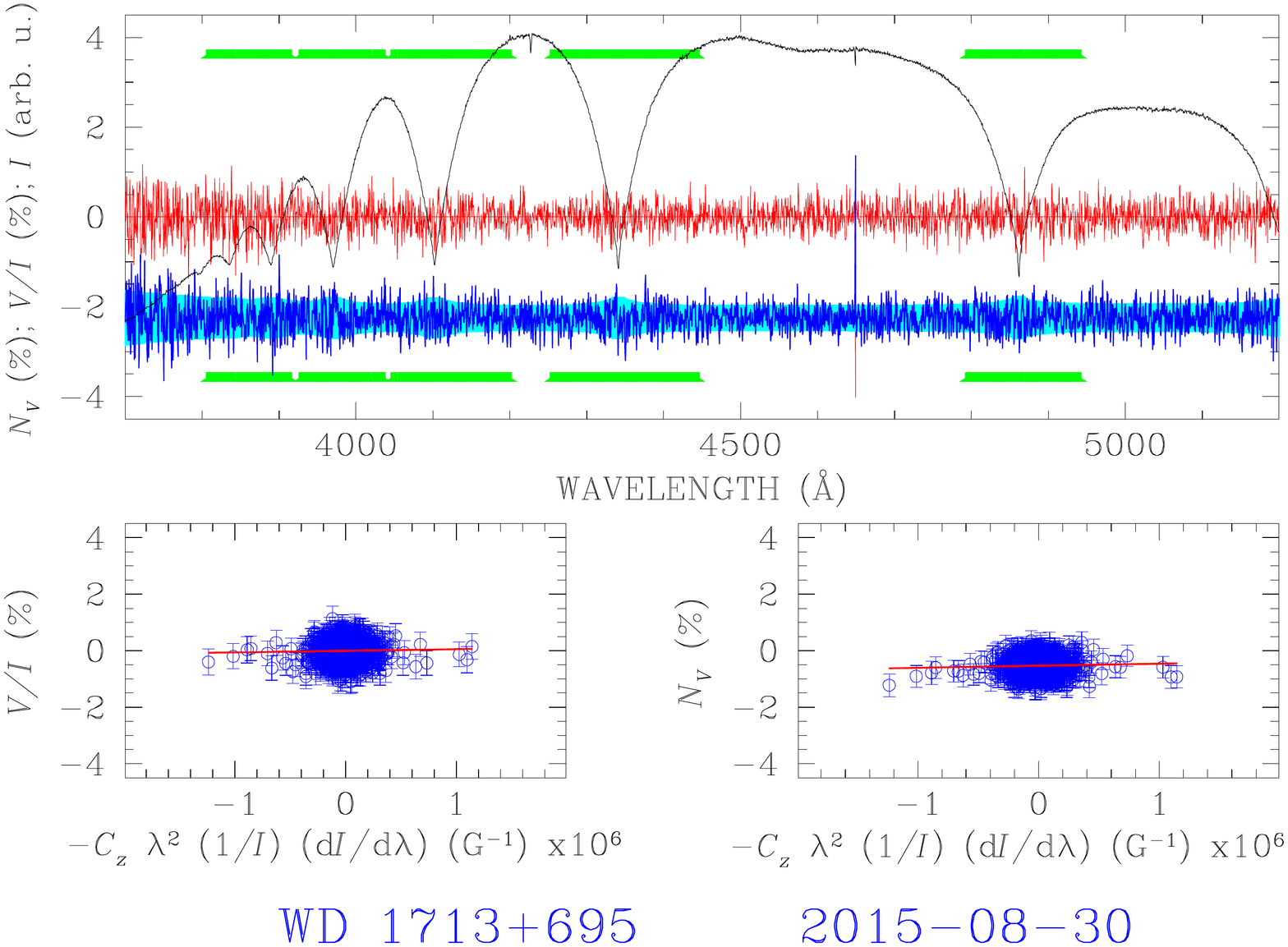}
\includegraphics*[angle=270,width=8.0cm,trim={0.90cm 0.0cm 0.1cm 1.0cm},clip]{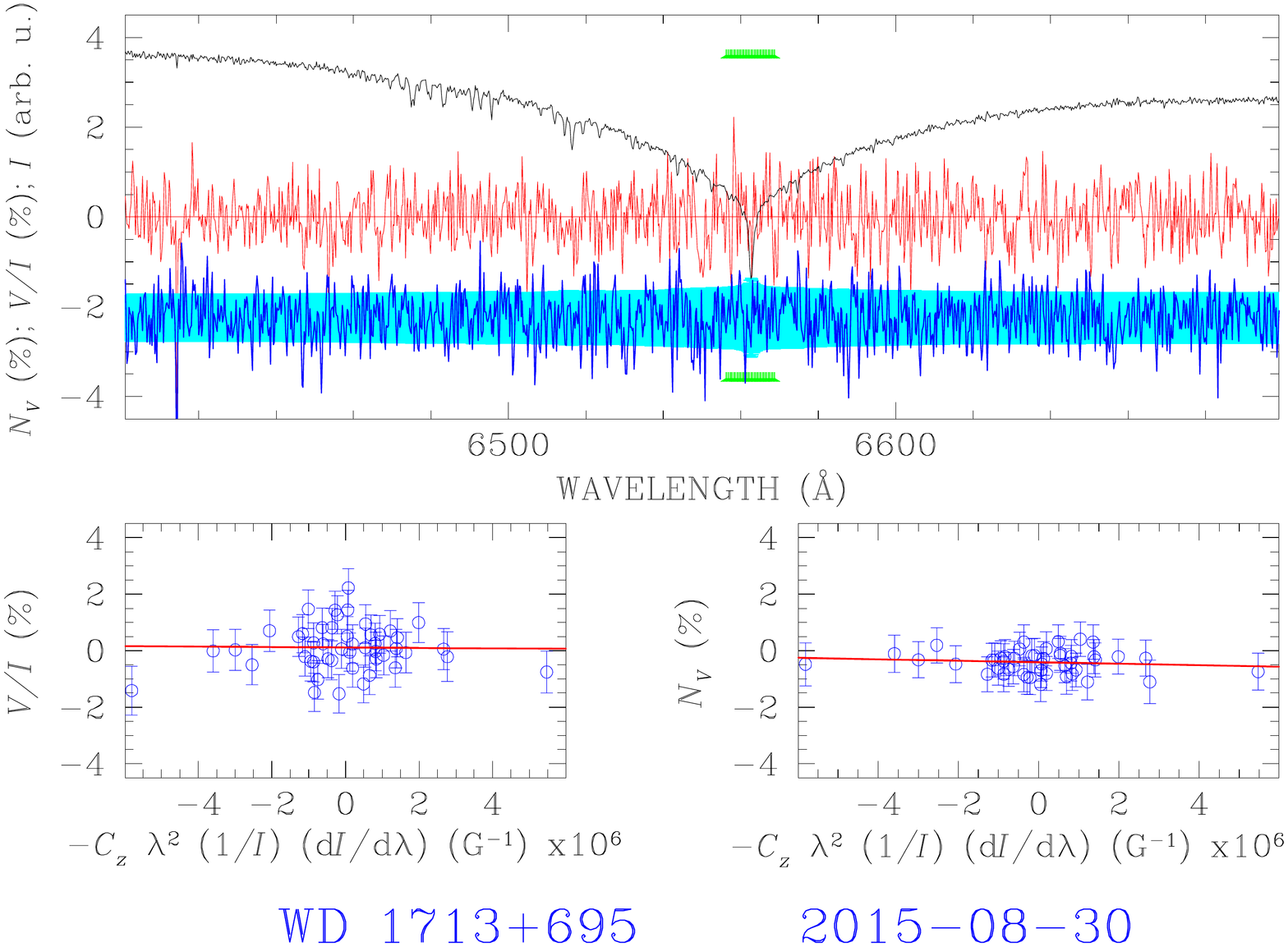} \\
\includegraphics*[angle=270,width=8.0cm,trim={0.90cm 0.0cm 0.1cm 1.0cm},clip]{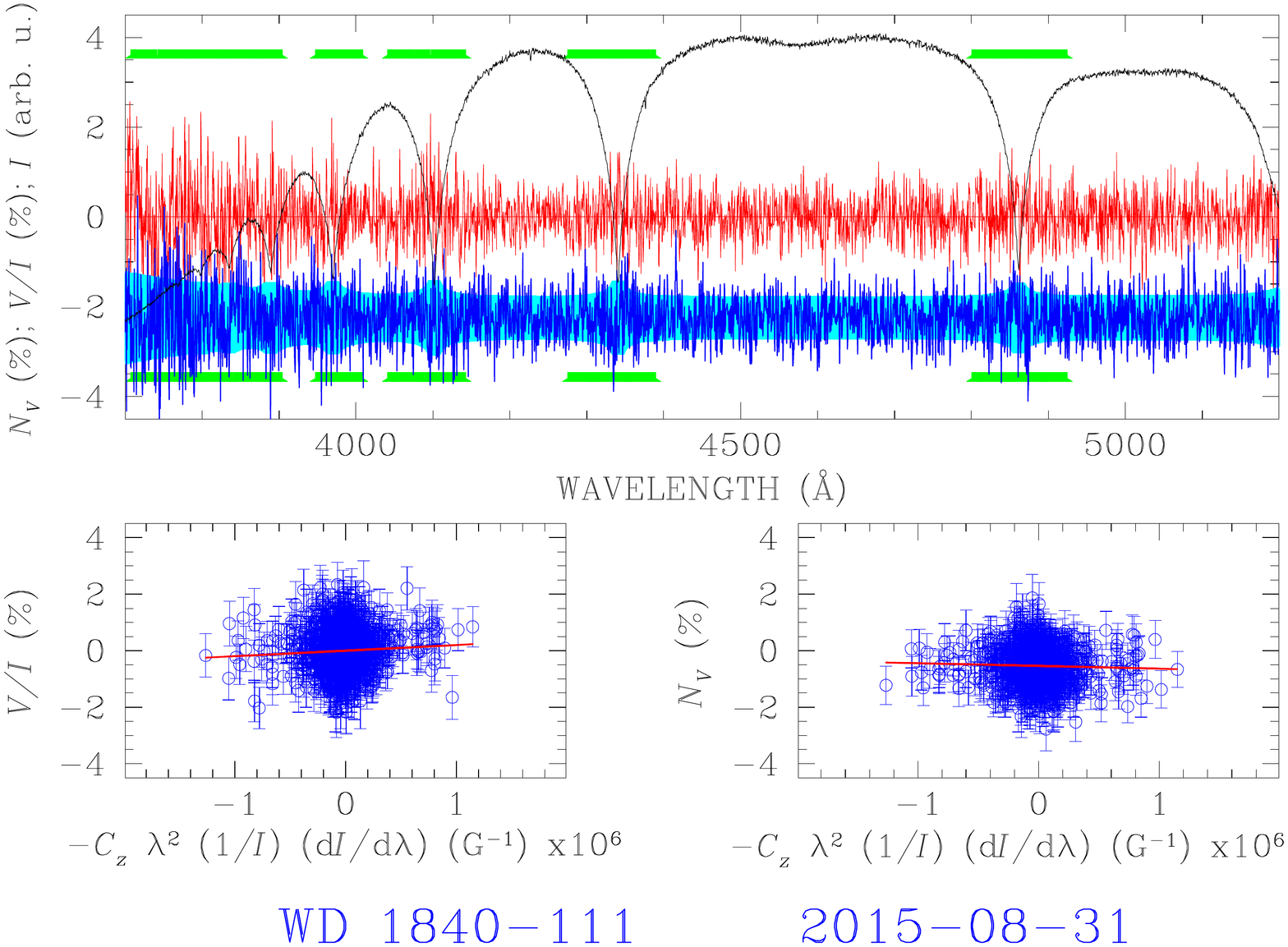}
\includegraphics*[angle=270,width=8.0cm,trim={0.90cm 0.0cm 0.1cm 1.0cm},clip]{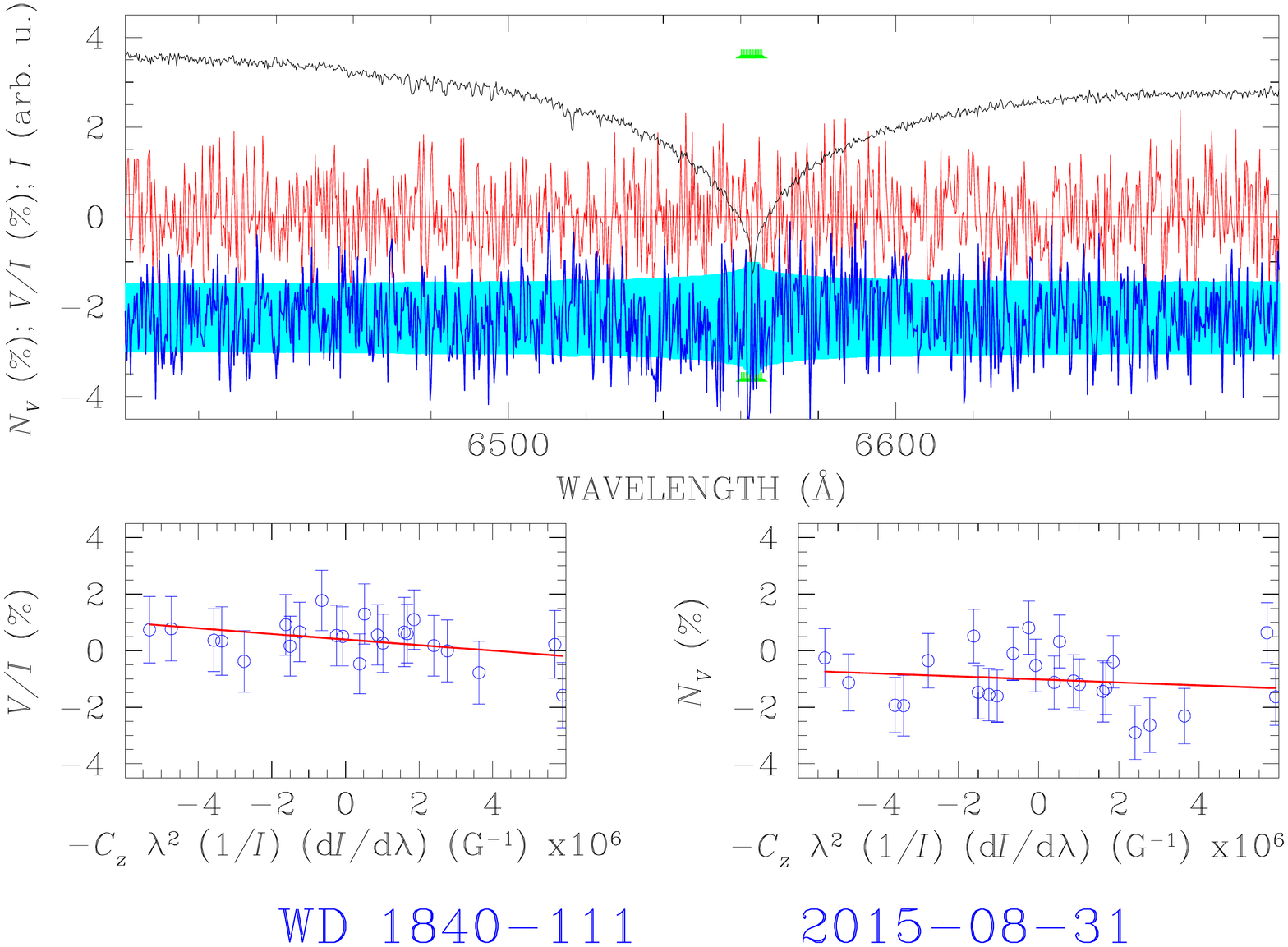} \\
\includegraphics*[angle=270,width=8.0cm,trim={0.90cm 0.0cm 0.1cm 1.0cm},clip]{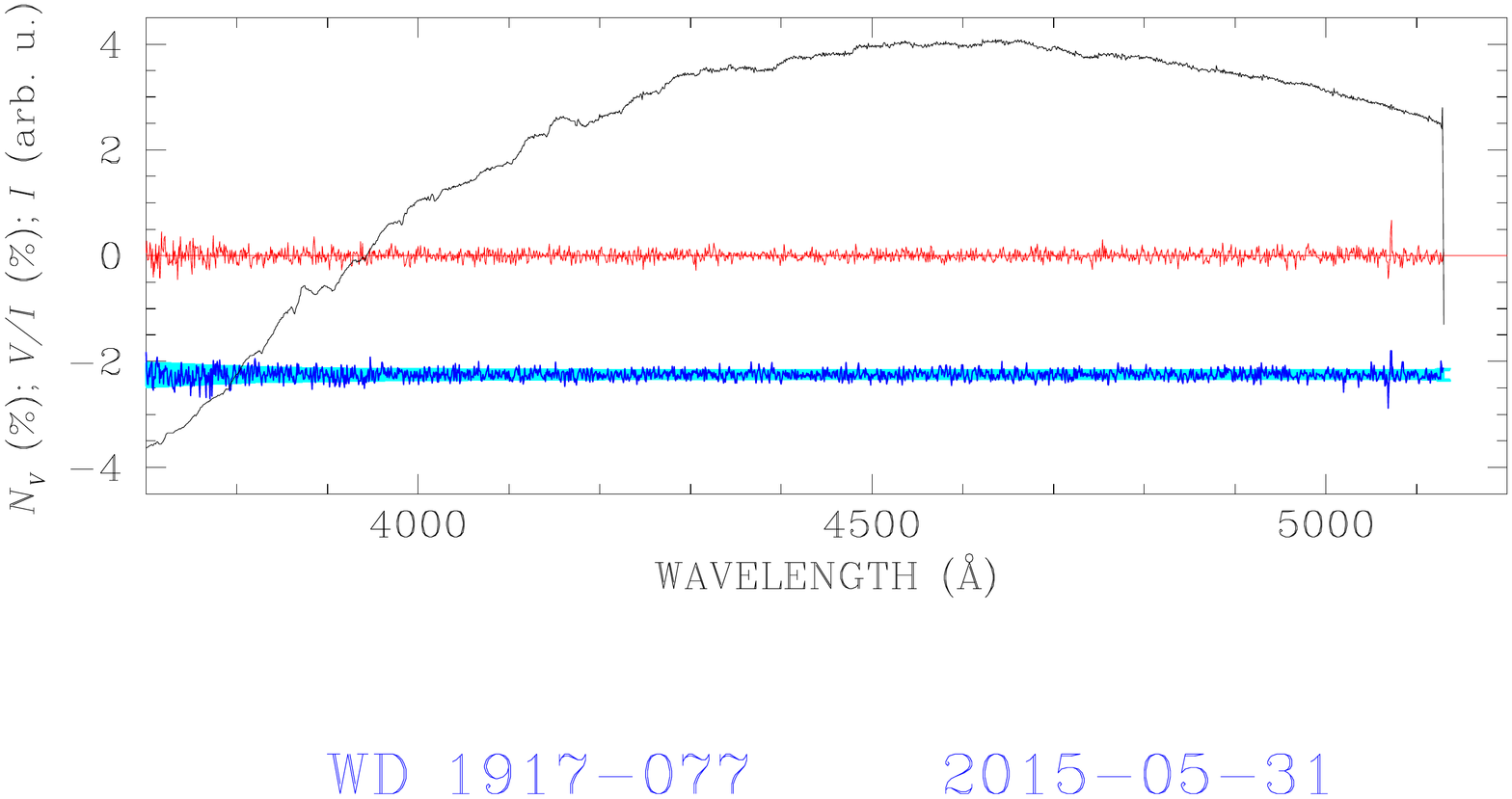} \\
\includegraphics*[angle=270,width=8.0cm,trim={0.90cm 0.0cm 0.1cm 1.0cm},clip]{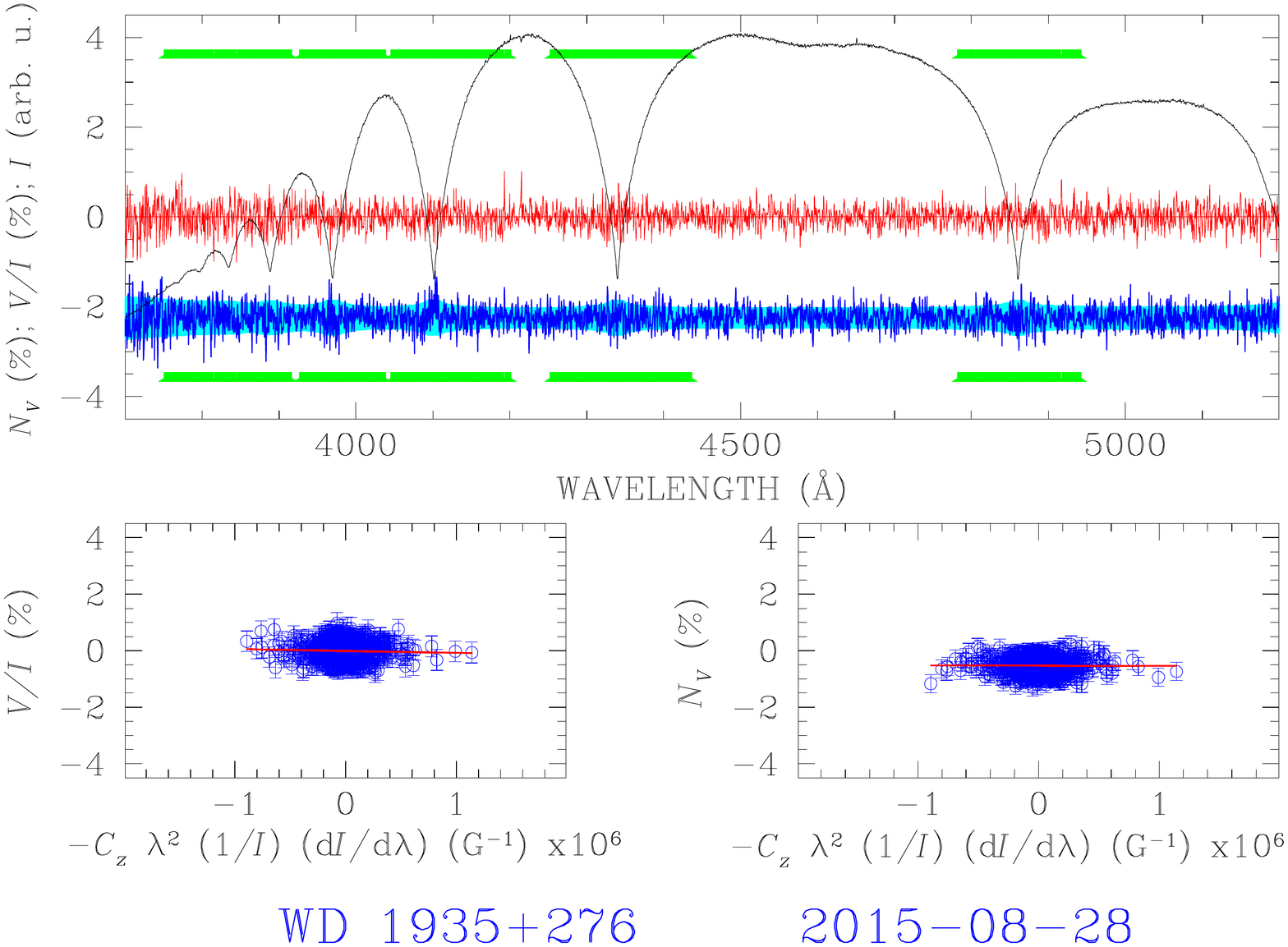}
\includegraphics*[angle=270,width=8.0cm,trim={0.90cm 0.0cm 0.1cm 1.0cm},clip]{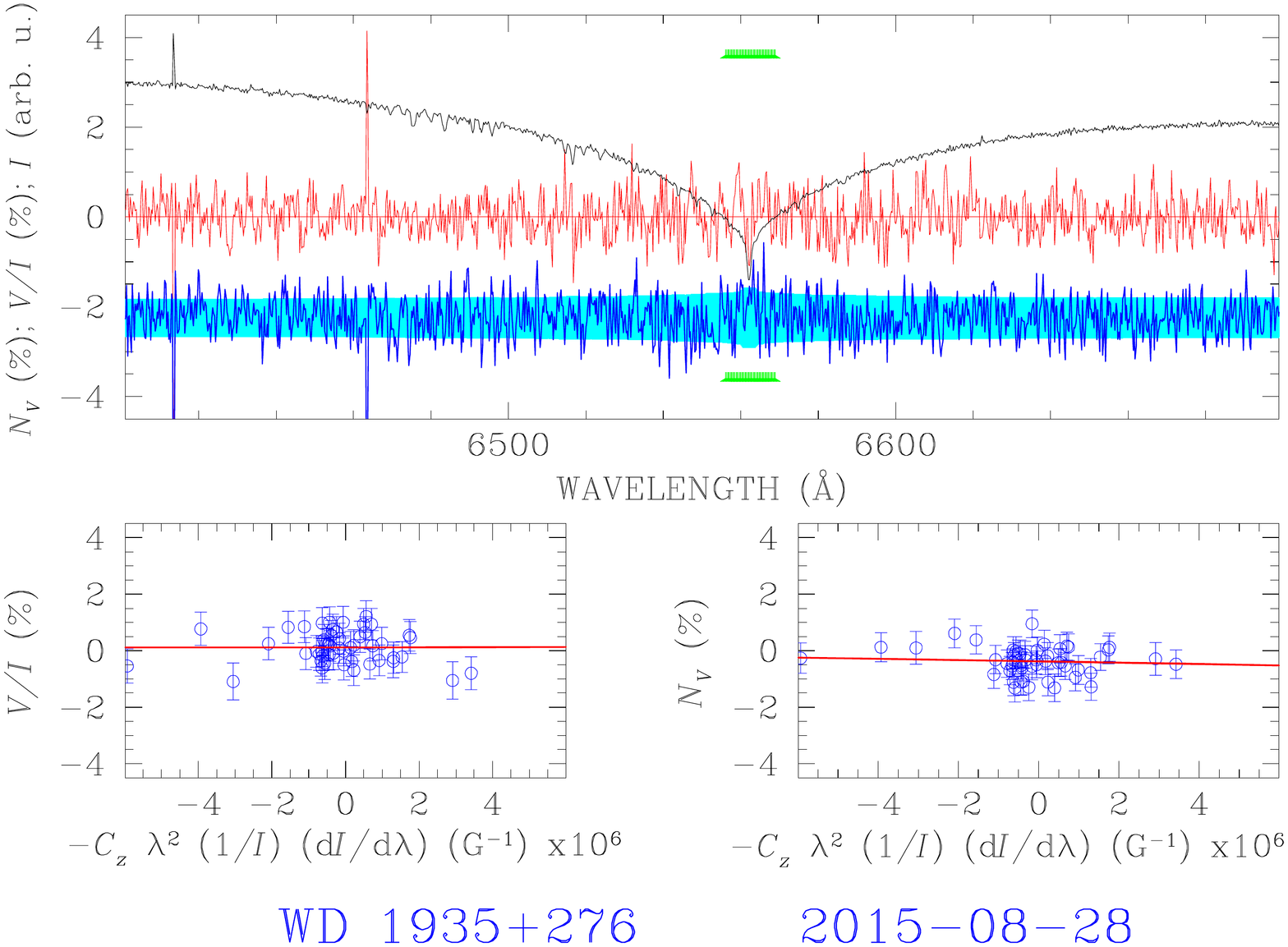} \\
\includegraphics*[angle=270,width=8.0cm,trim={0.90cm 0.0cm 0.1cm 1.0cm},clip]{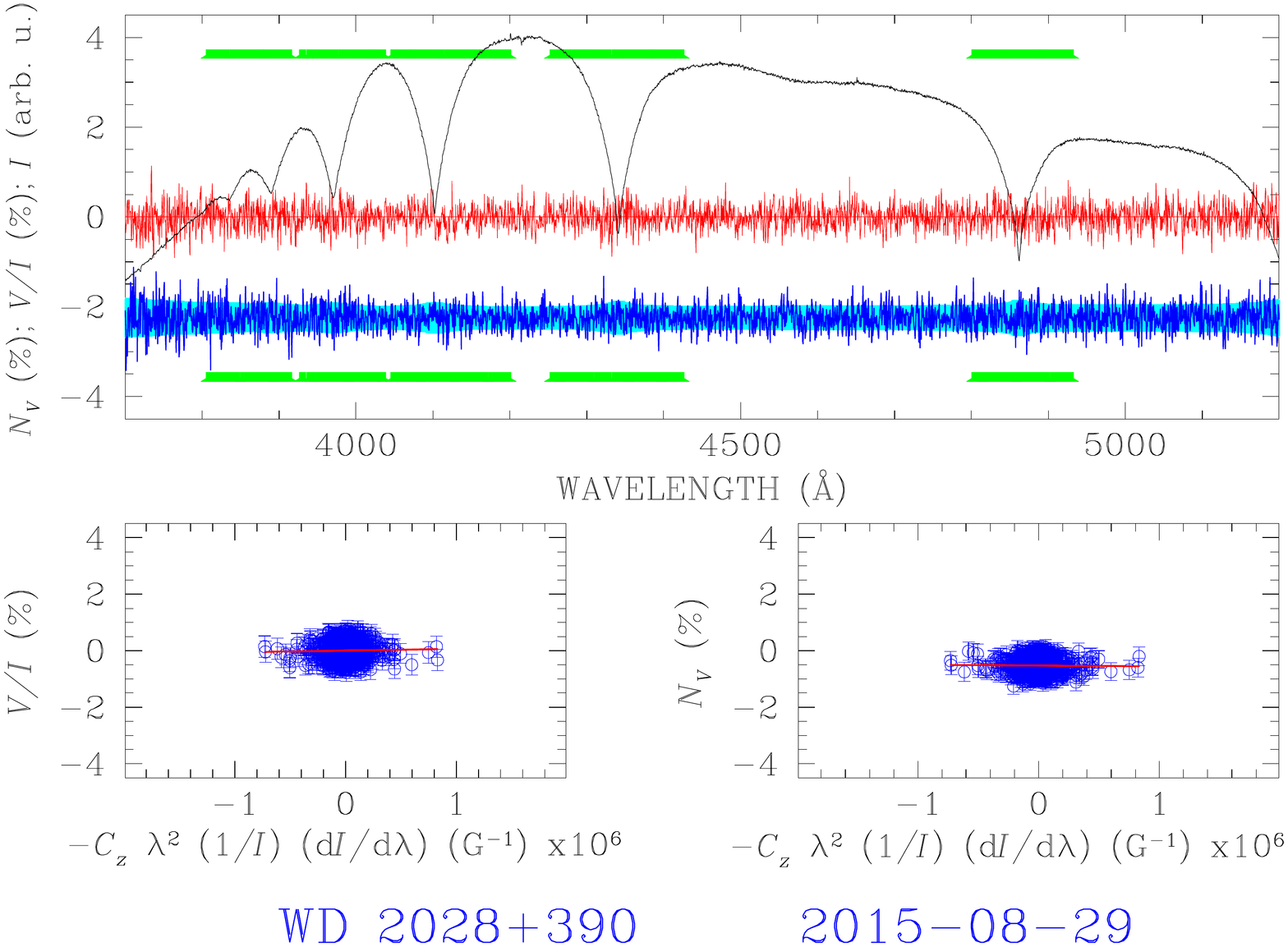}
\includegraphics*[angle=270,width=8.0cm,trim={0.90cm 0.0cm 0.1cm 1.0cm},clip]{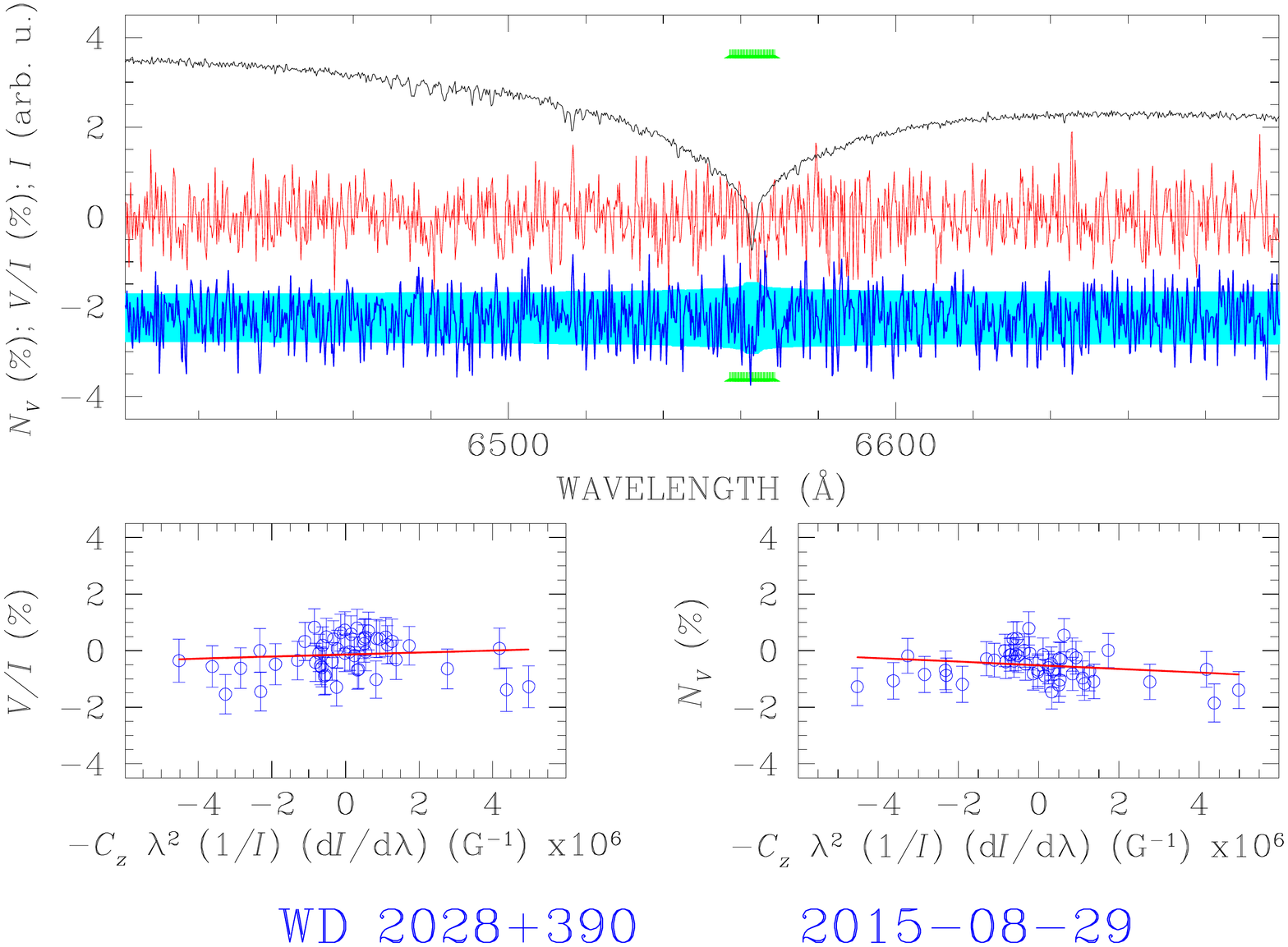} \\
\includegraphics*[angle=270,width=8.0cm,trim={0.90cm 0.0cm 0.1cm 1.0cm},clip]{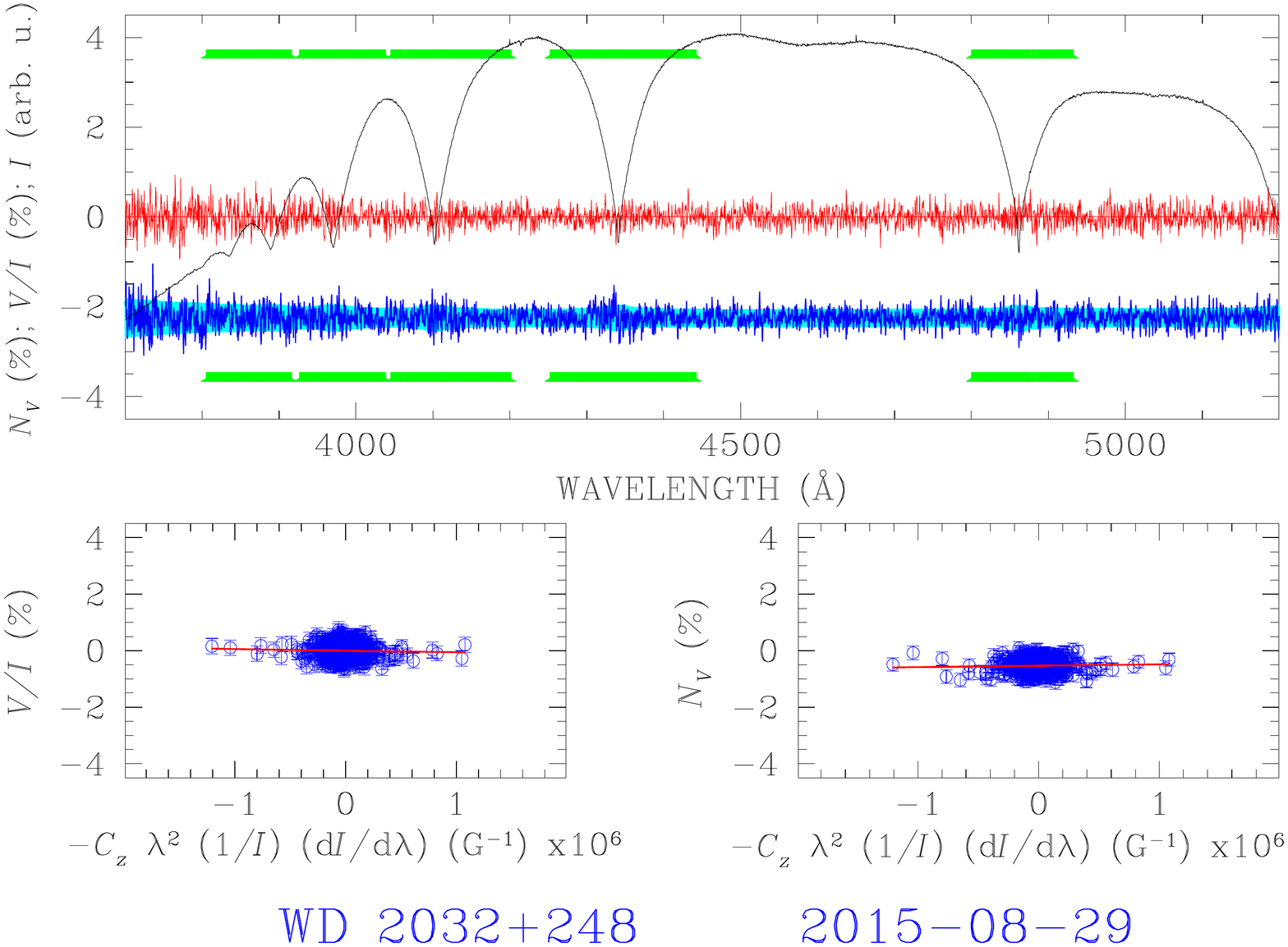}
\includegraphics*[angle=270,width=8.0cm,trim={0.90cm 0.0cm 0.1cm 1.0cm},clip]{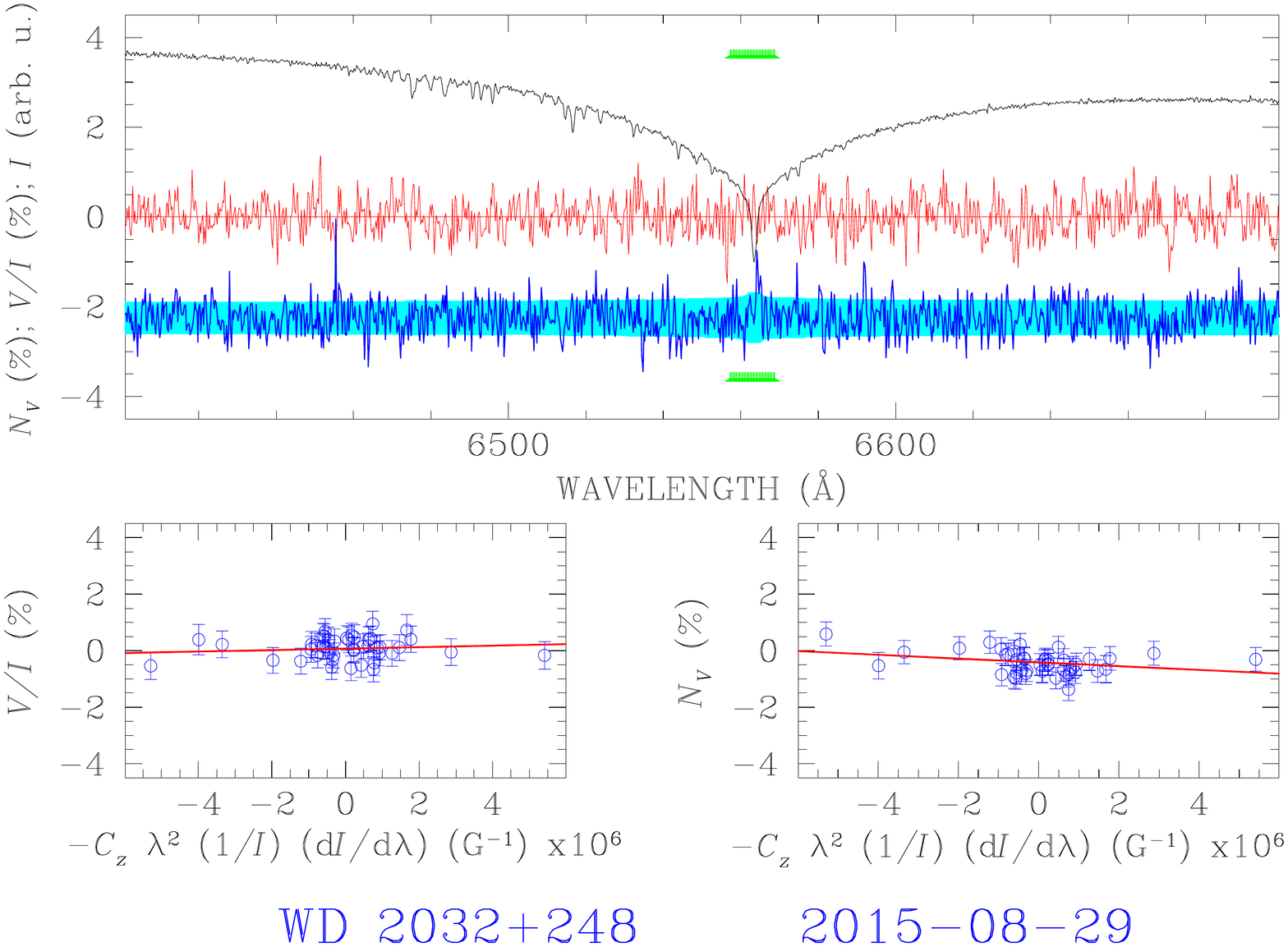} \\
\includegraphics*[angle=270,width=8.0cm,trim={0.90cm 0.0cm 0.1cm 1.0cm},clip]{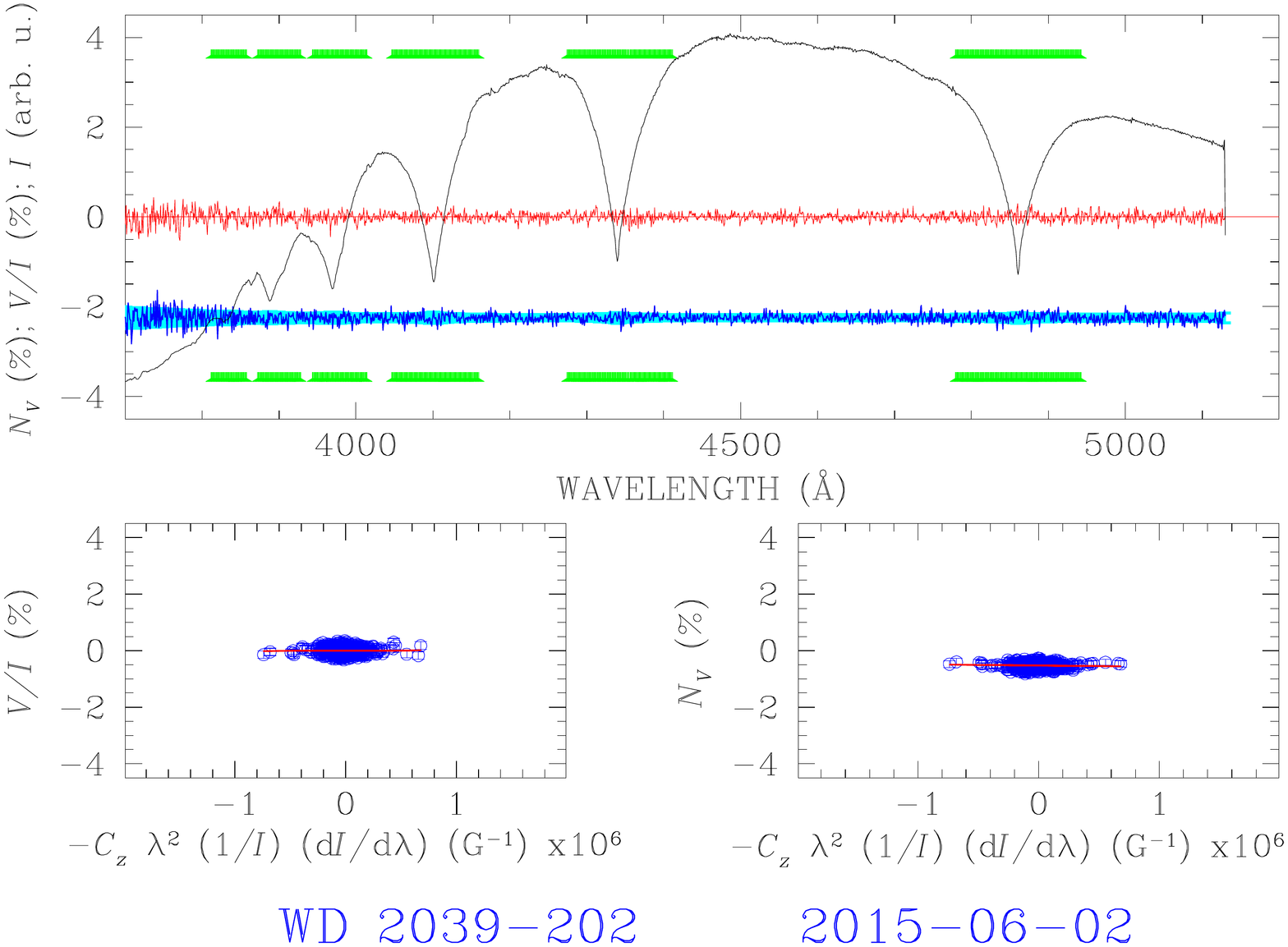} \\
\includegraphics*[angle=270,width=8.0cm,trim={0.90cm 0.0cm 0.1cm 1.0cm},clip]{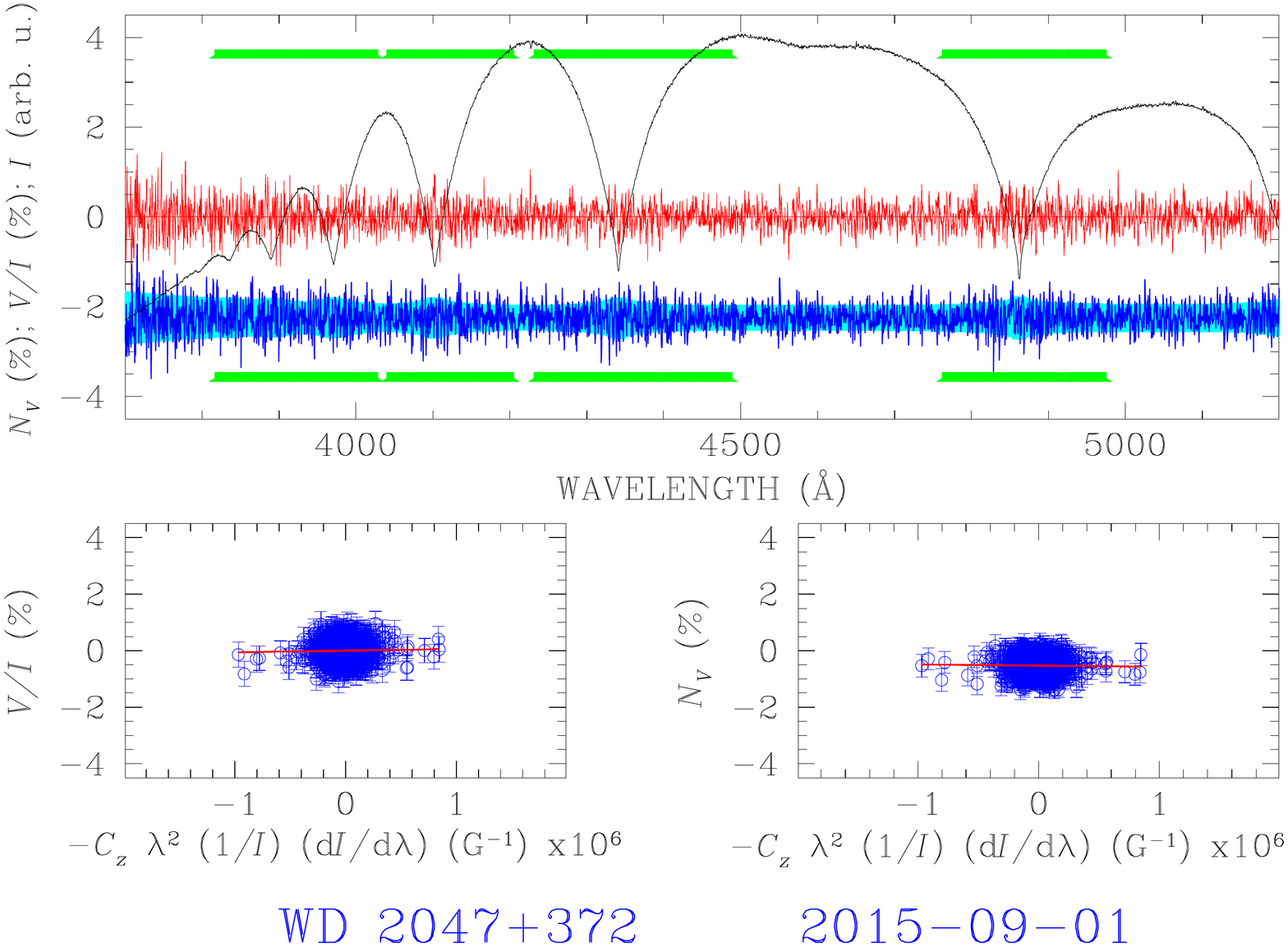}
\includegraphics*[angle=270,width=8.0cm,trim={0.90cm 0.0cm 0.1cm 1.0cm},clip]{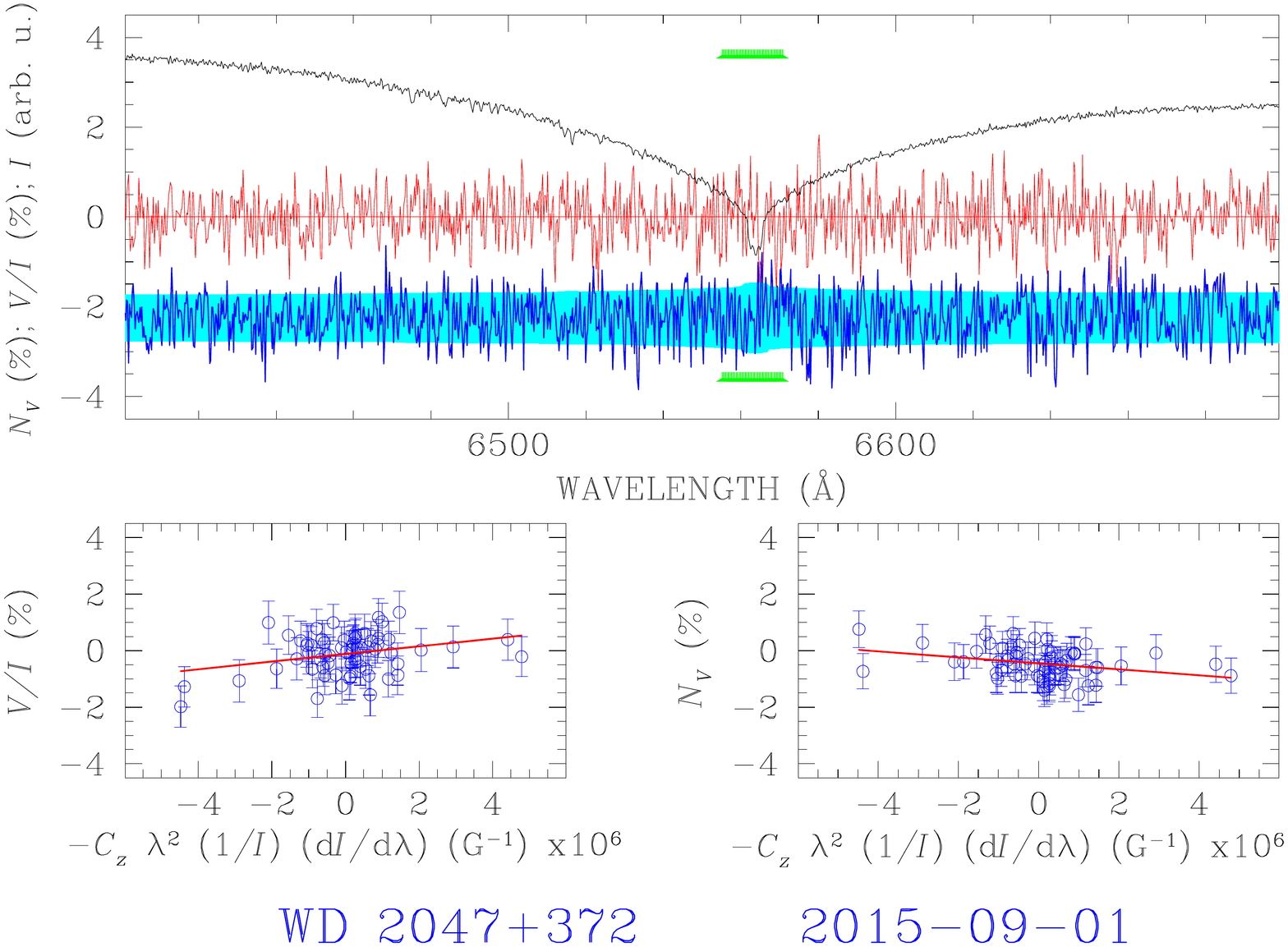} \\
\includegraphics*[angle=270,width=8.0cm,trim={0.90cm 0.0cm 0.1cm 1.0cm},clip]{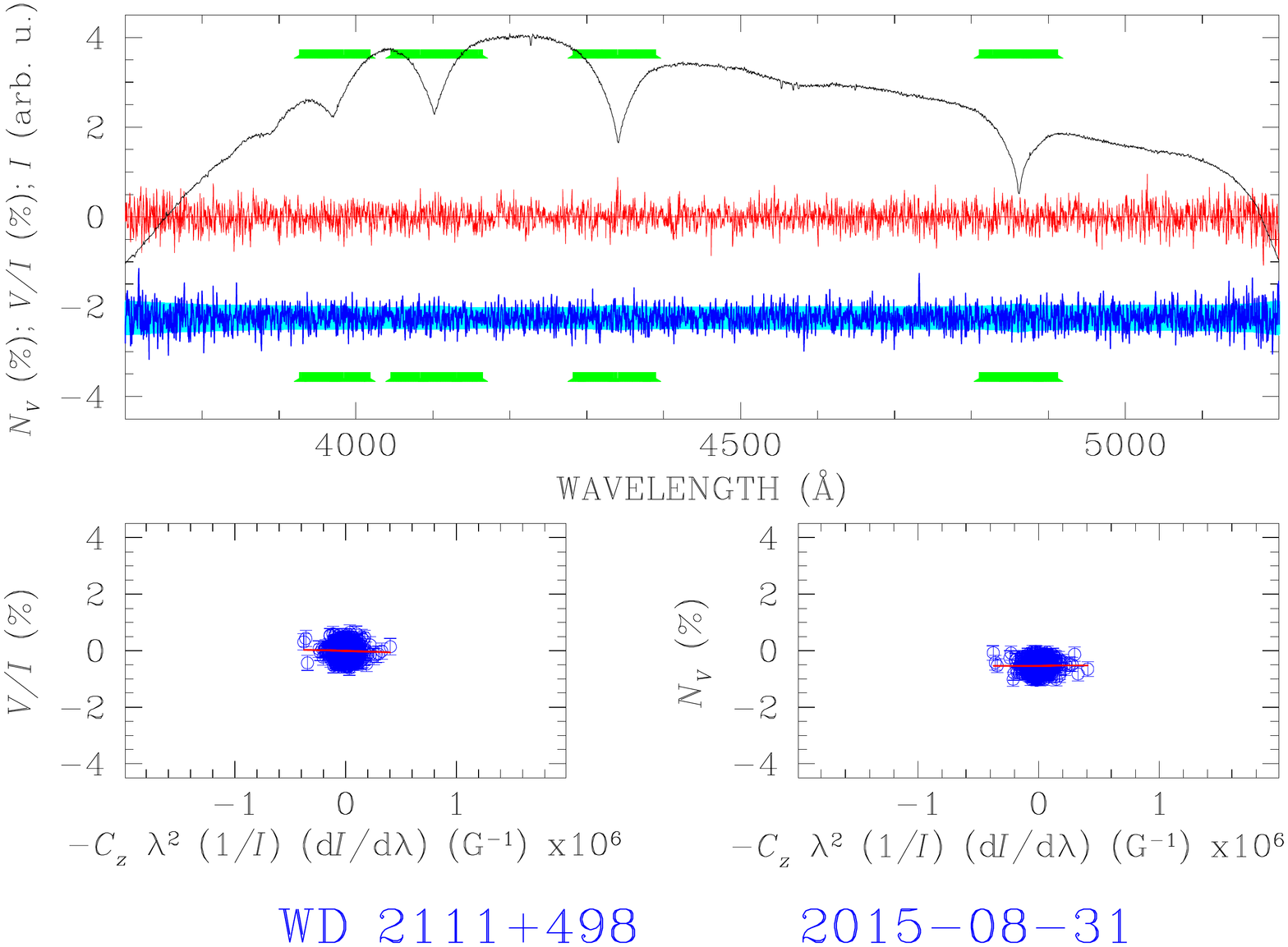}
\includegraphics*[angle=270,width=8.0cm,trim={0.90cm 0.0cm 0.1cm 1.0cm},clip]{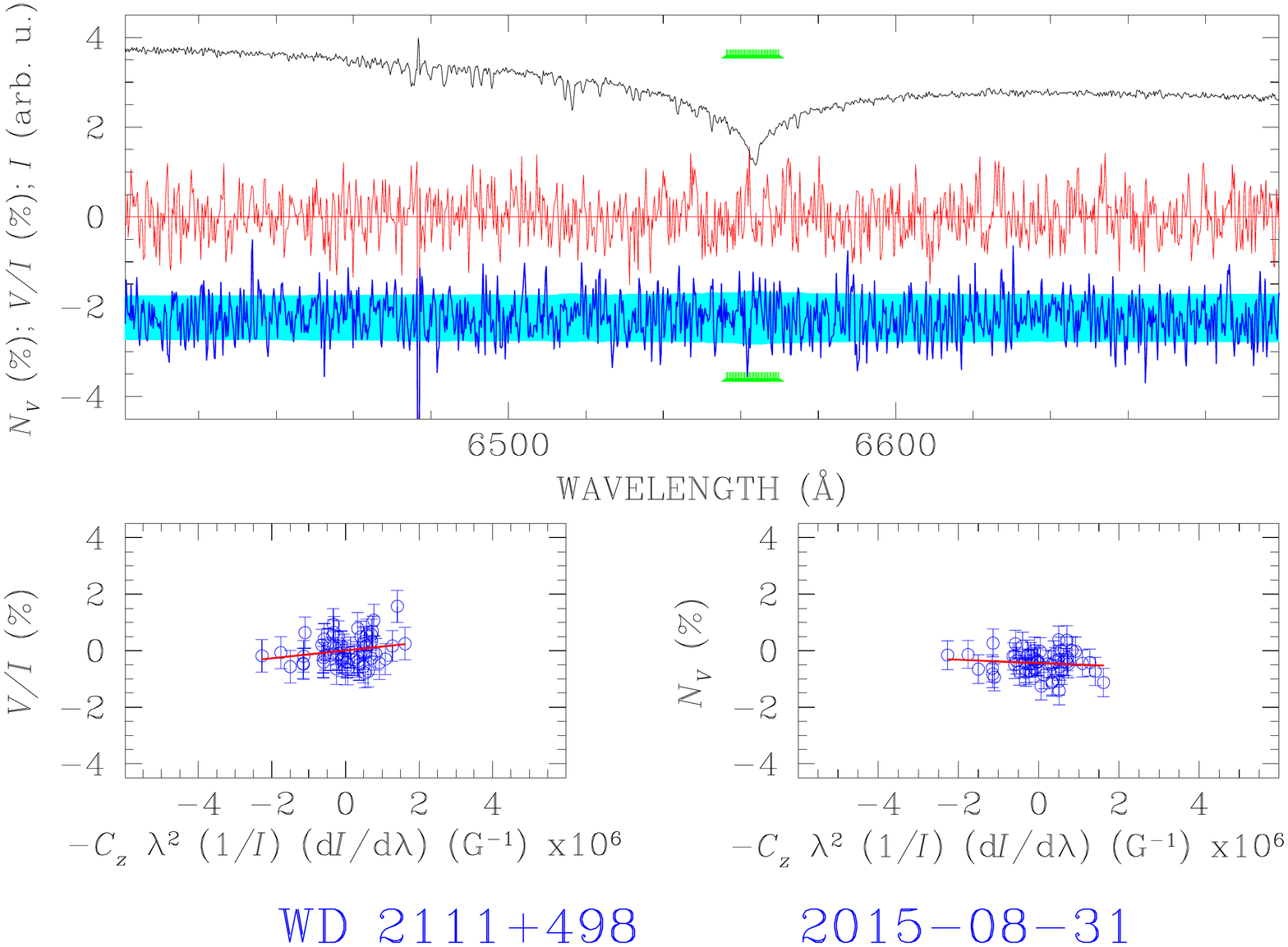} \\
\includegraphics*[angle=270,width=8.0cm,trim={0.90cm 0.0cm 0.1cm 1.0cm},clip]{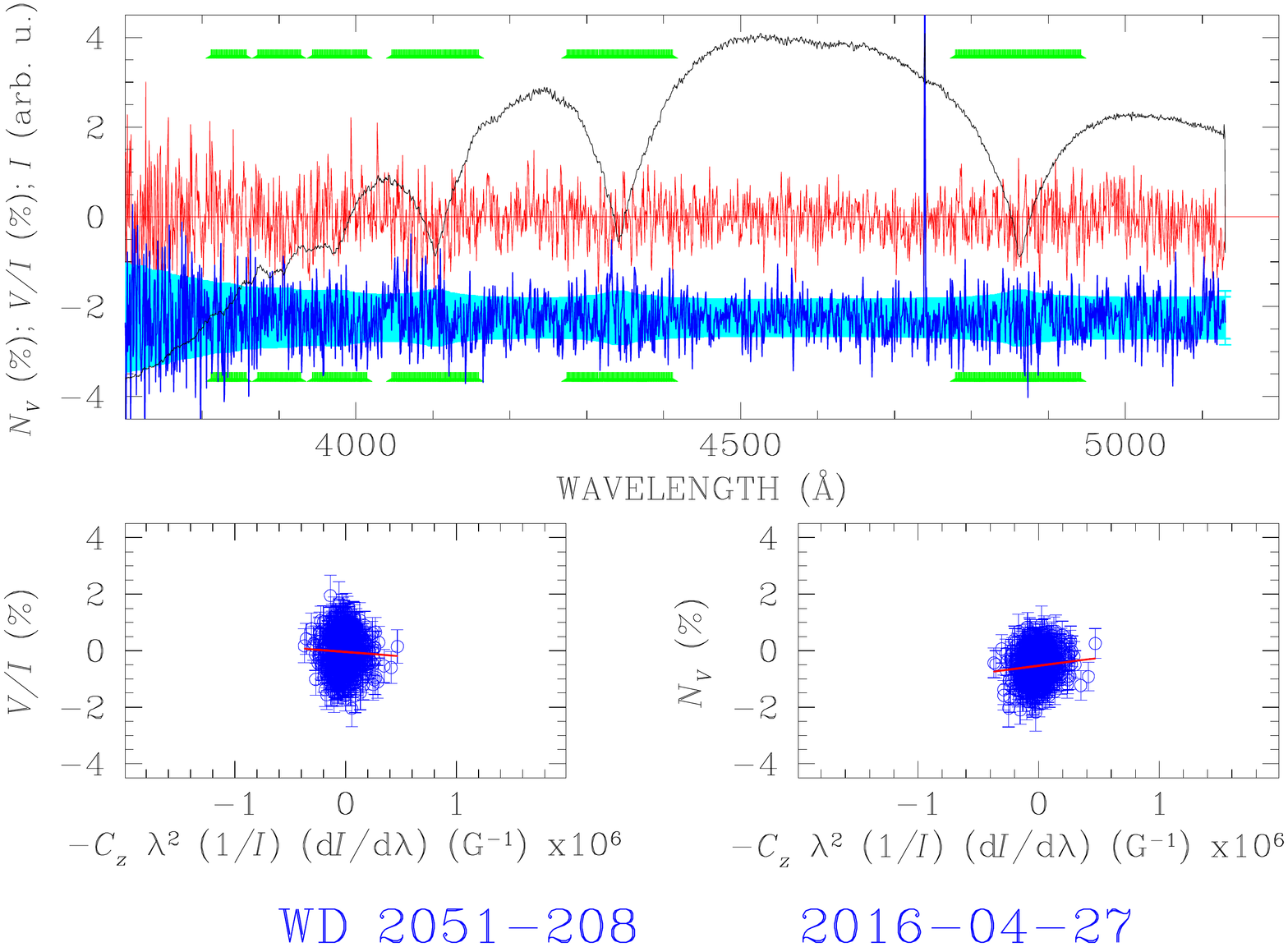} \\
\includegraphics*[angle=270,width=8.0cm,trim={0.90cm 0.0cm 0.1cm 1.0cm},clip]{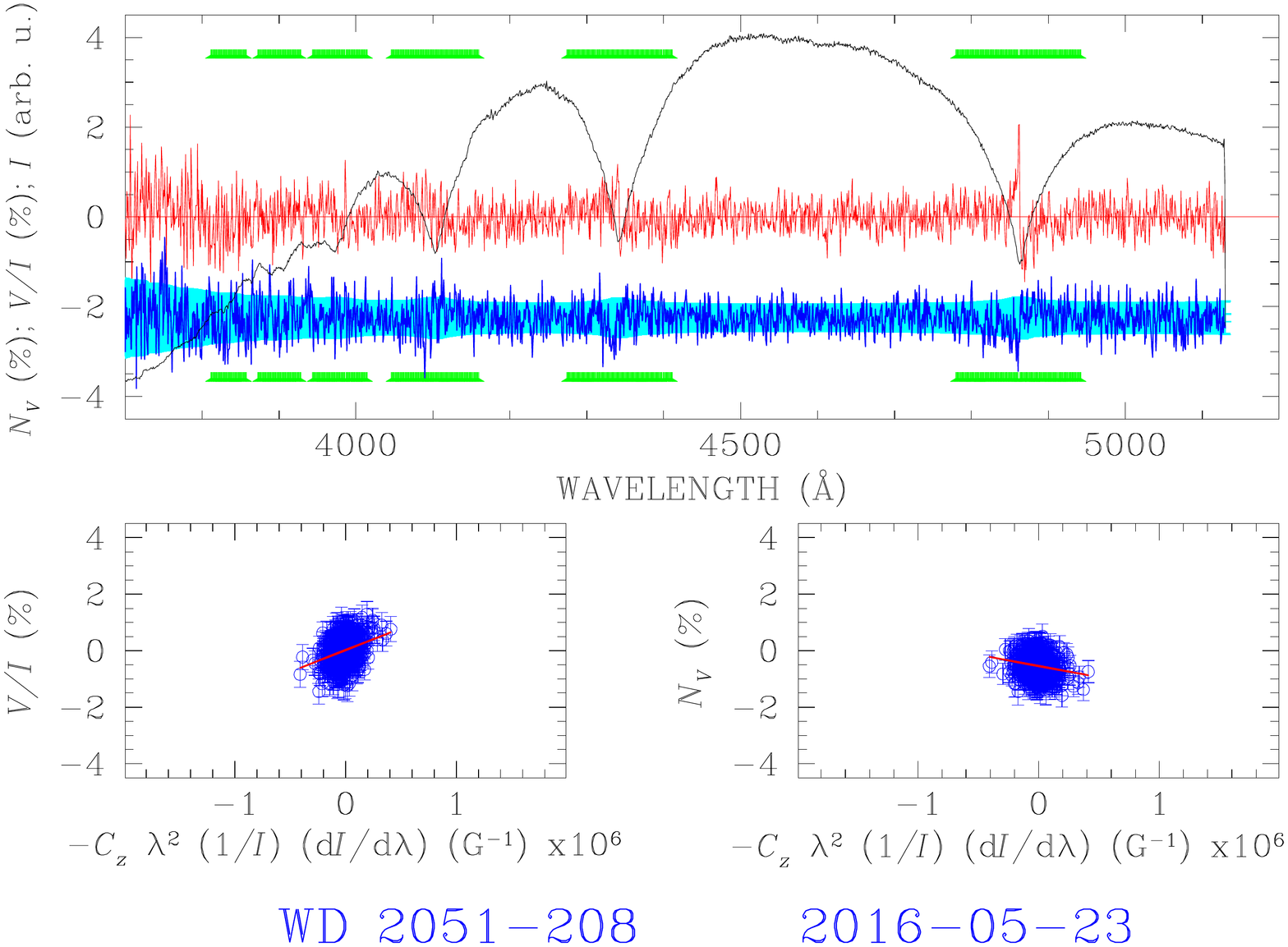} \\
\includegraphics*[angle=270,width=8.0cm,trim={0.90cm 0.0cm 0.1cm 1.0cm},clip]{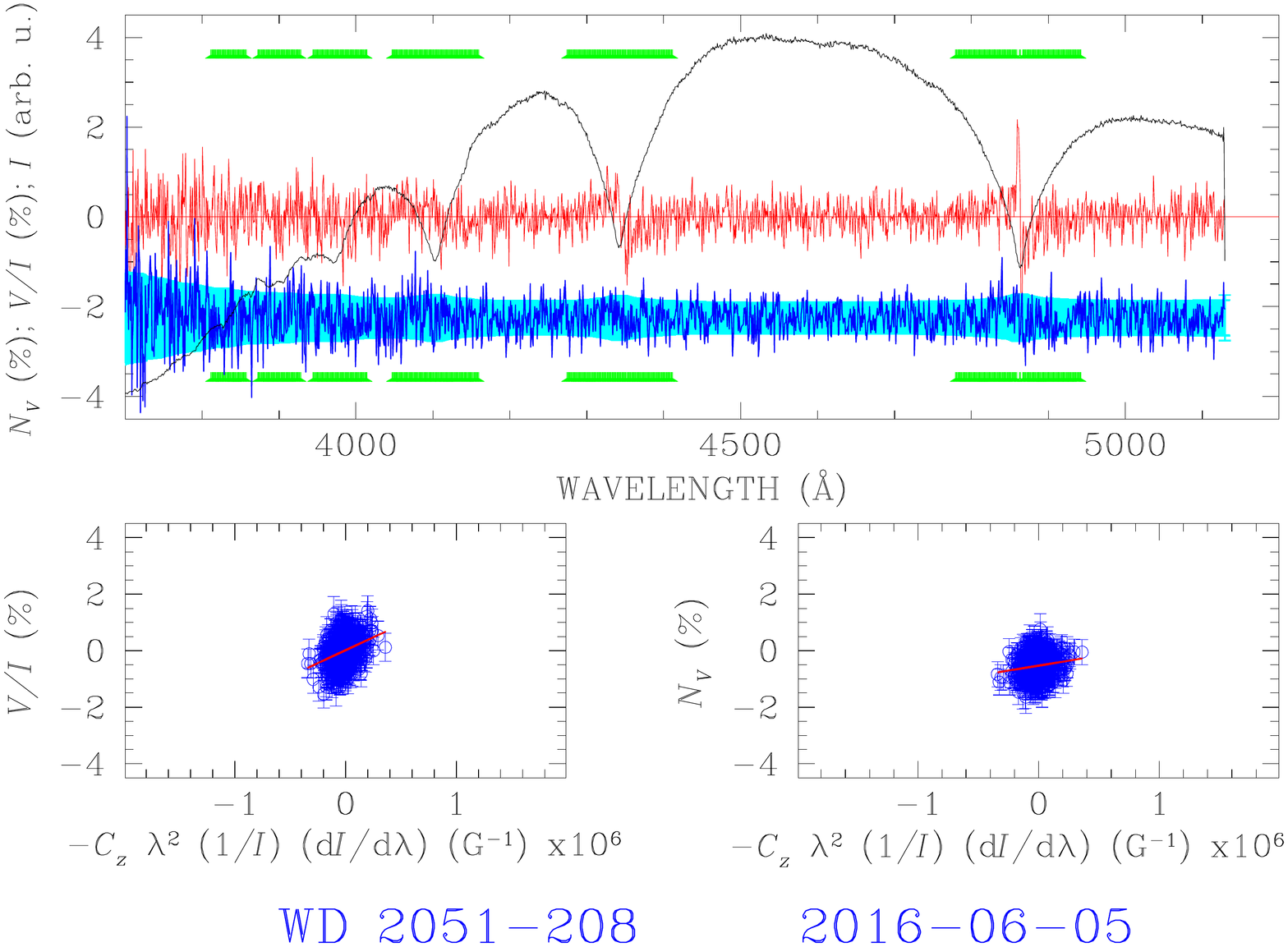} \\
\includegraphics*[angle=270,width=8.0cm,trim={0.90cm 0.0cm 0.1cm 1.0cm},clip]{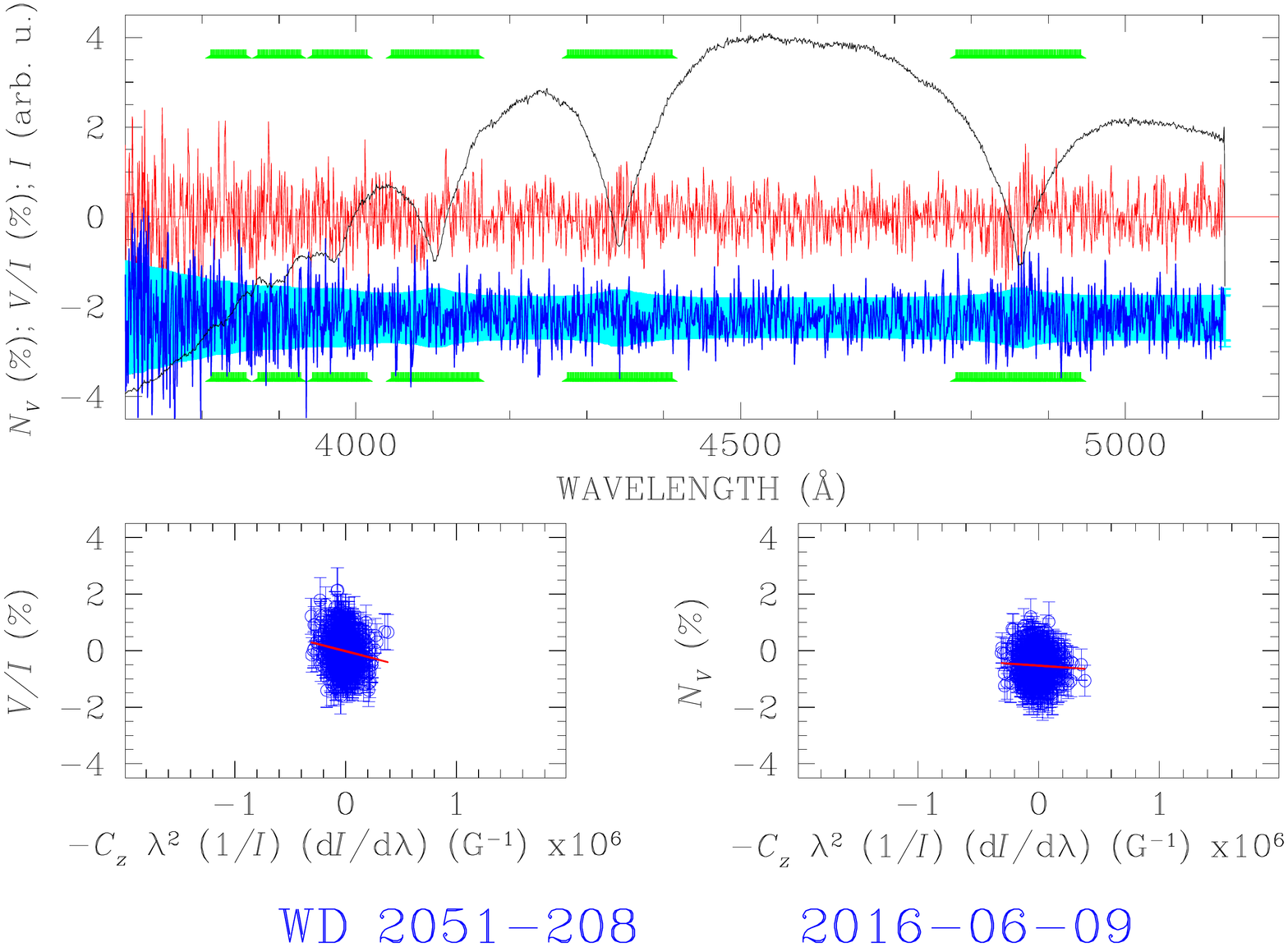} \\
\includegraphics*[angle=270,width=8.0cm,trim={0.90cm 0.0cm 0.1cm 1.0cm},clip]{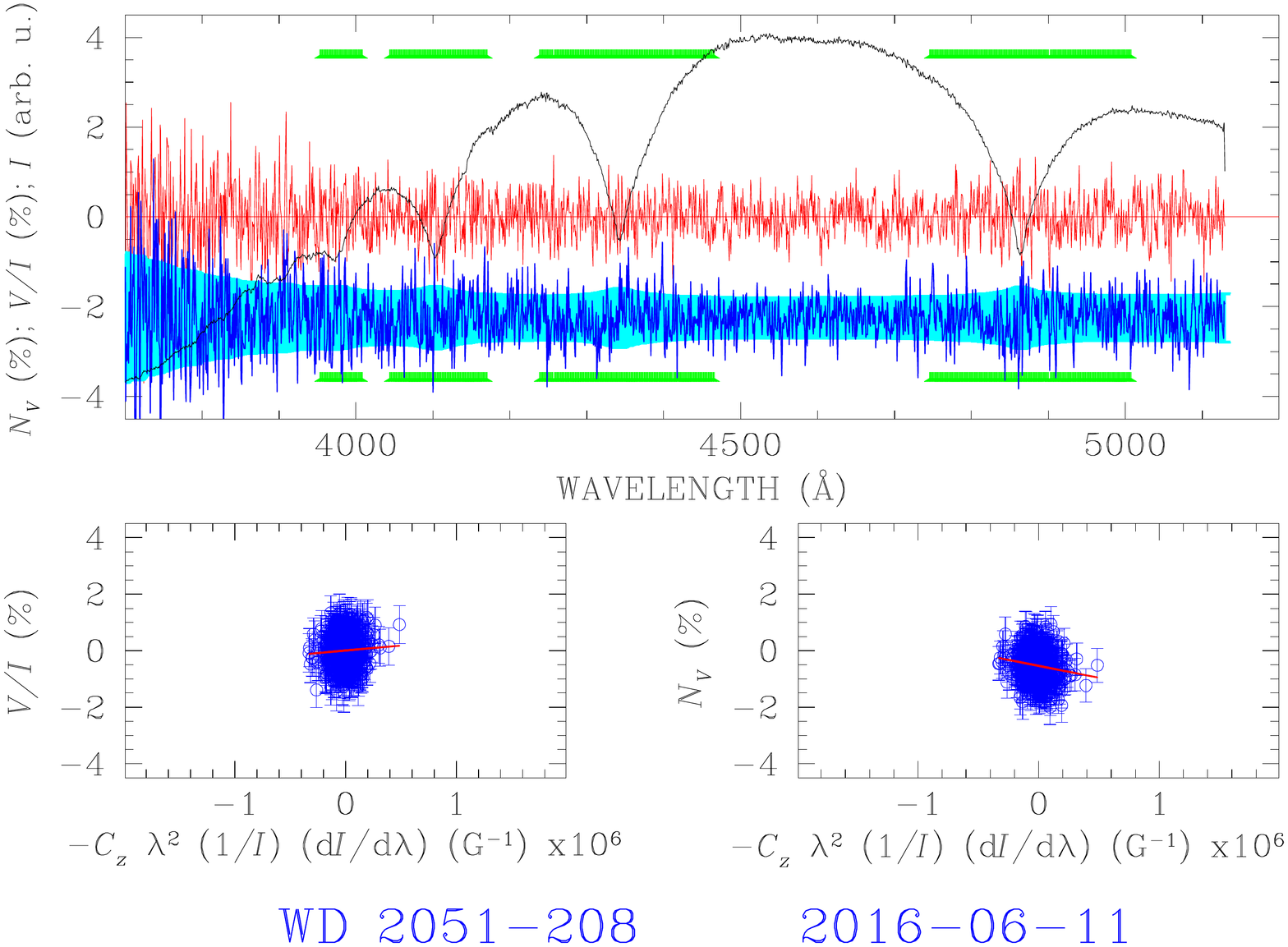} \\
\includegraphics*[angle=270,width=8.0cm,trim={0.90cm 0.0cm 0.1cm 1.0cm},clip]{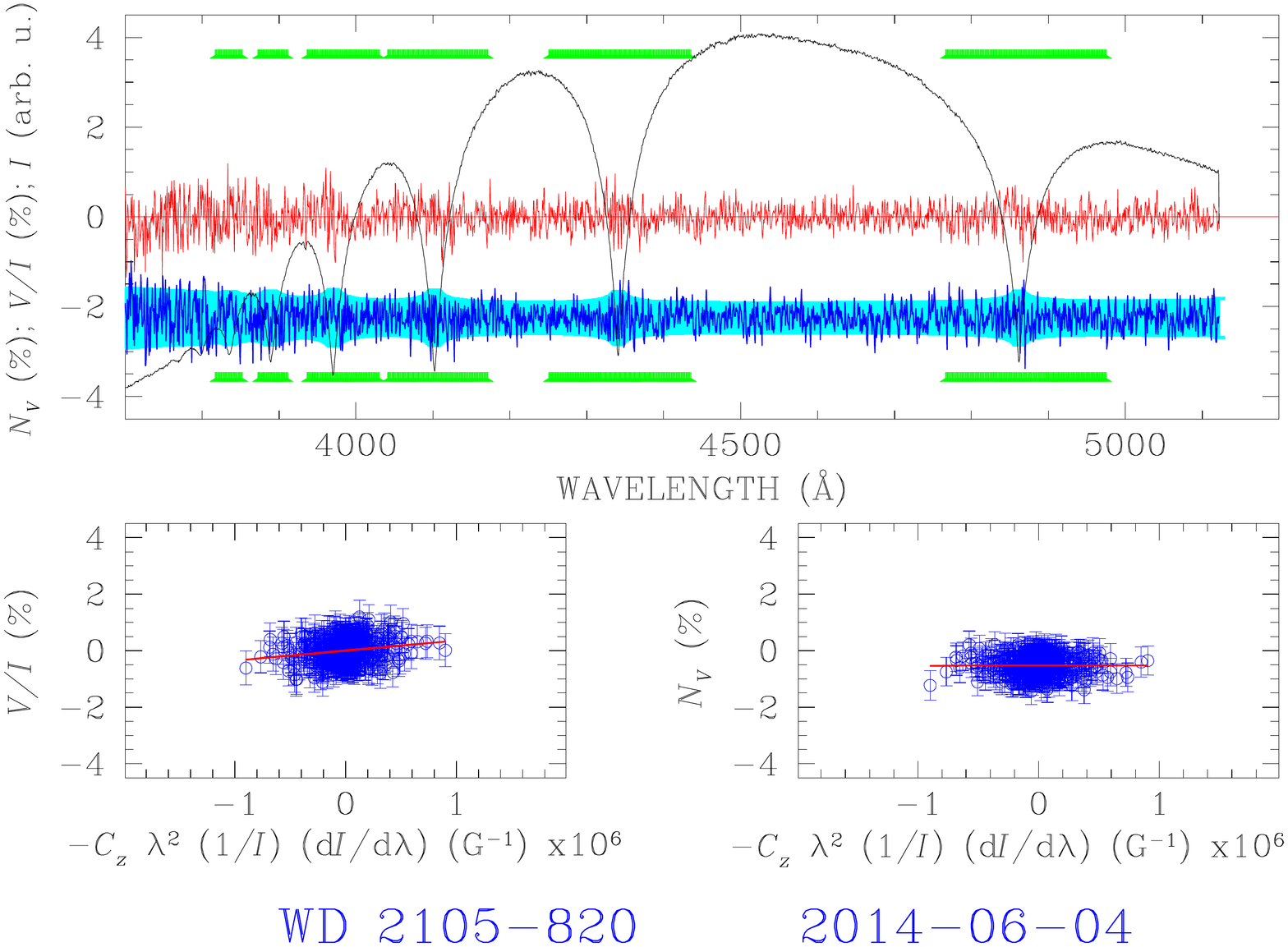} \\
\includegraphics*[angle=270,width=8.0cm,trim={0.90cm 0.0cm 0.1cm 1.0cm},clip]{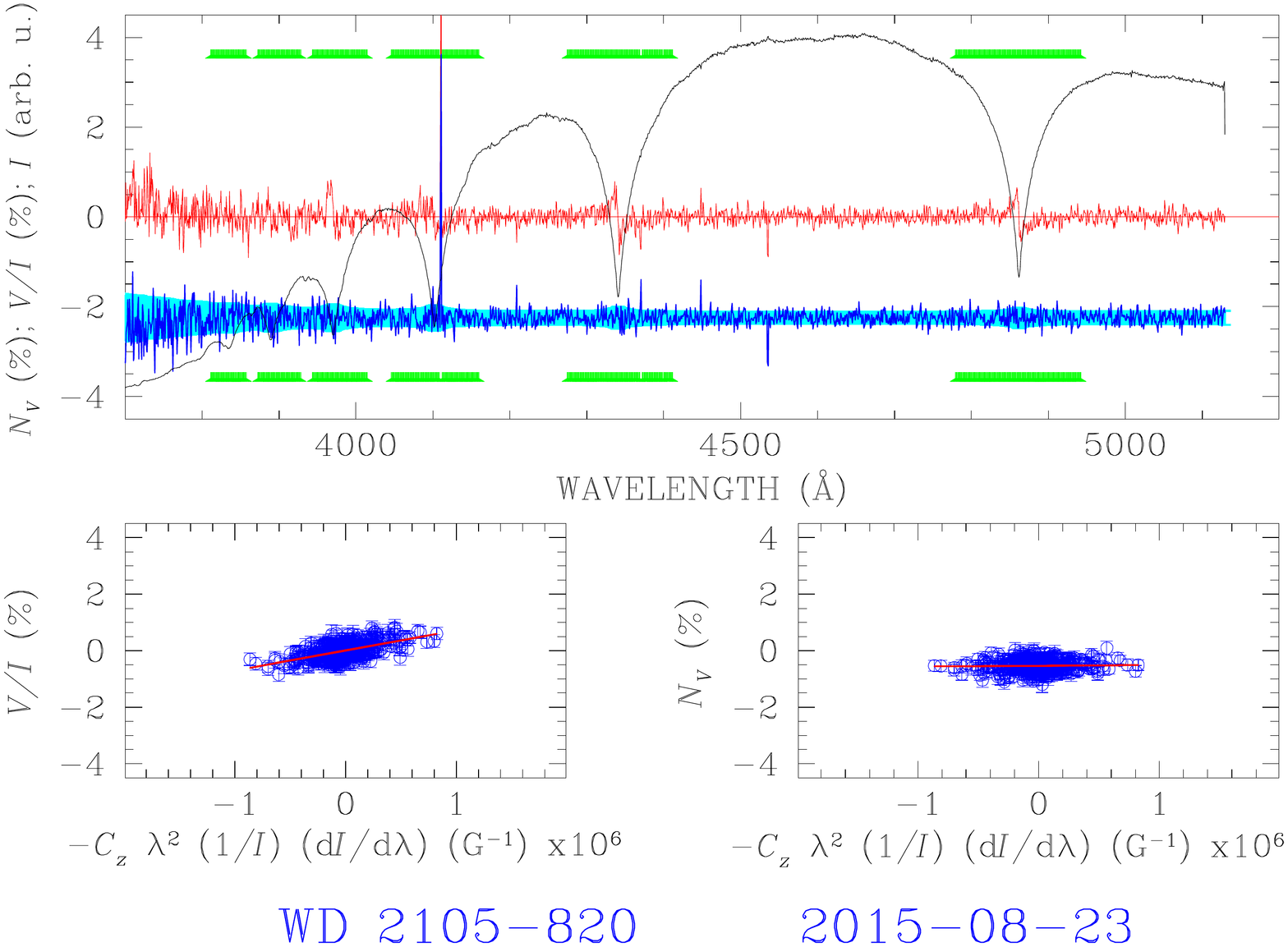} \\
\includegraphics*[angle=270,width=8.0cm,trim={0.90cm 0.0cm 0.1cm 1.0cm},clip]{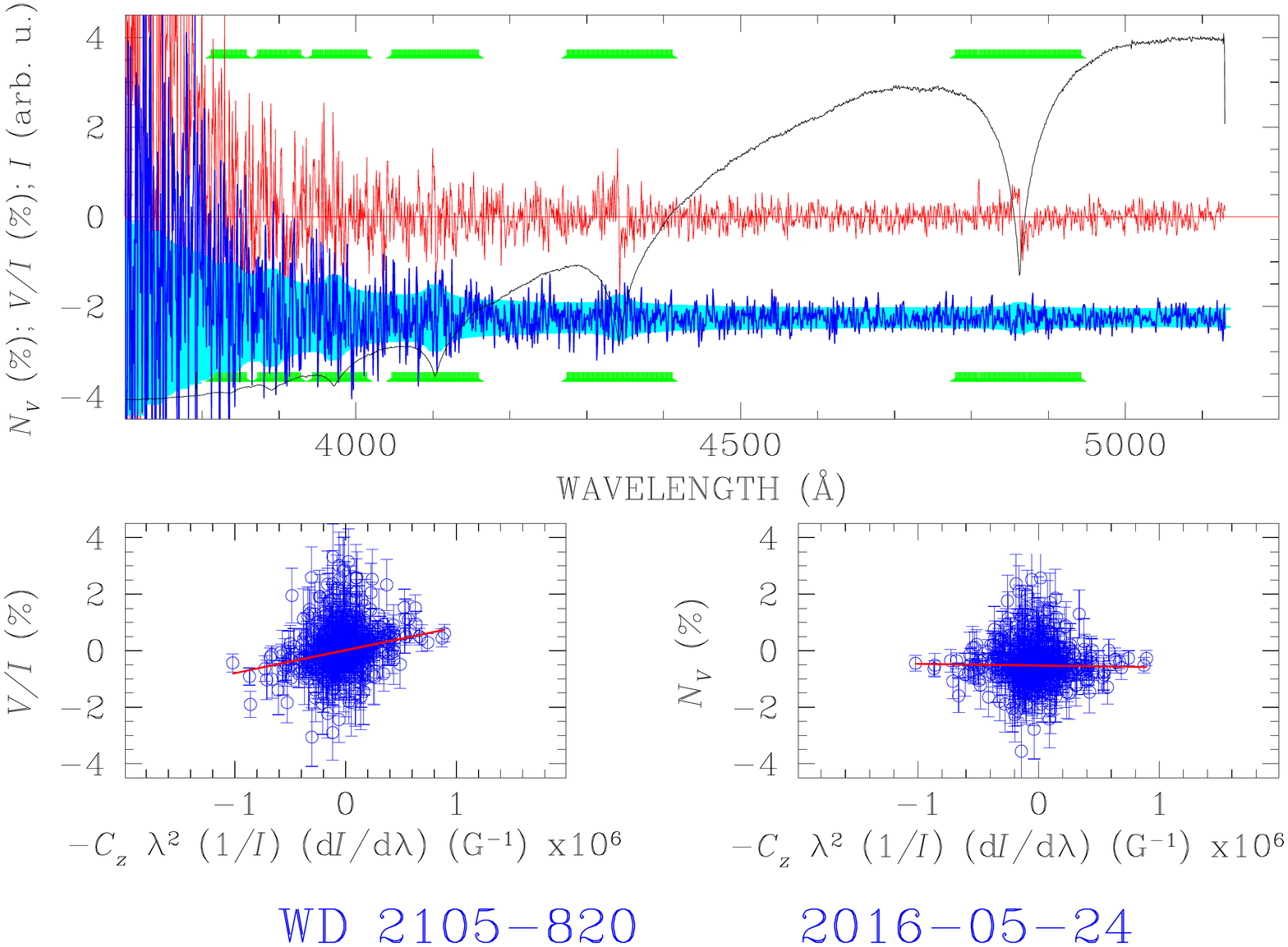} \\
\includegraphics*[angle=270,width=8.0cm,trim={0.90cm 0.0cm 0.1cm 1.0cm},clip]{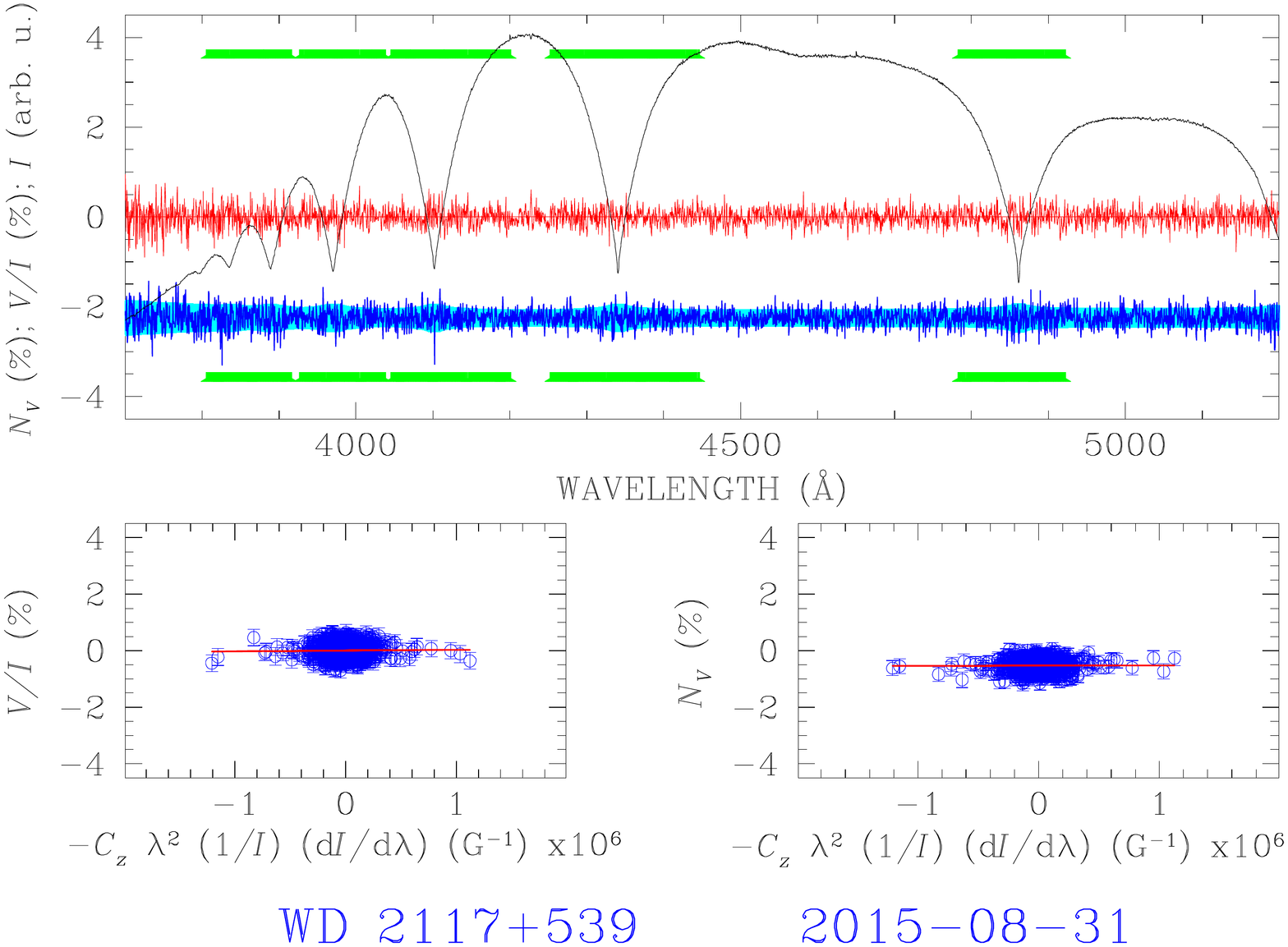}
\includegraphics*[angle=270,width=8.0cm,trim={0.90cm 0.0cm 0.1cm 1.0cm},clip]{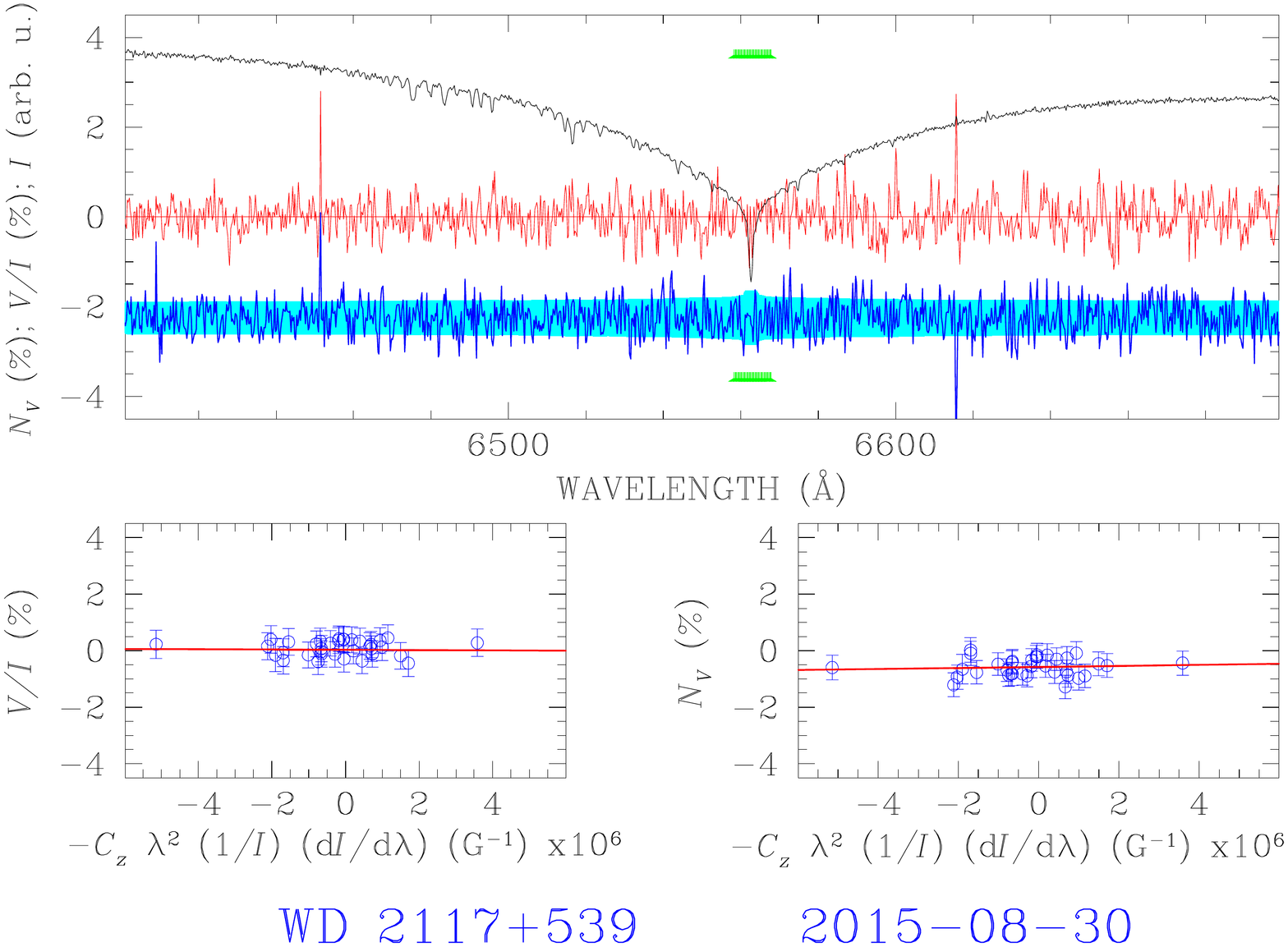} \\
\includegraphics*[angle=270,width=8.0cm,trim={0.90cm 0.0cm 0.1cm 1.0cm},clip]{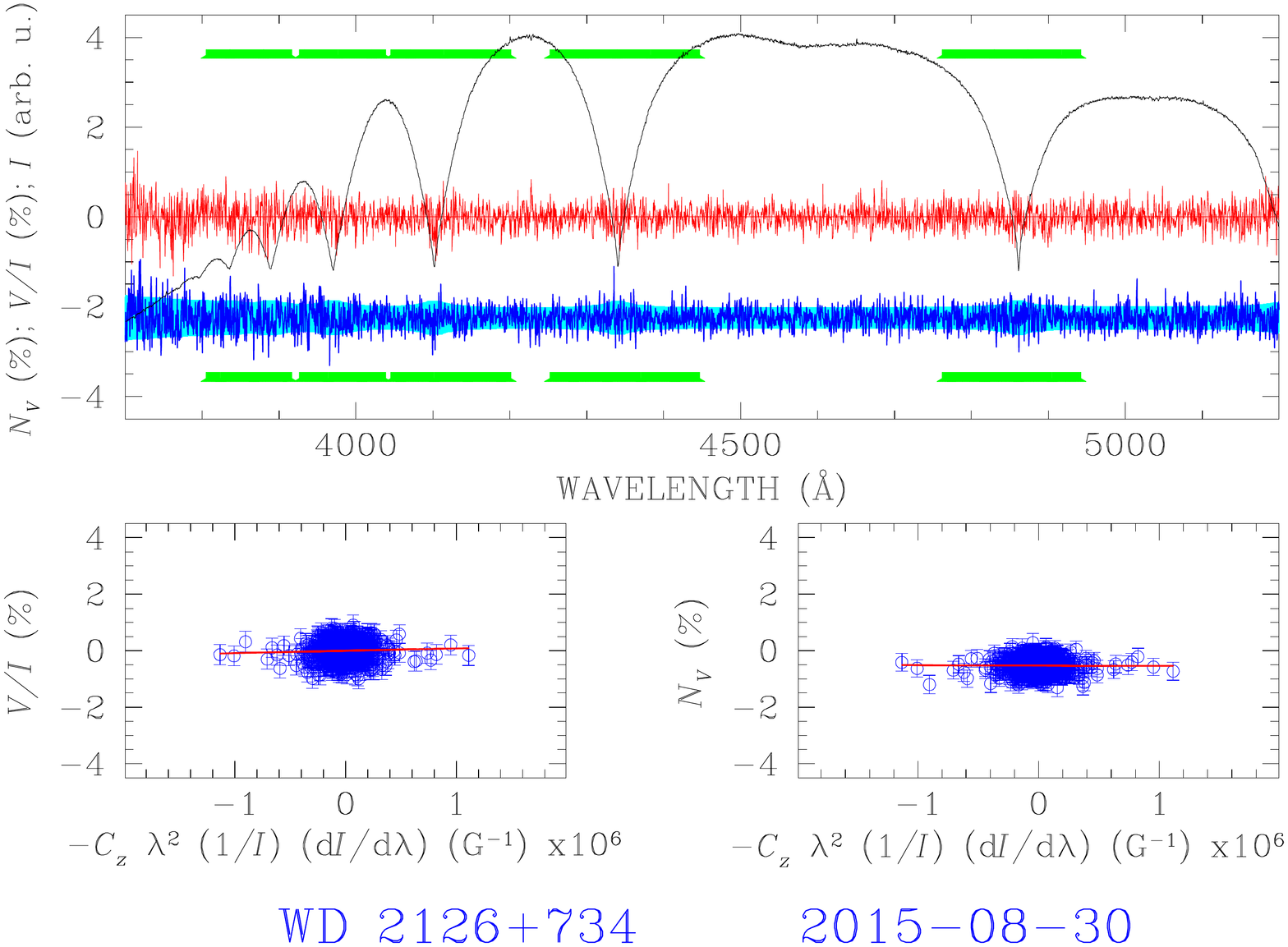}
\includegraphics*[angle=270,width=8.0cm,trim={0.90cm 0.0cm 0.1cm 1.0cm},clip]{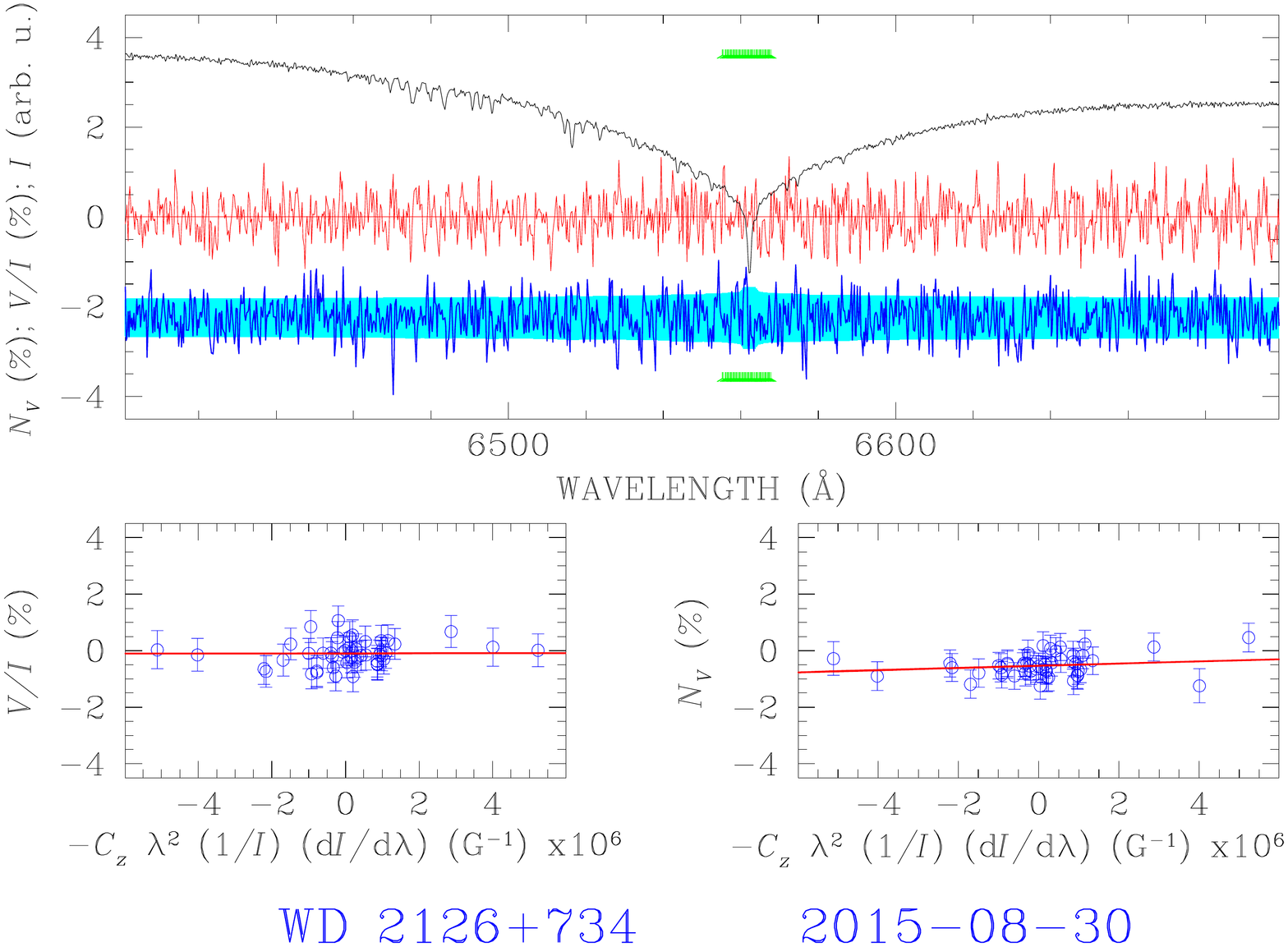} \\
\includegraphics*[angle=270,width=8.0cm,trim={0.90cm 0.0cm 0.1cm 1.0cm},clip]{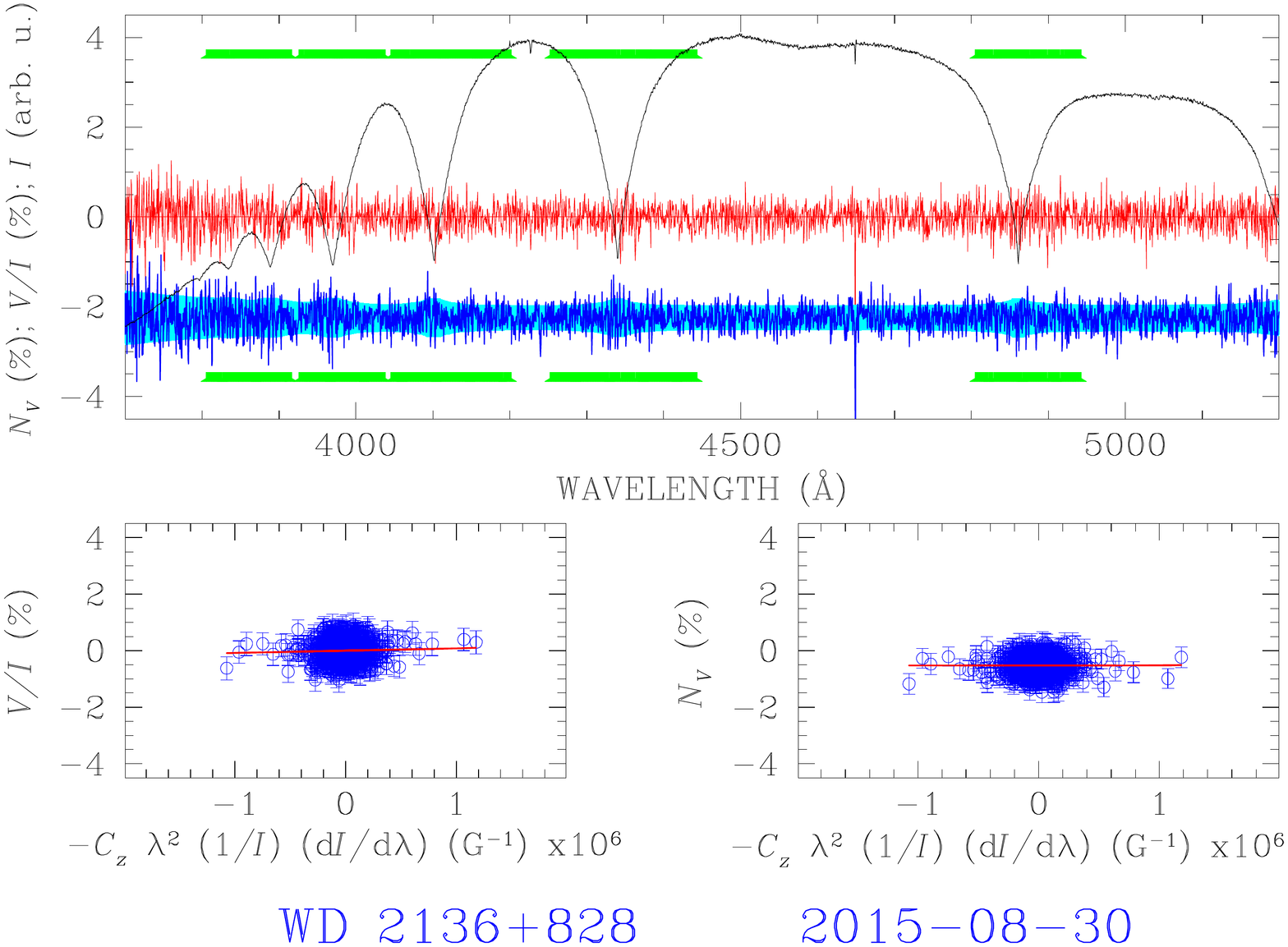}
\includegraphics*[angle=270,width=8.0cm,trim={0.90cm 0.0cm 0.1cm 1.0cm},clip]{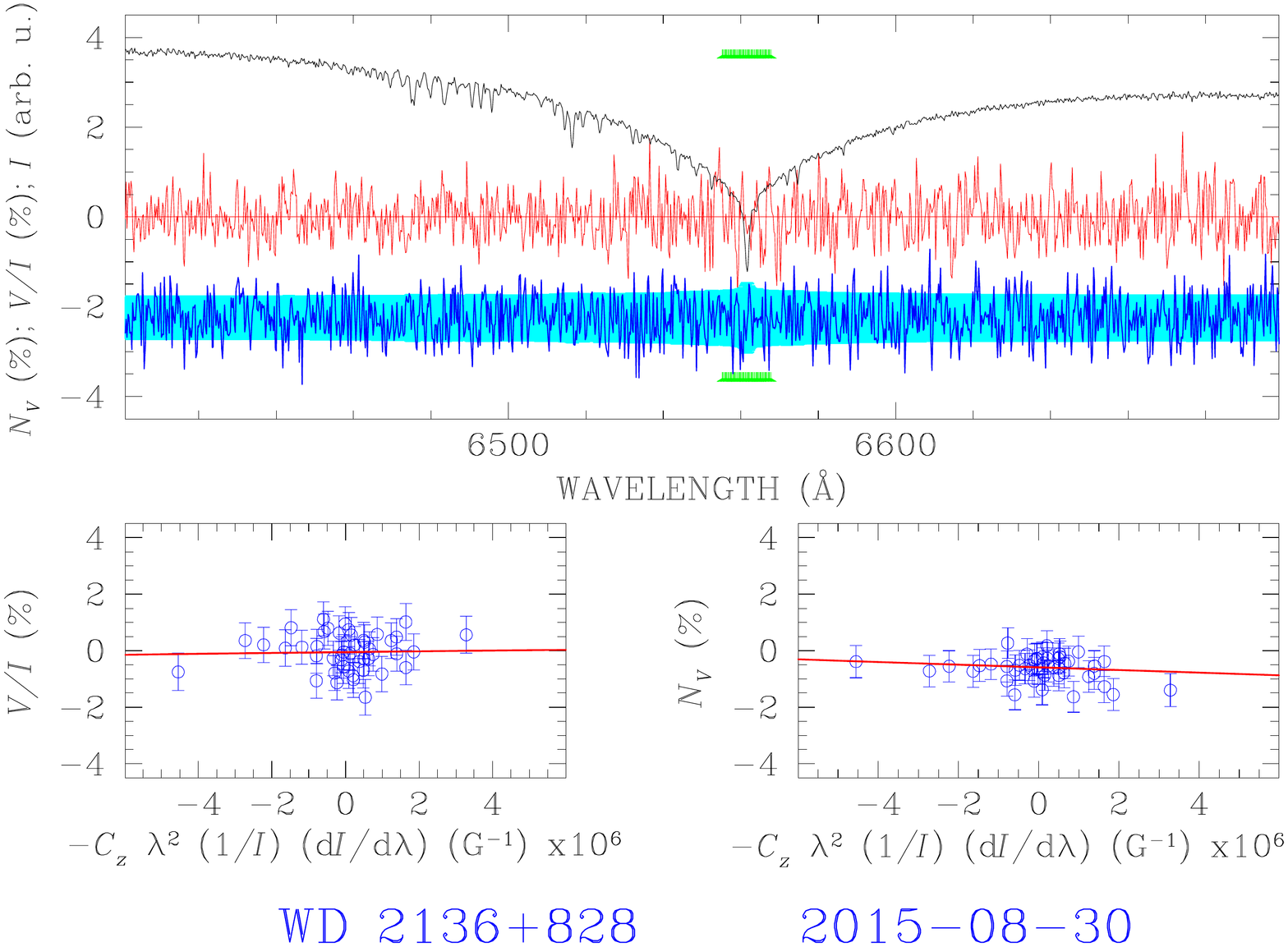} \\
\includegraphics*[angle=270,width=8.0cm,trim={0.90cm 0.0cm 0.1cm 1.0cm},clip]{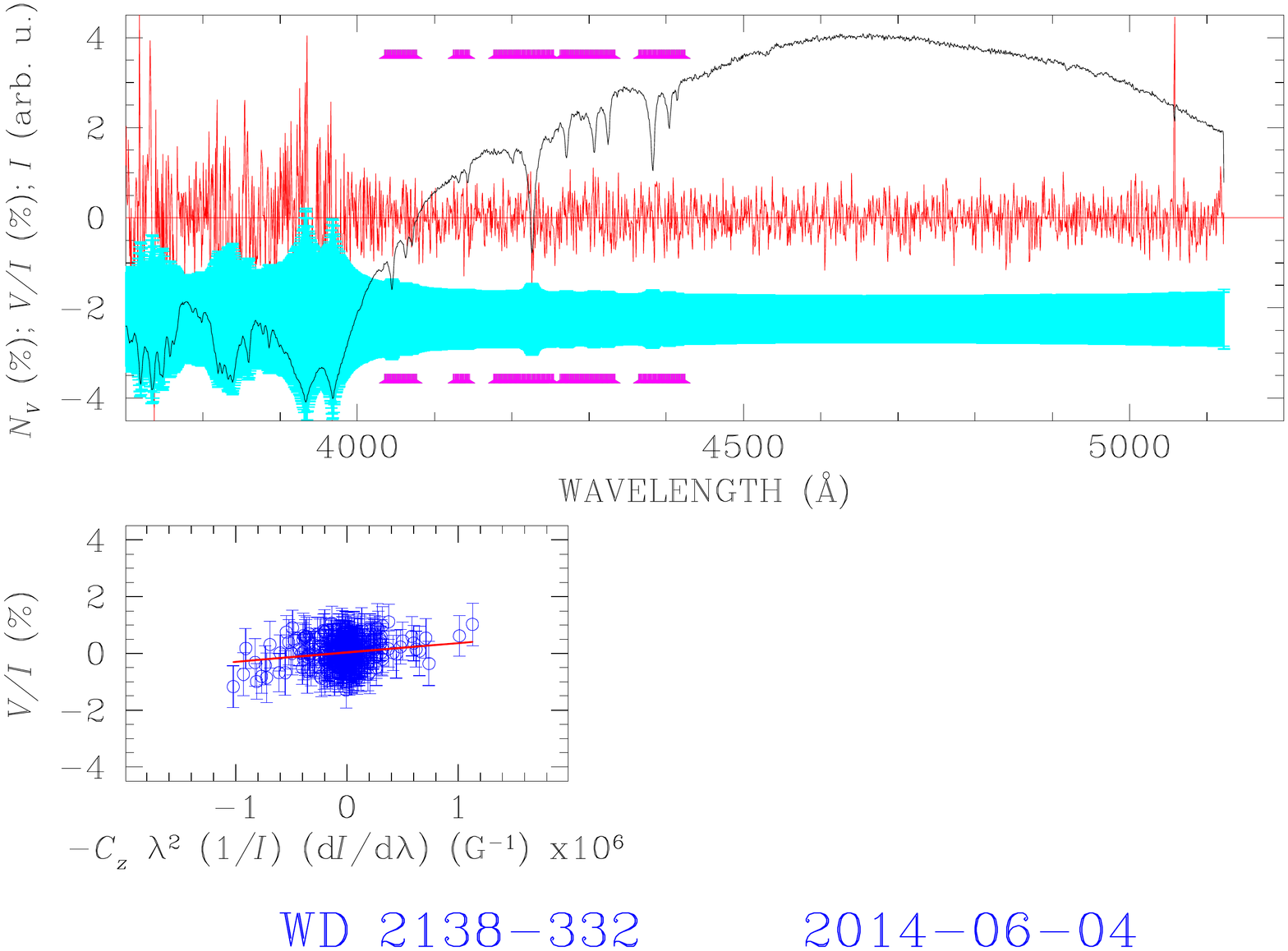}\\
\includegraphics*[angle=270,width=8.0cm,trim={0.90cm 0.0cm 0.1cm 1.0cm},clip]{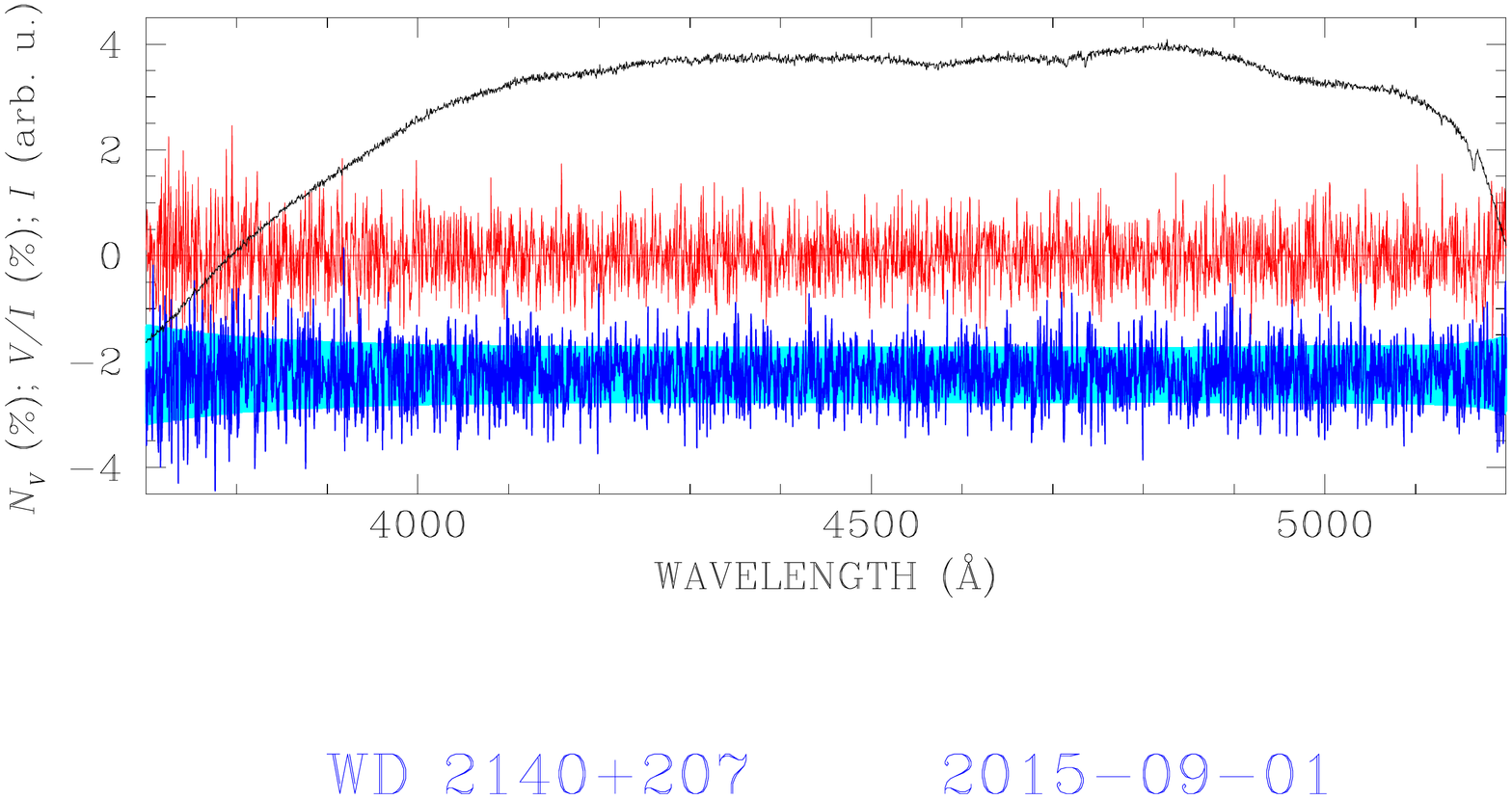}
\includegraphics*[angle=270,width=8.0cm,trim={0.90cm 0.0cm 0.1cm 1.0cm},clip]{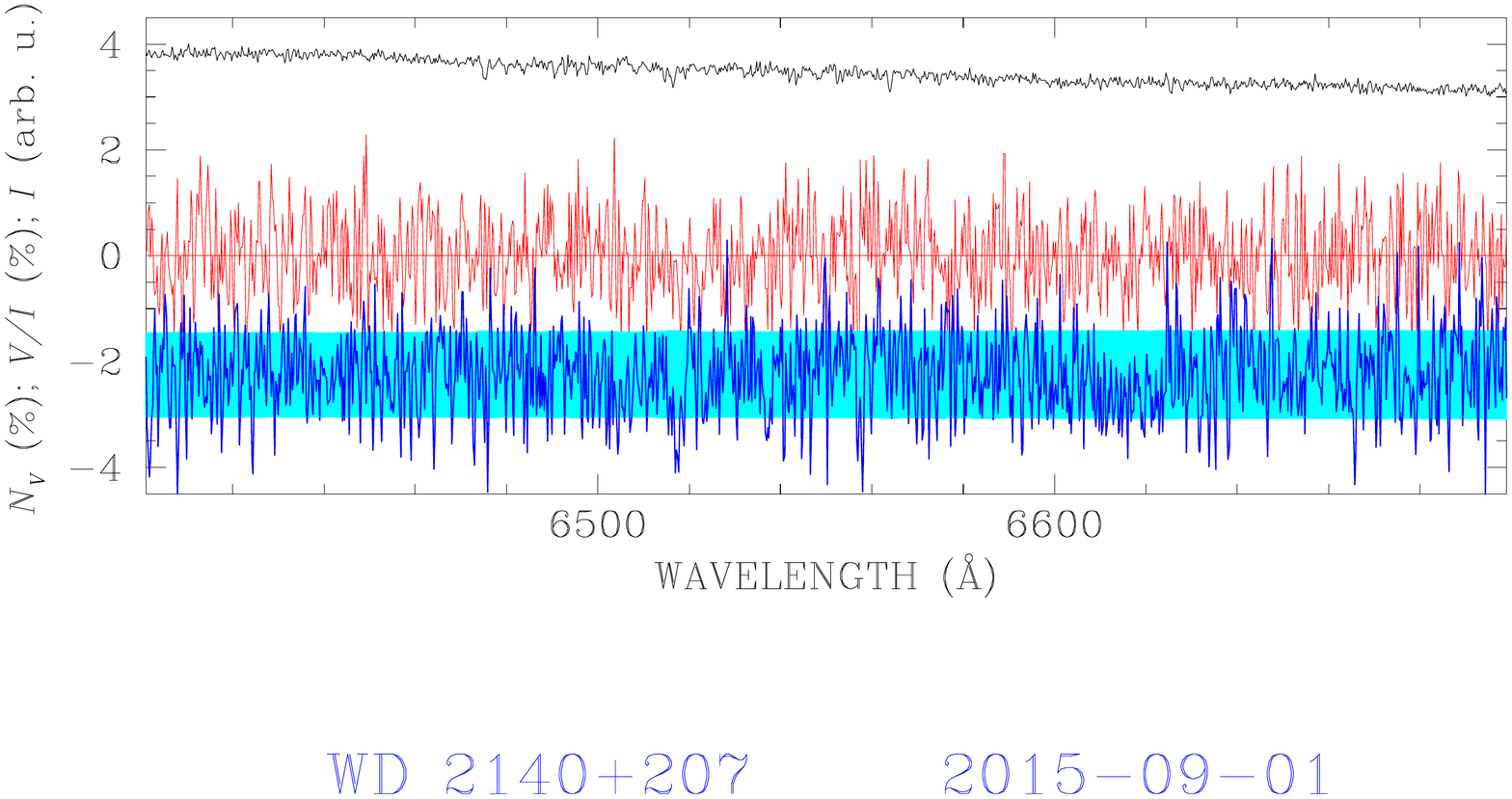} \\
\includegraphics*[angle=270,width=8.0cm,trim={0.90cm 0.0cm 0.1cm 1.0cm},clip]{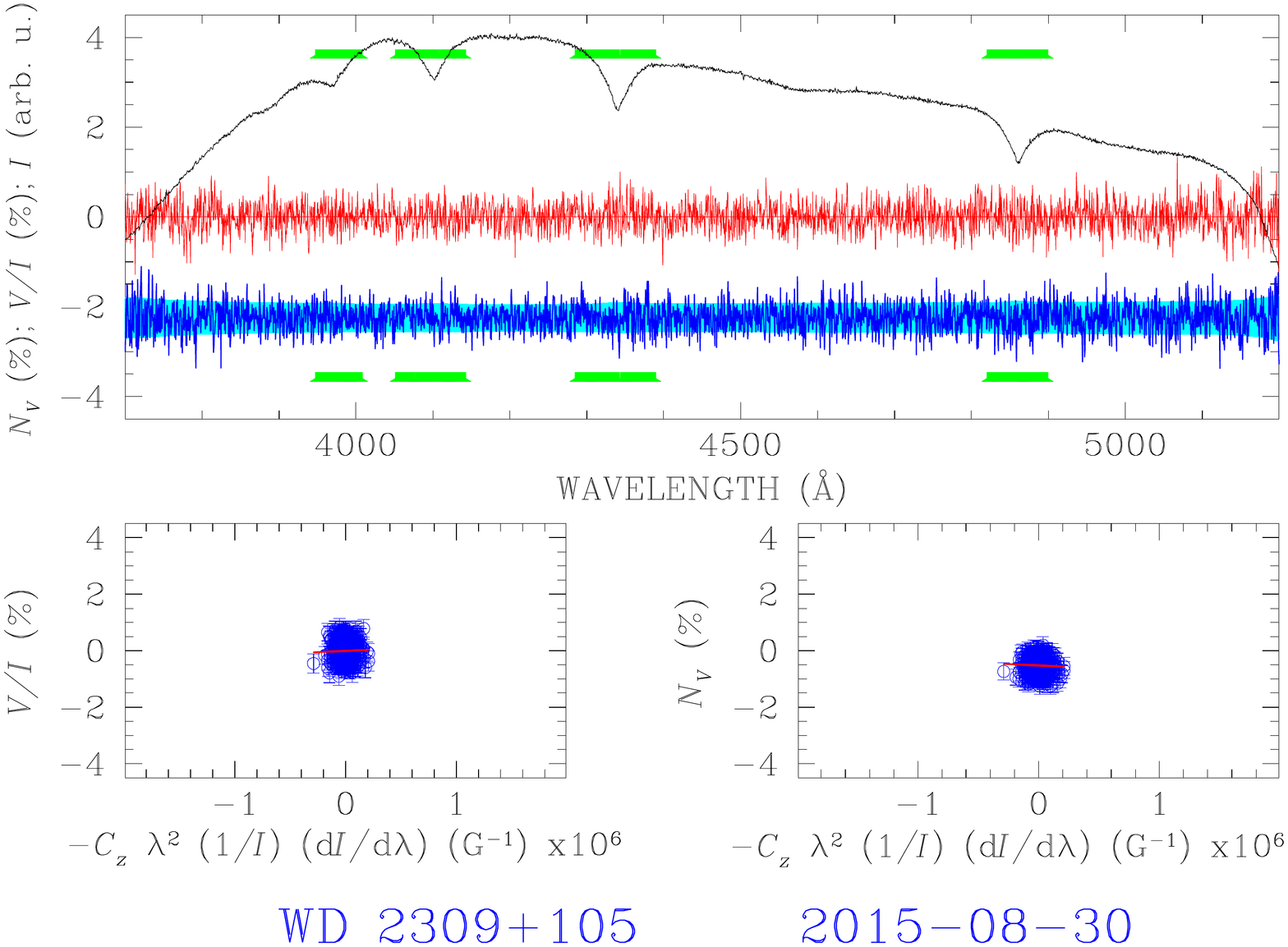}
\includegraphics*[angle=270,width=8.0cm,trim={0.90cm 0.0cm 0.1cm 1.0cm},clip]{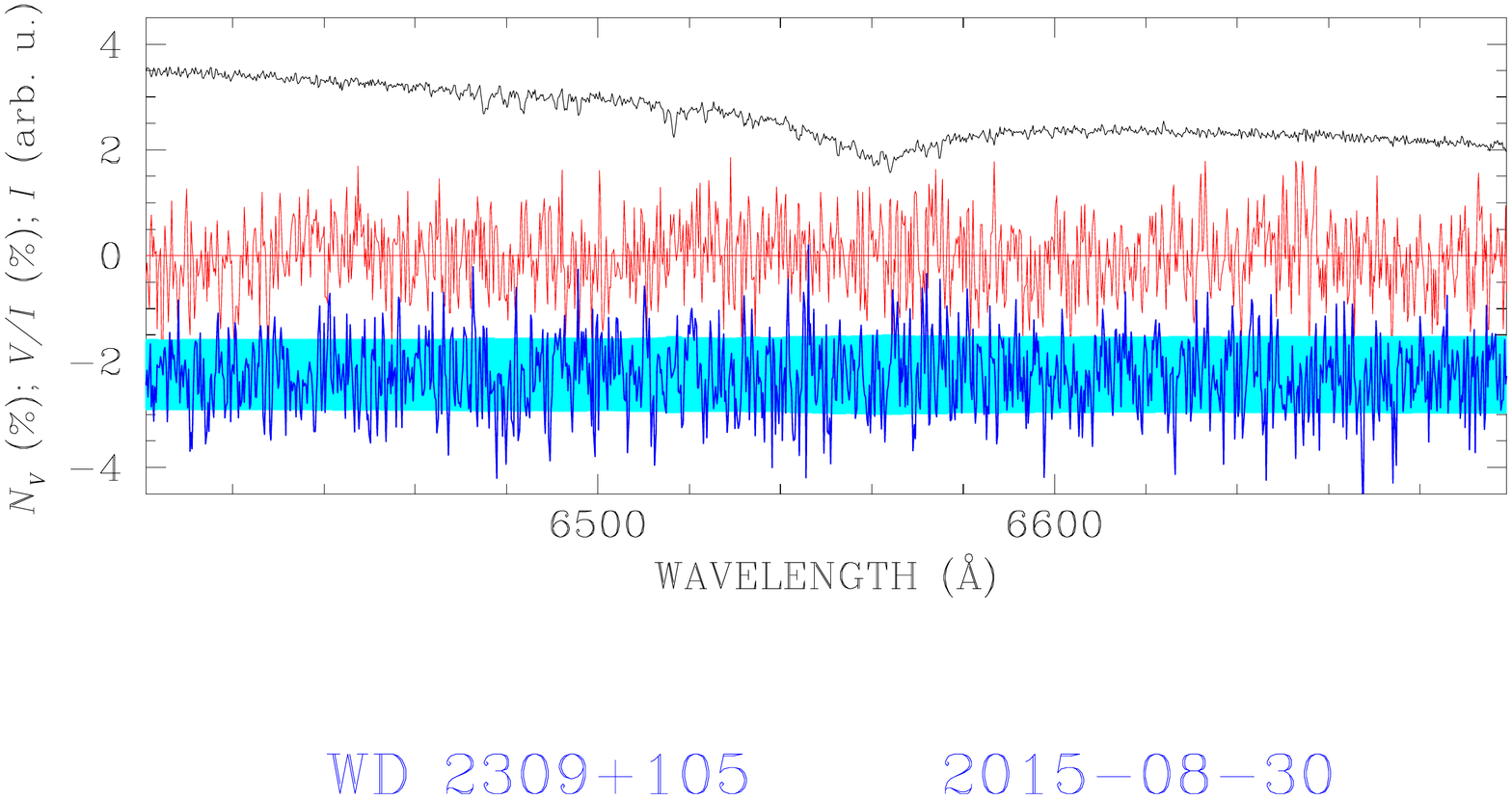} \\
\includegraphics*[angle=270,width=8.0cm,trim={0.90cm 0.0cm 0.1cm 1.0cm},clip]{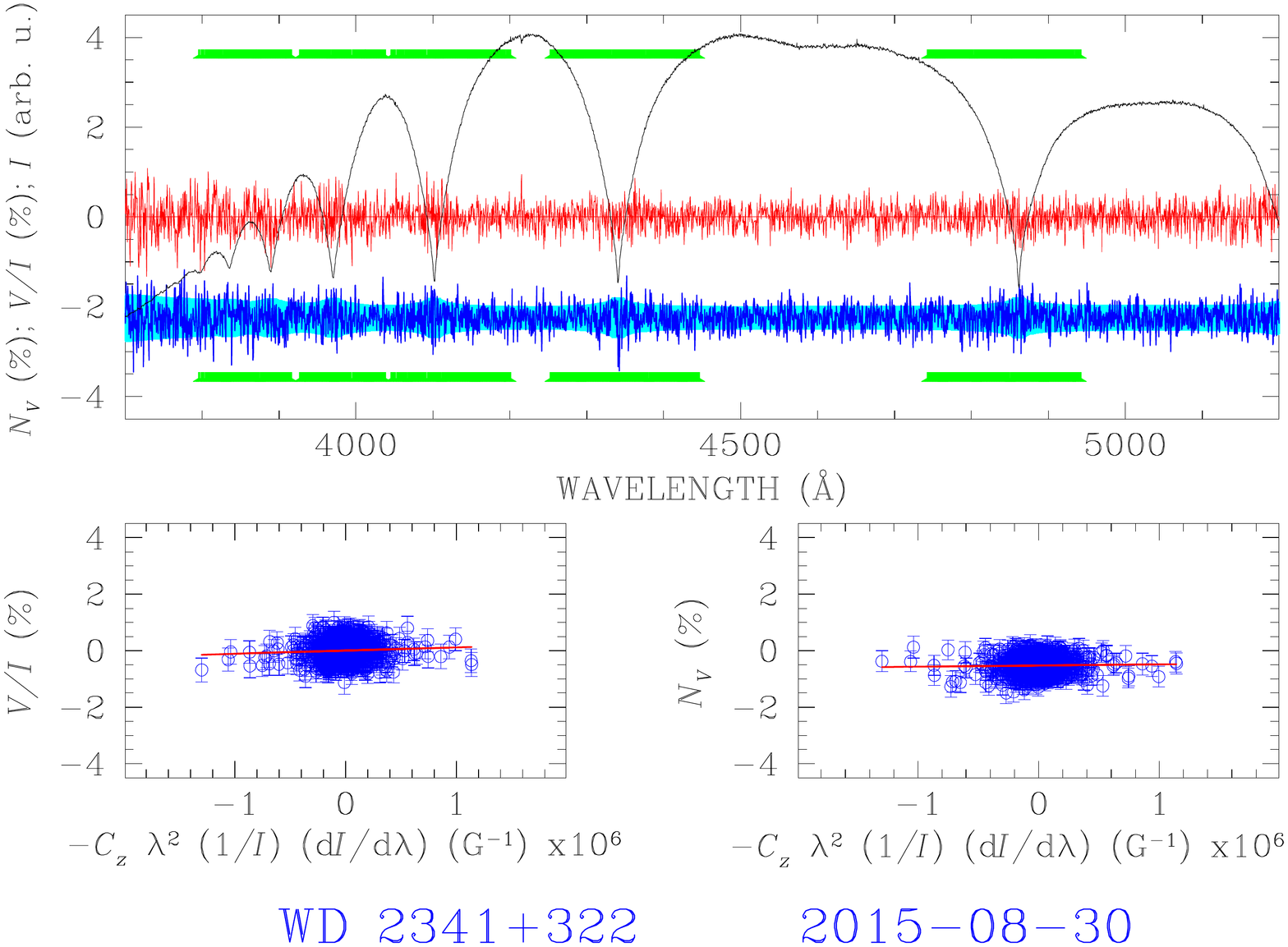}
\includegraphics*[angle=270,width=8.0cm,trim={0.90cm 0.0cm 0.1cm 1.0cm},clip]{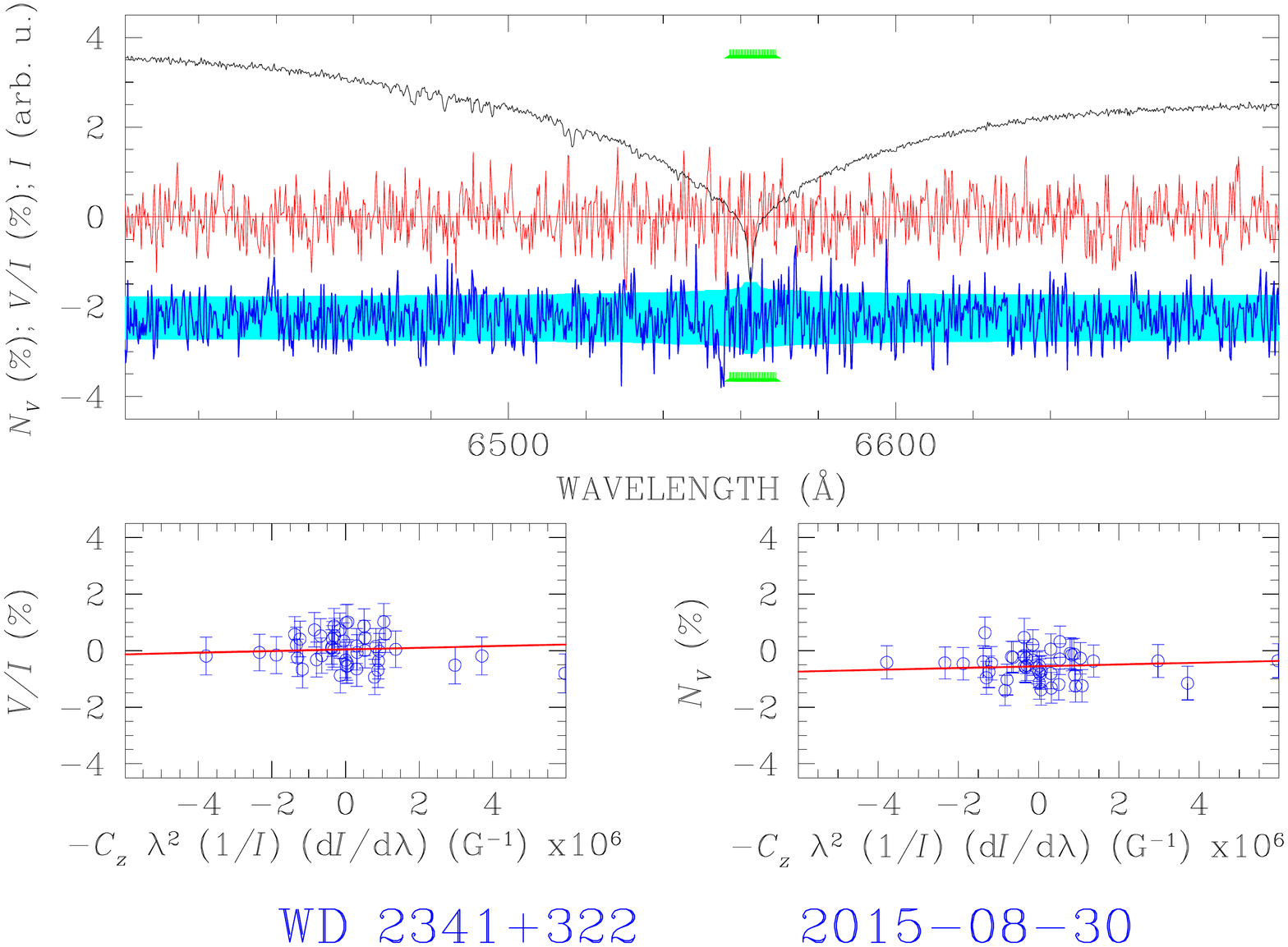} \\
\includegraphics*[angle=270,width=8.0cm,trim={0.90cm 0.0cm 0.1cm 1.0cm},clip]{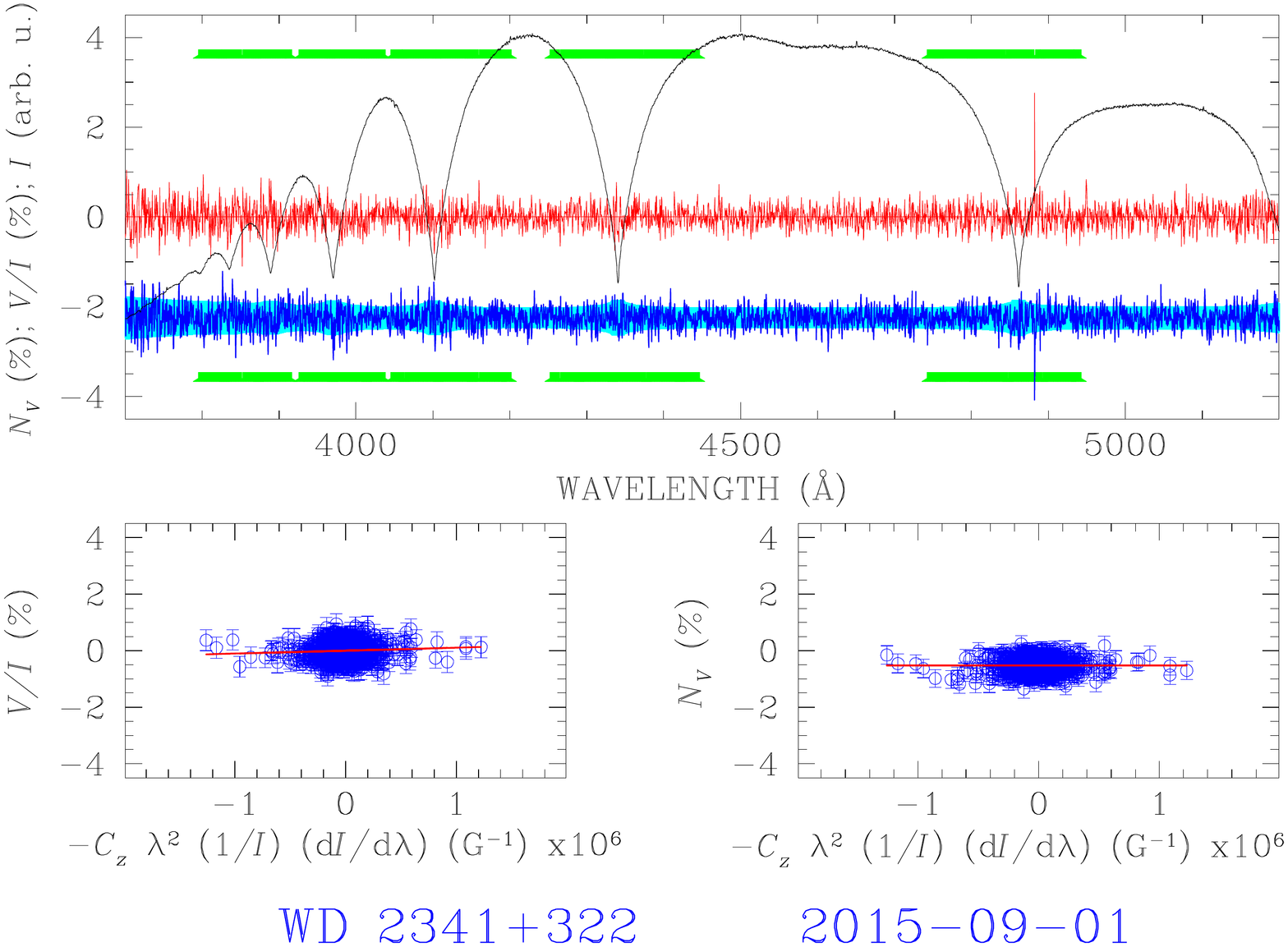}
\includegraphics*[angle=270,width=8.0cm,trim={0.90cm 0.0cm 0.1cm 1.0cm},clip]{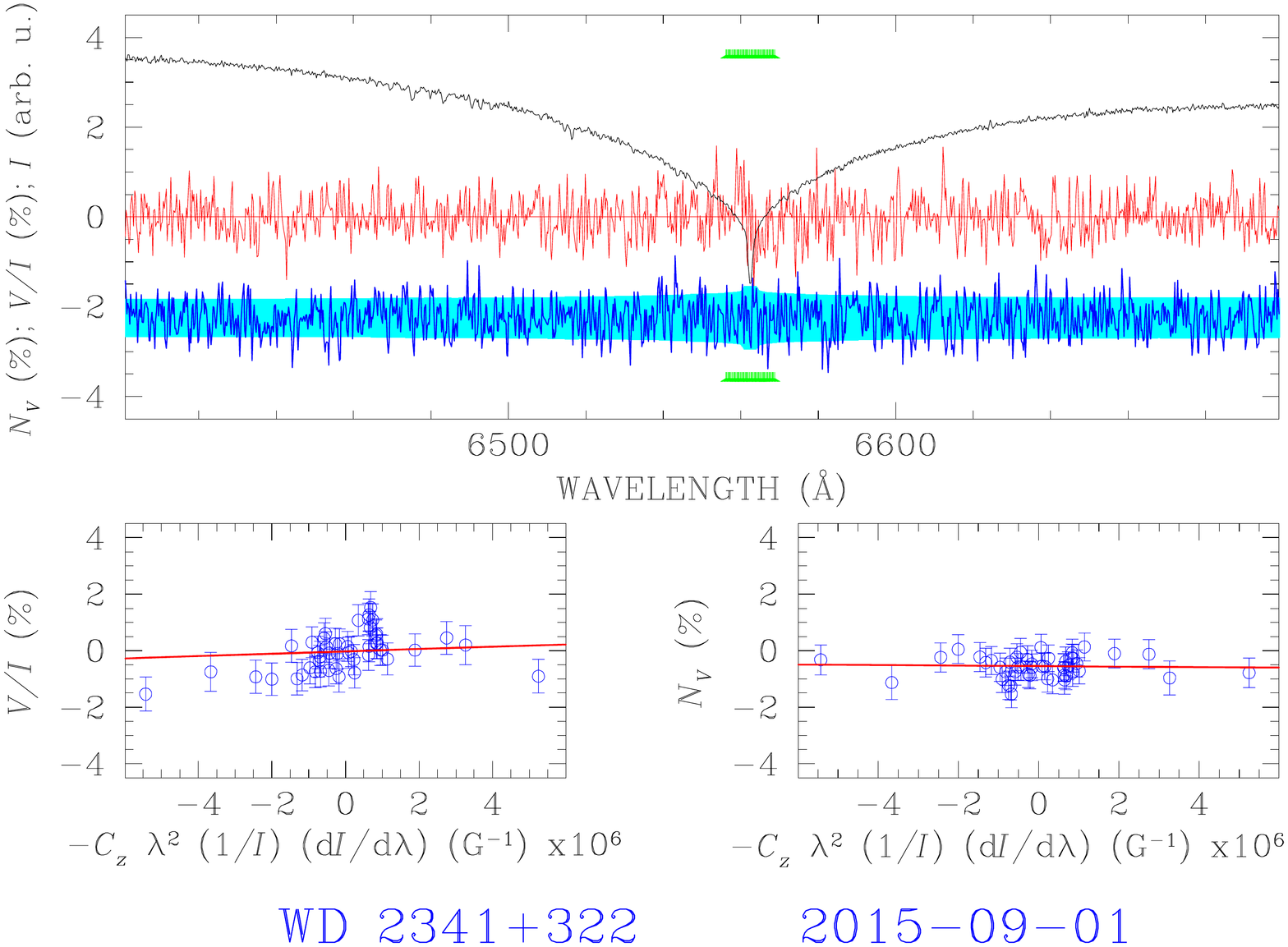} \\